\documentclass[11pt]{article}
\newcounter{ourcount}
\usepackage{comment}
\usepackage{amsmath,amsfonts,amssymb,graphicx,epsfig,pdflscape,multirow,gensymb,soul}
\usepackage{ntheorem}
\usepackage[matrix,arrow,curve]{xy}
\numberwithin{equation}{section}
\graphicspath{{figpdf/}}
\usepackage[nosort]{cite}
\usepackage[usenames]{color}
\usepackage{pstricks}
\usepackage[backref=false]{hyperref}
\hypersetup{
colorlinks=true,
citecolor=red,
linkcolor=darkblue,
urlcolor=darkblue
}
\usepackage[capitalise,noabbrev]{cleveref}

\theorembodyfont{\itshape} 
\theoremheaderfont{\scshape}
\theoremstyle{plain}

\newcommand{\thispic}[1]{#1}

\definecolor{darkblue}{rgb}{0,0,.8}
\definecolor{red}{rgb}{1,0,0}
\numberwithin{equation}{section}
\newtheorem{Proposition}{Proposition}[section]
\newtheorem{Lemma}[Proposition]{Lemma}
\newtheorem{Theorem}{Theorem}
\newtheorem{Corollary}[Proposition]{Corollary}

\numberwithin{equation}{section}

\newcommand{\nc}{\newcommand}

\crefname{Proposition}{Proposition}{Propositions}
\crefname{Lemma}{Lemma}{Lemmas}
\Crefname{Theorem}{Theorem}{Theorems}

\nc{\ir}{\mathrm{i}}
\nc{\dd}{\mathrm{d}} 
\nc{\eE}{\mathsf{e}}
\renewcommand{\ge}{\geqslant}
\renewcommand{\le}{\leqslant}
\renewcommand{\geq}{\geqslant}
\renewcommand{\leq}{\leqslant}

\nc{\mmod}{\,\textrm{mod}\,}
\nc{\pa}{\mathsf{a}}
\nc{\pb}{\mathsf{b}}
\nc{\pc}{\mathsf{c}}
\nc{\pd}{\mathsf{d}}
\nc{\ppi}{\mathsf{i}}
\nc{\ppj}{\mathsf{j}}
\nc{\g}{\mathfrak{g}}

\nc{\proof}{{\scshape Proof.\ }} 				
\nc{\eproof}{{\hfill \rule{0.5em}{0.5em}\medskip}}
\nc{\be}{\begin{equation}}
\nc{\ee}{\end{equation}}

\newcommand{\nn}{\nonumber}
\nc{\repB}{\mathsf{B}}
\nc{\repBa}{\mathsf{B}\mathrm{a}}
\nc{\repBd}{\mathsf{B}\mathrm{d}}
\nc{\repE}{\mathsf{E}}
\nc{\repI}{\mathsf{I}}
\nc{\repL}{\mathsf{L}}
\nc{\repM}{\mathsf{M}}
\nc{\repN}{\mathsf{N}}
\nc{\repQ}{\mathsf{Q}}
\nc{\repS}{\mathsf{S}}
\nc{\repV}{\mathsf{V}}
\nc{\repW}{\mathsf{W}}
\nc{\repX}{\mathsf{X}}
\nc{\tl}{\mathsf{TL}}
\nc{\eptl}{\mathsf{\mathcal EPTL}}
\nc{\bb}{\bar{b}}
\nc{\Fb}{\overline{F}}
\nc{\Cbb}{\mathbb{C}}
\nc{\Zbb}{\mathbb{Z}}
\nc{\id}{\mathbf{1}}
\nc{\timesf}{\times_{\!f}}
\nc{\timesfh}{\hspace{0.07cm} \widehat{\times}_{\!f}\hspace{0.07cm} }
\nc{\cL}{\mbox{$\mathcal L$}}
\nc{\cLa}{\mbox{$\mathcal L\mathrm{a}$}}
\nc{\cLd}{\mbox{$\mathcal L\mathrm{d}$}}
\nc{\cLaa}{\mbox{$\Lambda\mathrm{a}$}}
\nc{\cLad}{\mbox{$\Lambda\mathrm{d}$}}
\nc{\cLk}[1]{\mbox{$\mathcal L^{\tinyx {#1}}$}}
\nc{\cLkd}[1]{\mbox{$\mathcal L\mathrm{d}^{\tinyx {#1}}$}}
\nc{\tinyx}[1]{\textrm{\tiny$(#1)$}}
\nc{\tinyz}[1]{\textrm{\tiny$[#1]$}}

\newcommand{\ket}[1]{| {#1} \rangle}

\definecolor{lightblue}{rgb}{.7,.7,1}
\definecolor{lightestblue}{rgb}{.95,.95,1}
\definecolor{lightlightblue}{rgb}{.9,.9,1}
\definecolor{midblue}{rgb}{.7,.7,1}
\definecolor{babyblue}{rgb}{0.2, 0.75, 1}
\definecolor{purple}{rgb}{0.5,0,0.5}
\definecolor{peach}{rgb}{1, 0.854902, 0.72549}
\definecolor{pastelgreen}{rgb}{0.467, 0.867, 0.467}
\newrgbcolor{darkgreen}{0., 0.733333, 0.0621042}
\definecolor{darkpink}{rgb}{1,.2,1}
\definecolor{pastelpink}{rgb}{1, 0.82, 0.863}

\nc{\elegant}{1.5pt}
\nc{\moyen}{1.0pt}
\nc{\mince}{0.5pt}

\begin{document}

\topmargin -5mm
\oddsidemargin 5mm

\makeatletter 
\newcommand\Larger{\@setfontsize\semiHuge{19.00}{22.78}}
\makeatother 

\vspace*{-2cm}
\setcounter{page}{1}
\vspace{22mm}

\begin{center}
{\huge {\bf 
Fusion in the periodic Temperley--Lieb algebra: \\[0.2cm]
general definition of a bifunctor
}}

\end{center}

\vspace{0.6cm}

\begin{center}
{\vspace{-5mm}\Large Yacine Ikhlef$^{\,\dagger}$ \qquad Alexi Morin-Duchesne}
\\[.4cm]
{\em $^{\dagger}$Sorbonne Universit\'{e}, CNRS, Laboratoire de Physique Th\'{e}orique}
\\
{\em et Hautes \'{E}nergies, LPTHE, F-75005 Paris, France}
\\[.5cm] 
{\tt ikhlef\,@\,lpthe.jussieu.fr \qquad alexi.morin.duchesne\,@\,gmail.com}
\end{center}

\vspace{10pt}

%
%
 
\begin{abstract}

The periodic Temperley--Lieb category consists of connectivity diagrams drawn on a ring with $N$ and $N'$ nodes on the outer and inner boundary, respectively. We consider families of modules, namely sequences of modules $\repM(N)$ over the enlarged periodic Temperley--Lieb algebra for varying values of~$N$, endowed with an action $\repM(N') \to \repM(N)$ of the diagrams. Examples of modules that can be organised into families are those arising in the RSOS model and in the XXZ spin-$\frac12$ chain, as well as several others constructed from link states.\medskip

We construct a fusion product which outputs a family of modules from any pair of families. Its definition is inspired from connectivity diagrams drawn on a disc with two holes. It is thus defined in a way to describe intermediate states in lattice correlation functions. We prove that this fusion product is a bifunctor, and that it is distributive, commutative, and associative.

\end{abstract}
%
%
\tableofcontents
%

%
\section{Introduction}
%

The Temperley--Lieb algebras \cite{TL71,L91,MS93,GL98,G98,EG99} describe the underlying symmetry for a wide variety of two-dimensional lattice models of statistical mechanics: the Ising and $Q$-state Potts models, percolation, the O($n$) loop models, the model of critical dense polymers, etc. These algebras are particularly well-suited to address non-local observables, such as the connectivity probabilities of Ising domain walls or the hull of percolation clusters. One way to study the lattice correlation functions in the bulk, analogous to the radial quantisation of conformal field theory, is to consider the evolution operator (called the \emph{transfer matrix}) between concentric closed contours around a given point $r$. The transfer matrix is then typically an element of the {\it enlarged periodic} Temperley--Lieb algebra $\eptl_N(\beta)$, and the vectors of configuration probabilities on a closed contour around $r$ live in a module over this algebra. If an operator $\mathcal{O}(r)$ is inserted in the bulk of the lattice, then the contours enclosing $r$ carry a module $\repM(N)$ associated to $\mathcal{O}$. An important problem is then to understand the lattice operator product expansion: if one inserts the operators $\mathcal{O}_\pa(r_\pa)$ and $\mathcal{O}_\pb(r_\pb)$ associated to the modules $\repM_\pa(N_\pa)$ and $\repM_\pb(N_\pb)$, respectively, then what characterises the module carried by a contour enclosing both $r_\pa$ and $r_\pb$? The object of the present work is to answer this fundamental question, known as the \emph{fusion} of modules.
\medskip

In \cite{GL98}, Graham and Lehrer described a class of finite-dimensional indecomposable modules $\repW_{k,x}(N)$ over $\eptl_N(\beta)$. The vector space of $\repW_{k,x}(N)$ is spanned by link states with one marked point attached to $2k$ defects, and involves a complex variable $x$ associated to the non-contractible loops around the marked point for $k=0$, and to the winding of the defects for $k>0$. All the finite-dimensional irreducible modules of $\eptl_N(\beta)$ can be constructed from the modules $\repW_{k,x}(N)$ and their quotients. This fundamental result was an important progress in the representation theory of the $\eptl_N(\beta)$ algebras. However, the problem of fusion for these modules turns out to be significantly harder than its non-periodic analog, studied in \cite{PRZ06,RP07,NS07,GV13}. Indeed, although from an abstract viewpoint, the constructions of fusion products for $\eptl_N(\beta)$ proposed in the literature \cite{GS16,GJS18,BSA18} fulfill many of the expected algebraic properties, they do not appear to be related to the lattice operator product expansion, which is crucial for the study of lattice correlation functions in statistical models. In two previous papers \cite{IMD22,IMD24}, we studied the specific case where each of the modules $\repM_\pa(N_\pa)$ and $\repM_\pb(N_\pb)$ is a module $\repW_{k,x}(N)$ or a quotient thereof. In this context, we proposed a class of finite-dimensional modules defined by link states with \emph{two} marked points. We determined their decomposition in terms of irreducible modules, and argued why these were interesting candidates for the fusion of these modules. In the present work, following the same type of intuition, we address the general problem of fusing two \emph{arbitrary} modules of $\eptl_N(\beta)$.\medskip

The algebra $\eptl_N(\beta)$ is generated by connectivity diagrams on an ring, with non-intersecting loop segments connecting $N$ nodes on the inner boundary and $N$ nodes on the outer boundary. The product of diagrams is computed from concatenation of the diagrams, with each contractible closed loop removed and replaced by a factor of $\beta$. In the present work, we make no assumption on the value of the loop weight $\beta=-q-q^{-1}$, namely, our results are valid for all values of $q \in \Cbb^\times$.
\medskip

First, following \cite{GL98}, we consider the periodic Temperley--Lieb category, whose objects are the nonnegative integers. The morphisms $N' \to N$ between two objects are diagrams on a ring, with non-intersecting loop segments connecting $N$ nodes on the outer boundary and $N'$ nodes on the inner boundary. We denote by $\cL(N,N')$ the linear span of the corresponding diagrams, and set $N$ and $N'$ to have the same parity, so that this space of diagrams is non-empty. The composition of the morphisms is the product $\lambda\lambda'$ of two diagrams $\lambda \in \cL(N,N')$ and $\lambda' \in \cL(N',N'')$. It is defined diagrammatically from their concatenation and yields a diagram in $\cL(N,N'')$. We show that this definition of $\cL(N,N')$ is equivalent to an algebraic definition in terms of generators $c_j$ and $c^\dag_j$, subject to a set of algebraic relations. The operator $c_j$ is the diagram that inserts a loop segment connecting the inner nodes $j$ and $j+1$, whereas $c^\dag_j$ is the diagram inserting a loop segment connecting the outer nodes $j$ and $j+1$. The usual generators of $\eptl_N(\beta)=\cL(N,N)$ are recovered through the relations
\begin{equation} \label{eq:ej.Omega}
e_j=c_j^\dag c_j^{\phantom{\dag}}\,,
\qquad \Omega=c_1^{\phantom{\dag}} c_0^\dag = c_0^{\phantom{\dag}} c_{N+1}^\dag \,,
\qquad \Omega^{-1} = c_0^{\phantom{\dag}} c_1^\dag = c_{N+1}^{\phantom{\dag}} c_0^\dag \,.
\end{equation}

We then define a \emph{family of modules} as a set $\repM = \{\repM(N)\,|\,N=N_0,N_0+2, N_0+4, \dots\}$ of $\eptl_N(\beta)$-modules, with $N_0 \in \{0,1\}$, endowed with an action of the diagrams such that $\cL(N,N')$ acts on states in $\repM(N')$ to produce elements in $\repM(N)$, for every $N,N' \in \{N_0,N_0+2, N_0+4, \dots\}$. A family is thus a set of $\eptl_N(\beta)$-modules with $N$ taking only even or only odd values. Moreover, for a given pair $(\repM,\repM')$ of families of modules, we define a \emph{family of homomorphisms} to be a set $\phi = \{\phi_N\,|\,N=N_0,N_0+2, N_0+4, \dots\}$ of linear maps $\phi_N:\repM(N)\to\repM'(N)$ that preserve the action of the diagram spaces $\cL(N,N')$. We show that various examples of solvable lattice models commonly studied in the literature have an underlying description in terms of families of modules. In the terminology of \cite{GL98}, a family of modules is viewed as a \textit{functor} from the periodic Temperley--Lieb category to the category of complex vector spaces, and a family of homomorphisms as a \textit{natural transformation}.
\medskip

Next, we consider the spaces of diagram $\Lambda(N,N_\pa,N_\pb)$ drawn on a disc with two holes, thus generalising the diagrams in $\cL(N,N')$ which have just one hole. The elements of $\Lambda(N,N_\pa,N_\pb)$ are endowed with a natural left action of $\cL(N',N)$ and a right action of $\big(\cL(N_\pa,N_\pa') \otimes \cL(N_\pb,N_\pb')\big)$, defined in both cases from the concatenation of the diagrams. We show that the diagrammatic rules for these products of diagrams are equivalent to a set of algebraic relations satisfied by the operators $c_j$ and $c_j^\dag$.\medskip 

We use this inspiration to introduce a fusion map
\begin{equation} \label{eq:fusion.map}
\mathcal F: (\repM_\pa,\repM_\pb) \mapsto (\repM_\pa \timesf \repM_\pb) \,,
\end{equation}
where $\repM_\pa$ and $\repM_\pb$ are two arbitrary families of modules, and $(\repM_\pa \timesf \repM_\pb)$ denotes the family of fused modules.
Its modules are defined as the sum of vector spaces
\begin{equation} \label{eq:MxM}
(\repM_\pa \timesf \repM_\pb)(N)
= \sum_{N_\pa,N_\pb} \cL(N,N_\pa+N_\pb) \cdot \big(
\repM_\pa(N_\pa) \otimes \repM_\pb(N_\pb)
\big) \,,
\end{equation}
with the action $\lambda \cdot (u \otimes v)$, for $\lambda \in \cL(N,N_\pa+N_\pb)$, $u \in \repM_\pa(N_\pa)$ and $v \in \repM_\pb(N_\pb)$ subject to the relations
\begin{subequations} \label{eq:rules.MxM}
\begin{alignat}{2} 
&c_j \cdot (u \otimes v) = (c_j u \otimes v)\,, \qquad 
&& 1 \le j \le N_\pa-1\,,
\\[0.15cm]
&c_{j+N_\pa} \cdot (u \otimes v) = (u \otimes c_j v)\,, \qquad 
&& 1 \le j \le N_\pb-1\,,
\\[0.15cm]
&c_j^\dag \cdot (u \otimes v) = (c_j^\dag u \otimes v)\,, \qquad 
&& 1 \le j \le N_\pa + 1\,,
\\[0.15cm]
&c_{j+N_\pa}^\dag \cdot (u \otimes v) = (u \otimes c_j^\dag v)\,, \qquad 
&& 1 \le j \le N_\pb + 1\,,
\\[0.15cm]
&\Omega \cdot (u \otimes v) = c_{N_\pa} \cdot (\Omega\, u \otimes c_0^\dag v)\,, \qquad
&& \textrm{for }N_\pa>0 \,,
\end{alignat}
\end{subequations}
with $\Omega$ as in \eqref{eq:ej.Omega}.
Crucially, $\mathcal F$ takes as input families of modules over $\eptl_N(\beta)$, instead of individual modules. The relations \eqref{eq:rules.MxM} indeed allow for the dynamic change of $N_\pa$ and $N_\pb$, and the sum in \eqref{eq:MxM} indicates that the vector space is a sum over all values of $N_\pa$ and $N_\pb$ admissible in $\repM_\pa$ and $\repM_\pb$, respectively.\medskip 

We then prove a series of key properties satisfied by the fusion map $\mathcal F$:\footnote{Two families of modules $\repM$ and $\repM'$ are said to be isomorphic if and only if $\repM(N)$ and $\repM'(N)$ are isomorphic for every $N$. Similarly, we say that $\repM = \repM' \oplus \repM''$ if and only if $\repM(N) = \repM'(N) \oplus \repM''(N)$ for every $N$. Moreover, a family $\repQ$ is said to be a quotient family of another family $\repM$ if $\repQ(N)$ is a quotient of $\repM(N)$ for every $N$.}
\begin{enumerate}
\item[1)] We prove that the vector space \eqref{eq:MxM} with the relations \eqref{eq:rules.MxM} indeed defines a family of modules.
\item[2)] We complement $\mathcal F$ with a map $(\phi_\pa,\phi_\pb)\mapsto \phi_{\pa\pb}$ which preserves the families of identity homomorphisms and satisfies the rule of composition. This shows that $\mathcal F$ is a \emph{bifunctor} on the category of families of modules over $\eptl_N(\beta)$.
\item[3)] The fusion map respects the isomorphisms between families of modules: Let $\repM_\pa,\repM'_\pa,\repM_\pb$ be three families of modules with $\repM_\pa$ isomorphic to $\repM'_\pa$. Then $(\repM_\pa\timesf \repM_\pb)$ is isomorphic to $(\repM'_\pa\timesf \repM_\pb)$. This is in fact a direct consequence of the fact that $\mathcal F$ is a bifunctor.
\item[4)] The fusion map is distributive: Let $\repM_\pa,\repM_\pb$ be two families of modules, with $\repM_\pa$ decomposing in terms of two more families $\repM'_\pa$ and $\repM''_\pa$ as $\repM_\pa=\repM'_\pa\oplus \repM''_\pa$. Then $(\repM_\pa\timesf \repM_\pb)$ is isomorphic to $(\repM'_\pa\timesf \repM_\pb)\oplus (\repM''_\pa\timesf \repM_\pb)$.
\item[5)] The fusion map is commutative: Let $\repM_\pa,\repM_\pb$ be two families of modules. Then the fused families $(\repM_\pa\timesf \repM_\pb)$ and $(\repM_\pb\timesf \repM_\pa)$ are isomorphic.
\item[6)] Fusing with a quotient family produces a quotient family: Let $\repM_\pa,\repQ_\pa,\repM_\pb$ be three families of modules, such that $\repQ_\pa$ is a quotient family of $\repM_\pa$. Then $(\repQ_\pa\timesf \repM_\pb)$ is isomorphic to a quotient family of $(\repM_\pa\timesf \repM_\pb)$. This is a direct consequence of the fact that $\mathcal F$ is a bifunctor.
\item[7)] The vacuum family $\repV$, made of the modules $\repV(N)$ spanned by link states on drawn on a disc with $N$ nodes and no marked point, is the neutral element for $\mathcal F$: Let $\repM$ be a family of modules. The families of fused modules $(\repM\timesf \repV)$ and $(\repV\timesf \repM)$ are both isomorphic to $\repM$.
\item[8)] We study the effect of the fusion map on families of modules transformed by two $\eptl_N(\beta)$-automorphisms, namely the {\it parity sign flip} and the {\it reflection}, as well as on {\it adjoint families of modules}.
\item[9)] The fusion map is associative: Let $\repM_\pa,\repM_\pb,\repM_\pc$ be families of modules. Then the families of fused modules $\big((\repM_\pa\timesf \repM_\pb)\timesf \repM_\pc\big)$ and $\big(\repM_\pa\timesf (\repM_\pb\timesf \repM_\pc)\big)$ are isomorphic.
\end{enumerate}
For our proof of the associativity, we introduce a \emph{triple} fusion product, whose vector space is 
\begin{equation} \label{eq:MxMxM}
(\repM_\pa \timesf \repM_\pb \timesf \repM_\pc)(N)
= \sum_{N_\pa,N_\pb,N_\pc} \cL(N,N_\pa+N_\pb+N_\pc) \cdot \big(
\repM_\pa(N_\pa) \otimes \repM_\pb(N_\pb) \otimes \repM_\pc(N_\pc)
\big) \,,
\end{equation}
and is subject to algebraic relations similar to \eqref{eq:rules.MxM}. These are inspired from the algebraic rules for the concatenation of diagrams with three holes. We in fact prove that the families $\big((\repM_\pa\timesf \repM_\pb)\timesf \repM_\pc\big)$ and $\big(\repM_\pa\timesf (\repM_\pb\timesf \repM_\pc)\big)$ are both isomorphic to $(\repM_\pa \timesf \repM_\pb \timesf \repM_\pc)$. We show that this implies the equivalence of the algebraic and diagrammatic definitions for the spaces of diagrams with three holes.
\medskip

We remark that our definitions for diagrams with one, two or three holes are very similar to the planar tangles introduced by Jones in the context of planar algebras \cite{JonesPlanar} --- see also \cite{Rasmussen-Poncini1,Rasmussen-Poncini2} for recent developments.\medskip

The outline of this paper is as follows. In \cref{sec:diagrams.and.families}, we define the diagram spaces $\cL(N,N')$ and discuss some of their properties. The proof that their algebraic and diagrammatic definitions are equivalent is deferred to \cref{app:equiv.L}. We also introduce families of modules in \cref{sec:diagrams.and.families}, and give several examples. We describe two automorphisms of the diagram spaces and the transformations that they induce over families of modules. The action of the transformation on the example families of modules is described in \cref{app:mod.transf}. We also consider in \cref{sec:diagrams.and.families} families of right modules of the diagram spaces and how they are related to families of (left) modules by adjunction. In \cref{sec:fusion.modules}, we first discuss connectivity diagrams drawn on discs with two holes, and in particular their equivalent diagrammatic and algebraic definitions, established in \cref{app:equiv.Lambda}. We then define the fusion map, show that it is a bifunctor, and study its various properties listed above. The property of associativity of the bifunctor is proven separately in \cref{sec:associativity}, after a discussion of connectivity diagrams drawn on a disc with three holes. Concluding remarks are presented in \cref{sec:conclusion}.

%
\section{Diagram spaces and families of modules}
\label{sec:diagrams.and.families}
%

\subsection{Diagram spaces}
\label{sec:definitions-EPTL}

In this section, we define the diagram spaces $\cL(N,N')$, of which the enlarged periodic Temperley--Lieb algebra $\eptl_N(\beta)$ is the special case $N = N'$. These have equivalent definitions in terms of generators satisfying relations and in terms of diagrams. We describe both definitions and discuss their equivalence.

\paragraph{Diagrammatic definition of $\boldsymbol{\cL(N,N')}$.} 
Let $N, N' \in \Zbb_{\ge 0}$ with $N+N'$ even. We define the {\it diagram space} $\cL(N,N')$ as the vector space of {\it connectivity diagrams} drawn on a ring, where non-intersecting loop segments, drawn here as blue curves, connect pairwise $N$ nodes on the outer boundary and $N'$~nodes on the inner boundary. The nodes are numbered in the counter-clockwise direction, from $1$~to~$N$ and $1$~to~$N'$ for the outer and inner boundaries respectively. We place a first reference point $\pc'$ on the inner boundary between the nodes $N'$ and $1$, and a second reference point $\pc$ on the outer boundary between the nodes $N$ and $1$. We draw a dashed segment connecting $\pc$ and $\pc'$. A connectivity diagram~$\lambda$ in $\cL(N,N')$ thus consists in {\it arches} connecting nodes of the same boundary component, and {\it bridges} connecting nodes of the inner and outer boundaries. Diagrams with no bridges may also include {\it non-contractible} closed loops that encircle the inner boundary. Connectivity diagrams are defined up to continuous deformations of the loop segments inside the ring. Then $\cL(N,N')$ is the linear span of these connectivity diagrams. Here are examples of connectivity diagrams in $\cL(12,4)$ and $\cL(4,6)$ respectively:
\begin{equation}
\label{eq:first diagrams}
\psset{unit=0.8cm}
\lambda_1 = \
\thispic{\begin{pspicture}[shift=-1.7](-1.8,-1.8)(1.8,1.8)
\psarc[linecolor=black,linewidth=0.5pt,fillstyle=solid,fillcolor=lightlightblue]{-}(0,0){1.5}{0}{360}
\psarc[linecolor=black,linewidth=0.5pt,fillstyle=solid,fillcolor=white]{-}(0,0){0.7}{0}{360}
\rput{30}(0,0){\psbezier[linecolor=blue,linewidth=1.5pt]{-}(-0.388229, -1.44889)(-0.310583, -1.15911)(0.310583, -1.15911)(0.388229, -1.44889)
\psbezier[linecolor=blue,linewidth=1.5pt]{-}(-1.06066, -1.06066)(-0.848528, -0.948528)(0.848528, -0.948528)(1.06066, -1.06066)}
\rput{-90}(0,0){\psbezier[linecolor=blue,linewidth=1.5pt]{-}(-0.388229, -1.44889)(-0.310583, -1.15911)(0.310583, -1.15911)(0.388229, -1.44889)}
\rput{-150}(0,0){\psbezier[linecolor=blue,linewidth=1.5pt]{-}(-0.388229, -1.44889)(-0.310583, -1.15911)(0.310583, -1.15911)(0.388229, -1.44889)}
\rput{120}(0,0){\psbezier[linecolor=blue,linewidth=1.5pt]{-}(-0.388229, -1.44889)(-0.310583, -1.15911)(0.310583, -1.15911)(0.388229, -1.44889)}
\psline[linecolor=blue,linewidth=1.5pt]{-}(-1.06066, -1.06066)(-0.494975, -0.494975)
\psbezier[linecolor=blue,linewidth=1.5pt]{-}(0.494975, 0.494975)(0.989949, 0.989949)(-0.989949, 0.989949)(-0.494975, 0.494975)
\psbezier[linecolor=blue,linewidth=1.5pt]{-}(0.494975, -0.494975)(0.636396, -0.636396)(1.06252, -0.284701)(0.965926, 0.258819)
\psbezier[linecolor=blue,linewidth=1.5pt]{-}(0.968, 0.25)(0.777817, 0.777817)(0.310583, 1.15911)(0.388229, 1.44889)
\psline[linestyle=dashed, dash= 1.5pt 1.5pt,linewidth=0.5pt]{-}(0,-1.5)(0,-0.7)
\rput(0.452934, -1.69038){$_1$}
\rput(1.23744, -1.23744){$_2$}
\rput(1.69038, -0.452934){$_3$}
\rput(1.69038, 0.452934){$_4$}
\rput(1.23744, 1.23744){$_5$}
\rput(0.452934, 1.69038){$_6$}
\rput(-0.452934, 1.69038){$_7$}
\rput(-1.23744, 1.23744){$_8$}
\rput(-1.69038, 0.452934){$_9$}
\rput(-1.69038, -0.452934){$_{10}$}
\rput(-1.23744, -1.23744){$_{11}$}
\rput(-0.452934, -1.69038){$_{12}$}
\rput(0.353553, -0.353553){$_1$}
\rput(0.353553, 0.353553){$_2$}
\rput(-0.353553, 0.353553){$_3$}
\rput(-0.353553, -0.353553){$_4$}
\rput(0,-0.5){$_{\pc'}$}
\rput(0,-1.7){$_{\pc}$}
\end{pspicture}}
\ , \qquad \lambda_2 = \
\thispic{\begin{pspicture}[shift=-1.5](-1.6,-1.6)(1.6,1.6)
\psarc[linecolor=black,linewidth=0.5pt,fillstyle=solid,fillcolor=lightlightblue]{-}(0,0){1.5}{0}{360}
\psarc[linecolor=black,linewidth=0.5pt,fillstyle=solid,fillcolor=white]{-}(0,0){0.7}{0}{360}
\rput{180}(0,0){\psbezier[linecolor=blue,linewidth=1.5pt]{-}(-1.06066, -1.06066)(-0.848528, -0.948528)(0.848528, -0.948528)(1.06066, -1.06066)}
\psbezier[linecolor=blue,linewidth=1.5pt]{-}(0.7, 0.)(1.1, 0.)(0.55, 0.952628)(0.35, 0.606218)
\psbezier[linecolor=blue,linewidth=1.5pt]{-}(-0.7, 0.)(-1.1, 0.)(-0.55, -0.952628)(-0.35, -0.606218)
\psbezier[linecolor=blue,linewidth=1.5pt]{-}(-0.35, 0.606218)(-0.7, 1.3)(-1.0,0.3)(-1.0, 0)
\psbezier[linecolor=blue,linewidth=1.5pt]{-}(-1.0, 0)(-1.0,-0.3)(-1, -1)(-1.06066, -1.06066)
\psbezier[linecolor=blue,linewidth=1.5pt]{-}(0.35, -0.606218)(0.55, -0.952628)(0.777817, -0.777817)(1.06066, -1.06066)
\psline[linestyle=dashed, dash= 1.5pt 1.5pt,linewidth=0.5pt]{-}(0,-1.5)(0,-0.7)
\rput(1.23744, -1.23744){$_1$}
\rput(1.23744, 1.23744){$_2$}
\rput(-1.23744, 1.23744){$_3$}
\rput(-1.23744, -1.23744){$_{4}$}
\rput(0.25, -0.433013){$_1$}
\rput(0.5, 0.){$_2$}
\rput(0.25, 0.433013){$_3$}
\rput(-0.25, 0.433013){$_4$}
\rput(-0.5, 0.){$_5$}
\rput(-0.25, -0.433013){$_6$}
\rput(0,-0.5){$_{\pc'}$}
\rput(0,-1.7){$_{\pc}$}
\end{pspicture}}
\ .
\ee

We define the product $\lambda_1 \lambda_2$, for diagrams $\lambda_1 \in \cL(N,N')$ and $\lambda_2 \in \cL(N',N'')$. It is computed by drawing $\lambda_2$ inside $\lambda_1$, with the resulting diagram in $\cL(N,N'')$ read off from the connection of the outer and inner nodes of the new bigger ring. Closed loops may be formed in the process. These are said to be non-contractible if they encircle the inner boundary, and contractible otherwise. Each contractible loop is erased and replaced by a factor of $\beta$ multiplying the resulting connectivity diagram, which we parameterise as
\be 
\beta=-q-q^{-1}\,, \qquad q \in \Cbb^\times.
\ee
In contrast, non-contractible loops are not erased and instead remain in the connectivity diagram. For example, the product of the two connectivity diagrams in \eqref{eq:first diagrams} is\footnote{We usually omit the labels of the nodes in the diagrams.}
\be
\psset{unit=0.8cm}
\lambda_1 \lambda_2 = \
\psset{unit=1.2}
\thispic{\begin{pspicture}[shift=-1.4](-1.5,-1.5)(1.5,1.5)
\psarc[linecolor=black,linewidth=0.5pt,fillstyle=solid,fillcolor=lightlightblue]{-}(0,0){1.5}{0}{360}
\psarc[linecolor=black,linewidth=0.5pt,fillstyle=solid,fillcolor=white]{-}(0,0){0.7}{0}{360}
\rput{30}(0,0){\psbezier[linecolor=blue,linewidth=1.5pt]{-}(-0.388229, -1.44889)(-0.310583, -1.15911)(0.310583, -1.15911)(0.388229, -1.44889)
\psbezier[linecolor=blue,linewidth=1.5pt]{-}(-1.06066, -1.06066)(-0.848528, -0.948528)(0.848528, -0.948528)(1.06066, -1.06066)}
\rput{-90}(0,0){\psbezier[linecolor=blue,linewidth=1.5pt]{-}(-0.388229, -1.44889)(-0.310583, -1.15911)(0.310583, -1.15911)(0.388229, -1.44889)}
\rput{-150}(0,0){\psbezier[linecolor=blue,linewidth=1.5pt]{-}(-0.388229, -1.44889)(-0.310583, -1.15911)(0.310583, -1.15911)(0.388229, -1.44889)}
\rput{120}(0,0){\psbezier[linecolor=blue,linewidth=1.5pt]{-}(-0.388229, -1.44889)(-0.310583, -1.15911)(0.310583, -1.15911)(0.388229, -1.44889)}
\psline[linecolor=blue,linewidth=1.5pt]{-}(-1.06066, -1.06066)(-0.494975, -0.494975)
\psbezier[linecolor=blue,linewidth=1.5pt]{-}(0.494975, 0.494975)(0.989949, 0.989949)(-0.989949, 0.989949)(-0.494975, 0.494975)
\psbezier[linecolor=blue,linewidth=1.5pt]{-}(0.494975, -0.494975)(0.636396, -0.636396)(1.06252, -0.284701)(0.965926, 0.258819)
\psbezier[linecolor=blue,linewidth=1.5pt]{-}(0.968, 0.25)(0.777817, 0.777817)(0.310583, 1.15911)(0.388229, 1.44889)
\psline[linestyle=dashed, dash= 1.5pt 1.5pt,linewidth=0.5pt]{-}(0,-1.5)(0,-0.7)
\psset{unit=0.4666}
\psarc[linecolor=black,linewidth=0.5pt,fillstyle=solid,fillcolor=lightlightblue]{-}(0,0){1.5}{0}{360}
\psarc[linecolor=black,linewidth=0.5pt,fillstyle=solid,fillcolor=white]{-}(0,0){0.7}{0}{360}
\rput{180}(0,0){\psbezier[linecolor=blue,linewidth=1.5pt]{-}(-1.06066, -1.06066)(-0.848528, -0.948528)(0.848528, -0.948528)(1.06066, -1.06066)}
\psbezier[linecolor=blue,linewidth=1.5pt]{-}(0.7, 0.)(1.1, 0.)(0.55, 0.952628)(0.35, 0.606218)
\psbezier[linecolor=blue,linewidth=1.5pt]{-}(-0.7, 0.)(-1.1, 0.)(-0.55, -0.952628)(-0.35, -0.606218)
\psbezier[linecolor=blue,linewidth=1.5pt]{-}(-0.35, 0.606218)(-0.7, 1.3)(-1.0,0.3)(-1.0, 0)
\psbezier[linecolor=blue,linewidth=1.5pt]{-}(-1.0, 0)(-1.0,-0.3)(-1, -1)(-1.06066, -1.06066)
\psbezier[linecolor=blue,linewidth=1.5pt]{-}(0.35, -0.606218)(0.55, -0.952628)(0.777817, -0.777817)(1.06066, -1.06066)
\psline[linestyle=dashed, dash= 1.5pt 1.5pt,linewidth=0.5pt]{-}(0,-1.5)(0,-0.7)
\end{pspicture}}
\ = \beta \
\thispic{\begin{pspicture}[shift=-1.4](-1.6,-1.5)(1.6,1.5)
\psarc[linecolor=black,linewidth=0.5pt,fillstyle=solid,fillcolor=lightlightblue]{-}(0,0){1.5}{0}{360}
\psarc[linecolor=black,linewidth=0.5pt,fillstyle=solid,fillcolor=white]{-}(0,0){0.7}{0}{360}
\rput{30}(0,0){\psbezier[linecolor=blue,linewidth=1.5pt]{-}(-0.388229, -1.44889)(-0.310583, -1.15911)(0.310583, -1.15911)(0.388229, -1.44889)
\psbezier[linecolor=blue,linewidth=1.5pt]{-}(-1.06066, -1.06066)(-0.848528, -0.948528)(0.848528, -0.948528)(1.06066, -1.06066)}
\rput{-90}(0,0){\psbezier[linecolor=blue,linewidth=1.5pt]{-}(-0.388229, -1.44889)(-0.310583, -1.15911)(0.310583, -1.15911)(0.388229, -1.44889)}
\rput{-150}(0,0){\psbezier[linecolor=blue,linewidth=1.5pt]{-}(-0.388229, -1.44889)(-0.310583, -1.15911)(0.310583, -1.15911)(0.388229, -1.44889)}
\rput{120}(0,0){\psbezier[linecolor=blue,linewidth=1.5pt]{-}(-0.388229, -1.44889)(-0.310583, -1.15911)(0.310583, -1.15911)(0.388229, -1.44889)}
\psbezier[linecolor=blue,linewidth=1.5pt]{-}(0.7, 0.)(1.1, 0.)(0.55, 0.952628)(0.35, 0.606218)
\psbezier[linecolor=blue,linewidth=1.5pt]{-}(-0.7, 0.)(-1.1, 0.)(-0.55, -0.952628)(-0.35, -0.606218)
\psbezier[linecolor=blue,linewidth=1.5pt]{-}(0.35, -0.606218)(0.5, -0.866025)(1.06252, -0.284701)(0.965926, 0.258819)
\psbezier[linecolor=blue,linewidth=1.5pt]{-}(0.968, 0.25)(0.777817, 0.777817)(0.310583, 1.15911)(0.388229, 1.44889)
\psbezier[linecolor=blue,linewidth=1.5pt]{-}(-0.35, 0.606218)(-0.5, 0.866025)(-1.1, 0.284701)(-0.965926,-0.258819)
\psbezier[linecolor=blue,linewidth=1.5pt]{-}(-0.968,-0.25)(-0.76,-0.9)(-0.919239, -0.919239)(-1.06066, -1.06066)
\psline[linestyle=dashed, dash= 1.5pt 1.5pt,linewidth=0.5pt]{-}(0,-1.5)(0,-0.7)
\end{pspicture}}\ .
\ee
By construction, we have
\begin{equation} \label{eq:LLinL}
\cL(N,N') \,\cL(N',N'') \subseteq \cL(N,N'') \,.
\end{equation}
It is also easy to see that this product is associative.

\paragraph{Algebraic definition of $\boldsymbol{\cL(N,N')}$.}

Let us introduce the diagrams $c_{N,j} \in \cL(N-2,N)$ and $c^\dag_{N,j} \in \cL(N,N-2)$, defined as
\be
\label{eq:def.cj}
c_{N,j} = \ 
\psset{unit=0.8cm}
\thispic{\begin{pspicture}[shift=-1.25](-1.2,-1.4)(1.5,1.2)
\psarc[linecolor=black,linewidth=0.5pt,fillstyle=solid,fillcolor=lightlightblue]{-}(0,0){1.2}{0}{360}
\psarc[linecolor=black,linewidth=0.5pt,fillstyle=solid,fillcolor=white]{-}(0,0){0.7}{0}{360}
\psline[linecolor=blue,linewidth=1.5pt]{-}(0.676148, 0.181173)(1.15911, 0.310583)
\psbezier[linecolor=blue,linewidth=1.5pt]{-}(0.494975, 0.494975)(0.671751, 0.671751)(0.245878, 0.91763)(0.181173, 0.676148)
\psline[linecolor=blue,linewidth=1.5pt]{-}(-0.181173, 0.676148)(-0.310583, 1.15911)
\psline[linecolor=blue,linewidth=1.5pt]{-}(-0.494975, 0.494975)(-0.848528, 0.848528)
\psline[linecolor=blue,linewidth=1.5pt]{-}(-0.676148, 0.181173)(-1.15911, 0.310583)
\psline[linecolor=blue,linewidth=1.5pt]{-}(-0.676148, -0.181173)(-1.15911, -0.310583)
\psline[linecolor=blue,linewidth=1.5pt]{-}(-0.494975, -0.494975)(-0.848528, -0.848528)
\psline[linecolor=blue,linewidth=1.5pt]{-}(-0.181173, -0.676148)(-0.310583, -1.15911)
\psline[linecolor=blue,linewidth=1.5pt]{-}(0.181173, -0.676148)(0.310583, -1.15911)
\psline[linecolor=blue,linewidth=1.5pt]{-}(0.494975, -0.494975)(0.848528, -0.848528)
\psline[linecolor=blue,linewidth=1.5pt]{-}(0.676148, -0.181173)(1.15911, -0.310583)
\psline[linestyle=dashed, dash= 1.5pt 1.5pt,linewidth=0.5pt]{-}(0,-1.2)(0,-0.7)
\rput(0.362347, -1.3523){$_1$}
\rput(0.989949, -0.989949){$_2$}
\rput(1.3523, -0.362347){$_{...}$}
\rput(1.57, 0.362347){$_{j-1}$}
\rput(-0.362347, 1.4023){$_{j}$}
\rput(-1.13066, -1.13066){$_{N-3}$}
\rput(-0.362347, -1.4423){$_{N-2}$}
\end{pspicture}}\ , 
\qquad 
c_{N,j}^\dagger =\
\psset{unit=0.8cm}
\thispic{\begin{pspicture}[shift=-1.1](-1.2,-1.4)(1.4,1.2)
\psarc[linecolor=black,linewidth=0.5pt,fillstyle=solid,fillcolor=lightlightblue]{-}(0,0){1.2}{0}{360}
\psarc[linecolor=black,linewidth=0.5pt,fillstyle=solid,fillcolor=white]{-}(0,0){0.7}{0}{360}
\psline[linecolor=blue,linewidth=1.5pt]{-}(0.676148, 0.181173)(1.15911, 0.310583)
\psbezier[linecolor=blue,linewidth=1.5pt]{-}(0.848528, 0.848528)(0.671751, 0.671751)(0.245878, 0.91763)(0.310583, 1.15911)
\psline[linecolor=blue,linewidth=1.5pt]{-}(-0.181173, 0.676148)(-0.310583, 1.15911)
\psline[linecolor=blue,linewidth=1.5pt]{-}(-0.494975, 0.494975)(-0.848528, 0.848528)
\psline[linecolor=blue,linewidth=1.5pt]{-}(-0.676148, 0.181173)(-1.15911, 0.310583)
\psline[linecolor=blue,linewidth=1.5pt]{-}(-0.676148, -0.181173)(-1.15911, -0.310583)
\psline[linecolor=blue,linewidth=1.5pt]{-}(-0.494975, -0.494975)(-0.848528, -0.848528)
\psline[linecolor=blue,linewidth=1.5pt]{-}(-0.181173, -0.676148)(-0.310583, -1.15911)
\psline[linecolor=blue,linewidth=1.5pt]{-}(0.181173, -0.676148)(0.310583, -1.15911)
\psline[linecolor=blue,linewidth=1.5pt]{-}(0.494975, -0.494975)(0.848528, -0.848528)
\psline[linecolor=blue,linewidth=1.5pt]{-}(0.676148, -0.181173)(1.15911, -0.310583)
\psline[linestyle=dashed, dash= 1.5pt 1.5pt,linewidth=0.5pt]{-}(0,-1.2)(0,-0.7)
\rput(0.362347, -1.3523){$_1$}
\rput(0.989949, -0.989949){$_2$}
\rput(1.3523, -0.362347){$_3$}
\rput(1.3523, 0.362347){$_{...}$}
\rput(0.989949, 0.989949){$_j$}
\rput(0.362347, 1.3923){$_{j+1}$}
\rput(-1.13066, -1.13066){$_{N-1}$}
\rput(-0.362347, -1.3523){$_N$}
\end{pspicture}}\ ,
\qquad
j = 0, 1,\dots, N-1.
\ee
These operators satisfy the relations
\begingroup
\allowdisplaybreaks
\begin{subequations}
\label{eq:c.relations}
\begin{alignat}{2}
\label{eq:c.relations.a}
&c_{N,j}\, c_{N+2,k} = \left\{\begin{array}{ll}
c_{N,k-2}\, c_{N+2,j} & j\le k-2 \\[0.15cm]
c_{N,k}\, c_{N+2,j+2} & j \ge k \\[0.15cm]
\end{array}\right.
&& \hspace{-1.8cm} 1 \le j \le N-1\,,\quad 1 \le k \le N+1\,,
\\[0.15cm]
\label{eq:c.relations.b}
&c^\dagger_{N+2,j}\, c^\dagger_{N,k} = \left\{\begin{array}{ll}
c^\dagger_{N+2,k+2}\, c^\dagger_{N,j} & j\le k \\[0.15cm]
c^\dagger_{N+2,k}\, c^\dagger_{N,j-2} & j \ge k+2
\end{array}\right. 
&& \hspace{-1.8cm} 1 \le j \le N+1\,,\quad 1 \le k \le N-1\,,
\\[0.15cm]
\label{eq:c.relations.c}
&c^\dagger_{N,j}\, c_{N,k} = \left\{\begin{array}{ll}
c_{N+2,k+2}\, c^\dagger_{N+2,j} & j \le k -1 \\[0.15cm]
c_{N+2,j}\, c^\dagger_{N+2,j+2} = c_{N+2,j+2}\, c^\dagger_{N+2,j} & j = k \\[0.15cm]
c_{N+2,k}\, c^\dagger_{N+2,j+2} & j \ge k+1 \\[0.15cm]
\end{array}\right. 
&& \hspace{1.22cm} 1\le j,k\le N-1\,, 
\\[0.15cm]
\label{eq:c.relations.d}
&c_{N,j}\, c^\dagger_{N,k} = 
\left\{\begin{array}{lll}
c^\dagger_{N-2,k-2}\, c_{N-2,j} & j\le k-2 \\[0.15cm]
\id_{N-2} & j=k\pm 1 \\[0.15cm]
\beta\, \id_{N-2} & \quad j=k \\[0.15cm]
c^\dagger_{N-2,k}\, c_{N-2,j-2} & j \ge k+2 \\[0.15cm]
\end{array}\right. 
&& \hspace{-1.8cm} 1\le j,k\le N-1\,,
\end{alignat}
as well as
\begin{alignat}{2}
&c_{N,0} \, c_{N+2,j} = c_{N,j-1} \, c_{N+2,0}
&& \hspace{-1.5cm} 2 \le j \le N \,,
\label{eq:c.relations.e}
\\[0.15cm]
&c^\dagger_{N+2,0} \, c^\dagger_{N,j} = c^\dagger_{N+2,j+1} \, c^\dagger_{N,0}
&& \hspace{-1.5cm}1 \le j \le N-1\,,
\label{eq:c.relations.f}
\\[0.15cm]
&c^\dagger_{N,0} \, c_{N,j} = c_{N+2,j+1} \, c^\dagger_{N+2,0}
&& \hspace{-1.5cm}1 \le j \le N-1\,,
\label{eq:c.relations.g}
\\[0.15cm]
&c_{N,0} \, c^\dagger_{N,j} = 
\left\{\begin{array}{l}
\beta\,\id_{N-2}\\[0.15cm]
c_{N,N-1} c_{N,0}^\dagger \\[0.15cm]
c^\dagger_{N-2,j-1} \, c_{N-2,0} \\[0.15cm]
c_{N,1} c_{N,0}^\dagger
\end{array}\right. 
&& \hspace{-1.65cm} \begin{array}{l}
j=0\,, \\[0.15cm]
j=1\,, \\[0.15cm]
2 \le j \le N-2\,, \\[0.15cm]
j=N-1\,,
\end{array}
\label{eq:c.relations.h}
\\[0.15cm]
&c_{N,1}\,c^\dag_{N,0}\,c_{N,0}\,c^\dag_{N,1} = c_{N,0}\,c^\dag_{N,1}\,c_{N,1}\,c^\dag_{N,0} =\id_{N-2}\,.
\label{eq:c.relations.i}
\end{alignat}
\end{subequations}
\endgroup
We will often omit the label $N$ and write $c_{N,j}$ and $c^\dag_{N,j}$ as $c_j$ and $c_j^\dag$, respectively. \medskip

We define $\cL(N,N')$ algebraically as the vector space of words in $c_j$ and $c^\dag_j$, endowed with the relations~\eqref{eq:c.relations}. More precisely, a word $w$ in $\cL(N,N')$ is an ordered product of these operators, where the numbers $n(w)$ and $n^\dag(w)$ that respectively count the occurences of $c_j$ and $c^\dagger_j$ operators (independently of $j$) satisfy $n^\dag(w)-n(w) = \frac12(N-N')$. We often omit the size index of the operators $c_j$ and $c^\dag_j$ for simplicity, understanding that any word $w \in \cL(N,N')$ is either of the form $w=c_{N+2,j}w'$ for some shorter word $w' \in \cL(N+2,N')$ and with $0 \le j \le N+1$, or $w=c^\dag_{N,j}w'$ for some $w' \in \cL(N-2,N')$ with $0 \le j \le N-1$, or is the empty word $\id$ in $\cL(N,N)$. Thus, pairs of adjacent letters always have size indices that are equal or differ by two, as in \eqref{eq:c.relations}, and the indices $j$ are taken to have the lower bound $0$ and an upper bound that depends on the size index.

\begin{Theorem} \label{prop:equivalence.L}
The diagrammatic and algebraic definitions of $\cL(N,N')$ are equivalent.
\end{Theorem}

\noindent The proof is given in \cref{app:equiv.L}. In general, we say that two diagram spaces $\cL_1,\cL_2$ are isomorphic, and write $\cL_1 \simeq \cL_2$, if and only if there exists a linear map $\phi:\cL_1 \to \cL_2$ that is bijective and preserves the left and right actions of the diagrams. The above proposition thus states that the diagram spaces corresponding to the diagrammatic and algebraic definitions of $\cL$, which we respectively denote as $\cLd$ and $\cLa$, are isomorphic: $\cLd(N,N')\simeq \cLa(N,N')$.
\medskip

\paragraph{Algebraic definition of $\boldsymbol{\eptl_N(\beta)}$.}
We denote by $\eptl_N(\beta)$ the enlarged periodic Temperley--Lieb algebra on $N$ sites, with $N \in \Zbb_{\ge 0}$. For $N>2$, the algebra is defined in terms of generators as $\eptl_N(\beta) = \langle\Omega, \Omega^{-1}, e_0, e_1, \dots,e_{N-1}\rangle$ with the relations
\begin{subequations}
\label{eq:def.EPTL}
\begin{alignat}{4}
& e_j^2 = \beta \, e_j \,, \qquad &&e_j \,e_{j \pm 1 \, } e_j = e_j \,, \qquad &&e_i \,e_j = e_j \,e_i \qquad \text{for } |i-j|>1 \,, \\[0.1cm]
& \Omega \, e_j \, \Omega^{-1} = e_{j-1} \,, \qquad &&\Omega \, \Omega^{-1} = \Omega^{-1} \, \Omega = \id \,, \qquad &&\Omega^2 e_1 = e_{N-1}e_{N-2} \cdots e_2e_1\,.
\end{alignat}
\end{subequations}
Here $\id$ is the identity and the indices $i$ and $j$ are taken modulo~$N$. For $N=2$, the algebra is defined as $\eptl_2(\beta) = \langle\Omega, \Omega^{-1}, e_0, e_1\rangle$ with the relations 
\be
e_j^2 = \beta \, e_j\,, 
\qquad 
\Omega \, e_j \, \Omega^{-1} = e_{j-1}\,, 
\qquad
\Omega \, \Omega^{-1} = \Omega^{-1} \, \Omega = \id\,, 
\qquad 
\Omega^2 e_j = e_j\,.
\ee
For $N=1$, the algebra is defined as $\eptl_1(\beta)= \langle \Omega, \Omega^{-1}\rangle$ with $\Omega \, \Omega^{-1} = \Omega^{-1} \, \Omega = \id$. Finally for $N=0$, the algebra is defined in terms of a generator $f$ as $\eptl_0(\beta)= \langle f \rangle$, with no extra relation.\medskip

The ordinary Temperley--Lieb algebra is the subalgebra $\tl_N(\beta) = \langle e_1, e_2, \dots,e_{N-1}\rangle$.

\paragraph{Diagrammatic definition of $\boldsymbol{\eptl_N(\beta)}$.}

The algebra $\eptl_N(\beta)$ is defined diagrammatically as the vector space $\cL(N,N)$ endowed with the product of diagrams defined above, with $N=N'=N''$. Its generators are then assigned diagrams as follows:
\begin{subequations}
\begin{alignat}{4}
&\Omega = \ 
\psset{unit=0.8cm}
\thispic{\begin{pspicture}[shift=-1.3](-1.2,-1.4)(1.4,1.2)
\psarc[linecolor=black,linewidth=0.5pt,fillstyle=solid,fillcolor=lightlightblue]{-}(0,0){1.2}{0}{360}
\psarc[linecolor=black,linewidth=0.5pt,fillstyle=solid,fillcolor=white]{-}(0,0){0.7}{0}{360}
\psbezier[linecolor=blue,linewidth=1.5pt]{-}(0.676148, 0.181173)(0.91763, 0.245878)(0.91763, -0.245878)(1.15911, -0.310583)
\psbezier[linecolor=blue,linewidth=1.5pt]{-}(0.494975, 0.494975)(0.671751, 0.671751)(0.91763, 0.245878)(1.15911, 0.310583)
\psbezier[linecolor=blue,linewidth=1.5pt]{-}(0.181173, 0.676148)(0.245878, 0.91763)(0.671751, 0.671751)(0.848528, 0.848528)
\psbezier[linecolor=blue,linewidth=1.5pt]{-}(-0.181173, 0.676148)(-0.245878, 0.91763)(0.245878, 0.91763)(0.310583, 1.15911)
\psbezier[linecolor=blue,linewidth=1.5pt]{-}(-0.494975, 0.494975)(-0.671751, 0.671751)(-0.245878, 0.91763)(-0.310583, 1.15911)
\psbezier[linecolor=blue,linewidth=1.5pt]{-}(-0.676148, 0.181173)(-0.91763, 0.245878)(-0.671751, 0.671751)(-0.848528, 0.848528)
\psbezier[linecolor=blue,linewidth=1.5pt]{-}(-0.676148, -0.181173)(-0.91763, -0.245878)(-0.91763, 0.245878)(-1.15911, 0.310583)
\psbezier[linecolor=blue,linewidth=1.5pt]{-}(-0.494975, -0.494975)(-0.671751, -0.671751)(-0.91763, -0.245878)(-1.15911, -0.310583)
\psbezier[linecolor=blue,linewidth=1.5pt]{-}(-0.181173, -0.676148)(-0.245878, -0.91763)(-0.671751, -0.671751)(-0.848528, -0.848528)
\psbezier[linecolor=blue,linewidth=1.5pt]{-}(0.181173, -0.676148)(0.245878, -0.91763)(-0.245878, -0.91763)(-0.310583, -1.15911)
\psbezier[linecolor=blue,linewidth=1.5pt]{-}(0.494975, -0.494975)(0.671751, -0.671751)(0.245878, -0.91763)(0.310583, -1.15911)
\psbezier[linecolor=blue,linewidth=1.5pt]{-}(0.676148, -0.181173)(0.91763, -0.245878)(0.671751, -0.671751)(0.848528, -0.848528)
\psline[linestyle=dashed, dash= 1.5pt 1.5pt,linewidth=0.5pt]{-}(0,-1.2)(0,-0.7)
\rput(0.362347, -1.3523){$_1$}
\rput(0.989949, -0.989949){$_2$}
\rput(1.3523, -0.362347){$_3$}
\rput(1.3523, 0.362347){$_{...}$}
\rput(-1.13066, -1.13066){$_{N-1}$}
\rput(-0.362347, -1.3523){$_N$}
\end{pspicture}} \ \ ,
\qquad
&&\Omega^{-1}\,= \
\psset{unit=0.8cm}
\thispic{\begin{pspicture}[shift=-1.3](-1.2,-1.4)(1.4,1.2)
\psarc[linecolor=black,linewidth=0.5pt,fillstyle=solid,fillcolor=lightlightblue]{-}(0,0){1.2}{0}{360}
\psarc[linecolor=black,linewidth=0.5pt,fillstyle=solid,fillcolor=white]{-}(0,0){0.7}{0}{360}
\psbezier[linecolor=blue,linewidth=1.5pt]{-}(0.676148, 0.181173)(0.91763, 0.245878)(0.671751, 0.671751)(0.848528, 0.848528)
\psbezier[linecolor=blue,linewidth=1.5pt]{-}(0.494975, 0.494975)(0.671751, 0.671751)(0.245878, 0.91763)(0.310583, 1.15911)
\psbezier[linecolor=blue,linewidth=1.5pt]{-}(0.181173, 0.676148)(0.245878, 0.91763)(-0.245878, 0.91763)(-0.310583, 1.15911)
\psbezier[linecolor=blue,linewidth=1.5pt]{-}(-0.181173, 0.676148)(-0.245878, 0.91763)(-0.671751, 0.671751)(-0.848528, 0.848528)
\psbezier[linecolor=blue,linewidth=1.5pt]{-}(-0.494975, 0.494975)(-0.671751, 0.671751)(-0.91763, 0.245878)(-1.15911, 0.310583)
\psbezier[linecolor=blue,linewidth=1.5pt]{-}(-0.676148, 0.181173)(-0.91763, 0.245878)(-0.91763, -0.245878)(-1.15911, -0.310583)
\psbezier[linecolor=blue,linewidth=1.5pt]{-}(-0.676148, -0.181173)(-0.91763, -0.245878)(-0.671751, -0.671751)(-0.848528, -0.848528)
\psbezier[linecolor=blue,linewidth=1.5pt]{-}(-0.494975, -0.494975)(-0.671751, -0.671751)(-0.245878, -0.91763)(-0.310583, -1.15911)
\psbezier[linecolor=blue,linewidth=1.5pt]{-}(-0.181173, -0.676148)(-0.245878, -0.91763)(0.245878, -0.91763)(0.310583, -1.15911)
\psbezier[linecolor=blue,linewidth=1.5pt]{-}(0.181173, -0.676148)(0.245878, -0.91763)(0.671751, -0.671751)(0.848528, -0.848528)
\psbezier[linecolor=blue,linewidth=1.5pt]{-}(0.494975, -0.494975)(0.671751, -0.671751)(0.91763, -0.245878)(1.15911, -0.310583)
\psbezier[linecolor=blue,linewidth=1.5pt]{-}(0.676148, -0.181173)(0.91763, -0.245878)(0.91763, 0.245878)(1.15911, 0.310583)
\psline[linestyle=dashed, dash= 1.5pt 1.5pt,linewidth=0.5pt]{-}(0,-1.2)(0,-0.7)
\rput(0.362347, -1.3523){$_1$}
\rput(0.989949, -0.989949){$_2$}
\rput(1.3523, -0.362347){$_3$}
\rput(1.3523, 0.362347){$_{...}$}
\rput(-1.13066, -1.13066){$_{N-1}$}
\rput(-0.362347, -1.3523){$_N$}
\end{pspicture}}\ \ ,
\qquad 
\id = \ 
\psset{unit=0.8cm}
\thispic{\begin{pspicture}[shift=-1.3](-1.2,-1.4)(1.4,1.2)
\psarc[linecolor=black,linewidth=0.5pt,fillstyle=solid,fillcolor=lightlightblue]{-}(0,0){1.2}{0}{360}
\psarc[linecolor=black,linewidth=0.5pt,fillstyle=solid,fillcolor=white]{-}(0,0){0.7}{0}{360}
\psline[linecolor=blue,linewidth=1.5pt]{-}(0.676148, 0.181173)(1.15911, 0.310583)
\psline[linecolor=blue,linewidth=1.5pt]{-}(0.494975, 0.494975)(0.848528, 0.848528)
\psline[linecolor=blue,linewidth=1.5pt]{-}(0.181173, 0.676148)(0.310583, 1.15911)
\psline[linecolor=blue,linewidth=1.5pt]{-}(-0.181173, 0.676148)(-0.310583, 1.15911)
\psline[linecolor=blue,linewidth=1.5pt]{-}(-0.494975, 0.494975)(-0.848528, 0.848528)
\psline[linecolor=blue,linewidth=1.5pt]{-}(-0.676148, 0.181173)(-1.15911, 0.310583)
\psline[linecolor=blue,linewidth=1.5pt]{-}(-0.676148, -0.181173)(-1.15911, -0.310583)
\psline[linecolor=blue,linewidth=1.5pt]{-}(-0.494975, -0.494975)(-0.848528, -0.848528)
\psline[linecolor=blue,linewidth=1.5pt]{-}(-0.181173, -0.676148)(-0.310583, -1.15911)
\psline[linecolor=blue,linewidth=1.5pt]{-}(0.181173, -0.676148)(0.310583, -1.15911)
\psline[linecolor=blue,linewidth=1.5pt]{-}(0.494975, -0.494975)(0.848528, -0.848528)
\psline[linecolor=blue,linewidth=1.5pt]{-}(0.676148, -0.181173)(1.15911, -0.310583)
\psline[linestyle=dashed, dash= 1.5pt 1.5pt,linewidth=0.5pt]{-}(0,-1.2)(0,-0.7)
\rput(0.362347, -1.3523){$_1$}
\rput(0.989949, -0.989949){$_2$}
\rput(1.3523, -0.362347){$_3$}
\rput(1.3523, 0.362347){$_{...}$}
\rput(-1.13066, -1.13066){$_{N-1}$}
\rput(-0.362347, -1.3523){$_N$}
\end{pspicture}}\ \ ,
\\[0.5cm] 
& e_0\,= \
\psset{unit=0.8cm}
\thispic{\begin{pspicture}[shift=-1.3](-1.2,-1.4)(1.4,1.2)
\psarc[linecolor=black,linewidth=0.5pt,fillstyle=solid,fillcolor=lightlightblue]{-}(0,0){1.2}{0}{360}
\psarc[linecolor=black,linewidth=0.5pt,fillstyle=solid,fillcolor=white]{-}(0,0){0.7}{0}{360}
\psline[linecolor=blue,linewidth=1.5pt]{-}(0.676148, 0.181173)(1.15911, 0.310583)
\psline[linecolor=blue,linewidth=1.5pt]{-}(0.494975, 0.494975)(0.848528, 0.848528)
\psline[linecolor=blue,linewidth=1.5pt]{-}(0.181173, 0.676148)(0.310583, 1.15911)
\psline[linecolor=blue,linewidth=1.5pt]{-}(-0.181173, 0.676148)(-0.310583, 1.15911)
\psline[linecolor=blue,linewidth=1.5pt]{-}(-0.494975, 0.494975)(-0.848528, 0.848528)
\psline[linecolor=blue,linewidth=1.5pt]{-}(-0.676148, 0.181173)(-1.15911, 0.310583)
\psline[linecolor=blue,linewidth=1.5pt]{-}(-0.676148, -0.181173)(-1.15911, -0.310583)
\psline[linecolor=blue,linewidth=1.5pt]{-}(-0.494975, -0.494975)(-0.848528, -0.848528)
\psbezier[linecolor=blue,linewidth=1.5pt]{-}(-0.181173, -0.676148)(-0.245878, -0.91763)(0.245878, -0.91763)(0.181173, -0.676148)
\psbezier[linecolor=blue,linewidth=1.5pt]{-}(-0.310583, -1.15911)(-0.245878, -0.91763)(0.245878, -0.91763)(0.310583, -1.15911)
\psline[linecolor=blue,linewidth=1.5pt]{-}(0.494975, -0.494975)(0.848528, -0.848528)
\psline[linecolor=blue,linewidth=1.5pt]{-}(0.676148, -0.181173)(1.15911, -0.310583)
\psline[linestyle=dashed, dash= 1.5pt 1.5pt,linewidth=0.5pt]{-}(0,-1.2)(0,-0.7)
\rput(0.362347, -1.3523){$_1$}
\rput(0.989949, -0.989949){$_2$}
\rput(1.3523, -0.362347){$_3$}
\rput(1.3523, 0.362347){$_{...}$}
\rput(-1.13066, -1.13066){$_{N-1}$}
\rput(-0.362347, -1.3523){$_N$}
\end{pspicture}}\ \ , \qquad
&&e_j =\
\psset{unit=0.8cm}
\thispic{\begin{pspicture}[shift=-1.3](-1.2,-1.4)(1.4,1.2)
\psarc[linecolor=black,linewidth=0.5pt,fillstyle=solid,fillcolor=lightlightblue]{-}(0,0){1.2}{0}{360}
\psarc[linecolor=black,linewidth=0.5pt,fillstyle=solid,fillcolor=white]{-}(0,0){0.7}{0}{360}
\psline[linecolor=blue,linewidth=1.5pt]{-}(0.676148, 0.181173)(1.15911, 0.310583)
\psbezier[linecolor=blue,linewidth=1.5pt]{-}(0.494975, 0.494975)(0.671751, 0.671751)(0.245878, 0.91763)(0.181173, 0.676148)
\psbezier[linecolor=blue,linewidth=1.5pt]{-}(0.848528, 0.848528)(0.671751, 0.671751)(0.245878, 0.91763)(0.310583, 1.15911)
\psline[linecolor=blue,linewidth=1.5pt]{-}(-0.181173, 0.676148)(-0.310583, 1.15911)
\psline[linecolor=blue,linewidth=1.5pt]{-}(-0.494975, 0.494975)(-0.848528, 0.848528)
\psline[linecolor=blue,linewidth=1.5pt]{-}(-0.676148, 0.181173)(-1.15911, 0.310583)
\psline[linecolor=blue,linewidth=1.5pt]{-}(-0.676148, -0.181173)(-1.15911, -0.310583)
\psline[linecolor=blue,linewidth=1.5pt]{-}(-0.494975, -0.494975)(-0.848528, -0.848528)
\psline[linecolor=blue,linewidth=1.5pt]{-}(-0.181173, -0.676148)(-0.310583, -1.15911)
\psline[linecolor=blue,linewidth=1.5pt]{-}(0.181173, -0.676148)(0.310583, -1.15911)
\psline[linecolor=blue,linewidth=1.5pt]{-}(0.494975, -0.494975)(0.848528, -0.848528)
\psline[linecolor=blue,linewidth=1.5pt]{-}(0.676148, -0.181173)(1.15911, -0.310583)
\psline[linestyle=dashed, dash= 1.5pt 1.5pt,linewidth=0.5pt]{-}(0,-1.2)(0,-0.7)
\rput(0.362347, -1.3523){$_1$}
\rput(0.989949, -0.989949){$_2$}
\rput(1.3523, -0.362347){$_3$}
\rput(1.3523, 0.362347){$_{...}$}
\rput(0.989949, 0.989949){$_j$}
\rput(0.362347, 1.3523){$_{j+1}$}
\rput(-1.13066, -1.13066){$_{N-1}$}
\rput(-0.362347, -1.3523){$_N$}
\end{pspicture}}
\qquad 1\le j \le N-1\,.
\end{alignat}
\end{subequations}
For $N=0$, the generator $f$ corresponds to the diagram without nodes on any boundary and with a {\it non-contractible} closed loop, namely a loop encircling the inner boundary:
\be
\label{eq:def.f}
f = \ 
\psset{unit=0.8cm}
\thispic{\begin{pspicture}[shift=-1.1](-1.2,-1.2)(1.4,1.2)
\psarc[linecolor=black,linewidth=0.5pt,fillstyle=solid,fillcolor=lightlightblue]{-}(0,0){1.2}{0}{360}
\psarc[linecolor=black,linewidth=0.5pt,fillstyle=solid,fillcolor=white]{-}(0,0){0.7}{0}{360}
\psarc[linecolor=blue,linewidth=1.5pt]{-}(0,0){0.95}{0}{360}
\psline[linestyle=dashed, dash= 1.5pt 1.5pt,linewidth=0.5pt]{-}(0,-1.2)(0,-0.7)
\end{pspicture}}\ .
\ee
These generators can be written as
\begin{equation} \label{eq:Omega.and.c} 
e_{N,j} = c_{N,j}^{\dag} \, c_{N,j}^{\phantom{\dag}} \,,
\qquad 
\begin{array}{l}
\Omega_N = c_{N+2,1}\, c_{N+2,0}^\dag = c_{N+2,0}\, c_{N+2,N+1}^\dag \,,
\\[0.15cm]
\Omega_N^{-1} = c_{N+2,0}\, c_{N+2,1}^\dag = c_{N+2,N+1}\, c_{N+2,0}^\dag \,, 
\end{array}
\qquad 
f = c_{2,1}\, c_{2,0}^\dag = c_{2,0}\, c_{2,1}^\dag \,,
\end{equation}
where we write $e_j$, $\Omega$ and $\Omega^{-1}$ in $\eptl_N(\beta)$ as $e_{N,j}$, $\Omega_N$ and $\Omega^{-1}_N$ for clarity. Using the relations \eqref{eq:c.relations}, one can show that all the relations \eqref{eq:def.EPTL} are satisfied, as well as
\begin{equation} \label{eq:relations.c.Omega}
c_{N,0}\, \Omega_N = c_{N,1} \,, 
\qquad c_{N,N-1}\, \Omega_N = c_{N,0}\,, 
\qquad \Omega^{-1}_{N-2} \, c_{N,j}\, \Omega_N = c_{N,j+1} \,, 
\qquad 1 \le j \le N-2 \,.
\end{equation}

We note that the equivalence of the diagrammatic and algebraic definitions of $\eptl_N(\beta)$ was first proved in \cite{G98,EG99,FanGreen99}. \cref{prop:equivalence.L} can be seen as a generalisation of this result to the case $N\neq N'$.

\paragraph{Subspaces.}

We introduce the following subspaces of $\cL(N,N')$:
\begin{enumerate}
\item[1)] $\cLk{k}(N,N')$ is the subspace of $\cL(N,N')$ generated by the diagrams with exactly $2k$ bridges, where $k \in \frac12 \Zbb_{\ge 0}$ and $2k$ has the same parity as $N$ and $N'$. 
\item[2)] For $k>0$, $\cLk{k,0}(N,N')$ is the subspace of $\cLk{k}(N,N')$ generated by the diagrams without any bridge crossing the line $\pc'\pc$. For $k=0$, we instead define $\cLk{0,0}(N,N')$ to be the subspace of $\cLk{0}(N,N')$ made of connectivity diagrams without non-contractible loops. 
\item[3)] $\cL_0(N,N')$ is the subspace of $\cL(N,N')$ generated by the diagrams without any loop segment crossing the line $\pc'\pc$.
\end{enumerate}
With these definitions, one can check that $\cLk{k}(N,N')$ is infinite dimensional, whereas $\cLk{k,0}(N,N')$ and $\cL_0(N,N')$ are finite dimensional.

\paragraph{Basic properties.}

For each connectivity diagram $\lambda \in \cL(N,N')$, we define the {\it adjoint} diagram $\lambda^\dag \in \cL(N',N)$ obtained from~$\lambda$ by exchanging the inner and outer boundaries and reflecting the diagram radially. With this definition, the diagrams $c_{N,j}$ and $c_{N,j}^\dag$ defined in \eqref{eq:def.cj} are indeed the adjoints of each other. Moreover, we have
\begin{equation}
\id^\dag = \id\,, \qquad
e_j^\dag = e_j \,, \qquad 
\Omega^\dag = \Omega^{-1} \,, \qquad
f^\dag = f\,.
\end{equation}
On linear combinations of diagrams, we define this adjoint operation to act linearly. We now prove a basic property of the diagram spaces.

\begin{Proposition}\label{prop:LL=L}
Let $(N,N',N'')$ be a triplet of non-negative integers with the same parity and such that $N' \geq \mathrm{min}(N,N'')$. Then the following identity holds:
\begin{equation} \label{eq:L.L=L}
\cL(N,N') \,\cL(N',N'') = \cL(N,N'') \,. 
\end{equation}
\end{Proposition}
\proof We know from \eqref{eq:LLinL} that the left-hand side is a subspace of the right-hand side. We now proceed to prove the reverse inclusion in the case $N' \geq \mathrm{min}(N,N'')$. Let $\lambda$ be a diagram in $\cL(N,N'')$. We denote by $n'$ the number of bridges of $\lambda$. Clearly, $n'$ has the same parity as $N'$ and satisfies $n' \leq \mathrm{min}(N,N'')\leq N'$. We decompose $\lambda$ by drawing a closed non-contractible curve on the annulus that intersects each bridge of $\lambda$ exactly once and has no intersection with the arches. This decomposes $\lambda$ as $\lambda_1 \, \lambda_2$, where $\lambda_1 \in \cL(N,n')$ and $\lambda_2 \in \cL(n',N'')$ are respectively the diagram on the outside and the inside of the closed curve. Using the identity 
\begin{equation} \label{eq:identity.identity}
(c_2)^n (c_1^\dag)^n = \id_{n'} \,, 
\qquad
n \in \Zbb_{\geq 0} \,,
\end{equation}
we set $n = \frac12(N'-n')$ and write
\begin{equation}
\lambda = \big(\lambda_1 \,(c_2)^{n}\big) \big((c_1^\dag)^{n}\lambda_2\big)
\in \cL(N,N') \, \cL(N',N'') \,,
\end{equation}
which ends the proof.
\eproof

\begin{Proposition}
For all $\lambda \in \cL(N,N')$, we have $\Omega^N \,\lambda = \lambda\, \Omega^{N'}$.
\end{Proposition}
\proof
Using the relations \eqref{eq:relations.c.Omega} and their adjoints repeatedly, we find
\be
\label{eq:Omega.cj}
\Omega^{N-2} \,c_{N,j} = c_{N,j}\, \Omega^N\,,
\qquad 
\Omega^N \,c^\dag_{N,j}=c^\dag_{N,j} \, \Omega^{N-2}\,,
\qquad j = 0,1, \dots, N-1.
\ee
For an arbitrary diagram $\lambda \in \cL(N,N')$, we decompose $\lambda$ as a word in the generators $c_j$ and $c^\dag_j$ and, using \eqref{eq:Omega.cj}, we prove $\Omega^N \,\lambda = \lambda\, \Omega^{N'}$ inductively on the length of the word.
\eproof

\subsection{Braid generators and braid transfer matrices}

In this subsection, we recall the definition and properties of braid operators and braid transfer matrices. The braid generators are elements of $\eptl_N(\beta)$ defined as
\begin{equation}
b_{N,j} = b_j = q^{1/2} \id + q^{-1/2} e_j \,, \qquad 
\bb_{N,j} = \bb_j = q^{-1/2} \id + q^{1/2} e_j = b^{-1}_j\,.
\end{equation}
These are used to define the braid matrices $F_N,\Fb_N \in \eptl_N(\beta)$:
\begin{subequations}
\begin{alignat}{2}
F_N &= c_{N+2,1} b_{N+2,2} b_{N+2,3} \cdots b_{N+2,N+1}c_{N+2,0}^\dag \,,
\\
\Fb_N &= c_{N+2,1} \bb_{N+2,2} \bb_{N+2,3} \cdots \bb_{N+2,N+1} c_{N+2,0}^\dag \,.
\end{alignat}
\end{subequations}
This also holds for $N=0$, where we have $F_0=\Fb_0=c_{2,1}^{\phantom{\dag}} c_{2,0}^\dag = f$. For $N\geq 1$, we have
\begin{subequations}
\begin{alignat}{2}
F_N &= q^{1/2}\,b_{N,1}b_{N,2}\cdots b_{N,N-1}\Omega + q^{-1/2}\,\Omega^{-1}\bb_{N,N-1}\bb_{N,N-2}\cdots \bb_{N,1} \,, \\[0.1cm]
\Fb_N &= q^{-1/2}\,\bb_{N,1}\bb_{N,2}\cdots \bb_{N,N-1}\Omega + q^{1/2}\,\Omega^{-1}b_{N,N-1}b_{N,N-2}\cdots b_{N,1} \,.
\end{alignat}
\end{subequations}
The braid matrices satisfy the \textit{push-through relations}
\begin{subequations} \label{eq:F.push-through}
\begin{alignat}{3}
c_{N,j} \, F_N &= F_{N-2} \, c_{N,j} \,,
\qquad &F_{N} \, c^\dag_{N,j} &= c^\dag_{N,j} \, F_{N-2} \,,
\\[0.15cm]
c_{N,j} \, \Fb_N &= \Fb_{N-2} \, c_{N,j} \,,
\qquad &\Fb_{N} \, c^\dag_{N,j} &= c^\dag_{N,j} \, \Fb_{N-2} \,.
\end{alignat}
\end{subequations}
As a consequence, we have more generally
\begin{equation} \label{eq:F.push-through.L}
F_{N} \, \lambda = \lambda \, F_{N'} \,,
\qquad \Fb_{N} \, \lambda = \lambda \, \Fb_{N'} \,,
\end{equation}
for all $\lambda \in \cL(N,N')$. Hereafter, we usually omit the index $N$ and simply write the braid transfer matrices as $F$ and $\Fb$. The operators $F$, $\Fb$ and $\Omega^{\pm N}$ are in the center of $\eptl_N(\beta)$.\medskip

In terms of the diagrams, we depict $b_j$ and $\bb_j$ as
\be
b_j\,= \
\psset{unit=0.8cm}
\thispic{\begin{pspicture}[shift=-1.25](-1.2,-1.4)(1.4,1.3)
\psarc[linecolor=black,linewidth=0.5pt,fillstyle=solid,fillcolor=lightlightblue]{-}(0,0){1.2}{0}{360}
\psarc[linecolor=black,linewidth=0.5pt,fillstyle=solid,fillcolor=white]{-}(0,0){0.7}{0}{360}
\rput{150}(0,0){
\psline[linecolor=blue,linewidth=1.5pt]{-}(0.676148, 0.181173)(1.15911, 0.310583)
\psline[linecolor=blue,linewidth=1.5pt]{-}(0.494975, 0.494975)(0.848528, 0.848528)
\psline[linecolor=blue,linewidth=1.5pt]{-}(0.181173, 0.676148)(0.310583, 1.15911)
\psline[linecolor=blue,linewidth=1.5pt]{-}(-0.181173, 0.676148)(-0.310583, 1.15911)
\psline[linecolor=blue,linewidth=1.5pt]{-}(-0.494975, 0.494975)(-0.848528, 0.848528)
\psline[linecolor=blue,linewidth=1.5pt]{-}(-0.676148, 0.181173)(-1.15911, 0.310583)
\psline[linecolor=blue,linewidth=1.5pt]{-}(-0.676148, -0.181173)(-1.15911, -0.310583)
\psline[linecolor=blue,linewidth=1.5pt]{-}(-0.494975, -0.494975)(-0.848528, -0.848528)
\psbezier[linecolor=blue,linewidth=1.5pt]{-}(-0.181173, -0.676148)(-0.245878, -0.91763)(0.245878, -0.91763)(0.310583, -1.15911)
\psbezier[linecolor=lightlightblue,linewidth=3.5pt]{-}(-0.310583, -1.15911)(-0.245878, -0.91763)(0.245878, -0.91763)(0.181173, -0.676148)
\psbezier[linecolor=blue,linewidth=1.5pt]{-}(-0.310583, -1.15911)(-0.245878, -0.91763)(0.245878, -0.91763)(0.181173, -0.676148)
\psline[linecolor=blue,linewidth=1.5pt]{-}(0.494975, -0.494975)(0.848528, -0.848528)
\psline[linecolor=blue,linewidth=1.5pt]{-}(0.676148, -0.181173)(1.15911, -0.310583)
}
\psline[linestyle=dashed, dash= 1.5pt 1.5pt,linewidth=0.5pt]{-}(0,-1.2)(0,-0.7)
\psarc[linecolor=black,linewidth=0.5pt]{-}(0,0){1.2}{0}{360}
\psarc[linecolor=black,linewidth=0.5pt]{-}(0,0){0.7}{0}{360}
\rput(0.362347, -1.3523){$_1$}
\rput(0.989949, -0.989949){$_2$}
\rput(1.3523, -0.362347){$_3$}
\rput(1.3523, 0.362347){$_{...}$}
\rput(0.989949, 0.989949){$_j$}
\rput(0.362347, 1.3923){$_{j+1}$}
\rput(-1.13066, -1.13066){$_{N-1}$}
\rput(-0.362347, -1.3523){$_N$}
\end{pspicture}}\ \ ,
\qquad
\bb_j\,= \
\thispic{\begin{pspicture}[shift=-1.25](-1.2,-1.4)(1.4,1.3)
\psarc[linecolor=black,linewidth=0.5pt,fillstyle=solid,fillcolor=lightlightblue]{-}(0,0){1.2}{0}{360}
\psarc[linecolor=black,linewidth=0.5pt,fillstyle=solid,fillcolor=white]{-}(0,0){0.7}{0}{360}
\rput{150}(0,0){
\psline[linecolor=blue,linewidth=1.5pt]{-}(0.676148, 0.181173)(1.15911, 0.310583)
\psline[linecolor=blue,linewidth=1.5pt]{-}(0.494975, 0.494975)(0.848528, 0.848528)
\psline[linecolor=blue,linewidth=1.5pt]{-}(0.181173, 0.676148)(0.310583, 1.15911)
\psline[linecolor=blue,linewidth=1.5pt]{-}(-0.181173, 0.676148)(-0.310583, 1.15911)
\psline[linecolor=blue,linewidth=1.5pt]{-}(-0.494975, 0.494975)(-0.848528, 0.848528)
\psline[linecolor=blue,linewidth=1.5pt]{-}(-0.676148, 0.181173)(-1.15911, 0.310583)
\psline[linecolor=blue,linewidth=1.5pt]{-}(-0.676148, -0.181173)(-1.15911, -0.310583)
\psline[linecolor=blue,linewidth=1.5pt]{-}(-0.494975, -0.494975)(-0.848528, -0.848528)
\psbezier[linecolor=blue,linewidth=1.5pt]{-}(-0.310583, -1.15911)(-0.245878, -0.91763)(0.245878, -0.91763)(0.181173, -0.676148)
\psbezier[linecolor=lightlightblue,linewidth=3.5pt]{-}(-0.181173, -0.676148)(-0.245878, -0.91763)(0.245878, -0.91763)(0.310583, -1.15911)
\psbezier[linecolor=blue,linewidth=1.5pt]{-}(-0.181173, -0.676148)(-0.245878, -0.91763)(0.245878, -0.91763)(0.310583, -1.15911)
\psline[linecolor=blue,linewidth=1.5pt]{-}(0.494975, -0.494975)(0.848528, -0.848528)
\psline[linecolor=blue,linewidth=1.5pt]{-}(0.676148, -0.181173)(1.15911, -0.310583)
}
\psline[linestyle=dashed, dash= 1.5pt 1.5pt,linewidth=0.5pt]{-}(0,-1.2)(0,-0.7)
\psarc[linecolor=black,linewidth=0.5pt]{-}(0,0){1.2}{0}{360}
\psarc[linecolor=black,linewidth=0.5pt]{-}(0,0){0.7}{0}{360}
\rput(0.362347, -1.3523){$_1$}
\rput(0.989949, -0.989949){$_2$}
\rput(1.3523, -0.362347){$_3$}
\rput(1.3523, 0.362347){$_{...}$}
\rput(0.989949, 0.989949){$_j$}
\rput(0.362347, 1.3923){$_{j+1}$}
\rput(-1.13066, -1.13066){$_{N-1}$}
\rput(-0.362347, -1.3523){$_N$}
\end{pspicture}}\ \ .
\ee
The braid matrices are then depicted as
\be
F = \ 
\thispic{\begin{pspicture}[shift=-1.3](-1.2,-1.4)(1.4,1.2)
\psarc[linecolor=black,linewidth=0.5pt,fillstyle=solid,fillcolor=lightlightblue]{-}(0,0){1.2}{0}{360}
\psarc[linecolor=black,linewidth=0.5pt,fillstyle=solid,fillcolor=white]{-}(0,0){0.7}{0}{360}
\psline[linecolor=blue,linewidth=1.25pt]{-}(0.676148, 0.181173)(0.850015, 0.227761)\psline[linecolor=blue,linewidth=1.25pt]{-}(0.985244, 0.263995)(1.15911, 0.310583)
\psline[linecolor=blue,linewidth=1.25pt]{-}(0.494975, 0.494975)(0.622254, 0.622254)\psline[linecolor=blue,linewidth=1.25pt]{-}(0.721249, 0.721249)(0.848528, 0.848528)
\psline[linecolor=blue,linewidth=1.25pt]{-}(0.181173, 0.676148)(0.227761, 0.850015)\psline[linecolor=blue,linewidth=1.25pt]{-}(0.263995, 0.985244)(0.310583, 1.15911)
\psline[linecolor=blue,linewidth=1.25pt]{-}(-0.181173, 0.676148)(-0.227761, 0.850015)\psline[linecolor=blue,linewidth=1.25pt]{-}(-0.263995, 0.985244)(-0.310583, 1.15911)
\psline[linecolor=blue,linewidth=1.25pt]{-}(-0.494975, 0.494975)(-0.622254, 0.622254)\psline[linecolor=blue,linewidth=1.25pt]{-}(-0.721249, 0.721249)(-0.848528, 0.848528)
\psline[linecolor=blue,linewidth=1.25pt]{-}(-0.676148, 0.181173)(-0.850015, 0.227761)\psline[linecolor=blue,linewidth=1.25pt]{-}(-0.985244, 0.263995)(-1.15911, 0.310583)
\psline[linecolor=blue,linewidth=1.25pt]{-}(-0.676148, -0.181173)(-0.850015, -0.227761)\psline[linecolor=blue,linewidth=1.25pt]{-}(-0.985244, -0.263995)(-1.15911, -0.310583)
\psline[linecolor=blue,linewidth=1.25pt]{-}(-0.494975, -0.494975)(-0.622254, -0.622254)\psline[linecolor=blue,linewidth=1.25pt]{-}(-0.721249, -0.721249)(-0.848528, -0.848528)
\psline[linecolor=blue,linewidth=1.25pt]{-}(-0.181173, -0.676148)(-0.227761, -0.850015)\psline[linecolor=blue,linewidth=1.25pt]{-}(-0.263995, -0.985244)(-0.310583, -1.15911)
\psline[linecolor=blue,linewidth=1.25pt]{-}(0.181173, -0.676148)(0.227761, -0.850015)\psline[linecolor=blue,linewidth=1.25pt]{-}(0.263995, -0.985244)(0.310583, -1.15911)
\psline[linecolor=blue,linewidth=1.25pt]{-}(0.494975, -0.494975)(0.622254, -0.622254)\psline[linecolor=blue,linewidth=1.25pt]{-}(0.721249, -0.721249)(0.848528, -0.848528)
\psline[linecolor=blue,linewidth=1.25pt]{-}(0.676148, -0.181173)(0.850015, -0.227761)\psline[linecolor=blue,linewidth=1.25pt]{-}(0.985244, -0.263995)(1.15911, -0.310583)
\psarc[linecolor=blue,linewidth=1.25pt]{-}(0,0){0.95}{0}{360}
\psline[linestyle=dashed, dash= 1.5pt 1.5pt,linewidth=0.5pt]{-}(0,-1.2)(0,-0.7)
\rput(0.362347, -1.3523){$_1$}
\rput(0.989949, -0.989949){$_2$}
\rput(1.3523, -0.362347){$_3$}
\rput(1.3523, 0.362347){$_{...}$}
\rput(-1.06066, -1.06066){$_{N-1}$}
\rput(-0.362347, -1.3523){$_N$}
\end{pspicture}}\ \ ,
\qquad 
\Fb = \
\thispic{\begin{pspicture}[shift=-1.3](-1.2,-1.4)(1.4,1.2)
\psarc[linecolor=black,linewidth=0.5pt,fillstyle=solid,fillcolor=lightlightblue]{-}(0,0){1.2}{0}{360}
\psarc[linecolor=black,linewidth=0.5pt,fillstyle=solid,fillcolor=white]{-}(0,0){0.7}{0}{360}
\psline[linecolor=blue,linewidth=1.25pt]{-}(0.676148, 0.181173)(1.15911, 0.310583)
\psline[linecolor=blue,linewidth=1.25pt]{-}(0.494975, 0.494975)(0.848528, 0.848528)
\psline[linecolor=blue,linewidth=1.25pt]{-}(0.181173, 0.676148)(0.310583, 1.15911)
\psline[linecolor=blue,linewidth=1.25pt]{-}(-0.181173, 0.676148)(-0.310583, 1.15911)
\psline[linecolor=blue,linewidth=1.25pt]{-}(-0.494975, 0.494975)(-0.848528, 0.848528)
\psline[linecolor=blue,linewidth=1.25pt]{-}(-0.676148, 0.181173)(-1.15911, 0.310583)
\psline[linecolor=blue,linewidth=1.25pt]{-}(-0.676148, -0.181173)(-1.15911, -0.310583)
\psline[linecolor=blue,linewidth=1.25pt]{-}(-0.494975, -0.494975)(-0.848528, -0.848528)
\psline[linecolor=blue,linewidth=1.25pt]{-}(-0.181173, -0.676148)(-0.310583, -1.15911)
\psline[linecolor=blue,linewidth=1.25pt]{-}(0.181173, -0.676148)(0.310583, -1.15911)
\psline[linecolor=blue,linewidth=1.25pt]{-}(0.494975, -0.494975)(0.848528, -0.848528)
\psline[linecolor=blue,linewidth=1.25pt]{-}(0.676148, -0.181173)(1.15911, -0.310583)
\psarc[linecolor=blue,linewidth=1.25pt]{-}(0,0){0.95}{-10}{10}
\psarc[linecolor=blue,linewidth=1.25pt]{-}(0,0){0.95}{20}{40}
\psarc[linecolor=blue,linewidth=1.25pt]{-}(0,0){0.95}{50}{70}
\psarc[linecolor=blue,linewidth=1.25pt]{-}(0,0){0.95}{80}{100}
\psarc[linecolor=blue,linewidth=1.25pt]{-}(0,0){0.95}{110}{130}
\psarc[linecolor=blue,linewidth=1.25pt]{-}(0,0){0.95}{140}{160}
\psarc[linecolor=blue,linewidth=1.25pt]{-}(0,0){0.95}{170}{190}
\psarc[linecolor=blue,linewidth=1.25pt]{-}(0,0){0.95}{200}{220}
\psarc[linecolor=blue,linewidth=1.25pt]{-}(0,0){0.95}{230}{250}
\psarc[linecolor=blue,linewidth=1.25pt]{-}(0,0){0.95}{260}{280}
\psarc[linecolor=blue,linewidth=1.25pt]{-}(0,0){0.95}{290}{310}
\psarc[linecolor=blue,linewidth=1.25pt]{-}(0,0){0.95}{320}{340}
\psline[linestyle=dashed, dash= 1.5pt 1.5pt,linewidth=0.5pt]{-}(0,-1.2)(0,-0.7)
\rput(0.362347, -1.3523){$_1$}
\rput(0.989949, -0.989949){$_2$}
\rput(1.3523, -0.362347){$_3$}
\rput(1.3523, 0.362347){$_{...}$}
\rput(-1.06066, -1.06066){$_{N-1}$}
\rput(-0.362347, -1.3523){$_N$}
\end{pspicture}}\ \ .
\ee
We note that\footnote{For a diagram with braid crossings, the adjoint operation also interchanges the over and under crossings in addition to flipping the diagram inside out.}
\be
b_j^\dag = b_j\,, \qquad (\bb_j)^\dag = \bb_j\,, \qquad F^\dag = \Fb\,.
\ee

\subsection{Families of modules}

\paragraph{Definitions.}

We define a {\it family} $\repM$ of modules to be a set of $\eptl_N(\beta)$-modules indexed by increasing values of $N$,\footnote{Here, we use $\repM$ and $\repM(N)$ to denote a family of modules and a module in this family, respectively. This contrasts with the notation used in \cite{IMD22,IMD24} whereby $\repM$ and $\repM(N)$ were used interchangeably to denote the modules.} 
\begin{equation} \label{eq:module.family}
\repM = \{\repM(N)\,|\,N=N_0, N_0+2, N_0+4, \dots\} \,,
\qquad N_0 \in \{0,1\},
\end{equation}
and endowed with an action of the diagram spaces satisfying
\begin{subequations} \label{eq:family.properties}
\begin{alignat}{2}
& \lambda \cdot u \in \repM(N') 
\qquad &&\forall \lambda \in \cL(N',N), \, u \in \repM(N)\,,
\label{eq:family.stable} \\
& \lambda' \cdot (\lambda \cdot u) = (\lambda' \lambda) \cdot u
\qquad &&\forall \lambda' \in \cL(N'',N'),\, \lambda \in \cL(N',N),\,
u \in \repM(N)\,, \label{eq:family.consistency}
\end{alignat}
\end{subequations}
for all $N',N$ with the same parity as $N_0$. For a given family, the integer $N_0$ is uniquely fixed to either $N_0 = 0$ or $N_0 = 1$. We refer to the set $\{N_0, N_0+2, N_0+4,\dots\}$ as the set of all {\it admissible} values of $N$ for the family $\repM$. Moreover, we say that $\repM$ is an even or odd family of modules for $N_0=0$ and $N_0=1$, respectively.\medskip

Equivalently, a set $\repM$ of $\eptl_N(\beta)$-modules $\repM(N)$ is a family if we can construct operators $c_j: \repM(N) \to \repM(N-2)$ and $c^\dag_j: \repM(N-2) \to \repM(N)$ that respect the relations \eqref{eq:c.relations}, for each $N$. The next proposition gives a first set of examples of families of modules.\medskip

\begin{Proposition} \label{prop:L.family}
For all $N \in \Zbb_{\ge 0}$, the set 
\begin{equation} \label{eq:LN}
\repL_N=\{\cL(N',N)\,|\, N'=N_0, N_0+2, N_0+4, \dots\}\,,\qquad
N_0 = \left\{\begin{array}{ll}
0 & N \textrm{ even,}\\[0.1cm]
1 & N \textrm{ odd,}
\end{array}\right.
\end{equation}
is a family of modules.
\end{Proposition}
\proof This follows from \eqref{eq:LLinL} and from the associativity of the action in the diagrams spaces.
\eproof

Let $\repM,\repM'$ be two families of modules with the same parity. We define a \emph{family of homomorphisms}~$\phi$ from $\repM$ to $\repM'$ to be a set of $\eptl_N(\beta)$-homomorphisms $\{\phi_N:\repM(N) \to \repM'(N)\,|\,N=N_0, N_0+2, \dots\}$, satisfying
\begin{equation} \label{eq:def.family.morphism}
\phi_{N'}(\lambda \cdot u) = \lambda \cdot \phi_N(u)\,, \qquad
\forall u \in \repM(N) \,,\qquad 
\forall \lambda \in \cL(N',N) \,,
\end{equation}
for all admissible $N$ and $N'$. In the case where $\repM = \repM'$, we say that $\phi$ is a family of endomorphisms. Furthermore, we say that $\repM$ and $\repM'$ are isomorphic families if and only if there exists a family of homomorphisms~$\phi$ from $\repM$ to $\repM'$, such that each $\phi_N:\repM(N) \to \repM'(N)$ is an isomorphism.
\medskip

In the language of categories, a family of modules with $N_0=0$ or $N_0=1$ is a \textit{functor} from the subcategory of even or odd nonnegative integers to the category of complex vector spaces --- see \cite{GL98}. Similarly, a family of homomorphisms is a \textit{natural transformation} on a pair of families of modules.

\paragraph{Basic properties.}

We now describe some basic properties of families of modules.

\begin{Proposition}\label{prop:LM=M}
Let $\repM$ be a family of modules. Then for all admissible integers $N$ and $N'$ such that $N'\leq N$, we have
\begin{equation} \label{eq:family.nested}
\cL(N',N)\cdot \repM(N) = \repM(N') \,.
\end{equation}
\end{Proposition}
\proof
By \eqref{eq:family.properties}, it is clear that $\cL(N',N)\cdot \repM(N) \subseteq \repM(N')$. To show the reverse inclusion, we use the identity \eqref{eq:identity.identity} and write, for all $u \in \repM(N')$,
\begin{equation}
u = (c_2)^n \cdot \big((c_1^\dag)^n\cdot u \big)
\in \cL(N',N)\cdot \repM(N) \,, \qquad n=\frac12(N-N')\,,
\end{equation}
which ends the proof.
\eproof

The properties given below follow directly from the fact that a family of modules is a functor. We include the very simple proofs for the readers that are not familiar with theory of categories.

\begin{Proposition} \label{prop:sub.quot.families}
Let $\repM$ be a family and $\repM' = \{\repM'(N)\,|\,N = N_0, N_0+2, \dots\}$ be a set of submodules of the family $\repM$, namely $\repM'(N) \subseteq \repM(N)$ for all admissible $N$, satisfying the property
\begin{equation} \label{eq:subfamily}
\lambda \cdot u \in \repM'(N') \,,
\qquad \forall \lambda \in \cL(N',N) \,, \qquad \forall u \in \repM'(N) \,.
\end{equation}
Then the sets $\repM'$ and $\repM/\repM' = \{\repM(N)/\repM'(N)\,|\,N = N_0, N_0+2, \dots\}$ are both families of modules. 
\end{Proposition}
\proof It is obvious that $\repM'$ is a family. Indeed, \eqref{eq:subfamily} is precisely the first property \eqref{eq:family.stable} defining families. Moreover, the second property \eqref{eq:family.consistency} is guaranteed to hold because it holds for $\repM$, and $\repM'(N) \subseteq \repM(N)$ for each admissible $N$.
\medskip 

Let $u$ be a state in $\repM(N)$. We denote by $[u]$ the equivalence class of $u$ modulo $\repM'(N)$. Thus, the quotient reads
\begin{equation}
\repM(N)/\repM'(N) = \mathrm{span}\Big(\,[u]\,|\, u \in \repM(N)\Big) \,.
\end{equation}
For each pair of states $u,v \in \repM(N)$, we have $[u]=[v]$ if and only if $u-v\in \repM'(N)$. Therefore $\lambda\cdot(u-v)$ is also in $\repM'(N)$, which implies that $[\lambda \cdot u] = [\lambda \cdot v]$. We therefore define the action of $\cL(N',N)$ on $\repM(N)/\repM'(N)$ as $\lambda \cdot [u] = [\lambda \cdot u]$, for all $u \in \repM(N)$. This action satisfies 
\begin{equation}
\lambda' \cdot (\lambda \cdot [u])
= \lambda' \cdot [\lambda \cdot u]
= [\lambda' \cdot (\lambda \cdot u)]
= [(\lambda' \, \lambda) \cdot u]
= (\lambda'\, \lambda) \cdot [u] \,.
\end{equation}
Thus $\repM/\repM'$ is indeed a family of modules.
\eproof

\noindent We say that $\repM'$ and $\repM/\repM'$ are a {\it subfamily} and a {\it quotient family} of $\repM$, respectively.\medskip

Furthermore, it is straightforward to show that the sum of two subfamilies $\repM',\repM''$ is a subfamily. 

\begin{Proposition}
Let $\repM$ be a family and $\repM',\repM''$ be two subfamilies of $\repM$. Then the set of modules defined as $(\repM'+\repM'')(N)=\repM'(N)+\repM''(N)$ is a subfamily of $\repM$.\footnote{This result does not require that $\repM',\repM''$ be in direct sum, namely that the intersection $\repM'(N) \cap \repM''(N)$ be identically zero.}
\end{Proposition}

\subsection{Examples of families}\label{sec:example.fams}

\paragraph{Link state families.} 

We introduce three families of modules whose vector spaces are spanned by link states: $\repW_{k}$, $\repW_{k,x}$ and $\repV$.\medskip

First, we define the modules $\repW_{k}(N)$ for $k \in \frac12 \Zbb_{\geq 0}$, $N \in \Zbb_{\ge 0}$ and $N-2k$ even. For $2k \ge N$, we define $\repW_{k}(N)$ to be the trivial module: $\repW_{k}(N) = 0$. For $0 \le 2k \le N$, the module $\repW_k(N)$ is defined on the vector space $\cLk{k}(N,2k)$. These diagrams have all $2k$ nodes from the inner boundary of the ring connected to the outer boundary by bridges. For $k>0$, the bridges may wind around the ring an arbitrary number of times. For $k=0$, there can be an arbitrary number of non-contractible loops. In both cases, the vector spaces are infinite dimensional.\medskip

It is customary to draw the diagrams in $\repW_k(N)$ by shrinking the inner boundary to a point~$\pa$, called a {\it marked point}. For example, for $N=12$ and $k=2$, we have 
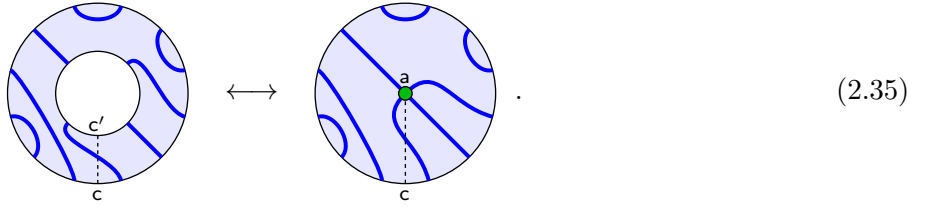
\be 
\label{eq:Wk.vs.L(N,2k)}
\psset{unit=0.8cm}
\thispic{\begin{pspicture}[shift=-1.5](-1.5,-1.6)(1.5,1.5)
\psarc[linecolor=black,linewidth=0.5pt,fillstyle=solid,fillcolor=lightlightblue]{-}(0,0){1.5}{0}{360}
\psarc[linecolor=black,linewidth=0.5pt,fillstyle=solid,fillcolor=white]{-}(0,0){0.7}{0}{360}
\rput{180}(0,0){
\rput{-60}(0,0){\psbezier[linecolor=blue,linewidth=1.5pt]{-}(-0.388229, -1.44889)(-0.310583, -1.15911)(0.310583, -1.15911)(0.388229, -1.44889)}
\psbezier[linecolor=blue,linewidth=1.5pt]{-}(-0.388229, -1.44889)(-0.310583, -1.15911)(0.310583, -1.15911)(0.388229, -1.44889)
\rput{120}(0,0){\psbezier[linecolor=blue,linewidth=1.5pt]{-}(-0.388229, -1.44889)(-0.310583, -1.15911)(0.310583, -1.15911)(0.388229, -1.44889)
\psbezier[linecolor=blue,linewidth=1.5pt]{-}(-1.06066, -1.06066)(-0.848528, -0.948528)(0.848528, -0.948528)(1.06066, -1.06066)}
\psline[linecolor=blue,linewidth=1.5pt]{-}(-1.06066, 1.06066)(-0.494975, 0.494975)
\psbezier[linecolor=blue,linewidth=1.5pt]{-}(-0.388229, 1.44889)(-0.284701, 1.06252)(0.777817, 0.777817)(0.494975, 0.494975)
\psbezier[linecolor=blue,linewidth=1.5pt]{-}(-1.44889, 0.388229)(-1.06252, 0.284701)(-0.777817, -0.777817)(-0.494975, -0.494975)
\psline[linecolor=blue,linewidth=1.5pt]{-}(1.06066, -1.06066)(0.494975, -0.494975)
}
\psline[linestyle=dashed, dash= 1.5pt 1.5pt,linewidth=0.5pt]{-}(0,-1.5)(0,-0.7)
\rput(0,-0.5){$_{\pc'}$}
\rput(0,-1.7){$_\pc$}
\end{pspicture}}
\quad
\longleftrightarrow
\quad
\thispic{\begin{pspicture}[shift=-1.5](-1.5,-1.6)(1.5,1.5)
\psarc[linecolor=black,linewidth=0.5pt,fillstyle=solid,fillcolor=lightlightblue]{-}(0,0){1.5}{0}{360}
\rput{180}(0,0){
\rput{-60}(0,0){\psbezier[linecolor=blue,linewidth=1.5pt]{-}(-0.388229, -1.44889)(-0.310583, -1.15911)(0.310583, -1.15911)(0.388229, -1.44889)}
\psbezier[linecolor=blue,linewidth=1.5pt]{-}(-0.388229, -1.44889)(-0.310583, -1.15911)(0.310583, -1.15911)(0.388229, -1.44889)
\rput{120}(0,0){\psbezier[linecolor=blue,linewidth=1.5pt]{-}(-0.388229, -1.44889)(-0.310583, -1.15911)(0.310583, -1.15911)(0.388229, -1.44889)
\psbezier[linecolor=blue,linewidth=1.5pt]{-}(-1.06066, -1.06066)(-0.848528, -0.948528)(0.848528, -0.948528)(1.06066, -1.06066)}
\psline[linecolor=blue,linewidth=1.5pt]{-}(-1.06066, 1.06066)(0,0)
\psline[linecolor=blue,linewidth=1.5pt]{-}(1.06066, -1.06066)(0,0)
}
\psbezier[linecolor=blue,linewidth=1.5pt]{-}(0.388229,-1.44889)(0.194114, -0.724444)(-0.53033, -0.53033)(0,0)
\psbezier[linecolor=blue,linewidth=1.5pt]{-}(1.44889, -0.388229)(0.724444, -0.194114)(0.53033, 0.53033)(0, 0)
\psline[linestyle=dashed, dash=1.5pt 1.5pt,linewidth=0.5pt]{-}(0,-1.5)(0,0)
\psarc[linecolor=black,linewidth=0.5pt,fillstyle=solid,fillcolor=darkgreen]{-}(0,0){0.1125}{0}{360}
\rput(0,0.25){$_{\pa}$}
\rput(0,-1.7){$_\pc$}
\end{pspicture}}\ \ . 
\ee
The resulting diagram drawn on a disc with a marked point is called a {\it link state}, and its bridges attached to $\pa$ are called {\it defects}. A link state thus consists of $\frac N2-k$ arches connecting the nodes on the boundary of the disc, and $2k$ defects connecting $\pa$ to the boundary. There may be an arbitrary number of defects that cross the line~$\pa\pc$. For $k=0$, the link states may also contain non-contractible loops, namely loops encircling the marked point.
\medskip

The action $\lambda \cdot u$ of a diagram $\lambda\in\cL(N',N)$ on a link state $u \in \repW_k(N)$ follows the action of $\cL(N',N)$ on $\cL(N,2k)$, but with an extra rule for $k>0$. In this case, any diagram produced in $\cL(N',2k)$ with less than $2k$ bridges is set to zero. For $k=0$, there is no extra rule. Each $\repW_k(N)$ is an infinite-dimensional module of $\eptl_N(\beta)$. We note that $\repW_k(N)$ can be seen as the quotient
\be
\repW_k(N) = \frac{\repL_{2k}(N)}{\{E_{k-1}=0\}}\,,
\qquad
\textrm{where}
\qquad
E_m = e_2 e_4 \cdots e_{N-2m}\,, \qquad m \ge 0\,,
\ee
and $E_{-1} = E_{-1/2}= 0$.\medskip

Second, we define the modules $\repW_{k,x}(N)$ for $k \in \frac12 \Zbb_{\geq 0}$, $x\in \Cbb^\times$, $N \in \Zbb_{\geq 0}$ and $N-2k$ even. For $2k \ge N$, we define $\repW_{k,x}(N) = 0$. For $0 \le 2k \le N$, the module $\repW_{k,x}(N)$ is defined on the vector space $\cLk{k,0}(N,2k)$. Equivalently, using the same map as in \eqref{eq:Wk.vs.L(N,2k)}, we see that the module $\repW_{k,x}(N)$ for $k>0$ is spanned by the subset of link states of $\repW_{k}(N)$ that have no defects crossing the line $\pa\pc$. For $k=0$, the link states have no non-contractible loops. The resulting module $\repW_{k,x}(N)$ is finite dimensional, namely
\begin{equation} \label{eq:dim.W}
\mathrm{dim} \, \repW_{k,x}(N) = \binom{N}{\frac N2-k} \,.
\end{equation}
As examples, the bases of $\repW_{0,x}(4)$ and $\repW_{1,x}(4)$ are made of the following link states:
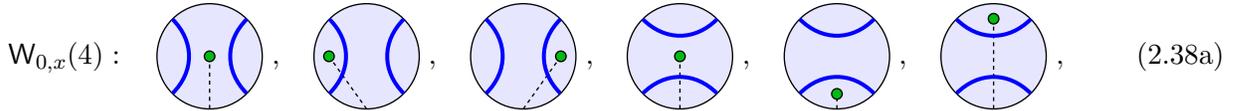
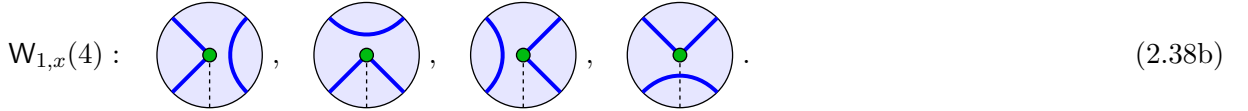
\begin{subequations}
\begin{alignat}{2}
&\repW_{0,x}(4): \quad
\thispic{\begin{pspicture}[shift=-0.6](-0.7,-0.7)(0.7,0.7)
\psarc[linecolor=black,linewidth=0.5pt,fillstyle=solid,fillcolor=lightlightblue]{-}(0,0){0.7}{0}{360}
\psline[linestyle=dashed, dash= 1.5pt 1.5pt, linewidth=0.5pt]{-}(0,0)(0,-0.7)
\psarc[linecolor=black,linewidth=0.5pt,fillstyle=solid,fillcolor=darkgreen]{-}(0,0){0.07}{0}{360}
\psbezier[linecolor=blue,linewidth=1.5pt]{-}(0.494975, 0.494975)(0.20,0.20)(0.20,-0.20)(0.494975, -0.494975)
\psbezier[linecolor=blue,linewidth=1.5pt]{-}(-0.494975, 0.494975)(-0.20,0.20)(-0.20,-0.20)(-0.494975, -0.494975)
\end{pspicture}}\ ,
\quad
\thispic{\begin{pspicture}[shift=-0.6](-0.7,-0.7)(0.7,0.7)
\psarc[linecolor=black,linewidth=0.5pt,fillstyle=solid,fillcolor=lightlightblue]{-}(0,0){0.7}{0}{360}
\psline[linestyle=dashed, dash=1.5pt 1.5pt, linewidth=0.5pt]{-}(-0.5,0)(0,-0.7)
\psarc[linecolor=black,linewidth=0.5pt,fillstyle=solid,fillcolor=darkgreen]{-}(-0.5,0){0.07}{0}{360}
\psbezier[linecolor=blue,linewidth=1.5pt]{-}(0.494975, 0.494975)(0.20,0.20)(0.20,-0.20)(0.494975, -0.494975)
\psbezier[linecolor=blue,linewidth=1.5pt]{-}(-0.494975, 0.494975)(-0.20,0.20)(-0.20,-0.20)(-0.494975, -0.494975)
\end{pspicture}}\ ,
\quad
\thispic{\begin{pspicture}[shift=-0.6](-0.7,-0.7)(0.7,0.7)
\psarc[linecolor=black,linewidth=0.5pt,fillstyle=solid,fillcolor=lightlightblue]{-}(0,0){0.7}{0}{360}
\psline[linestyle=dashed, dash=1.5pt 1.5pt, linewidth=0.5pt]{-}(0.5,0)(0,-0.7)
\psarc[linecolor=black,linewidth=0.5pt,fillstyle=solid,fillcolor=darkgreen]{-}(0.5,0){0.07}{0}{360}
\psbezier[linecolor=blue,linewidth=1.5pt]{-}(0.494975, 0.494975)(0.20,0.20)(0.20,-0.20)(0.494975, -0.494975)
\psbezier[linecolor=blue,linewidth=1.5pt]{-}(-0.494975, 0.494975)(-0.20,0.20)(-0.20,-0.20)(-0.494975, -0.494975)
\end{pspicture}}\ ,
\quad
\thispic{\begin{pspicture}[shift=-0.6](-0.7,-0.7)(0.7,0.7)
\psarc[linecolor=black,linewidth=0.5pt,fillstyle=solid,fillcolor=lightlightblue]{-}(0,0){0.7}{0}{360}
\psline[linestyle=dashed, dash= 1.5pt 1.5pt, linewidth=0.5pt]{-}(0,0)(0,-0.7)
\psarc[linecolor=black,linewidth=0.5pt,fillstyle=solid,fillcolor=darkgreen]{-}(0,0){0.07}{0}{360}
\psbezier[linecolor=blue,linewidth=1.5pt]{-}(0.494975, 0.494975)(0.20,0.20)(-0.20,0.20)(-0.494975, 0.494975)
\psbezier[linecolor=blue,linewidth=1.5pt]{-}(0.494975, -0.494975)(0.20,-0.20)(-0.20,-0.20)(-0.494975, -0.494975)
\end{pspicture}}\ ,
\quad
\thispic{\begin{pspicture}[shift=-0.6](-0.7,-0.7)(0.7,0.7)
\psarc[linecolor=black,linewidth=0.5pt,fillstyle=solid,fillcolor=lightlightblue]{-}(0,0){0.7}{0}{360}
\psline[linestyle=dashed, dash=1.5pt 1.5pt, linewidth=0.5pt]{-}(0,-0.45)(0,-0.7)
\psarc[linecolor=black,linewidth=0.5pt,fillstyle=solid,fillcolor=darkgreen]{-}(0,-0.5){0.07}{0}{360}
\psbezier[linecolor=blue,linewidth=1.5pt]{-}(0.494975, 0.494975)(0.20,0.20)(-0.20,0.20)(-0.494975, 0.494975)
\psbezier[linecolor=blue,linewidth=1.5pt]{-}(0.494975, -0.494975)(0.20,-0.20)(-0.20,-0.20)(-0.494975, -0.494975)
\end{pspicture}}\ ,
\quad
\thispic{\begin{pspicture}[shift=-0.6](-0.7,-0.7)(0.7,0.7)
\psarc[linecolor=black,linewidth=0.5pt,fillstyle=solid,fillcolor=lightlightblue]{-}(0,0){0.7}{0}{360}
\psline[linestyle=dashed, dash=1.5pt 1.5pt, linewidth=0.5pt]{-}(0,0.5)(0,-0.7)
\psarc[linecolor=black,linewidth=0.5pt,fillstyle=solid,fillcolor=darkgreen]{-}(0,0.5){0.07}{0}{360}
\psbezier[linecolor=blue,linewidth=1.5pt]{-}(0.494975, 0.494975)(0.20,0.20)(-0.20,0.20)(-0.494975, 0.494975)
\psbezier[linecolor=blue,linewidth=1.5pt]{-}(0.494975, -0.494975)(0.20,-0.20)(-0.20,-0.20)(-0.494975, -0.494975)
\end{pspicture}}\ ,
\\[0.2cm]
\label{eq:W14.basis}
&\repW_{1,x}(4): \quad
\thispic{\begin{pspicture}[shift=-0.6](-0.7,-0.7)(0.7,0.7)
\psarc[linecolor=black,linewidth=0.5pt,fillstyle=solid,fillcolor=lightlightblue]{-}(0,0){0.7}{0}{360}
\psline[linestyle=dashed, dash= 1.5pt 1.5pt, linewidth=0.5pt]{-}(0,0)(0,-0.7)
\psbezier[linecolor=blue,linewidth=1.5pt]{-}(0.494975, 0.494975)(0.20,0.20)(0.20,-0.20)(0.494975, -0.494975)
\psline[linecolor=blue,linewidth=1.5pt]{-}(-0.494975, 0.494975)(0,0)
\psline[linecolor=blue,linewidth=1.5pt]{-}(-0.494975, -0.494975)(0,0)
\psarc[linecolor=black,linewidth=0.5pt,fillstyle=solid,fillcolor=darkgreen]{-}(0,0){0.09}{0}{360}
\end{pspicture}}\ ,
\quad
\thispic{\begin{pspicture}[shift=-0.6](-0.7,-0.7)(0.7,0.7)
\psarc[linecolor=black,linewidth=0.5pt,fillstyle=solid,fillcolor=lightlightblue]{-}(0,0){0.7}{0}{360}
\psline[linestyle=dashed, dash= 1.5pt 1.5pt, linewidth=0.5pt]{-}(0,0)(0,-0.7)
\psbezier[linecolor=blue,linewidth=1.5pt]{-}(0.494975, 0.494975)(0.20,0.20)(-0.20,0.20)(-0.494975, 0.494975)
\psline[linecolor=blue,linewidth=1.5pt]{-}(0.494975, -0.494975)(0,0)
\psline[linecolor=blue,linewidth=1.5pt]{-}(-0.494975, -0.494975)(0,0)
\psarc[linecolor=black,linewidth=0.5pt,fillstyle=solid,fillcolor=darkgreen]{-}(0,0){0.09}{0}{360}
\end{pspicture}}\ ,
\quad
\thispic{\begin{pspicture}[shift=-0.6](-0.7,-0.7)(0.7,0.7)
\psarc[linecolor=black,linewidth=0.5pt,fillstyle=solid,fillcolor=lightlightblue]{-}(0,0){0.7}{0}{360}
\psline[linestyle=dashed, dash= 1.5pt 1.5pt, linewidth=0.5pt]{-}(0,0)(0,-0.7)
\psbezier[linecolor=blue,linewidth=1.5pt]{-}(-0.494975, -0.494975)(-0.20,-0.20)(-0.20,0.20)(-0.494975, 0.494975)
\psline[linecolor=blue,linewidth=1.5pt]{-}(0.494975, -0.494975)(0,0)
\psline[linecolor=blue,linewidth=1.5pt]{-}(0.494975, 0.494975)(0,0)
\psarc[linecolor=black,linewidth=0.5pt,fillstyle=solid,fillcolor=darkgreen]{-}(0,0){0.09}{0}{360}
\end{pspicture}}\ ,
\quad
\thispic{\begin{pspicture}[shift=-0.6](-0.7,-0.7)(0.7,0.7)
\psarc[linecolor=black,linewidth=0.5pt,fillstyle=solid,fillcolor=lightlightblue]{-}(0,0){0.7}{0}{360}
\psline[linestyle=dashed, dash= 1.5pt 1.5pt, linewidth=0.5pt]{-}(0,0)(0,-0.7)
\psbezier[linecolor=blue,linewidth=1.5pt]{-}(-0.494975, -0.494975)(-0.20,-0.20)(0.20,-0.20)(0.494975, -0.494975)
\psline[linecolor=blue,linewidth=1.5pt]{-}(-0.494975, 0.494975)(0,0)
\psline[linecolor=blue,linewidth=1.5pt]{-}(0.494975, 0.494975)(0,0)
\psarc[linecolor=black,linewidth=0.5pt,fillstyle=solid,fillcolor=darkgreen]{-}(0,0){0.09}{0}{360}
\end{pspicture}}\ .
\end{alignat}
\end{subequations}

The action $\lambda \cdot u$ of $\lambda\in\cL(N',N)$ on a state $u \in \repW_{k,x}(N)$ follows the action of $\cL(N',N)$ on $\cL(N,2k)$, but with some extra rules. For $k=0$, each non-contractible loop is removed and replaced by a multiplicative factor $\alpha = x+x^{-1}$. For $k>0$, if the action of the diagrams produces a link state with less than $2k$ defects, the result is set to zero. Moreover, if the action of the diagrams produces a state with a defect crossing the line $\pa\pc$, then this defect is unwound and the resulting link state is multiplied by a twist factor. If in the original diagram, a walker traveling on the defect from $\pa$ to $\pc$ crosses the dashed line $\pa\pc$ with $\pa$ to its right, the twist factor is $x$. It is $x^{-1}$ if $\pa$ is instead to the left of the walker. If more than one defect cross the line $\pa\pc$, then the resulting twist factor is the product of the individual twist factors. The diagrammatic rules for the action of $\eptl_N(\beta)$ in these modules are summarised as
\be
\label{eq:Wkx.rules}
\psset{unit=0.6cm}
\thispic{\begin{pspicture}[shift=-0.6](-0.7,-0.7)(0.7,0.7)
\psarc[linecolor=black,linewidth=0.5pt,fillstyle=solid,fillcolor=lightlightblue]{-}(0,0){0.7}{0}{360}
\psline[linestyle=dashed, dash= 1.5pt 1.5pt, linewidth=0.5pt]{-}(0,0)(0,-0.7)
\psarc[linecolor=black,linewidth=0.5pt,fillstyle=solid,fillcolor=darkgreen]{-}(0,0){0.07}{0}{360}
\psarc[linecolor=blue,linewidth=1.5pt]{-}(0,0){0.27}{0}{360}
\end{pspicture}}
\ \, = \alpha \ \,
\thispic{\begin{pspicture}[shift=-0.6](-0.7,-0.7)(0.7,0.7)
\psarc[linecolor=black,linewidth=0.5pt,fillstyle=solid,fillcolor=lightlightblue]{-}(0,0){0.7}{0}{360}
\psline[linestyle=dashed, dash= 1.5pt 1.5pt, linewidth=0.5pt]{-}(0,0)(0,-0.7)
\psarc[linecolor=black,linewidth=0.5pt,fillstyle=solid,fillcolor=darkgreen]{-}(0,0){0.07}{0}{360}
\end{pspicture}}
\ ,
\qquad
\psset{unit=0.6cm}
\thispic{\begin{pspicture}[shift=-0.6](-0.7,-0.7)(0.7,0.7)
\psarc[linecolor=black,linewidth=0.5pt,fillstyle=solid,fillcolor=lightlightblue]{-}(0,0){0.7}{0}{360}
\psline[linestyle=dashed, dash= 1.5pt 1.5pt, linewidth=0.5pt]{-}(0,0)(0,-0.7)
\psbezier[linecolor=blue,linewidth=1.5pt]{-}(0,0)(0.4,-0.4)(-0.4,-0.4)(-0.495,-0.495)
\psarc[linecolor=black,linewidth=0.5pt,fillstyle=solid,fillcolor=darkgreen]{-}(0,0){0.09}{0}{360}
\end{pspicture}}
\ \, = x \ \,
\thispic{\begin{pspicture}[shift=-0.6](-0.7,-0.7)(0.7,0.7)
\psarc[linecolor=black,linewidth=0.5pt,fillstyle=solid,fillcolor=lightlightblue]{-}(0,0){0.7}{0}{360}
\psline[linestyle=dashed, dash= 1.5pt 1.5pt, linewidth=0.5pt]{-}(0,0)(0,-0.7)
\psline[linecolor=blue,linewidth=1.5pt]{-}(0,0)(-0.495,-0.495)
\psarc[linecolor=black,linewidth=0.5pt,fillstyle=solid,fillcolor=darkgreen]{-}(0,0){0.09}{0}{360}
\end{pspicture}}\ ,
\qquad
\thispic{\begin{pspicture}[shift=-0.6](-0.7,-0.7)(0.7,0.7)
\psarc[linecolor=black,linewidth=0.5pt,fillstyle=solid,fillcolor=lightlightblue]{-}(0,0){0.7}{0}{360}
\psline[linestyle=dashed, dash= 1.5pt 1.5pt, linewidth=0.5pt]{-}(0,0)(0,-0.7)
\psbezier[linecolor=blue,linewidth=1.5pt]{-}(0,0)(-0.4,-0.4)(0.4,-0.4)(0.495,-0.495)
\psarc[linecolor=black,linewidth=0.5pt,fillstyle=solid,fillcolor=darkgreen]{-}(0,0){0.09}{0}{360}
\end{pspicture}}
\ \, = x^{-1} \ \,
\thispic{\begin{pspicture}[shift=-0.6](-0.7,-0.7)(0.7,0.7)
\psarc[linecolor=black,linewidth=0.5pt,fillstyle=solid,fillcolor=lightlightblue]{-}(0,0){0.7}{0}{360}
\psline[linestyle=dashed, dash= 1.5pt 1.5pt, linewidth=0.5pt]{-}(0,0)(0,-0.7)
\psline[linecolor=blue,linewidth=1.5pt]{-}(0,0)(0.495,-0.495)
\psarc[linecolor=black,linewidth=0.5pt,fillstyle=solid,fillcolor=darkgreen]{-}(0,0){0.09}{0}{360}
\end{pspicture}}\ ,
\qquad
\thispic{\begin{pspicture}[shift=-0.6](-0.7,-0.7)(0.7,0.7)
\psarc[linecolor=black,linewidth=0.5pt,fillstyle=solid,fillcolor=lightlightblue]{-}(0,0){0.7}{0}{360}
\psline[linestyle=dashed, dash= 1.5pt 1.5pt, linewidth=0.5pt]{-}(0,0)(0,-0.7)
\psline[linecolor=blue,linewidth=1.5pt]{-}(0,0)(0.2,0.2)
\psline[linecolor=blue,linewidth=1.5pt]{-}(0,0)(-0.2,0.2)
\psarc[linecolor=black,linewidth=0.5pt,fillstyle=solid,fillcolor=darkgreen]{-}(0,0){0.09}{0}{360}
\psbezier[linecolor=blue,linewidth=1.5pt]{-}(0.2,0.2)(0.5,0.5)(-0.5,0.5)(-0.2,0.2)
\end{pspicture}}
\ \, = 0\,,
\ee
in addition to the rule
\be
\label{eq:beta.rule}
\psset{unit=0.6cm}
\thispic{\begin{pspicture}[shift=-0.6](-0.7,-0.7)(0.7,0.7)
\psarc[linecolor=black,linewidth=0.5pt,fillstyle=solid,fillcolor=lightlightblue]{-}(0,0){0.7}{0}{360}
\psline[linestyle=dashed, dash= 1.5pt 1.5pt, linewidth=0.5pt]{-}(0,0)(0,-0.7)
\psarc[linecolor=black,linewidth=0.5pt,fillstyle=solid,fillcolor=darkgreen]{-}(0,0){0.07}{0}{360}
\psarc[linecolor=blue,linewidth=1.5pt]{-}(0.3, 0.3){0.17}{0}{360}
\end{pspicture}}
\ \, = \beta \ \,
\thispic{\begin{pspicture}[shift=-0.6](-0.7,-0.7)(0.7,0.7)
\psarc[linecolor=black,linewidth=0.5pt,fillstyle=solid,fillcolor=lightlightblue]{-}(0,0){0.7}{0}{360}
\psline[linestyle=dashed, dash= 1.5pt 1.5pt, linewidth=0.5pt]{-}(0,0)(0,-0.7)
\psarc[linecolor=black,linewidth=0.5pt,fillstyle=solid,fillcolor=darkgreen]{-}(0,0){0.07}{0}{360}
\end{pspicture}}
\ee
that applies generally for the product in the diagram spaces $\cL(N',N)$. We note that the module $\repW_{k,x}(N)$ can be obtained from the quotient 
\begin{equation}
\repW_{k,x}(N) = \frac{\repW_{k}(N)}
{\{\Omega^N = x^{2k} \id, F = (q^{k} x + q^{-k} x^{-1}) \id\}} \,.
\end{equation}

Third, we define the vacuum modules $\repV(N)$ for $N \in 2 \Zbb_{\ge 0}$. The vector space for $\repV(N)$ is spanned by link states drawn on a disc with no defects and no marked point. Its dimension is given by the Catalan number:
\be
\dim \repV(N) = C_{N/2} = \frac1{N/2+1} \binom{N}{N/2}\,. 
\ee
Here is an example of the vector space for $N=6$:
\be
\repV(6): \ \
\thispic{\begin{pspicture}[shift=-0.7](-0.8,-0.8)(0.8,0.8)
\psarc[linecolor=black,linewidth=0.5pt,fillstyle=solid,fillcolor=lightlightblue]{-}(0,0){0.7}{0}{360}
\psbezier[linecolor=blue,linewidth=1.5pt](0.35, -0.606218)(0.175, -0.303109)(0.35, 0)(0.7, 0.)
\psbezier[linecolor=blue,linewidth=1.5pt](0.35, 0.606218)(0.175, 0.303109)(-0.175, 0.303109)(-0.35, 0.606218)
\psbezier[linecolor=blue,linewidth=1.5pt](-0.7, 0.)(-0.35, 0.)(-0.175, -0.303109)(-0.35, -0.606218)
\end{pspicture}}\,,
\quad
\thispic{\begin{pspicture}[shift=-0.7](-0.8,-0.8)(0.8,0.8)
\psarc[linecolor=black,linewidth=0.5pt,fillstyle=solid,fillcolor=lightlightblue]{-}(0,0){0.7}{0}{360}
\psbezier[linecolor=blue,linewidth=1.5pt](0.35, -0.606218)(0.175, -0.303109)(-0.175, -0.303109)(-0.35, -0.606218)
\psbezier[linecolor=blue,linewidth=1.5pt](0.35, 0.606218)(0.175, 0.303109)(0.35, 0)(0.7, 0.)
\psbezier[linecolor=blue,linewidth=1.5pt](-0.7, 0.)(-0.35, 0.)(-0.175, 0.303109)(-0.35, 0.606218)
\end{pspicture}}\,,
\quad
\thispic{\begin{pspicture}[shift=-0.7](-0.8,-0.8)(0.8,0.8)
\psarc[linecolor=black,linewidth=0.5pt,fillstyle=solid,fillcolor=lightlightblue]{-}(0,0){0.7}{0}{360}
\psbezier[linecolor=blue,linewidth=1.5pt](0.35, -0.606218)(0.175, -0.303109)(0.35, 0)(0.7, 0.)
\psbezier[linecolor=blue,linewidth=1.5pt](0.35, 0.606218)(0.175, 0.303109)(-0.175, -0.303109)(-0.35, -0.606218)
\psbezier[linecolor=blue,linewidth=1.5pt](-0.7, 0.)(-0.35, 0.)(-0.175, 0.303109)(-0.35, 0.606218)
\end{pspicture}}\,,
\quad
\thispic{\begin{pspicture}[shift=-0.7](-0.8,-0.8)(0.8,0.8)
\psarc[linecolor=black,linewidth=0.5pt,fillstyle=solid,fillcolor=lightlightblue]{-}(0,0){0.7}{0}{360}
\psbezier[linecolor=blue,linewidth=1.5pt](0.35, -0.606218)(0.175, -0.303109)(-0.175, 0.303109)(-0.35, 0.606218)
\psbezier[linecolor=blue,linewidth=1.5pt](0.35, 0.606218)(0.175, 0.303109)(0.35, 0)(0.7, 0.)
\psbezier[linecolor=blue,linewidth=1.5pt](-0.7, 0.)(-0.35, 0.)(-0.175, -0.303109)(-0.35, -0.606218)
\end{pspicture}}\,,
\quad
\thispic{\begin{pspicture}[shift=-0.7](-0.8,-0.8)(0.8,0.8)
\psarc[linecolor=black,linewidth=0.5pt,fillstyle=solid,fillcolor=lightlightblue]{-}(0,0){0.7}{0}{360}
\psbezier[linecolor=blue,linewidth=1.5pt](0.35, -0.606218)(0.175, -0.303109)(-0.175, -0.303109)(-0.35, -0.606218)
\psbezier[linecolor=blue,linewidth=1.5pt](0.35, 0.606218)(0.175, 0.303109)(-0.175, 0.303109)(-0.35, 0.606218)
\psbezier[linecolor=blue,linewidth=1.5pt](-0.7, 0.)(-0.35, 0.)(0.35, 0)(0.7, 0.)
\end{pspicture}}\ .
\ee
The action $\lambda \cdot u$ of a diagram $\lambda \in \cL(N',N)$ on a link state $u \in \repV(N)$ is computed as usual, by drawing $u$ inside $\lambda$. In this case, all closed loops are assigned a weight $\beta$. As discussed in \cite{IMD24}, the module $\repV(N)$ is the quotient of $\repW_{0,-q}(N)$ by its submodule isomorphic to $\repW_{1,-1}(N)$.\medskip

In all three cases, the modules are endowed with an action of the diagram spaces $\cL(N',N)$. This readily yields the following proposition.

\begin{Proposition} \label{prop:Wk.families}
Let $N_0 \in \{0,1\}$, $k \in \frac12\Zbb_{\geq 0}$, $N_0-2k \in 2\Zbb$ and $x \in \Cbb^\times$. Each of the sets
\begin{subequations}
\begin{alignat}{2}
\repW_k &= \{\repW_{k}(N)\,|\, N = N_0, N_0+2, N_0+4, \dots \}\,, \\[0.1cm]
\repW_{k,x} &=\{\repW_{k,x}(N)\,|\, N = N_0, N_0+2, N_0+4, \dots \}\,, \\[0.1cm]
\repV &=\{\repV(N)\,|\, N = 0, 2, 4, \dots \}\,,
\end{alignat}
\end{subequations}
is a family of modules. 
\end{Proposition}
We note that the fact that $\repW_{k,x}$ is a family of modules was proven in \cite{GL98}.

\paragraph{XXZ families.}

The vector space $\repX_{\phi}(N)$ for the periodic spin-$\frac12$ XXZ chain of length $N$ with anisotropy $\Delta = \frac12(q+q^{-1})$ and a diagonal twist $\phi \in \Cbb$ is $(\Cbb^2)^{\otimes N}$. We use the canonical basis 
$\ket{+} = \left(\begin{smallmatrix}1\\0\end{smallmatrix}\right)$, 
$\ket{-} = \left(\begin{smallmatrix}0\\1\end{smallmatrix}\right)$
so that spin configurations for the chain of length $N$ take the form $\ket{s_1, s_2, \dots, s_N}$ with each $s_i \in \{+,-\}$. This vector space is endowed with a representation of $\eptl_N(\beta=-q-q^{-1})$ defined by
\begin{equation}
\Omega = t \exp(\ir \phi \sigma^z_1)\,,
\qquad e_j = \underbrace{\mathbb I_2 \otimes \dots \otimes \mathbb I_2}_{j-1}\otimes 
\,e \otimes \underbrace{\mathbb I_2 \otimes \dots \otimes \mathbb I_2}_{N-j-1}\,, \qquad 1 \le j \le N-1\,,
\end{equation}
where 
$\mathbb I_2$ is the $2\times2$ identity matrix, $\sigma^a_j$ is the Pauli matrix $\sigma^a$ with $a \in \{x,y,z\}$ acting on the $j$-th tensorand of $(\Cbb^2)^{\otimes N}$, $t$ is the shift operator
\be
t\, \ket{s_1,s_2, \dots, s_N} = \ket{s_2, s_3, \dots, s_N, s_1} \,,
\ee
and
\be
e = \begin{pmatrix}
0 & 0 & 0 & 0\\
0 & -q & 1 & 0\\
0 & 1 & -q^{-1} & 0\\
0 & 0 & 0 & 0\\
\end{pmatrix}\,.
\ee
The action of $e_0$ is defined from the relation $e_0 = \Omega\, e_1\, \Omega^{-1}$. The XXZ Hamiltonian is given by
\begin{equation}
H = - \sum_{j=0}^{N-1} e_j\,
= \frac{N \Delta} 2- \frac12 \sum_{j=0}^N \Big(\sigma^x_j \sigma^x_{j+1} +\sigma^y_j \sigma^y_{j+1} + \Delta\, \sigma^z_j \sigma^z_{j+1}\Big)
\end{equation}
with twisted boundary conditions
\begin{equation}
\sigma^x_{N+1} = \cos (2\phi) \, \sigma^x_1 + \sin (2\phi) \,\sigma^y_1\,,\qquad
\sigma^y_{N+1} = - \sin (2\phi) \,\sigma^x_1 + \cos (2\phi) \, \sigma^y_1 \,,\qquad
\sigma^z_{N+1} = \sigma^z_1\,.
\end{equation}
The magnetisation $S^z = \frac12 \sum_{j=1}^N \sigma^z_j$ is preserved by the action of $\eptl_N(\beta)$. As a result, $\repX_{\phi}(N)$ splits as a direct sum of the modules $\repX_{m,\phi}(N)$ where $S^z = m$:
\begin{equation}
\repX_{\phi}(N) = \bigoplus_{m=-N/2}^{N/2} \repX_{m,\phi}(N) \,.
\end{equation}

The generators can be factored in terms of operators $c_j: \repX_{\phi}(N) \to \repX_{\phi}(N-2)$ and $c_j^\dag: \repX_{\phi}(N-2) \to \repX_{\phi}(N)$, defined as
\begin{equation}
c_j = \underbrace{\mathbb I_2 \otimes \dots \otimes \mathbb I_2}_{j-1}\otimes \,c
\otimes \underbrace{\mathbb I_2 \otimes \dots \otimes \mathbb I_2}_{N-j-1}\,,
\qquad
c^\dag_j = \underbrace{\mathbb I_2 \otimes \dots \otimes \mathbb I_2}_{j-1}\otimes \,c^\dag
\otimes \underbrace{\mathbb I_2 \otimes \dots \otimes \mathbb I_2}_{N-j-1}\,,
\end{equation}
for $1 \le j \le N-1$, where $c$ is the row-vector $(0,\ir q^{1/2},-\ir q^{-1/2},0)$ and $c^\dag$ is its transpose. The generators $c_0$ and $c_0^\dag$ are defined as $c_0 = c_1\, \Omega^{-1}$ and $c^\dag_0 = \Omega\, c^\dag_1$. One can check that these operators satisfy all the relations \eqref{eq:c.relations}. Moreover, it is easy to see that $c_j$ and $c^\dag_j$ commute with $S^z$. This yields the following proposition.
\begin{Proposition}
Let $N_0 \in \{0,1\}$, $m \in \frac12 \Zbb$, $N_0-2m \in 2 \Zbb$ and $\phi \in \Cbb$. Each of the sets
\begin{subequations}
\begin{alignat}{2}
\repX_{\phi} &= \{\repX_{\phi}(N)\,|\,N=N_0,N_0+2,N_0+4, \dots \}\,, \\[0.1cm]
\repX_{m,\phi} &= \{\repX_{m,\phi}(N)\,|\,N=N_0,N_0+2,N_0+4, \dots \}\,,
\end{alignat}
\end{subequations}
is a family of modules.
\end{Proposition}

\paragraph{Unitary RSOS families.}

We use the construction of unitary ADE lattice models in \cite{Pasquier87} and
consider the Dynkin diagram $\mathcal G$ with $n$ nodes associated to a Lie algebra $\g=A_n, D_n, E_6, E_7$, or $E_8$. The adjacency matrix of the graph $\mathcal G$ is denoted $A$. The eigenvalues of $A$ are of the form
\begin{equation}
\beta_\mu = 2\cos \frac{\pi m_\mu}{\ell} \,,
\end{equation}
where $\ell$ is the Coxeter number associated to the algebra $\g$, and $m_\mu$ is an integer satisfying $1 \leq m_\mu \leq \ell-1$ called an \textit{exponent}. The values of $\ell$ and the possible exponents $m_1,m_2,\dots,m_n$ are given in \cref{tab:RSOS}.\medskip
\begin{table}
\begin{center}
\begin{tabular}{c|c|c}
Algebra $\g$ & Coxeter number $\ell$ & Exponents $m_1,m_2,\dots,m_n$\\ 
\hline
$A_n$ & $n+1$ & $1,2,\dots,n$ \\
$D_n$ & $2n-2$ & $1,3,\dots,2n-3,n-1$ \\
$E_6$ & $12$ & $1,4,5,7,8,11$ \\
$E_7$ & $18$ & $1,5,7,9,11,13,17$ \\
$E_8$ & $30$ & $1,7,11,13,17,19,23,29$
\end{tabular}
\end{center}
\caption{The Lie algebra, Coxeter number and exponents for the ADE series}
\label{tab:RSOS}
\end{table}

An automorphism of $\cal G$ is a permuation $K:a \mapsto K(a)$ of the nodes, such that any pair of adjacent vertices is mapped to a pair of adjacent vertices. The matrix $K_{ab}=\delta_{K(a),b}$ then commutes with $A$, implying that the two matrices can be diagonalised in the same basis $\{S_{a\mu}\}$. Because $A$ and $K$ are real symmetric matrices, this basis can be chosen such that each $S_{a\mu}$ is real.\medskip

The unitary RSOS model $(\g,K)$ is defined by a choice of an algebra $\g$, and an automorphism $K$ of $\cal G$. The Perron--Frobenius theorem ensures that the leading eigenvalue $\beta_{1}$ has multiplicity one, and that the corresponding eigenvector has real positive entries: $S_{a1}>0$. We denote the components $S_{a1}$ of this eigenvector as $S_a$.\footnote{We investigate the non-unitary RSOS models in \cite{IMD26}.} The $\eptl_N(\beta)$-module $\repM_{\g,K}(N)$ with twisted periodic boundary conditions is constructed as follows. The allowed configurations are of the form $\ket{a_0,a_1,a_2,\dots,a_N}$, where (i) $a_j \in \mathcal G$, (ii) $a_j$ and $a_{j+1}$ are adjacent on $\mathcal G$, and
(iii) $a_N=K(a_0)$.
The dimension of $\repM_{\g,K}(N)$ is
\begin{equation}
\dim \, \repM_{\g,K}(N)
= \textrm{tr}(K A^N) = \sum_{\nu=1}^n \kappa_\nu \, (\beta_{\nu})^N \,,
\end{equation}
where $\kappa_\mu$ is the eigenvalue of $K$ associated to the eigenvector $S_{a\mu}$.
The Dynkin diagrams of type $A,D,E$ are all bipartite, namely their nodes can be coloured in black and white, so that any edge connects two nodes of opposite colours. Since $\cal G$ is connected, the automorphism $K$ either conserves or inverts the colour of all the nodes. In the former case (for instance for $K=\id$), the module $\repM_{\g,K}(N)$ is nonzero only for $N$ even. In the latter case, $\repM_{\g,K}(N)$ is nonzero only for $N$ odd.
\medskip

For $j=1,2,\dots,N-1$, we define $c_j:\repM_{\g,K}(N) \to \repM_{\g,K}(N-2)$ and $c^\dag_j:\repM_{\g,K}(N-2) \to \repM_{\g,K}(N)$ as
\begin{subequations} \label{eq:def.RSOS}
\begin{alignat}{2}
& c_j \cdot \ket{a_0, a_1, \dots, a_N}
= \delta_{a_{j-1},a_{j+1}} \, \sqrt{\frac{S_{a_j}}{S_{a_{j+1}}}}\, \ket{a_0, a_1, \dots, a_{j-1},a_{j+2}, a_{j+3},\dots, a_N} \,, 
\\
& c^\dag_j \cdot \ket{a_0, a_1, \dots, a_{N-2}} = \sum_{a'_j=1}^n A_{a_{j-1},a'_j} \sqrt{\frac{S_{a'_j}}{S_{a_{j-1}}}}
\, \ket{a_0, a_1, \dots, a_{j-1}, a'_{j} ,a_{j-1}, a_j, \dots, a_{N-2}} \,.
\end{alignat}
The operators $c_0$ and $c_0^\dag$ are similarly defined as
\begin{alignat}{2}
& c_0 \cdot \ket{a_0, a_1, \dots, a_N}
= \delta_{a_{N-1},K(a_1)} \sqrt{\frac{S_{a_0}}{S_{a_1}}} \,\ket{a_1,a_2,\dots,a_{N-1}} \,, 
\\
& c_0^\dag \cdot \ket{a_0, a_1, \dots, a_{N-2}}
= \sum_{a'_0=1}^n A_{a_0,a'_0}
\sqrt{\frac{S_{a'_0}}{S_{a_0}}} \,\ket{a'_0,a_0,a_1,\dots,a_{N-2},K(a'_0)} \,.
\end{alignat}
\end{subequations}
It is tedious though straightforward to check that the operators $c_j$ and $c^\dag_j$ defined by this action satisfy the relations \eqref{eq:c.relations}, with the weight of contractible loops set to $\beta = \beta_1 = 2 \cos(\pi/\ell)$. For instance, we have
\begin{alignat}{2}
c_{j+1} c_{j}^\dag \cdot \ket{a_0, a_1, \dots, a_{N-2}}
&= \sum_{a'_{j}=1}^n A_{a_{j-1},a'_{j}} \sqrt{\frac{S_{a'_{j}}}{S_{a_{j-1}}}} \delta_{a'_{j},a_{j}}
\sqrt{\frac{S_{a_{j-1}}}{S_{a_{j}}}} \ket{a_0,a_1, \dots, a_{j-1}, a'_{j}, a_{j+1}, \dots, a_{N-2}} \nn \\
&= \ket{a_0, a_1, \dots, a_{N-2}} \,,
\end{alignat}
and
\begin{alignat}{2}
c_j c_j^\dag \cdot \ket{a_0, a_1, \dots, a_{N-2}}
= \sum_{a'_j=1}^n A_{a_{j-1},a'_j} \frac{S_{a'_j}}{S_{a_{j-1}}} \, \ket{a_0, a_1, \dots, a_{N-2}}
= \beta_1 \, \ket{a_0, a_1,\dots, a_{N-2}} \,.
\end{alignat}
Verifying that all the relations \eqref{eq:c.relations} hold leads to the following proposition.
\begin{Proposition}
Let $\g$ be a Lie algebra of type $A_n$, $D_n$ or $E_n$, and $K$ be an automorphism of the Dynkin diagram of $\g$. The set
\begin{equation}
\repM_{\g,K} = \{\repM_{\g,K}(N)\,|\,N=N_0,N_0+2,N_0+4, \dots \}\,,
\qquad N_0 \in \{0,1\}\,,
\end{equation}
is a family of modules.
\end{Proposition}
The action of the generators of $\eptl_N(\beta)$ on $\repM_{\g,K}(N)$ is as follows.
For $N=0$, the operator $f=c^{\phantom{\dag}}_0 c_1^\dag=c^{\phantom{\dag}}_1 c_0^\dag$ acts as
\begin{equation}
f \ket{b}
= \sum_{a=1}^n \delta_{K(a),a} \, A_{ab} \, \ket{a} \,.
\end{equation}
For $N\geq 1$, the translation operators $\Omega=c^{\phantom{\dag}}_1 c_0^\dag$ and $\Omega^{-1} =c^{\phantom{\dag}}_0 c_1^\dag$ act as
\begin{subequations}
\begin{alignat}{2} 
\label{eq:OmegaRSOS}
&\Omega \cdot \ket{a_0,a_1, \dots,a_N} = \ket{a_1,a_2,\dots,a_N, K(a_1)} \,, \\[0.1cm]
&\Omega^{-1} \cdot \ket{a_0,\dots,a_{N-1},a_N} =
\ket{K^{-1}(a_{N-1}),a_0,a_1,\dots,a_{N-1}} \,.
\end{alignat}
\end{subequations}
For $N\geq 2$, the action of the generators $e_j=c^\dag_j c_j^{\phantom{\dag}}$ is
\begin{subequations}
\label{eq:ejRSOS}
\begin{alignat}{2}
e_0 \cdot \ket{a_0,a_1,\dots,a_N}
&= \sum_{a'_j=1}^n \delta_{a_{N-1},K(a_1)} A_{a'_0,a_1} \sqrt{\frac{S_{a_0}S_{a'_0}}{(S_{a_1})^2}} \, \ket{a'_0,a_1, \dots, K(a'_0)}\,,
\\ 
e_j \cdot \ket{a_0,a_1,\dots,a_N}
&= \sum_{a'_j=1}^n \delta_{a_{j-1}a_{j+1}} A_{a'_j,a_{j-1}} \sqrt{\frac{S_{a_j}S_{a'_j}}{S_{a_{j-1}}S_{a_{j+1}}}} \, \ket{a_0,a_1,\dots, a_{j-1},a'_j,a_{j+1}, \dots, a_N} \,, 
\end{alignat}
\end{subequations}
for $j=1,2,\dots, N-1$. 

\subsection{Automorphisms and module transformations}\label{sec:automorphs}

We now describe two automorphisms of the diagram spaces that lead to non-trivial transformations of the families of modules. The first is the {\it parity sign flip automorphism} $C$. For each diagram $\lambda$ in $\cL(N',N)$, we define its parity $\sigma(\lambda)$ to be $\sigma(\lambda)=1$ or $\sigma(\lambda)= -1$ if the loop segments of $\lambda$ cross the dashed line $\pc\pc'$ an even or odd number of times, respectively. For a pair of diagrams $\lambda' \in \cL(N'',N')$ and $\lambda \in \cL(N',N)$, the parity satisfies the property
\begin{equation}
\sigma(\lambda' \lambda) =\sigma(\lambda') \, \sigma(\lambda) \,, 
\qquad \sigma(\lambda^\dag) = \sigma(\lambda) \,,
\end{equation}
For the generators, we have
\begin{equation}
\sigma(\id) = 1\,, \qquad 
\sigma(c_{N,0}) = \sigma(c_{N,0}^\dag) = -1\,, \qquad
\sigma(c_{N,j}) = \sigma(c_{N,j}^\dag) = 1 \,, \qquad 1 \le j \le N-1 \,.
\end{equation}
The map defined as $\lambda \mapsto C(\lambda) = \sigma(\lambda)\, \lambda$ for each diagram $\lambda \in \cL(N',N)$, is an automorphism of the diagram spaces, as it satisfies $C(\lambda' \lambda)=C(\lambda')C(\lambda)$ for all $\lambda' \in \cL(N'',N')$ and $\lambda \in \cL(N',N)$.\medskip

Let $\repM$ be a family of modules. For each admissible $N$, we define the module $\repM^-(N)$ as follows. The vector spaces of $\repM(N)$ and $\repM^-(N)$ are the same. To avoid confusion, for each $u \in \repM(N)$, we denote by $u^-$ the corresponding state in $\repM^-(N)$. The action of any diagram $\lambda \in \cL(N',N)$ on $\repM^-(N)$ is defined as
\begin{equation} \label{eq:L.on.M-}
\lambda \cdot u^- = \big(C(\lambda) \cdot u\big)^-= \sigma(\lambda) \, (\lambda \cdot u)^- \,,
\qquad 
\forall u^- \in \repM^-(N) \,.
\end{equation}
The actions on $\repM(N)$ and $\repM^-(N)$ thus differ only by extra minus signs for the odd diagrams. We then have
\begin{equation} \label{eq:consistency.M-}
\lambda' \cdot (\lambda\cdot u^-)
= \sigma(\lambda)\,\lambda' \cdot (\lambda \cdot u)^-
= \sigma(\lambda)\,\sigma(\lambda')\, (\lambda' \lambda \cdot u)^-
= (\lambda' \lambda) \cdot u^- \,,
\end{equation}
for all $\lambda' \in \cL(N'',N')$, $\lambda \in \cL(N',N)$ and $u \in \repM^-(N)$.
The relations \eqref{eq:L.on.M-} and \eqref{eq:consistency.M-} confirm that the conditions \eqref{eq:family.properties} are satisfied, and thus $\repM^- = \{\repM^-(N) \,|\, N = N_0, N_0+2, \dots \}$ is a family of modules. Moreover, it is easy to see that the map $\repM \mapsto \repM^-$ is an involution, namely $(\repM^-)^- = \repM$.
\medskip

The second automorphism is the {\it reflection automorphism} $R$, described for instance in \cite{PinetStAubin23}. For each diagram $\lambda \in \cL(N',N)$, we define the reflected diagram $R(\lambda)$ as the diagram obtained by exchanging the nodes $j$ and $(N+1-j)$ on the inner boundary, and the nodes $j$ and $(N'+1-j)$ on the outer boundary. This linear map satisfies
\begin{equation} \label{eq:R.on.c}
R(\id)=\id \,, \qquad
\left\{\begin{array}{l}
R(c_{N,0})=c_{N,0} \,, \\[0.15cm]
R(c^\dag_{N,0})=c^\dag_{N,0} \,, 
\end{array}\right.
\qquad
\left\{\begin{array}{l}
R(c_{N,j})=c_{N,N-j} \,, \\[0.15cm]
R(c^\dag_{N,j})=c^\dag_{N,N-j} \,, 
\end{array}\right.
\quad
1 \le j \le N-1 \,.
\end{equation}
It is again easy to show that $R(\lambda'\lambda) = R(\lambda')R(\lambda)$, thus confirming that this is indeed an automorphism. If diagrams are drawn in a rectangular strip with periodic boundary conditions, whereby one cuts along the dashed line $\pc\pc'$ and deforms the resulting shape into a rectangle, this transformation is simply a left-right reflection. For example, for the diagram $\lambda_1$ in \eqref{eq:first diagrams}, we have
\be
\lambda_1 = \
\thispic{\begin{pspicture}[shift=-0.45](0,-0.55)(4.8,0.55)
\rput(0.2,-0.55){$_1$}
\rput(0.6,-0.55){$_2$}\rput(0.6,0.55){$_1$}
\rput(1.0,-0.55){$_3$}
\rput(1.4,-0.55){$_4$}
\rput(1.8,-0.55){$_5$}\rput(1.8,0.55){$_2$}
\rput(2.2,-0.55){$_6$}
\rput(2.6,-0.55){$_7$}
\rput(3.0,-0.55){$_8$}\rput(3.0,0.55){$_3$}
\rput(3.4,-0.55){$_9$}
\rput(3.8,-0.55){$_{10}$}
\rput(4.2,-0.55){$_{11}$}\rput(4.2,0.55){$_4$}
\rput(4.6,-0.55){$_{12}$}
\pspolygon[fillstyle=solid,fillcolor=lightlightblue,linecolor=lightlightblue,linewidth=0pt](0,-0.35)(4.8,-0.35)(4.8,0.35)(0,0.35)
\psline[linewidth=1pt, linestyle=dashed, dash= 1.5pt 1.5pt](0,-0.35)(0,0.35)
\psline[linewidth=1pt, linestyle=dashed, dash= 1.5pt 1.5pt](4.8,-0.35)(4.8,0.35)
\psarc[linecolor=blue,linewidth=1.5pt]{-}(0.4,-0.35){0.2}{0}{180}
\psarc[linecolor=blue,linewidth=1.5pt]{-}(1.6,-0.35){0.2}{0}{180}
\psarc[linecolor=blue,linewidth=1.5pt]{-}(2.8,-0.35){0.2}{0}{180}
\psarc[linecolor=blue,linewidth=1.5pt]{-}(3.6,-0.35){0.2}{0}{180}
\psline[linecolor=blue,linewidth=1.5pt]{-}(4.2,-0.35)(4.2,0.35)
\psbezier[linecolor=blue,linewidth=1.5pt]{-}(1.8,0.35)(1.8,0)(3.0,0)(3.0,0.35)
\psbezier[linecolor=blue,linewidth=1.5pt]{-}(2.2,-0.35)(2.2,0.1)(0.6,-0.1)(0.6,0.35)
\psbezier[linecolor=blue,linewidth=1.5pt]{-}(-0.02,-0.08)(0.1,0.09)(1.0,0.09)(1.0,-0.35)
\psbezier[linecolor=blue,linewidth=1.5pt]{-}(4.6,-0.35)(4.6,-0.17)(4.75,-0.07)(4.82,-0.05)
\psframe[fillstyle=solid,linecolor=white,linewidth=0pt](-0.1,-0.4)(-0.005,0.4)
\psframe[fillstyle=solid,linecolor=white,linewidth=0pt](4.805,-0.4)(4.9,0.4)
\end{pspicture}} 
\quad \mapsto \quad 
R(\lambda_1) = \
\thispic{\begin{pspicture}[shift=-0.45](0,-0.55)(4.8,0.55)
\rput(0.2,-0.55){$_1$}
\rput(0.6,-0.55){$_2$}\rput(0.6,0.55){$_1$}
\rput(1.0,-0.55){$_3$}
\rput(1.4,-0.55){$_4$}
\rput(1.8,-0.55){$_5$}\rput(1.8,0.55){$_2$}
\rput(2.2,-0.55){$_6$}
\rput(2.6,-0.55){$_7$}
\rput(3.0,-0.55){$_8$}\rput(3.0,0.55){$_3$}
\rput(3.4,-0.55){$_9$}
\rput(3.8,-0.55){$_{10}$}
\rput(4.2,-0.55){$_{11}$}\rput(4.2,0.55){$_4$}
\rput(4.6,-0.55){$_{12}$}
\pspolygon[fillstyle=solid,fillcolor=lightlightblue,linecolor=lightlightblue,linewidth=0pt](0,-0.35)(4.8,-0.35)(4.8,0.35)(0,0.35)
\psline[linewidth=1pt, linestyle=dashed, dash= 1.5pt 1.5pt](0,-0.35)(0,0.35)
\psline[linewidth=1pt, linestyle=dashed, dash= 1.5pt 1.5pt](4.8,-0.35)(4.8,0.35)
\psarc[linecolor=blue,linewidth=1.5pt]{-}(4.4,-0.35){0.2}{0}{180}
\psarc[linecolor=blue,linewidth=1.5pt]{-}(3.2,-0.35){0.2}{0}{180}
\psarc[linecolor=blue,linewidth=1.5pt]{-}(2.0,-0.35){0.2}{0}{180}
\psarc[linecolor=blue,linewidth=1.5pt]{-}(1.2,-0.35){0.2}{0}{180}
\psline[linecolor=blue,linewidth=1.5pt]{-}(0.6,-0.35)(0.6,0.35)
\psbezier[linecolor=blue,linewidth=1.5pt]{-}(3.0,0.35)(3.0,0)(1.8,0)(1.8,0.35)
\psbezier[linecolor=blue,linewidth=1.5pt]{-}(2.6,-0.35)(2.6,0.1)(4.2,-0.1)(4.2,0.35)
\psbezier[linecolor=blue,linewidth=1.5pt]{-}(4.82,-0.08)(4.7,0.09)(3.8,0.09)(3.8,-0.35)
\psbezier[linecolor=blue,linewidth=1.5pt]{-}(0.2,-0.35)(0.2,-0.17)(0.05,-0.07)(-0.02,-0.05)
\psframe[fillstyle=solid,linecolor=white,linewidth=0pt](-0.1,-0.4)(-0.005,0.4)
\psframe[fillstyle=solid,linecolor=white,linewidth=0pt](4.805,-0.4)(4.9,0.4)
\end{pspicture}}\ \, .
\ee

Let $\repM$ be a family of modules. We define the reflected family of modules $\repM^{r}$ as follows. The vector spaces of $\repM(N)$ and $\repM^r(N)$ are the same, and we use $u$ and $u^r$ to distinguish between the identical states in these two modules. The action of $\cL(N',N)$ on $\repM^r(N)$ is given by
\begin{equation} \label{eq:L.on.Mr}
\lambda \cdot u^r = (R(\lambda) \cdot u)^r \,,
\qquad \forall u^r \in \repM^r(N) \,.
\end{equation}
We then have
\begin{equation} \label{eq:consistency.Mr}
\lambda' \cdot (\lambda \cdot u^r)
= \lambda' \cdot \big(R(\lambda) \cdot u\big)^r
= (R(\lambda')R(\lambda) \cdot u)^r
= (R(\lambda'\lambda) \cdot u)^r
= (\lambda'\lambda) \cdot u^r \,.
\end{equation}
Thus \eqref{eq:L.on.Mr} and \eqref{eq:consistency.Mr} confirm that the conditions \eqref{eq:family.properties} are satisfied, and
that $\repM^r=\{\repM^r(N)\,|\, N = N_0, N_0+2, \dots\}$ is a family of modules. Moreover, it is easy to see that $(\repM^r)^r = \repM$.\medskip

The action of these transformations on the families defined in \cref{sec:example.fams} is
\begin{subequations} \label{eq:transformed}
\begin{alignat}{5}
&\repW^-_{k} \simeq \repW_{k} \,, \qquad
&&\repW^-_{k,x} \simeq \repW_{k,-x} \,,\qquad
&&\qquad
&&\repX_{m,\phi}^- \simeq \repX_{m,\phi+\pi}\,, \qquad
&&\repM^-_{\g,K} \simeq \repM_{\g,K} \,, 
\\[0.1cm]
&\repW^r_{k} \simeq \repW_{k} \,, \qquad 
&&\repW^r_{k,x} \simeq \repW_{k,x^{-1}} \,, \qquad 
&&\repV^r \simeq \repV\,,\qquad
&&\repX_{m,\phi}^r \simeq \repX_{m,-\phi}\,, \qquad 
&&\repM^r_{\g,K} \simeq \repM_{\g,K^{-1}}\,.
\end{alignat}
\end{subequations}
For the family $\repX_{m,\phi}$, these results follow straightforwardly from the definitions of the operators $c_j$ and~$c^\dag_j$. For the other families, the proofs are given in \cref{app:mod.transf}. We note that the results for $\repW_{k,x}^r$ and $\repX_{m,\phi}^r$ were previously obtained in \cite{PinetStAubin23}. Lastly, we note that the vacuum family $\repV$ has a partner family $\repV^-$ which is not isomorphic to $\repV$.

\subsection{Right modules and adjoint modules}

Up to now, our discussion has focused on families of left modules\footnote{Throughout, we use the term {\it family of modules} to mean a family of left modules.}, namely of modules on which the diagram spaces act from the left. We now describe right modules, for which the diagrams act from the right. A family of right modules is a set 
\begin{equation}
\repM =\big\{ \repM(N)\,|\, N = N_0, N_0+2, N_0+4,\dots\big\}\,,
\qquad N_0 \in \{0,1\}\,,
\end{equation}
where each $\repM(N)$ is a right module of $\eptl_N(\beta)$, and
subject to the conditions
\begin{subequations}
\begin{alignat}{2}
& u \cdot \lambda \in \repM(N') 
\qquad &&\forall \lambda \in \cL(N,N'), \, u \in \repM(N)\,, \\[0.1cm]
& (u \cdot \lambda) \cdot \lambda' = u \cdot (\lambda \lambda')
\qquad &&\forall \lambda \in \cL(N,N'),\, \lambda' \in \cL(N',N''),\,
u \in \repM(N)\,. 
\end{alignat}
\end{subequations}
In the case where the families have modules spanned by link states, we view the link states of the right modules as drawn in the complement of the disc, with their nodes numbered anticlockwise on the circle.\medskip

The {\it adjoint operation} $\cL(N,N') \to \cL(N',N)$, defined by $\lambda \mapsto \lambda^\dagger$ for all $\lambda \in \cL(N,N')$, is an anti-automorphism of the diagram spaces, since $(\lambda'\lambda)^\dagger = \lambda^\dagger (\lambda')^\dagger$. Given a family of left modules $\repM$, we define the {\it adjoint family} of right modules $\repM^\dag$ as follows. For each admissible $N$, the vector space of $\repM^\dag(N)$ is the same as that of $\repM(N)$. For each $u\in \repM(N)$, we denote by $u^\dag$ the corresponding state in $\repM^\dag(N)$. The action of $\cL(N,N')$ on $\repM^\dag(N)$ is defined by
\begin{equation}
u^\dag \cdot \lambda = (\lambda^\dag \cdot u)^\dag \,.
\end{equation}
We then have
\begin{equation}
(u^\dag \cdot \lambda) \cdot \lambda' 
= (\lambda^\dag \cdot u)^\dag \cdot \lambda' 
= \big((\lambda')^\dag \lambda^\dag \cdot u\big)^\dag
= \big((\lambda\lambda')^\dag \cdot u\big)^\dag
= u^\dag \cdot (\lambda \lambda')\,,
\end{equation}
which confirms that $\repM^\dag = \{\repM^\dag(N)\,|\,N=N_0,N_0+2, \dots\}$ is a family of right modules.\medskip

This construction of right modules adjoint to left modules is similar but different from the duality associated to the anti-involution $\lambda \mapsto \lambda^\dag$ used in \cite{PinetStAubin23}. Indeed, in contrast with \cite{PinetStAubin23}, we anticipate on the applications to correlation functions and find it more convenient to work with left and right modules, related by adjunction.

%
\section{Fusion of arbitrary modules}
\label{sec:fusion.modules}
%

In this section, we first introduce the space of diagrams with two holes. We then define the fusion of arbitrary pairs of families of modules. We show that this defines a bifunctor over the category of families of modules of $\eptl_N(\beta)$, and we study some of its properties.

\subsection{Diagrams with two holes}\label{sec:diagrams.2holes}

\paragraph{Diagrammatic definition of $\boldsymbol{\Lambda(N,N_\pa,N_\pb)}$.}

Let $N$, $N_\pa$ and $N_\pb$ be positive integers such that $N$ and $N_{\pa\pb}=N_\pa + N_\pb$ have the same parity. We define the diagram space $\Lambda(N,N_\pa,N_\pb)$ as the vector space of connectivity diagrams on a disc with two holes, referred to as hole $\pa$ and hole $\pb$. In the diagrams, we draw hole $\pa$ in green on the right, and hole $\pb$ in purple on the left. A connectivity diagram in $\Lambda(N,N_\pa,N_\pb)$ consists of non-intersecting loop segments connecting $N$ nodes on the outer boundary, $N_\pa$~nodes on hole~$\pa$ and $N_\pb$~nodes on hole~$\pb$. It may also contain three types of non-contractible loops, namely loops encircling $\pa$, $\pb$ and $\pa\pb$. We insert reference points $\pa$, $\pb$ and $\pc$ on the three boundaries and draw the dashed lines $\pa\pc$ and $\pb\pc$. We assign the nodes on the outside of the disc, on hole $\pa$ and on hole~$\pb$ the labels $\{1,2, \dots, N\}$, $\{1,2,\dots, N_\pa\}$ and $\{1,2, \dots, N_\pb\}$, respectively, starting from the reference point and increasing in the counter-clockwise direction. Here is an example of a connectivity diagram in $\Lambda(12,6,4)$:
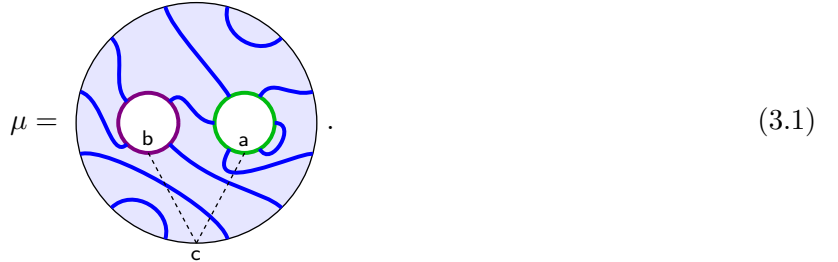
\be
\psset{unit=0.8cm}
\mu = \
\thispic{\begin{pspicture}[shift=-2.1](-2,-2.2)(2,2.2)
\psarc[linecolor=black,linewidth=0.5pt,fillstyle=solid,fillcolor=lightlightblue]{-}(0,0){2.0}{0}{360}
\psarc[linecolor=black,linewidth=0.5pt,fillstyle=solid,fillcolor=white]{-}(-0.8,0){0.5}{0}{360}
\psarc[linecolor=black,linewidth=0.5pt,fillstyle=solid,fillcolor=white]{-}(0.8,0){0.5}{0}{360}
\psarc[linecolor=purple,linewidth=1.5pt]{-}(-0.8,0){0.5}{0}{360}
\psarc[linecolor=darkgreen,linewidth=1.5pt]{-}(0.8,0){0.5}{0}{360}
\psbezier[linecolor=blue,linewidth=1.5pt]{-}(1.05, 0.433013)(1.25, 0.779423)(1.54548, 0.41411)(1.93185, 0.517638)
\psbezier[linecolor=blue,linewidth=1.5pt]{-}(0.55, -0.433013)(0.05, -1.29904)(1.73867, -0.465874)(1.93185, -0.517638)
\psbezier[linecolor=blue,linewidth=1.5pt]{-}(-0.446447, -0.353553)(0.26066, -1.06066)(1.13137, -1.13137)(1.41421, -1.41421)
\psbezier[linecolor=blue,linewidth=1.5pt]{-}(-0.446447, 0.353553)(-0.163604, 0.636396)(-0.1, 0)(0.3, 0)
\psbezier[linecolor=blue,linewidth=1.5pt]{-}(-0.517638, 1.93185)(-0.41411, 1.54548)(0.35, 0.779423)(0.55, 0.433013)
\psbezier[linecolor=blue,linewidth=1.5pt]{-}(-1.41421, 1.41421)(-1.13137, 1.13137)(-1.4364, 0.636396)(-1.15355, 0.353553)
\psbezier[linecolor=blue,linewidth=1.5pt]{-}(-1.93185, 0.517638)(-1.54548, 0.41411)(-1.4364, -0.636396)(-1.15355, -0.353553)
\rput{-30}(0,0){\psbezier[linecolor=blue,linewidth=1.5pt]{-}(0.517638, -1.93185)(0.388229, -1.44889)(-0.388229, -1.44889)(-0.517638, -1.93185)
\psbezier[linecolor=blue,linewidth=1.5pt]{-}(-1.41421, -1.41421)(-1.06066, -1.06066)(1.06066, -1.06066)(1.41421, -1.41421)
}
\rput{150}(0,0){\psbezier[linecolor=blue,linewidth=1.5pt]{-}(0.517638, -1.93185)(0.388229, -1.44889)(-0.388229, -1.44889)(-0.517638, -1.93185)
}
\psbezier[linecolor=blue,linewidth=1.5pt]{-}(1.05, -0.433013)(1.25, -0.779423)(1.7, 0)(1.3, 0)
\psline[linestyle=dashed, dash= 1.5pt 1.5pt,linewidth=0.5pt]{-}(0,-2)(0.8,-0.5)
\psline[linestyle=dashed, dash= 1.5pt 1.5pt,linewidth=0.5pt]{-}(0,-2)(-0.8,-0.5)
\rput(0,-2.2){$_\pc$}
\rput(0.8,-0.3){$_\pa$}
\rput(-0.8,-0.25){$_\pb$}
\end{pspicture}}\ .
\ee
The vector space $\Lambda(N,N_\pa,N_\pb)$ is equipped with the diagrammatic actions
\begin{subequations}
\begin{alignat}{4}
&\lambda\, \mu \in \Lambda(N,N_\pa,N_\pb) \qquad &&\forall \mu \in \Lambda(N',N_\pa,N_\pb) \,,\, \ \ &&\forall\lambda \in \cL(N,N')\,,
\\[0.15cm]
&\mu\, (\lambda_\pa \otimes \lambda_\pb)\in \Lambda(N,N_\pa,N_\pb) \qquad &&\forall \mu \in \Lambda(N,N_\pa',N_\pb') \,,\, \ \ &&\forall\lambda_\pa \in \cL(N'_\pa, N_\pa) \,,\, \ \ &&\forall\lambda_\pb \in \cL(N'_\pb, N_\pb)\,,
\end{alignat}
\end{subequations}
where the product $\lambda\, \mu$ is computed by drawing $\mu$ inside $\lambda$, and the product $\mu\, (\lambda_\pa \otimes \lambda_\pb)$ by drawing $\lambda_\pa$ and $\lambda_\pb$ in the holes $\pa$ and $\pb$, respectively. We equivalently write this as
\begin{subequations}
\begin{alignat}{2}
&\cL(N,N')\,\Lambda(N',N_\pa,N_\pb) \subseteq \Lambda(N,N_\pa,N_\pb)\,,
\label{eq:LLa=L}
\\[0.15cm]
&\Lambda(N,N'_\pa,N_\pb')\,\big(\cL(N'_\pa, N_\pa)\otimes\cL(N'_\pb, N_\pb) \big) \subseteq \Lambda(N,N_\pa,N_\pb)\,.
\label{eq:LaLL=La}
\end{alignat}
\end{subequations}
Thus $\Lambda(N,N_\pa,N_\pb)$ is the vector space of diagrams with two holes subject to this left action of $\cL(N',N)$ and right action of $\big(\cL(N'_\pa, N_\pa)\otimes\cL(N'_\pb, N_\pb) \big)$. This is the diagrammatic definition of $\Lambda(N,N_\pa,N_\pb)$.

\paragraph{Algebraic definition of $\boldsymbol{\Lambda(N,N_\pa,N_\pb)}$.}

To define $\Lambda(N,N_\pa,N_\pb)$ algebraically, we remark that any diagram $\mu \in \Lambda(N,N_\pa,N_\pb)$ can be decomposed in terms of a triple of elements $\lambda \in \cL(N,N'_{\pa\pb})$, $\lambda_\pa \in \cL(N'_\pa,N_\pa)$ and $\lambda_\pb \in \cL(N'_\pb,N_\pb)$, for some integers $N'_\pa$, $N'_\pb$ and $N'_{\pa\pb} = N'_{\pa} + N'_{\pb}$. We denote it as $\lambda(\lambda_\pa \otimes \lambda_\pb)$, where
\be
\label{eq:mu.diag}
\psset{unit=0.5cm}
\mu = \lambda(\lambda_\pa \otimes \lambda_\pb) = \
\thispic{\begin{pspicture}[shift=-4.4](-4.5,-4.5)(4.5,4.5)
\psarc[linecolor=black,linewidth=0.5pt,fillstyle=solid,fillcolor=lightlightblue]{-}(0,0){4.5}{0}{360}
\psarc[linecolor=black,linewidth=0.5pt,fillstyle=solid,fillcolor=white]{-}(-1.6,0){0.5}{0}{360}
\psarc[linecolor=black,linewidth=0.5pt,fillstyle=solid,fillcolor=white]{-}(1.6,0){0.5}{0}{360}
\psarc[linecolor=black,linewidth=0.5pt]{-}(-1.6,0){1.2}{0}{360}
\psarc[linecolor=black,linewidth=0.5pt]{-}(1.6,0){1.2}{0}{360}
\psarc[linecolor=purple,linewidth=1.5pt]{-}(-1.6,0){0.5}{0}{360}
\psarc[linecolor=darkgreen,linewidth=1.5pt]{-}(1.6,0){0.5}{0}{360}
\psarc[linecolor=black,linewidth=0.5pt]{-}(0,0){3.5}{0}{360}
\psline[linestyle=dashed, dash= 1.5pt 1.5pt,linewidth=0.5pt]{-}(-1.1,0)(-0.4,0)
\psline[linestyle=dashed, dash= 1.5pt 1.5pt,linewidth=0.5pt]{-}(1.1,0)(0.4,0)
\psline[linestyle=dashed, dash= 1.5pt 1.5pt,linewidth=0.5pt]{-}(-0.4,0)(0,-3.5)
\psline[linestyle=dashed, dash= 1.5pt 1.5pt,linewidth=0.5pt]{-}(0.4,0)(0,-3.5)
\psline[linestyle=dashed, dash= 1.5pt 1.5pt,linewidth=0.5pt]{-}(0,-3.5)(0,-4.5)
\psbezier[linecolor=blue,linewidth=1.5pt]{-}(0.491345, -0.45922)(0.121793, -0.612293)(0.689815, -3.02228)(0.778823, -3.41225)
\psbezier[linecolor=blue,linewidth=1.5pt]{-}(1.14078, -1.10866)(0.987707, -1.47821)(1.93282, -2.42368)(2.18221, -2.73641)
\psbezier[linecolor=blue,linewidth=1.5pt]{-}(2.05922, -1.10866)(2.21229, -1.47821)(2.793, -1.34504)(3.15339, -1.51859)
\psbezier[linecolor=blue,linewidth=1.5pt]{-}(2.70866, -0.45922)(3.07821, -0.612293)(3.1, 0.)(3.5, 0.)
\psbezier[linecolor=blue,linewidth=1.5pt]{-}(2.70866, 0.45922)(3.07821, 0.612293)(2.793, 1.34504)(3.15339, 1.51859)
\psbezier[linecolor=blue,linewidth=1.5pt]{-}(2.05922, 1.10866)(2.21229, 1.47821)(1.93282, 2.42368)(2.18221, 2.73641)
\psbezier[linecolor=blue,linewidth=1.5pt]{-}(1.14078, 1.10866)(0.987707, 1.47821)(0.689815, 3.02228)(0.778823, 3.41225)
\psbezier[linecolor=blue,linewidth=1.5pt]{-}(0.491345, 0.45922)(0.121793, 0.612293)(-0.689815, 3.02228)(-0.778823, 3.41225)
\psbezier[linecolor=blue,linewidth=1.5pt]{-}(-2.182, 2.736)(-1.964, 2.463)(0.115, 0.99)(-0.561, 0.6)
\psbezier[linecolor=blue,linewidth=1.5pt]{-}(-3.153, 1.519)(-2.838, 1.367)(-1.6, 1.8)(-1.6, 1.2)
\psbezier[linecolor=blue,linewidth=1.5pt]{-}(-2.63923, 0.6)(-2.98564, 0.8)(-3.1, 0.)(-3.5, 0.)
\psbezier[linecolor=blue,linewidth=1.5pt]{-}(-2.63923, -0.6)(-2.98564, -0.8)(-2.793, -1.34504)(-3.15339, -1.51859)
\psbezier[linecolor=blue,linewidth=1.5pt]{-}(-1.6, -1.2)(-1.6, -1.6)(-1.93282, -2.42368)(-2.18221, -2.73641)
\psbezier[linecolor=blue,linewidth=1.5pt]{-}(-0.779, -3.412)(-0.701, -3.071)(-0.041, -0.9)(-0.561, -0.6)
\rput(0.3,-4){$_\lambda$}
\rput(1.7,-0.85){$_{\lambda_\pa}$}
\rput(-1.6,-0.85){$_{\lambda_\pb}$}
\end{pspicture}}\ \, .
\ee
We say that $\lambda_\pa$, $\lambda_\pb$ and $\lambda$ are diagrams belonging to the {\it subsystems} $\pa$, $\pb$ and $\pa\pb$, respectively. This decomposition of diagrams with two holes in terms of three diagrams with one hole is typically not unique, namely there is usually more than one choice of $\lambda$, $\lambda_\pa$, $\lambda_\pb$ and $N'_\pa$, $N'_\pb$ leading to a same diagram $\mu \in \Lambda(N,N_\pa,N_\pb)$. The space of diagrams of $\Lambda(N,N_\pa,N_\pb)$ can then be written as a sum over triples of diagram spaces
\begin{equation} \label{eq:Lambda.sumsLLL}
\Lambda(N,N_\pa,N_\pb) =
\sum_{N'_\pa,N'_\pb} \cL(N,N'_{\pa\pb})
\big(\cL(N'_\pa,N_\pa) \otimes \cL(N'_\pb,N_\pb) \big) \,.
\end{equation}
Here, the values of $N'_\pa$ and $N'_\pb$ run over all non-negative integers with the same parities as $N_\pa$ and $N_\pb$, respectively, and $N'_{\pa\pb}=N'_\pa+N'_\pb$. We also note that \eqref{eq:Lambda.sumsLLL} expresses $\Lambda(N,N_\pa,N_\pb)$ as a sum of vector spaces, and not as a direct sum, as some of the diagrams from the spaces on the right-hand side are identical. These equalities of diagrams can be translated into relations satisfied by the generators $c_j$, $c_j^\dagger$ and $\Omega$. These relations are
\begin{subequations} \label{eq:Lambda.equations}
\begin{alignat}{2}
&\lambda\, c_j\, (\lambda_\pa \otimes \lambda_\pb) = 
\lambda\, (c_j \lambda_\pa \otimes \lambda_\pb)\,, \qquad 
&& 1 \le j \le N_\pa -1,
\label{eq:Lambda.equations.a}\\[0.15cm]
&\lambda\, c_{j+N_\pa}\, (\lambda_\pa \otimes \lambda_\pb) = 
\lambda\, (\lambda_\pa \otimes c_j\,\lambda_\pb)\,, \qquad 
&& 1 \le j \le N_\pb -1,
\label{eq:Lambda.equations.b}\\[0.15cm]
&\lambda\, c_j^\dagger (\lambda_\pa \otimes \lambda_\pb) = 
\lambda\, (c_j^\dagger \lambda_\pa \otimes \lambda_\pb)\,, \qquad 
&& 1 \le j \le N_\pa +1,
\label{eq:Lambda.equations.c}\\[0.15cm]
&\lambda\, c^\dagger_{j+N_\pa} (\lambda_\pa \otimes \lambda_\pb) = 
\lambda\, (\lambda_\pa \otimes c^\dagger_j\lambda_\pb)\,, \qquad 
&& 1 \le j \le N_\pb +1,
\label{eq:Lambda.equations.d}\\[0.15cm]
&\lambda\, \Omega\,(\lambda_\pa \otimes \lambda_\pb) = \lambda\, c_{N_\pa}\,(\Omega \,\lambda_\pa \otimes c_0^\dag \lambda_\pb)\,,
&& \textrm{for } N_\pa>0 \,,
\label{eq:Lambda.equations.e}
\end{alignat}
\end{subequations}
for all $\lambda \in \cL(N',N^{\tinyx \alpha}_{\pa\pb})$, $\lambda_\pa \in \cL(N_\pa,N'_\pa)$ and $\lambda_\pb \in \cL(N_\pb,N'_\pb)$, where $N_{\pa\pb}^{\tinyx \alpha}$ in (\ref{eq:Lambda.equations}$\alpha$) takes the values
\be
\label{eq:N.alpha}
N_{\pa\pb}^{\tinyx \alpha} = \left\{\begin{array}{cl}
N_{\pa\pb}-2 & \alpha \in \{a,b\}\,,\\[0.15cm]
N_{\pa\pb}+2 & \alpha \in \{c,d\}\,,\\[0.15cm]
N_{\pa\pb} & \alpha \in \{e\}\,.
\end{array}\right.
\ee
These identities allow us to move operators $c_j$ and $c_j^\dag$ between the different subsystems, and are central in the following. Diagrammatically, they take the form
\begin{subequations}
\be
\psset{unit=0.35cm}
\thispic{\begin{pspicture}[shift=-4.4](-4.5,-4.5)(4.5,4.5)
\psarc[linecolor=black,linewidth=0.5pt,fillstyle=solid,fillcolor=lightlightblue]{-}(0,0){4.5}{0}{360}
\psarc[linecolor=black,linewidth=0.5pt,fillstyle=solid,fillcolor=white]{-}(-1.6,0){0.5}{0}{360}
\psarc[linecolor=black,linewidth=0.5pt,fillstyle=solid,fillcolor=white]{-}(1.6,0){0.5}{0}{360}
\psarc[linecolor=black,linewidth=0.5pt]{-}(-1.6,0){1.2}{0}{360}
\psarc[linecolor=black,linewidth=0.5pt]{-}(1.6,0){1.2}{0}{360}
\psline[linestyle=dashed, dash= 1.5pt 1.5pt,linewidth=0.5pt]{-}(-1.1,0)(-0.4,0)
\psline[linestyle=dashed, dash= 1.5pt 1.5pt,linewidth=0.5pt]{-}(1.1,0)(0.4,0)
\psline[linestyle=dashed, dash= 1.5pt 1.5pt,linewidth=0.5pt]{-}(-0.4,0)(0,-3.5)
\psline[linestyle=dashed, dash= 1.5pt 1.5pt,linewidth=0.5pt]{-}(0.4,0)(0,-3.5)
\psline[linestyle=dashed, dash= 1.5pt 1.5pt,linewidth=0.5pt]{-}(0,-3.5)(0,-4.5)
\rput(1.6,0){
\psline[linecolor=blue,linewidth=1.5pt]{-}(-0.476, -0.155)(-1.141, -0.371)
\psline[linecolor=blue,linewidth=1.5pt]{-}(-0.294, -0.405)(-0.705, -0.971)
\psline[linecolor=blue,linewidth=1.5pt]{-}(0., -0.5)(0., -1.2)
\psline[linecolor=blue,linewidth=1.5pt]{-}(0.294, -0.405)(0.705, -0.971)
\psline[linecolor=blue,linewidth=1.5pt]{-}(0.476, -0.155)(1.141, -0.371)
\psline[linecolor=blue,linewidth=1.5pt]{-}(0.476, 0.155)(1.141, 0.371)
\psline[linecolor=blue,linewidth=1.5pt]{-}(0.294, 0.405)(0.705, 0.971)
\psline[linecolor=blue,linewidth=1.5pt]{-}(0., 0.5)(0., 1.2)
\psline[linecolor=blue,linewidth=1.5pt]{-}(-0.294, 0.405)(-0.705, 0.971)
\psline[linecolor=blue,linewidth=1.5pt]{-}(-0.476, 0.155)(-1.141, 0.371)
}
\rput(-1.6,0){
\psline[linecolor=blue,linewidth=1.5pt]{-}(0, -0.5)(0., -1.2)
\psline[linecolor=blue,linewidth=1.5pt]{-}(0.433013, -0.25)(1.03923, -0.6)
\psline[linecolor=blue,linewidth=1.5pt]{-}(0.433013, 0.25)(1.03923, 0.6)
\psline[linecolor=blue,linewidth=1.5pt]{-}(0, 0.5)(0., 1.2)
\psline[linecolor=blue,linewidth=1.5pt]{-}(-0.433013, 0.25)(-1.03923, 0.6)
\psline[linecolor=blue,linewidth=1.5pt]{-}(-0.433013, -0.25)(-1.03923, -0.6)
}
\psline[linecolor=blue,linewidth=1.5pt]{-}(0.683, -3.433)(0.878, -4.414)
\psline[linecolor=blue,linewidth=1.5pt]{-}(1.944, -2.91)(2.5, -3.742)
\psline[linecolor=blue,linewidth=1.5pt]{-}(2.91, -1.944)(3.742, -2.5)
\psbezier[linecolor=blue,linewidth=1.5pt]{-}(3.433, -0.683)(4.087, -0.813)(4.087, 0.813)(3.433, 0.683)
\psline[linecolor=blue,linewidth=1.5pt]{-}(2.91, 1.944)(3.742, 2.5)
\psline[linecolor=blue,linewidth=1.5pt]{-}(1.944, 2.91)(2.5, 3.742)
\psline[linecolor=blue,linewidth=1.5pt]{-}(0.683, 3.433)(0.878, 4.414)
\psline[linecolor=blue,linewidth=1.5pt]{-}(-0.683, 3.433)(-0.878, 4.414)
\psline[linecolor=blue,linewidth=1.5pt]{-}(-1.944, 2.91)(-2.5, 3.742)
\psline[linecolor=blue,linewidth=1.5pt]{-}(-2.91, 1.944)(-3.742, 2.5)
\psline[linecolor=blue,linewidth=1.5pt]{-}(-3.433, 0.683)(-4.414, 0.878)
\psline[linecolor=blue,linewidth=1.5pt]{-}(-3.433, -0.683)(-4.414, -0.878)
\psline[linecolor=blue,linewidth=1.5pt]{-}(-2.91, -1.944)(-3.742, -2.5)
\psline[linecolor=blue,linewidth=1.5pt]{-}(-1.944, -2.91)(-2.5, -3.742)
\psline[linecolor=blue,linewidth=1.5pt]{-}(-0.683, -3.433)(-0.878, -4.414)
\psbezier[linecolor=blue,linewidth=1.5pt]{-}(0.683, -3.433)(0.546, -2.746)(-0.112, -0.556)(0.459, -0.371)
\psbezier[linecolor=blue,linewidth=1.5pt]{-}(1.944, -2.91)(1.556, -2.328)(0.542, -1.456)(0.895, -0.971)
\psbezier[linecolor=blue,linewidth=1.5pt]{-}(2.91, -1.944)(2.328, -1.556)(1.6, -1.8)(1.6, -1.2)
\psbezier[linecolor=blue,linewidth=1.5pt]{-}(3.433, -0.683)(2.746, -0.546)(2.658, -1.456)(2.305, -0.971)
\psbezier[linecolor=blue,linewidth=1.5pt]{-}(3.433, 0.683)(2.746, 0.546)(3.312, -0.556)(2.741, -0.371)
\psbezier[linecolor=blue,linewidth=1.5pt]{-}(2.91, 1.944)(2.328, 1.556)(3.312, 0.556)(2.741, 0.371)
\psbezier[linecolor=blue,linewidth=1.5pt]{-}(1.944, 2.91)(1.556, 2.328)(2.658, 1.456)(2.305, 0.971)
\psbezier[linecolor=blue,linewidth=1.5pt]{-}(0.683, 3.433)(0.546, 2.746)(1.6, 1.8)(1.6, 1.2)
\psbezier[linecolor=blue,linewidth=1.5pt]{-}(-0.683, 3.433)(-0.546, 2.746)(0.542, 1.456)(0.895, 0.971)
\psbezier[linecolor=blue,linewidth=1.5pt]{-}(-1.944, 2.91)(-1.556, 2.528)(-0.112, 0.556)(0.459, 0.371)
\psbezier[linecolor=blue,linewidth=1.5pt]{-}(-2.91, 1.944)(-2.619, 1.75)(0.271, 1.08)(-0.561, 0.6)
\psbezier[linecolor=blue,linewidth=1.5pt]{-}(-3.433, 0.683)(-3.089, 0.615)(-1.6, 1.8)(-1.6, 1.2)
\psbezier[linecolor=blue,linewidth=1.5pt]{-}(-3.433, -0.683)(-3.089, -0.615)(-3.159, 0.9)(-2.639, 0.6)
\psbezier[linecolor=blue,linewidth=1.5pt]{-}(-2.91, -1.944)(-2.619, -1.75)(-3.159, -0.9)(-2.639, -0.6)
\psbezier[linecolor=blue,linewidth=1.5pt]{-}(-1.944, -2.91)(-1.75, -2.619)(-1.6, -1.8)(-1.6, -1.2)
\psbezier[linecolor=blue,linewidth=1.5pt]{-}(-0.683, -3.433)(-0.615, -3.089)(-0.041, -0.9)(-0.561, -0.6)
\psarc[linecolor=purple,linewidth=1.5pt]{-}(-1.6,0){0.5}{0}{360}
\psarc[linecolor=darkgreen,linewidth=1.5pt]{-}(1.6,0){0.5}{0}{360}
\psarc[linecolor=black,linewidth=0.5pt]{-}(0,0){3.5}{0}{360}
\end{pspicture}}
\ = \
\thispic{\begin{pspicture}[shift=-4.4](-4.5,-4.5)(4.5,4.5)
\psarc[linecolor=black,linewidth=0.5pt,fillstyle=solid,fillcolor=lightlightblue]{-}(0,0){4.5}{0}{360}
\psarc[linecolor=black,linewidth=0.5pt,fillstyle=solid,fillcolor=white]{-}(-1.6,0){0.5}{0}{360}
\psarc[linecolor=black,linewidth=0.5pt,fillstyle=solid,fillcolor=white]{-}(1.6,0){0.5}{0}{360}
\psarc[linecolor=black,linewidth=0.5pt]{-}(-1.6,0){1.2}{0}{360}
\psarc[linecolor=black,linewidth=0.5pt]{-}(1.6,0){1.2}{0}{360}
\psline[linestyle=dashed, dash= 1.5pt 1.5pt,linewidth=0.5pt]{-}(-1.1,0)(-0.4,0)
\psline[linestyle=dashed, dash= 1.5pt 1.5pt,linewidth=0.5pt]{-}(1.1,0)(0.4,0)
\psline[linestyle=dashed, dash= 1.5pt 1.5pt,linewidth=0.5pt]{-}(-0.4,0)(0,-3.5)
\psline[linestyle=dashed, dash= 1.5pt 1.5pt,linewidth=0.5pt]{-}(0.4,0)(0,-3.5)
\psline[linestyle=dashed, dash= 1.5pt 1.5pt,linewidth=0.5pt]{-}(0,-3.5)(0,-4.5)
\rput(1.6,0){
\psline[linecolor=blue,linewidth=1.5pt]{-}(-0.476, -0.155)(-1.141, -0.371)
\psline[linecolor=blue,linewidth=1.5pt]{-}(-0.294, -0.405)(-0.705, -0.971)
\psline[linecolor=blue,linewidth=1.5pt]{-}(0., -0.5)(0., -1.2)
\psbezier[linecolor=blue,linewidth=1.5pt]{-}(0.294, -0.405)(0.5, -0.688)(0.808, -0.263)(0.476, -0.155)
\psline[linecolor=blue,linewidth=1.5pt]{-}(0.476, 0.155)(1.141, 0.371)
\psline[linecolor=blue,linewidth=1.5pt]{-}(0.294, 0.405)(0.705, 0.971)
\psline[linecolor=blue,linewidth=1.5pt]{-}(0., 0.5)(0., 1.2)
\psline[linecolor=blue,linewidth=1.5pt]{-}(-0.294, 0.405)(-0.705, 0.971)
\psline[linecolor=blue,linewidth=1.5pt]{-}(-0.476, 0.155)(-1.141, 0.371)
}
\rput(-1.6,0){
\psline[linecolor=blue,linewidth=1.5pt]{-}(0, -0.5)(0., -1.2)
\psline[linecolor=blue,linewidth=1.5pt]{-}(0.433013, -0.25)(1.03923, -0.6)
\psline[linecolor=blue,linewidth=1.5pt]{-}(0.433013, 0.25)(1.03923, 0.6)
\psline[linecolor=blue,linewidth=1.5pt]{-}(0, 0.5)(0., 1.2)
\psline[linecolor=blue,linewidth=1.5pt]{-}(-0.433013, 0.25)(-1.03923, 0.6)
\psline[linecolor=blue,linewidth=1.5pt]{-}(-0.433013, -0.25)(-1.03923, -0.6)
}
\psline[linecolor=blue,linewidth=1.5pt]{-}(0.778823, -3.41225)(1.00134, -4.38718)
\psline[linecolor=blue,linewidth=1.5pt]{-}(2.18221, -2.73641)(2.8057, -3.51824)
\psline[linecolor=blue,linewidth=1.5pt]{-}(3.15339, -1.51859)(4.05436,-1.95248)
\psline[linecolor=blue,linewidth=1.5pt]{-}(3.5, 0.)(4.5, 0)
\psline[linecolor=blue,linewidth=1.5pt]{-}(3.15339, 1.51859)(4.05436, 1.95248)
\psline[linecolor=blue,linewidth=1.5pt]{-}(2.18221, 2.73641)(2.8057, 3.51824)
\psline[linecolor=blue,linewidth=1.5pt]{-}(0.778823, 3.41225)(1.00134, 4.38718)
\psline[linecolor=blue,linewidth=1.5pt]{-}(-0.778823, 3.41225)(-1.00134, 4.38718)
\psline[linecolor=blue,linewidth=1.5pt]{-}(-2.18221, 2.73641)(-2.8057, 3.51824)
\psline[linecolor=blue,linewidth=1.5pt]{-}(-3.15339, 1.51859)(-4.05436, 1.95248)
\psline[linecolor=blue,linewidth=1.5pt]{-}(-3.5, 0.)(-4.5, 0)
\psline[linecolor=blue,linewidth=1.5pt]{-}(-3.15339, -1.51859)(-4.05436, -1.95248)
\psline[linecolor=blue,linewidth=1.5pt]{-}(-2.18221, -2.73641)(-2.8057, -3.5182)
\psline[linecolor=blue,linewidth=1.5pt]{-}(-0.778823, -3.41225)(-1.00134, -4.38718)
\psbezier[linecolor=blue,linewidth=1.5pt]{-}(0.779, -3.412)(0.623, -2.73)(0.116, -0.482)(0.459, -0.371)
\psbezier[linecolor=blue,linewidth=1.5pt]{-}(2.182, -2.736)(1.746, -2.189)(0.683, -1.262)(0.895, -0.971)
\psbezier[linecolor=blue,linewidth=1.5pt]{-}(3.153, -1.519)(2.523, -1.215)(1.6, -1.8)(1.6, -1.2)
\psbezier[linecolor=blue,linewidth=1.5pt]{-}(3.5, 0.)(3.15, 0.)(2.97, 0.445)(2.741, 0.371)
\psbezier[linecolor=blue,linewidth=1.5pt]{-}(3.153, 1.519)(2.523, 1.215)(2.517, 1.262)(2.305, 0.971)
\psbezier[linecolor=blue,linewidth=1.5pt]{-}(2.182, 2.736)(1.746, 2.189)(1.6, 1.56)(1.6, 1.2)
\psbezier[linecolor=blue,linewidth=1.5pt]{-}(0.779, 3.412)(0.623, 2.73)(0.683, 1.262)(0.895, 0.971)
\psbezier[linecolor=blue,linewidth=1.5pt]{-}(-0.779, 3.412)(-0.623, 2.73)(0.116, 0.482)(0.459, 0.371)
\psbezier[linecolor=blue,linewidth=1.5pt]{-}(-0.56077, 0.6)(-0.214359, 0.8)(-1.93282, 2.42368)(-2.18221, 2.73641)
\psbezier[linecolor=blue,linewidth=1.5pt]{-}(-1.6, 1.2)(-1.6, 1.6)(-2.793, 1.34504)(-3.15339, 1.51859)
\psbezier[linecolor=blue,linewidth=1.5pt]{-}(-2.63923, 0.6)(-2.98564, 0.8)(-3.1, 0.)(-3.5, 0.)
\psbezier[linecolor=blue,linewidth=1.5pt]{-}(-2.63923, -0.6)(-2.98564, -0.8)(-2.793, -1.34504)(-3.15339, -1.51859)
\psbezier[linecolor=blue,linewidth=1.5pt]{-}(-1.6, -1.2)(-1.6, -1.6)(-1.93282, -2.42368)(-2.18221, -2.73641)
\psbezier[linecolor=blue,linewidth=1.5pt]{-}(-0.56077, -0.6)(-0.214359, -0.8)(-0.689815, -3.02228)(-0.778823, -3.41225)
\psarc[linecolor=purple,linewidth=1.5pt]{-}(-1.6,0){0.5}{0}{360}
\psarc[linecolor=darkgreen,linewidth=1.5pt]{-}(1.6,0){0.5}{0}{360}
\psarc[linecolor=black,linewidth=0.5pt]{-}(0,0){3.5}{0}{360}
\end{pspicture}}\ \,,
\qquad
\thispic{\begin{pspicture}[shift=-4.4](-4.5,-4.5)(4.5,4.5)
\psarc[linecolor=black,linewidth=0.5pt,fillstyle=solid,fillcolor=lightlightblue]{-}(0,0){4.5}{0}{360}
\psarc[linecolor=black,linewidth=0.5pt,fillstyle=solid,fillcolor=white]{-}(-1.6,0){0.5}{0}{360}
\psarc[linecolor=black,linewidth=0.5pt,fillstyle=solid,fillcolor=white]{-}(1.6,0){0.5}{0}{360}
\psarc[linecolor=black,linewidth=0.5pt]{-}(-1.6,0){1.2}{0}{360}
\psarc[linecolor=black,linewidth=0.5pt]{-}(1.6,0){1.2}{0}{360}
\psline[linestyle=dashed, dash= 1.5pt 1.5pt,linewidth=0.5pt]{-}(-1.1,0)(-0.4,0)
\psline[linestyle=dashed, dash= 1.5pt 1.5pt,linewidth=0.5pt]{-}(1.1,0)(0.4,0)
\psline[linestyle=dashed, dash= 1.5pt 1.5pt,linewidth=0.5pt]{-}(-0.4,0)(0,-3.5)
\psline[linestyle=dashed, dash= 1.5pt 1.5pt,linewidth=0.5pt]{-}(0.4,0)(0,-3.5)
\psline[linestyle=dashed, dash= 1.5pt 1.5pt,linewidth=0.5pt]{-}(0,-3.5)(0,-4.5)
\rput(1.6,0){
\psline[linecolor=blue,linewidth=1.5pt]{-}(0.191342, -0.46194)(0.45922, -1.10866)
\psline[linecolor=blue,linewidth=1.5pt]{-}(0.46194, -0.191342)(1.10866, -0.45922)
\psline[linecolor=blue,linewidth=1.5pt]{-}(0.46194, 0.191342)(1.10866, 0.45922)
\psline[linecolor=blue,linewidth=1.5pt]{-}(0.191342, 0.46194)(0.45922, 1.10866)
\psline[linecolor=blue,linewidth=1.5pt]{-}(-0.191342, 0.46194)(-0.45922, 1.10866)
\psline[linecolor=blue,linewidth=1.5pt]{-}(-0.46194, 0.191342)(-1.10866, 0.45922)
\psline[linecolor=blue,linewidth=1.5pt]{-}(-0.46194, -0.191342)(-1.10866, -0.45922)
\psline[linecolor=blue,linewidth=1.5pt]{-}(-0.191342, -0.46194)(-0.45922, -1.10866)
}
\rput(-1.6,0){
\psline[linecolor=blue,linewidth=1.5pt]{-}(0.462, 0.191)(1.109, 0.459)
\psline[linecolor=blue,linewidth=1.5pt]{-}(0.191, 0.462)(0.459, 1.109)
\psline[linecolor=blue,linewidth=1.5pt]{-}(-0.191, 0.462)(-0.459, 1.109)
\psline[linecolor=blue,linewidth=1.5pt]{-}(-0.462, 0.191)(-1.109, 0.459)
\psline[linecolor=blue,linewidth=1.5pt]{-}(-0.462, -0.191)(-1.109, -0.459)
\psline[linecolor=blue,linewidth=1.5pt]{-}(-0.191, -0.462)(-0.459, -1.109)
\psline[linecolor=blue,linewidth=1.5pt]{-}(0.191, -0.462)(0.459, -1.109)
\psline[linecolor=blue,linewidth=1.5pt]{-}(0.462, -0.191)(1.109, -0.459)
}
\psline[linecolor=blue,linewidth=1.5pt]{-}(0.683, -3.433)(0.878, -4.414)
\psline[linecolor=blue,linewidth=1.5pt]{-}(1.944, -2.91)(2.5, -3.742)
\psline[linecolor=blue,linewidth=1.5pt]{-}(2.91, -1.944)(3.742, -2.5)
\psline[linecolor=blue,linewidth=1.5pt]{-}(3.433, -0.683)(4.414, -0.878)
\psline[linecolor=blue,linewidth=1.5pt]{-}(3.433, 0.683)(4.414, 0.878)
\psline[linecolor=blue,linewidth=1.5pt]{-}(2.91, 1.944)(3.742, 2.5)
\psline[linecolor=blue,linewidth=1.5pt]{-}(1.944, 2.91)(2.5, 3.742)
\psline[linecolor=blue,linewidth=1.5pt]{-}(0.683, 3.433)(0.878, 4.414)
\psline[linecolor=blue,linewidth=1.5pt]{-}(-0.683, 3.433)(-0.878, 4.414)
\psline[linecolor=blue,linewidth=1.5pt]{-}(-1.944, 2.91)(-2.5, 3.742)
\psbezier[linecolor=blue,linewidth=1.5pt]{-}(-2.91, 1.944)(-3.464, 2.315)(-4.087, 0.813)(-3.433, 0.683)
\psline[linecolor=blue,linewidth=1.5pt]{-}(-3.433, -0.683)(-4.414, -0.878)
\psline[linecolor=blue,linewidth=1.5pt]{-}(-2.91, -1.944)(-3.742, -2.5)
\psline[linecolor=blue,linewidth=1.5pt]{-}(-1.944, -2.91)(-2.5, -3.742)
\psline[linecolor=blue,linewidth=1.5pt]{-}(-0.683, -3.433)(-0.878, -4.414)
\psbezier[linecolor=blue,linewidth=1.5pt]{-}(0.683, -3.433)(0.546, -2.746)(-0.063, -0.689)(0.491, -0.459)
\psbezier[linecolor=blue,linewidth=1.5pt]{-}(1.944, -2.91)(1.556, -2.328)(0.911, -1.663)(1.141, -1.109)
\psbezier[linecolor=blue,linewidth=1.5pt]{-}(2.91, -1.944)(2.328, -1.556)(2.289, -1.663)(2.059, -1.109)
\psbezier[linecolor=blue,linewidth=1.5pt]{-}(3.433, -0.683)(2.746, -0.546)(3.263, -0.689)(2.709, -0.459)
\psbezier[linecolor=blue,linewidth=1.5pt]{-}(3.433, 0.683)(2.746, 0.546)(3.263, 0.689)(2.709, 0.459)
\psbezier[linecolor=blue,linewidth=1.5pt]{-}(2.91, 1.944)(2.328, 1.556)(2.289, 1.663)(2.059, 1.109)
\psbezier[linecolor=blue,linewidth=1.5pt]{-}(1.944, 2.91)(1.556, 2.328)(0.911, 1.663)(1.141, 1.109)
\psbezier[linecolor=blue,linewidth=1.5pt]{-}(0.683, 3.433)(0.546, 2.746)(-0.063, 0.689)(0.491, 0.459)
\psbezier[linecolor=blue,linewidth=1.5pt]{-}(-0.683, 3.433)(-0.615, 3.089)(0.063, 0.689)(-0.491, 0.459)
\psbezier[linecolor=blue,linewidth=1.5pt]{-}(-1.944, 2.91)(-1.75, 2.619)(-0.911, 1.663)(-1.141, 1.109)
\psbezier[linecolor=blue,linewidth=1.5pt]{-}(-2.91, 1.944)(-2.619, 1.75)(-2.289, 1.663)(-2.059, 1.109)
\psbezier[linecolor=blue,linewidth=1.5pt]{-}(-3.433, 0.683)(-3.089, 0.615)(-3.263, 0.689)(-2.709, 0.459)
\psbezier[linecolor=blue,linewidth=1.5pt]{-}(-3.433, -0.683)(-3.089, -0.615)(-3.263, -0.689)(-2.709, -0.459)
\psbezier[linecolor=blue,linewidth=1.5pt]{-}(-2.91, -1.944)(-2.619, -1.75)(-2.289, -1.663)(-2.059, -1.109)
\psbezier[linecolor=blue,linewidth=1.5pt]{-}(-1.944, -2.91)(-1.75, -2.619)(-0.911, -1.663)(-1.141, -1.109)
\psbezier[linecolor=blue,linewidth=1.5pt]{-}(-0.683, -3.433)(-0.615, -3.089)(0.063, -0.689)(-0.491, -0.459)
\psarc[linecolor=purple,linewidth=1.5pt]{-}(-1.6,0){0.5}{0}{360}
\psarc[linecolor=darkgreen,linewidth=1.5pt]{-}(1.6,0){0.5}{0}{360}
\psarc[linecolor=black,linewidth=0.5pt]{-}(0,0){3.5}{0}{360}
\end{pspicture}}
\ = \
\thispic{\begin{pspicture}[shift=-4.4](-4.5,-4.5)(4.5,4.5)
\psarc[linecolor=black,linewidth=0.5pt,fillstyle=solid,fillcolor=lightlightblue]{-}(0,0){4.5}{0}{360}
\psarc[linecolor=black,linewidth=0.5pt,fillstyle=solid,fillcolor=white]{-}(-1.6,0){0.5}{0}{360}
\psarc[linecolor=black,linewidth=0.5pt,fillstyle=solid,fillcolor=white]{-}(1.6,0){0.5}{0}{360}
\psarc[linecolor=black,linewidth=0.5pt]{-}(-1.6,0){1.2}{0}{360}
\psarc[linecolor=black,linewidth=0.5pt]{-}(1.6,0){1.2}{0}{360}
\psline[linestyle=dashed, dash= 1.5pt 1.5pt,linewidth=0.5pt]{-}(-1.1,0)(-0.4,0)
\psline[linestyle=dashed, dash= 1.5pt 1.5pt,linewidth=0.5pt]{-}(1.1,0)(0.4,0)
\psline[linestyle=dashed, dash= 1.5pt 1.5pt,linewidth=0.5pt]{-}(-0.4,0)(0,-3.5)
\psline[linestyle=dashed, dash= 1.5pt 1.5pt,linewidth=0.5pt]{-}(0.4,0)(0,-3.5)
\psline[linestyle=dashed, dash= 1.5pt 1.5pt,linewidth=0.5pt]{-}(0,-3.5)(0,-4.5)
\rput(1.6,0){
\psline[linecolor=blue,linewidth=1.5pt]{-}(0.191342, -0.46194)(0.45922, -1.10866)
\psline[linecolor=blue,linewidth=1.5pt]{-}(0.46194, -0.191342)(1.10866, -0.45922)
\psline[linecolor=blue,linewidth=1.5pt]{-}(0.46194, 0.191342)(1.10866, 0.45922)
\psline[linecolor=blue,linewidth=1.5pt]{-}(0.191342, 0.46194)(0.45922, 1.10866)
\psline[linecolor=blue,linewidth=1.5pt]{-}(-0.191342, 0.46194)(-0.45922, 1.10866)
\psline[linecolor=blue,linewidth=1.5pt]{-}(-0.46194, 0.191342)(-1.10866, 0.45922)
\psline[linecolor=blue,linewidth=1.5pt]{-}(-0.46194, -0.191342)(-1.10866, -0.45922)
\psline[linecolor=blue,linewidth=1.5pt]{-}(-0.191342, -0.46194)(-0.45922, -1.10866)
}
\rput(-1.6,0){
\psline[linecolor=blue,linewidth=1.5pt]{-}(0.462, 0.191)(1.109, 0.459)
\psline[linecolor=blue,linewidth=1.5pt]{-}(0.191, 0.462)(0.459, 1.109)
\psline[linecolor=blue,linewidth=1.5pt]{-}(-0.462, -0.191)(-1.109, -0.459)
\psline[linecolor=blue,linewidth=1.5pt]{-}(-0.191, -0.462)(-0.459, -1.109)
\psline[linecolor=blue,linewidth=1.5pt]{-}(0.191, -0.462)(0.459, -1.109)
\psline[linecolor=blue,linewidth=1.5pt]{-}(0.462, -0.191)(1.109, -0.459)
\psbezier[linecolor=blue,linewidth=1.5pt]{-}(-0.191, 0.462)(-0.325, 0.785)(-0.785, 0.325)(-0.462, 0.191)
}
\psline[linecolor=blue,linewidth=1.5pt]{-}(0.778823, -3.41225)(1.00134, -4.38718)
\psline[linecolor=blue,linewidth=1.5pt]{-}(2.18221, -2.73641)(2.8057, -3.51824)
\psline[linecolor=blue,linewidth=1.5pt]{-}(3.15339, -1.51859)(4.05436,-1.95248)
\psline[linecolor=blue,linewidth=1.5pt]{-}(3.5, 0.)(4.5, 0)
\psline[linecolor=blue,linewidth=1.5pt]{-}(3.15339, 1.51859)(4.05436, 1.95248)
\psline[linecolor=blue,linewidth=1.5pt]{-}(2.18221, 2.73641)(2.8057, 3.51824)
\psline[linecolor=blue,linewidth=1.5pt]{-}(0.778823, 3.41225)(1.00134, 4.38718)
\psline[linecolor=blue,linewidth=1.5pt]{-}(-0.778823, 3.41225)(-1.00134, 4.38718)
\psline[linecolor=blue,linewidth=1.5pt]{-}(-2.18221, 2.73641)(-2.8057, 3.51824)
\psline[linecolor=blue,linewidth=1.5pt]{-}(-3.15339, 1.51859)(-4.05436, 1.95248)
\psline[linecolor=blue,linewidth=1.5pt]{-}(-3.5, 0.)(-4.5, 0)
\psline[linecolor=blue,linewidth=1.5pt]{-}(-3.15339, -1.51859)(-4.05436, -1.95248)
\psline[linecolor=blue,linewidth=1.5pt]{-}(-2.18221, -2.73641)(-2.8057, -3.5182)
\psline[linecolor=blue,linewidth=1.5pt]{-}(-0.778823, -3.41225)(-1.00134, -4.38718)
\psbezier[linecolor=blue,linewidth=1.5pt]{-}(0.491345, -0.45922)(0.121793, -0.612293)(0.689815, -3.02228)(0.778823, -3.41225)
\psbezier[linecolor=blue,linewidth=1.5pt]{-}(1.14078, -1.10866)(0.987707, -1.47821)(1.93282, -2.42368)(2.18221, -2.73641)
\psbezier[linecolor=blue,linewidth=1.5pt]{-}(2.05922, -1.10866)(2.21229, -1.47821)(2.793, -1.34504)(3.15339, -1.51859)
\psbezier[linecolor=blue,linewidth=1.5pt]{-}(2.70866, -0.45922)(3.07821, -0.612293)(3.1, 0.)(3.5, 0.)
\psbezier[linecolor=blue,linewidth=1.5pt]{-}(2.70866, 0.45922)(3.07821, 0.612293)(2.793, 1.34504)(3.15339, 1.51859)
\psbezier[linecolor=blue,linewidth=1.5pt]{-}(2.05922, 1.10866)(2.21229, 1.47821)(1.93282, 2.42368)(2.18221, 2.73641)
\psbezier[linecolor=blue,linewidth=1.5pt]{-}(1.14078, 1.10866)(0.987707, 1.47821)(0.689815, 3.02228)(0.778823, 3.41225)
\psbezier[linecolor=blue,linewidth=1.5pt]{-}(0.491345, 0.45922)(0.121793, 0.612293)(-0.689815, 3.02228)(-0.778823, 3.41225)
\psbezier[linecolor=blue,linewidth=1.5pt]{-}(-2.182, 2.736)(-1.964, 2.463)(0.063, 0.689)(-0.491, 0.459)
\psbezier[linecolor=blue,linewidth=1.5pt]{-}(-3.153, 1.519)(-2.838, 1.367)(-0.911, 1.663)(-1.141, 1.109)
\psbezier[linecolor=blue,linewidth=1.5pt]{-}(-3.5, 0.)(-3.15, 0.)(-3.263, -0.689)(-2.709, -0.459)
\psbezier[linecolor=blue,linewidth=1.5pt]{-}(-3.153, -1.519)(-2.838, -1.367)(-2.289, -1.663)(-2.059, -1.109)
\psbezier[linecolor=blue,linewidth=1.5pt]{-}(-2.182, -2.736)(-1.964, -2.463)(-0.911, -1.663)(-1.141, -1.109)
\psbezier[linecolor=blue,linewidth=1.5pt]{-}(-0.779, -3.412)(-0.701, -3.071)(0.063, -0.689)(-0.491, -0.459)
\psarc[linecolor=purple,linewidth=1.5pt]{-}(-1.6,0){0.5}{0}{360}
\psarc[linecolor=darkgreen,linewidth=1.5pt]{-}(1.6,0){0.5}{0}{360}
\psarc[linecolor=black,linewidth=0.5pt]{-}(0,0){3.5}{0}{360}
\end{pspicture}}\ \, ,
\ee\vspace{0.1cm}
\be
\psset{unit=0.35cm}
\thispic{\begin{pspicture}[shift=-4.4](-4.5,-4.5)(4.5,4.5)
\psarc[linecolor=black,linewidth=0.5pt,fillstyle=solid,fillcolor=lightlightblue]{-}(0,0){4.5}{0}{360}
\psarc[linecolor=black,linewidth=0.5pt,fillstyle=solid,fillcolor=white]{-}(-1.6,0){0.5}{0}{360}
\psarc[linecolor=black,linewidth=0.5pt,fillstyle=solid,fillcolor=white]{-}(1.6,0){0.5}{0}{360}
\psarc[linecolor=black,linewidth=0.5pt]{-}(-1.6,0){1.2}{0}{360}
\psarc[linecolor=black,linewidth=0.5pt]{-}(1.6,0){1.2}{0}{360}
\psline[linestyle=dashed, dash= 1.5pt 1.5pt,linewidth=0.5pt]{-}(-1.1,0)(-0.4,0)
\psline[linestyle=dashed, dash= 1.5pt 1.5pt,linewidth=0.5pt]{-}(1.1,0)(0.4,0)
\psline[linestyle=dashed, dash= 1.5pt 1.5pt,linewidth=0.5pt]{-}(-0.4,0)(0,-3.5)
\psline[linestyle=dashed, dash= 1.5pt 1.5pt,linewidth=0.5pt]{-}(0.4,0)(0,-3.5)
\psline[linestyle=dashed, dash= 1.5pt 1.5pt,linewidth=0.5pt]{-}(0,-3.5)(0,-4.5)
\rput(1.6,0){
\psline[linecolor=blue,linewidth=1.5pt]{-}(-0.433, -0.25)(-1.039, -0.6)
\psline[linecolor=blue,linewidth=1.5pt]{-}(0., -0.5)(0., -1.2)
\psline[linecolor=blue,linewidth=1.5pt]{-}(0.433, -0.25)(1.039, -0.6)
\psline[linecolor=blue,linewidth=1.5pt]{-}(0.433, 0.25)(1.039, 0.6)
\psline[linecolor=blue,linewidth=1.5pt]{-}(0., 0.5)(0., 1.2)
\psline[linecolor=blue,linewidth=1.5pt]{-}(-0.433, 0.25)(-1.039, 0.6)
}
\rput(-1.6,0){
\psline[linecolor=blue,linewidth=1.5pt]{-}(0, -0.5)(0., -1.2)
\psline[linecolor=blue,linewidth=1.5pt]{-}(0.433013, -0.25)(1.03923, -0.6)
\psline[linecolor=blue,linewidth=1.5pt]{-}(0.433013, 0.25)(1.03923, 0.6)
\psline[linecolor=blue,linewidth=1.5pt]{-}(0, 0.5)(0., 1.2)
\psline[linecolor=blue,linewidth=1.5pt]{-}(-0.433013, 0.25)(-1.03923, 0.6)
\psline[linecolor=blue,linewidth=1.5pt]{-}(-0.433013, -0.25)(-1.03923, -0.6)
}
\psline[linecolor=blue,linewidth=1.5pt]{-}(0.778823, -3.41225)(1.00134, -4.38718)
\psline[linecolor=blue,linewidth=1.5pt]{-}(2.18221, -2.73641)(2.8057, -3.51824)
\psline[linecolor=blue,linewidth=1.5pt]{-}(3.15339, -1.51859)(4.05436,-1.95248)
\psline[linecolor=blue,linewidth=1.5pt]{-}(3.5, 0.)(4.5, 0)
\psbezier[linecolor=blue,linewidth=1.5pt]{-}(4.054, 1.952)(3.604, 1.736)(2.494, 3.127)(2.806, 3.518)
\psline[linecolor=blue,linewidth=1.5pt]{-}(0.778823, 3.41225)(1.00134, 4.38718)
\psline[linecolor=blue,linewidth=1.5pt]{-}(-0.778823, 3.41225)(-1.00134, 4.38718)
\psline[linecolor=blue,linewidth=1.5pt]{-}(-2.18221, 2.73641)(-2.8057, 3.51824)
\psline[linecolor=blue,linewidth=1.5pt]{-}(-3.15339, 1.51859)(-4.05436, 1.95248)
\psline[linecolor=blue,linewidth=1.5pt]{-}(-3.5, 0.)(-4.5, 0)
\psline[linecolor=blue,linewidth=1.5pt]{-}(-3.15339, -1.51859)(-4.05436, -1.95248)
\psline[linecolor=blue,linewidth=1.5pt]{-}(-2.18221, -2.73641)(-2.8057, -3.5182)
\psline[linecolor=blue,linewidth=1.5pt]{-}(-0.778823, -3.41225)(-1.00134, -4.38718)
\psbezier[linecolor=blue,linewidth=1.5pt]{-}(0.779, -3.412)(0.623, -2.73)(0.041, -0.9)(0.561, -0.6)
\psbezier[linecolor=blue,linewidth=1.5pt]{-}(2.182, -2.736)(1.746, -2.189)(1.6, -1.8)(1.6, -1.2)
\psbezier[linecolor=blue,linewidth=1.5pt]{-}(3.153, -1.519)(2.523, -1.215)(3.159, -0.9)(2.639, -0.6)
\psbezier[linecolor=blue,linewidth=1.5pt]{-}(3.5, 0.)(2.8, 0.)(3.159, 0.9)(2.639, 0.6)
\psbezier[linecolor=blue,linewidth=1.5pt]{-}(0.779, 3.412)(0.623, 2.73)(1.6, 1.8)(1.6, 1.2)
\psbezier[linecolor=blue,linewidth=1.5pt]{-}(-0.779, 3.412)(-0.623, 2.73)(0.041, 0.9)(0.561, 0.6)
\psbezier[linecolor=blue,linewidth=1.5pt]{-}(-2.182, 2.736)(-1.746, 2.189)(-0.041, 0.9)(-0.561, 0.6)
\psbezier[linecolor=blue,linewidth=1.5pt]{-}(-3.153, 1.519)(-2.523, 1.215)(-1.6, 1.8)(-1.6, 1.2)
\psbezier[linecolor=blue,linewidth=1.5pt]{-}(-3.5, 0.)(-2.8, 0.)(-3.159, 0.9)(-2.639, 0.6)
\psbezier[linecolor=blue,linewidth=1.5pt]{-}(-3.153, -1.519)(-2.523, -1.215)(-3.159, -0.9)(-2.639, -0.6)
\psbezier[linecolor=blue,linewidth=1.5pt]{-}(-2.182, -2.736)(-1.746, -2.189)(-1.6, -1.8)(-1.6, -1.2)
\psbezier[linecolor=blue,linewidth=1.5pt]{-}(-0.779, -3.412)(-0.623, -2.73)(-0.041, -0.9)(-0.561, -0.6)
\psarc[linecolor=purple,linewidth=1.5pt]{-}(-1.6,0){0.5}{0}{360}
\psarc[linecolor=darkgreen,linewidth=1.5pt]{-}(1.6,0){0.5}{0}{360}
\psarc[linecolor=black,linewidth=0.5pt]{-}(0,0){3.5}{0}{360}
\end{pspicture}}
\ = \
\thispic{\begin{pspicture}[shift=-4.4](-4.5,-4.5)(4.5,4.5)
\psarc[linecolor=black,linewidth=0.5pt,fillstyle=solid,fillcolor=lightlightblue]{-}(0,0){4.5}{0}{360}
\psarc[linecolor=black,linewidth=0.5pt,fillstyle=solid,fillcolor=white]{-}(-1.6,0){0.5}{0}{360}
\psarc[linecolor=black,linewidth=0.5pt,fillstyle=solid,fillcolor=white]{-}(1.6,0){0.5}{0}{360}
\psarc[linecolor=black,linewidth=0.5pt]{-}(-1.6,0){1.2}{0}{360}
\psarc[linecolor=black,linewidth=0.5pt]{-}(1.6,0){1.2}{0}{360}
\psline[linestyle=dashed, dash= 1.5pt 1.5pt,linewidth=0.5pt]{-}(-1.1,0)(-0.4,0)
\psline[linestyle=dashed, dash= 1.5pt 1.5pt,linewidth=0.5pt]{-}(1.1,0)(0.4,0)
\psline[linestyle=dashed, dash= 1.5pt 1.5pt,linewidth=0.5pt]{-}(-0.4,0)(0,-3.5)
\psline[linestyle=dashed, dash= 1.5pt 1.5pt,linewidth=0.5pt]{-}(0.4,0)(0,-3.5)
\psline[linestyle=dashed, dash= 1.5pt 1.5pt,linewidth=0.5pt]{-}(0,-3.5)(0,-4.5)
\rput(1.6,0){
\psline[linecolor=blue,linewidth=1.5pt]{-}(0.191342, -0.46194)(0.45922, -1.10866)
\psline[linecolor=blue,linewidth=1.5pt]{-}(0.46194, -0.191342)(1.10866, -0.45922)
\psline[linecolor=blue,linewidth=1.5pt]{-}(-0.191342, 0.46194)(-0.45922, 1.10866)
\psline[linecolor=blue,linewidth=1.5pt]{-}(-0.46194, 0.191342)(-1.10866, 0.45922)
\psbezier[linecolor=blue,linewidth=1.5pt]{-}(1.109, 0.459)(0.785, 0.325)(0.325, 0.785)(0.459, 1.109)
\psline[linecolor=blue,linewidth=1.5pt]{-}(-0.46194, -0.191342)(-1.10866, -0.45922)
\psline[linecolor=blue,linewidth=1.5pt]{-}(-0.191342, -0.46194)(-0.45922, -1.10866)
}
\rput(-1.6,0){
\psline[linecolor=blue,linewidth=1.5pt]{-}(0, -0.5)(0., -1.2)
\psline[linecolor=blue,linewidth=1.5pt]{-}(0.433013, -0.25)(1.03923, -0.6)
\psline[linecolor=blue,linewidth=1.5pt]{-}(0.433013, 0.25)(1.03923, 0.6)
\psline[linecolor=blue,linewidth=1.5pt]{-}(0, 0.5)(0., 1.2)
\psline[linecolor=blue,linewidth=1.5pt]{-}(-0.433013, 0.25)(-1.03923, 0.6)
\psline[linecolor=blue,linewidth=1.5pt]{-}(-0.433013, -0.25)(-1.03923, -0.6)
}
\psline[linecolor=blue,linewidth=1.5pt]{-}(0.778823, -3.41225)(1.00134, -4.38718)
\psline[linecolor=blue,linewidth=1.5pt]{-}(2.18221, -2.73641)(2.8057, -3.51824)
\psline[linecolor=blue,linewidth=1.5pt]{-}(3.15339, -1.51859)(4.05436,-1.95248)
\psline[linecolor=blue,linewidth=1.5pt]{-}(3.5, 0.)(4.5, 0)
\psline[linecolor=blue,linewidth=1.5pt]{-}(3.15339, 1.51859)(4.05436, 1.95248)
\psline[linecolor=blue,linewidth=1.5pt]{-}(2.18221, 2.73641)(2.8057, 3.51824)
\psline[linecolor=blue,linewidth=1.5pt]{-}(0.778823, 3.41225)(1.00134, 4.38718)
\psline[linecolor=blue,linewidth=1.5pt]{-}(-0.778823, 3.41225)(-1.00134, 4.38718)
\psline[linecolor=blue,linewidth=1.5pt]{-}(-2.18221, 2.73641)(-2.8057, 3.51824)
\psline[linecolor=blue,linewidth=1.5pt]{-}(-3.15339, 1.51859)(-4.05436, 1.95248)
\psline[linecolor=blue,linewidth=1.5pt]{-}(-3.5, 0.)(-4.5, 0)
\psline[linecolor=blue,linewidth=1.5pt]{-}(-3.15339, -1.51859)(-4.05436, -1.95248)
\psline[linecolor=blue,linewidth=1.5pt]{-}(-2.18221, -2.73641)(-2.8057, -3.5182)
\psline[linecolor=blue,linewidth=1.5pt]{-}(-0.778823, -3.41225)(-1.00134, -4.38718)
\psbezier[linecolor=blue,linewidth=1.5pt]{-}(0.491345, -0.45922)(0.121793, -0.612293)(0.689815, -3.02228)(0.778823, -3.41225)
\psbezier[linecolor=blue,linewidth=1.5pt]{-}(1.14078, -1.10866)(0.987707, -1.47821)(1.93282, -2.42368)(2.18221, -2.73641)
\psbezier[linecolor=blue,linewidth=1.5pt]{-}(2.05922, -1.10866)(2.21229, -1.47821)(2.793, -1.34504)(3.15339, -1.51859)
\psbezier[linecolor=blue,linewidth=1.5pt]{-}(2.70866, -0.45922)(3.07821, -0.612293)(3.1, 0.)(3.5, 0.)
\psbezier[linecolor=blue,linewidth=1.5pt]{-}(2.70866, 0.45922)(3.07821, 0.612293)(2.793, 1.34504)(3.15339, 1.51859)
\psbezier[linecolor=blue,linewidth=1.5pt]{-}(2.05922, 1.10866)(2.21229, 1.47821)(1.93282, 2.42368)(2.18221, 2.73641)
\psbezier[linecolor=blue,linewidth=1.5pt]{-}(1.14078, 1.10866)(0.987707, 1.47821)(0.689815, 3.02228)(0.778823, 3.41225)
\psbezier[linecolor=blue,linewidth=1.5pt]{-}(0.491345, 0.45922)(0.121793, 0.612293)(-0.689815, 3.02228)(-0.778823, 3.41225)
\psbezier[linecolor=blue,linewidth=1.5pt]{-}(-0.56077, 0.6)(-0.214359, 0.8)(-1.93282, 2.42368)(-2.18221, 2.73641)
\psbezier[linecolor=blue,linewidth=1.5pt]{-}(-1.6, 1.2)(-1.6, 1.6)(-2.793, 1.34504)(-3.15339, 1.51859)
\psbezier[linecolor=blue,linewidth=1.5pt]{-}(-2.63923, 0.6)(-2.98564, 0.8)(-3.1, 0.)(-3.5, 0.)
\psbezier[linecolor=blue,linewidth=1.5pt]{-}(-2.63923, -0.6)(-2.98564, -0.8)(-2.793, -1.34504)(-3.15339, -1.51859)
\psbezier[linecolor=blue,linewidth=1.5pt]{-}(-1.6, -1.2)(-1.6, -1.6)(-1.93282, -2.42368)(-2.18221, -2.73641)
\psbezier[linecolor=blue,linewidth=1.5pt]{-}(-0.56077, -0.6)(-0.214359, -0.8)(-0.689815, -3.02228)(-0.778823, -3.41225)
\psarc[linecolor=purple,linewidth=1.5pt]{-}(-1.6,0){0.5}{0}{360}
\psarc[linecolor=darkgreen,linewidth=1.5pt]{-}(1.6,0){0.5}{0}{360}
\psarc[linecolor=black,linewidth=0.5pt]{-}(0,0){3.5}{0}{360}
\end{pspicture}} \ \, ,
\qquad
\thispic{\begin{pspicture}[shift=-4.4](-4.5,-4.5)(4.5,4.5)
\psarc[linecolor=black,linewidth=0.5pt,fillstyle=solid,fillcolor=lightlightblue]{-}(0,0){4.5}{0}{360}
\psarc[linecolor=black,linewidth=0.5pt,fillstyle=solid,fillcolor=white]{-}(-1.6,0){0.5}{0}{360}
\psarc[linecolor=black,linewidth=0.5pt,fillstyle=solid,fillcolor=white]{-}(1.6,0){0.5}{0}{360}
\psarc[linecolor=black,linewidth=0.5pt]{-}(-1.6,0){1.2}{0}{360}
\psarc[linecolor=black,linewidth=0.5pt]{-}(1.6,0){1.2}{0}{360}
\psline[linestyle=dashed, dash= 1.5pt 1.5pt,linewidth=0.5pt]{-}(-1.1,0)(-0.4,0)
\psline[linestyle=dashed, dash= 1.5pt 1.5pt,linewidth=0.5pt]{-}(1.1,0)(0.4,0)
\psline[linestyle=dashed, dash= 1.5pt 1.5pt,linewidth=0.5pt]{-}(-0.4,0)(0,-3.5)
\psline[linestyle=dashed, dash= 1.5pt 1.5pt,linewidth=0.5pt]{-}(0.4,0)(0,-3.5)
\psline[linestyle=dashed, dash= 1.5pt 1.5pt,linewidth=0.5pt]{-}(0,-3.5)(0,-4.5)
\rput(1.6,0){
\psline[linecolor=blue,linewidth=1.5pt]{-}(0.191342, -0.46194)(0.45922, -1.10866)
\psline[linecolor=blue,linewidth=1.5pt]{-}(0.46194, -0.191342)(1.10866, -0.45922)
\psline[linecolor=blue,linewidth=1.5pt]{-}(0.46194, 0.191342)(1.10866, 0.45922)
\psline[linecolor=blue,linewidth=1.5pt]{-}(0.191342, 0.46194)(0.45922, 1.10866)
\psline[linecolor=blue,linewidth=1.5pt]{-}(-0.191342, 0.46194)(-0.45922, 1.10866)
\psline[linecolor=blue,linewidth=1.5pt]{-}(-0.46194, 0.191342)(-1.10866, 0.45922)
\psline[linecolor=blue,linewidth=1.5pt]{-}(-0.46194, -0.191342)(-1.10866, -0.45922)
\psline[linecolor=blue,linewidth=1.5pt]{-}(-0.191342, -0.46194)(-0.45922, -1.10866)
}
\rput(-1.6,0){
\psline[linecolor=blue,linewidth=1.5pt]{-}(0.354, 0.354)(0.849, 0.849)
\psline[linecolor=blue,linewidth=1.5pt]{-}(-0.354, 0.354)(-0.849, 0.849)
\psline[linecolor=blue,linewidth=1.5pt]{-}(-0.354, -0.354)(-0.849, -0.849)
\psline[linecolor=blue,linewidth=1.5pt]{-}(0.354, -0.354)(0.849, -0.849)
}
\psline[linecolor=blue,linewidth=1.5pt]{-}(0.778823, -3.41225)(1.00134, -4.38718)
\psline[linecolor=blue,linewidth=1.5pt]{-}(2.18221, -2.73641)(2.8057, -3.51824)
\psline[linecolor=blue,linewidth=1.5pt]{-}(3.15339, -1.51859)(4.05436,-1.95248)
\psline[linecolor=blue,linewidth=1.5pt]{-}(3.5, 0.)(4.5, 0)
\psline[linecolor=blue,linewidth=1.5pt]{-}(3.15339, 1.51859)(4.05436, 1.95248)
\psline[linecolor=blue,linewidth=1.5pt]{-}(2.18221, 2.73641)(2.8057, 3.51824)
\psline[linecolor=blue,linewidth=1.5pt]{-}(0.778823, 3.41225)(1.00134, 4.38718)
\psline[linecolor=blue,linewidth=1.5pt]{-}(-0.778823, 3.41225)(-1.00134, 4.38718)
\psline[linecolor=blue,linewidth=1.5pt]{-}(-2.18221, 2.73641)(-2.8057, 3.51824)
\psline[linecolor=blue,linewidth=1.5pt]{-}(-3.15339, 1.51859)(-4.05436, 1.95248)
\psbezier[linecolor=blue,linewidth=1.5pt]{-}(-4.5, 0.)(-4., 0.)(-3.604, -1.736)(-4.054, -1.952)
\psline[linecolor=blue,linewidth=1.5pt]{-}(-2.18221, -2.73641)(-2.8057, -3.5182)
\psline[linecolor=blue,linewidth=1.5pt]{-}(-0.778823, -3.41225)(-1.00134, -4.38718)
\psbezier[linecolor=blue,linewidth=1.5pt]{-}(0.491345, -0.45922)(0.121793, -0.612293)(0.689815, -3.02228)(0.778823, -3.41225)
\psbezier[linecolor=blue,linewidth=1.5pt]{-}(1.14078, -1.10866)(0.987707, -1.47821)(1.93282, -2.42368)(2.18221, -2.73641)
\psbezier[linecolor=blue,linewidth=1.5pt]{-}(2.05922, -1.10866)(2.21229, -1.47821)(2.793, -1.34504)(3.15339, -1.51859)
\psbezier[linecolor=blue,linewidth=1.5pt]{-}(2.70866, -0.45922)(3.07821, -0.612293)(3.1, 0.)(3.5, 0.)
\psbezier[linecolor=blue,linewidth=1.5pt]{-}(2.70866, 0.45922)(3.07821, 0.612293)(2.793, 1.34504)(3.15339, 1.51859)
\psbezier[linecolor=blue,linewidth=1.5pt]{-}(2.05922, 1.10866)(2.21229, 1.47821)(1.93282, 2.42368)(2.18221, 2.73641)
\psbezier[linecolor=blue,linewidth=1.5pt]{-}(1.14078, 1.10866)(0.987707, 1.47821)(0.689815, 3.02228)(0.778823, 3.41225)
\psbezier[linecolor=blue,linewidth=1.5pt]{-}(0.491345, 0.45922)(0.121793, 0.612293)(-0.689815, 3.02228)(-0.778823, 3.41225)
\psbezier[linecolor=blue,linewidth=1.5pt]{-}(-2.182, 2.736)(-1.746, 2.189)(-0.327, 1.273)(-0.751, 0.849)
\psbezier[linecolor=blue,linewidth=1.5pt]{-}(-3.153, 1.519)(-2.838, 1.367)(-2.703, 1.103)(-2.449, 0.849)
\psbezier[linecolor=blue,linewidth=1.5pt]{-}(-2.182, -2.736)(-1.746, -2.189)(-2.873, -1.273)(-2.449, -0.849)
\psbezier[linecolor=blue,linewidth=1.5pt]{-}(-0.779, -3.412)(-0.623, -2.73)(-0.327, -1.273)(-0.751, -0.849)
\psarc[linecolor=purple,linewidth=1.5pt]{-}(-1.6,0){0.5}{0}{360}
\psarc[linecolor=darkgreen,linewidth=1.5pt]{-}(1.6,0){0.5}{0}{360}
\psarc[linecolor=black,linewidth=0.5pt]{-}(0,0){3.5}{0}{360}
\end{pspicture}}
\ = \
\thispic{\begin{pspicture}[shift=-4.4](-4.5,-4.5)(4.5,4.5)
\psarc[linecolor=black,linewidth=0.5pt,fillstyle=solid,fillcolor=lightlightblue]{-}(0,0){4.5}{0}{360}
\psarc[linecolor=black,linewidth=0.5pt,fillstyle=solid,fillcolor=white]{-}(-1.6,0){0.5}{0}{360}
\psarc[linecolor=black,linewidth=0.5pt,fillstyle=solid,fillcolor=white]{-}(1.6,0){0.5}{0}{360}
\psarc[linecolor=black,linewidth=0.5pt]{-}(-1.6,0){1.2}{0}{360}
\psarc[linecolor=black,linewidth=0.5pt]{-}(1.6,0){1.2}{0}{360}
\psline[linestyle=dashed, dash= 1.5pt 1.5pt,linewidth=0.5pt]{-}(-1.1,0)(-0.4,0)
\psline[linestyle=dashed, dash= 1.5pt 1.5pt,linewidth=0.5pt]{-}(1.1,0)(0.4,0)
\psline[linestyle=dashed, dash= 1.5pt 1.5pt,linewidth=0.5pt]{-}(-0.4,0)(0,-3.5)
\psline[linestyle=dashed, dash= 1.5pt 1.5pt,linewidth=0.5pt]{-}(0.4,0)(0,-3.5)
\psline[linestyle=dashed, dash= 1.5pt 1.5pt,linewidth=0.5pt]{-}(0,-3.5)(0,-4.5)
\rput(1.6,0){
\psline[linecolor=blue,linewidth=1.5pt]{-}(0.191342, -0.46194)(0.45922, -1.10866)
\psline[linecolor=blue,linewidth=1.5pt]{-}(0.46194, -0.191342)(1.10866, -0.45922)
\psline[linecolor=blue,linewidth=1.5pt]{-}(0.46194, 0.191342)(1.10866, 0.45922)
\psline[linecolor=blue,linewidth=1.5pt]{-}(0.191342, 0.46194)(0.45922, 1.10866)
\psline[linecolor=blue,linewidth=1.5pt]{-}(-0.191342, 0.46194)(-0.45922, 1.10866)
\psline[linecolor=blue,linewidth=1.5pt]{-}(-0.46194, 0.191342)(-1.10866, 0.45922)
\psline[linecolor=blue,linewidth=1.5pt]{-}(-0.46194, -0.191342)(-1.10866, -0.45922)
\psline[linecolor=blue,linewidth=1.5pt]{-}(-0.191342, -0.46194)(-0.45922, -1.10866)
}
\rput(-1.6,0){
\psline[linecolor=blue,linewidth=1.5pt]{-}(0, -0.5)(0., -1.2)
\psline[linecolor=blue,linewidth=1.5pt]{-}(0.433013, -0.25)(1.03923, -0.6)
\psline[linecolor=blue,linewidth=1.5pt]{-}(0.433013, 0.25)(1.03923, 0.6)
\psline[linecolor=blue,linewidth=1.5pt]{-}(0, 0.5)(0., 1.2)
\psbezier[linecolor=blue,linewidth=1.5pt]{-}(-1.039, 0.6)(-0.736, 0.425)(-0.736, -0.425)(-1.039, -0.6)
}
\psline[linecolor=blue,linewidth=1.5pt]{-}(0.778823, -3.41225)(1.00134, -4.38718)
\psline[linecolor=blue,linewidth=1.5pt]{-}(2.18221, -2.73641)(2.8057, -3.51824)
\psline[linecolor=blue,linewidth=1.5pt]{-}(3.15339, -1.51859)(4.05436,-1.95248)
\psline[linecolor=blue,linewidth=1.5pt]{-}(3.5, 0.)(4.5, 0)
\psline[linecolor=blue,linewidth=1.5pt]{-}(3.15339, 1.51859)(4.05436, 1.95248)
\psline[linecolor=blue,linewidth=1.5pt]{-}(2.18221, 2.73641)(2.8057, 3.51824)
\psline[linecolor=blue,linewidth=1.5pt]{-}(0.778823, 3.41225)(1.00134, 4.38718)
\psline[linecolor=blue,linewidth=1.5pt]{-}(-0.778823, 3.41225)(-1.00134, 4.38718)
\psline[linecolor=blue,linewidth=1.5pt]{-}(-2.18221, 2.73641)(-2.8057, 3.51824)
\psline[linecolor=blue,linewidth=1.5pt]{-}(-3.15339, 1.51859)(-4.05436, 1.95248)
\psline[linecolor=blue,linewidth=1.5pt]{-}(-3.5, 0.)(-4.5, 0)
\psline[linecolor=blue,linewidth=1.5pt]{-}(-3.15339, -1.51859)(-4.05436, -1.95248)
\psline[linecolor=blue,linewidth=1.5pt]{-}(-2.18221, -2.73641)(-2.8057, -3.5182)
\psline[linecolor=blue,linewidth=1.5pt]{-}(-0.778823, -3.41225)(-1.00134, -4.38718)
\psbezier[linecolor=blue,linewidth=1.5pt]{-}(0.491345, -0.45922)(0.121793, -0.612293)(0.689815, -3.02228)(0.778823, -3.41225)
\psbezier[linecolor=blue,linewidth=1.5pt]{-}(1.14078, -1.10866)(0.987707, -1.47821)(1.93282, -2.42368)(2.18221, -2.73641)
\psbezier[linecolor=blue,linewidth=1.5pt]{-}(2.05922, -1.10866)(2.21229, -1.47821)(2.793, -1.34504)(3.15339, -1.51859)
\psbezier[linecolor=blue,linewidth=1.5pt]{-}(2.70866, -0.45922)(3.07821, -0.612293)(3.1, 0.)(3.5, 0.)
\psbezier[linecolor=blue,linewidth=1.5pt]{-}(2.70866, 0.45922)(3.07821, 0.612293)(2.793, 1.34504)(3.15339, 1.51859)
\psbezier[linecolor=blue,linewidth=1.5pt]{-}(2.05922, 1.10866)(2.21229, 1.47821)(1.93282, 2.42368)(2.18221, 2.73641)
\psbezier[linecolor=blue,linewidth=1.5pt]{-}(1.14078, 1.10866)(0.987707, 1.47821)(0.689815, 3.02228)(0.778823, 3.41225)
\psbezier[linecolor=blue,linewidth=1.5pt]{-}(0.491345, 0.45922)(0.121793, 0.612293)(-0.689815, 3.02228)(-0.778823, 3.41225)
\psbezier[linecolor=blue,linewidth=1.5pt]{-}(-0.56077, 0.6)(-0.214359, 0.8)(-1.93282, 2.42368)(-2.18221, 2.73641)
\psbezier[linecolor=blue,linewidth=1.5pt]{-}(-1.6, 1.2)(-1.6, 1.6)(-2.793, 1.34504)(-3.15339, 1.51859)
\psbezier[linecolor=blue,linewidth=1.5pt]{-}(-2.63923, 0.6)(-2.98564, 0.8)(-3.1, 0.)(-3.5, 0.)
\psbezier[linecolor=blue,linewidth=1.5pt]{-}(-2.63923, -0.6)(-2.98564, -0.8)(-2.793, -1.34504)(-3.15339, -1.51859)
\psbezier[linecolor=blue,linewidth=1.5pt]{-}(-1.6, -1.2)(-1.6, -1.6)(-1.93282, -2.42368)(-2.18221, -2.73641)
\psbezier[linecolor=blue,linewidth=1.5pt]{-}(-0.56077, -0.6)(-0.214359, -0.8)(-0.689815, -3.02228)(-0.778823, -3.41225)
\psarc[linecolor=purple,linewidth=1.5pt]{-}(-1.6,0){0.5}{0}{360}
\psarc[linecolor=darkgreen,linewidth=1.5pt]{-}(1.6,0){0.5}{0}{360}
\psarc[linecolor=black,linewidth=0.5pt]{-}(0,0){3.5}{0}{360}
\end{pspicture}} \ \, ,
\ee\vspace{0.1cm}
\be
\psset{unit=0.35cm}
\thispic{\begin{pspicture}[shift=-4.4](-4.5,-4.5)(4.5,4.5)
\psarc[linecolor=black,linewidth=0.5pt,fillstyle=solid,fillcolor=lightlightblue]{-}(0,0){4.5}{0}{360}
\psarc[linecolor=black,linewidth=0.5pt,fillstyle=solid,fillcolor=white]{-}(-1.6,0){0.5}{0}{360}
\psarc[linecolor=black,linewidth=0.5pt,fillstyle=solid,fillcolor=white]{-}(1.6,0){0.5}{0}{360}
\psarc[linecolor=black,linewidth=0.5pt]{-}(-1.6,0){1.2}{0}{360}
\psarc[linecolor=black,linewidth=0.5pt]{-}(1.6,0){1.2}{0}{360}
\psline[linestyle=dashed, dash= 1.5pt 1.5pt,linewidth=0.5pt]{-}(-1.1,0)(-0.4,0)
\psline[linestyle=dashed, dash= 1.5pt 1.5pt,linewidth=0.5pt]{-}(1.1,0)(0.4,0)
\psline[linestyle=dashed, dash= 1.5pt 1.5pt,linewidth=0.5pt]{-}(-0.4,0)(0,-3.5)
\psline[linestyle=dashed, dash= 1.5pt 1.5pt,linewidth=0.5pt]{-}(0.4,0)(0,-3.5)
\psline[linestyle=dashed, dash= 1.5pt 1.5pt,linewidth=0.5pt]{-}(0,-3.5)(0,-4.5)
\rput(1.6,0){
\psline[linecolor=blue,linewidth=1.5pt]{-}(0.191342, -0.46194)(0.45922, -1.10866)
\psline[linecolor=blue,linewidth=1.5pt]{-}(0.46194, -0.191342)(1.10866, -0.45922)
\psline[linecolor=blue,linewidth=1.5pt]{-}(0.46194, 0.191342)(1.10866, 0.45922)
\psline[linecolor=blue,linewidth=1.5pt]{-}(0.191342, 0.46194)(0.45922, 1.10866)
\psline[linecolor=blue,linewidth=1.5pt]{-}(-0.191342, 0.46194)(-0.45922, 1.10866)
\psline[linecolor=blue,linewidth=1.5pt]{-}(-0.46194, 0.191342)(-1.10866, 0.45922)
\psline[linecolor=blue,linewidth=1.5pt]{-}(-0.46194, -0.191342)(-1.10866, -0.45922)
\psline[linecolor=blue,linewidth=1.5pt]{-}(-0.191342, -0.46194)(-0.45922, -1.10866)
}
\rput(-1.6,0){
\psline[linecolor=blue,linewidth=1.5pt]{-}(0, -0.5)(0., -1.2)
\psline[linecolor=blue,linewidth=1.5pt]{-}(0.433013, -0.25)(1.03923, -0.6)
\psline[linecolor=blue,linewidth=1.5pt]{-}(0.433013, 0.25)(1.03923, 0.6)
\psline[linecolor=blue,linewidth=1.5pt]{-}(0, 0.5)(0., 1.2)
\psline[linecolor=blue,linewidth=1.5pt]{-}(-0.433013, 0.25)(-1.03923, 0.6)
\psline[linecolor=blue,linewidth=1.5pt]{-}(-0.433013, -0.25)(-1.03923, -0.6)
}
\psbezier[linecolor=blue,linewidth=1.5pt]{-}(0.779, -3.412)(0.89, -3.9)(-0.89, -3.9)(-1.001, -4.387)
\psbezier[linecolor=blue,linewidth=1.5pt]{-}(2.182, -2.736)(2.494, -3.127)(0.89, -3.9)(1.001, -4.387)
\psbezier[linecolor=blue,linewidth=1.5pt]{-}(3.153, -1.519)(3.604, -1.736)(2.494, -3.127)(2.806, -3.518)
\psbezier[linecolor=blue,linewidth=1.5pt]{-}(3.5, 0.)(4., 0.)(3.604, -1.736)(4.054, -1.952)
\psbezier[linecolor=blue,linewidth=1.5pt]{-}(3.153, 1.519)(3.604, 1.736)(4., 0.)(4.5, 0)
\psbezier[linecolor=blue,linewidth=1.5pt]{-}(2.182, 2.736)(2.494, 3.127)(3.604, 1.736)(4.054, 1.952)
\psbezier[linecolor=blue,linewidth=1.5pt]{-}(0.779, 3.412)(0.89, 3.9)(2.494, 3.127)(2.806, 3.518)
\psbezier[linecolor=blue,linewidth=1.5pt]{-}(-0.779, 3.412)(-0.89, 3.9)(0.89, 3.9)(1.001, 4.387)
\psbezier[linecolor=blue,linewidth=1.5pt]{-}(-2.182, 2.736)(-2.494, 3.127)(-0.89, 3.9)(-1.001, 4.387)
\psbezier[linecolor=blue,linewidth=1.5pt]{-}(-3.153, 1.519)(-3.604, 1.736)(-2.494, 3.127)(-2.806, 3.518)
\psbezier[linecolor=blue,linewidth=1.5pt]{-}(-3.5, 0.)(-4., 0.)(-3.604, 1.736)(-4.054, 1.952)
\psbezier[linecolor=blue,linewidth=1.5pt]{-}(-3.153, -1.519)(-3.604, -1.736)(-4., 0.)(-4.5, 0.)
\psbezier[linecolor=blue,linewidth=1.5pt]{-}(-2.182, -2.736)(-2.494, -3.127)(-3.604, -1.736)(-4.054, -1.952)
\psbezier[linecolor=blue,linewidth=1.5pt]{-}(-0.779, -3.412)(-0.89, -3.9)(-2.494, -3.127)(-2.806, -3.518)
\psbezier[linecolor=blue,linewidth=1.5pt]{-}(0.491345, -0.45922)(0.121793, -0.612293)(0.689815, -3.02228)(0.778823, -3.41225)
\psbezier[linecolor=blue,linewidth=1.5pt]{-}(1.14078, -1.10866)(0.987707, -1.47821)(1.93282, -2.42368)(2.18221, -2.73641)
\psbezier[linecolor=blue,linewidth=1.5pt]{-}(2.05922, -1.10866)(2.21229, -1.47821)(2.793, -1.34504)(3.15339, -1.51859)
\psbezier[linecolor=blue,linewidth=1.5pt]{-}(2.70866, -0.45922)(3.07821, -0.612293)(3.1, 0.)(3.5, 0.)
\psbezier[linecolor=blue,linewidth=1.5pt]{-}(2.70866, 0.45922)(3.07821, 0.612293)(2.793, 1.34504)(3.15339, 1.51859)
\psbezier[linecolor=blue,linewidth=1.5pt]{-}(2.05922, 1.10866)(2.21229, 1.47821)(1.93282, 2.42368)(2.18221, 2.73641)
\psbezier[linecolor=blue,linewidth=1.5pt]{-}(1.14078, 1.10866)(0.987707, 1.47821)(0.689815, 3.02228)(0.778823, 3.41225)
\psbezier[linecolor=blue,linewidth=1.5pt]{-}(0.491345, 0.45922)(0.121793, 0.612293)(-0.689815, 3.02228)(-0.778823, 3.41225)
\psbezier[linecolor=blue,linewidth=1.5pt]{-}(-2.182, 2.736)(-1.964, 2.463)(0.115, 0.99)(-0.561, 0.6)
\psbezier[linecolor=blue,linewidth=1.5pt]{-}(-3.153, 1.519)(-2.838, 1.367)(-1.6, 1.8)(-1.6, 1.2)
\psbezier[linecolor=blue,linewidth=1.5pt]{-}(-2.63923, 0.6)(-2.98564, 0.8)(-3.1, 0.)(-3.5, 0.)
\psbezier[linecolor=blue,linewidth=1.5pt]{-}(-2.63923, -0.6)(-2.98564, -0.8)(-2.793, -1.34504)(-3.15339, -1.51859)
\psbezier[linecolor=blue,linewidth=1.5pt]{-}(-1.6, -1.2)(-1.6, -1.6)(-1.93282, -2.42368)(-2.18221, -2.73641)
\psbezier[linecolor=blue,linewidth=1.5pt]{-}(-0.779, -3.412)(-0.701, -3.071)(-0.041, -0.9)(-0.561, -0.6)
\psarc[linecolor=purple,linewidth=1.5pt]{-}(-1.6,0){0.5}{0}{360}
\psarc[linecolor=darkgreen,linewidth=1.5pt]{-}(1.6,0){0.5}{0}{360}
\psarc[linecolor=black,linewidth=0.5pt]{-}(0,0){3.5}{0}{360}
\end{pspicture}}
\ = \
\thispic{\begin{pspicture}[shift=-4.4](-4.5,-4.5)(4.5,4.5)
\psarc[linecolor=black,linewidth=0.5pt,fillstyle=solid,fillcolor=lightlightblue]{-}(0,0){4.5}{0}{360}
\psarc[linecolor=black,linewidth=0.5pt,fillstyle=solid,fillcolor=white]{-}(-1.6,0){0.5}{0}{360}
\psarc[linecolor=black,linewidth=0.5pt,fillstyle=solid,fillcolor=white]{-}(1.6,0){0.5}{0}{360}
\psarc[linecolor=black,linewidth=0.5pt]{-}(-1.6,0){1.2}{0}{360}
\psarc[linecolor=black,linewidth=0.5pt]{-}(1.6,0){1.2}{0}{360}
\psline[linestyle=dashed, dash= 1.5pt 1.5pt,linewidth=0.5pt]{-}(-1.1,0)(-0.4,0)
\psline[linestyle=dashed, dash= 1.5pt 1.5pt,linewidth=0.5pt]{-}(1.1,0)(0.4,0)
\psline[linestyle=dashed, dash= 1.5pt 1.5pt,linewidth=0.5pt]{-}(-0.4,0)(0,-3.5)
\psline[linestyle=dashed, dash= 1.5pt 1.5pt,linewidth=0.5pt]{-}(0.4,0)(0,-3.5)
\psline[linestyle=dashed, dash= 1.5pt 1.5pt,linewidth=0.5pt]{-}(0,-3.5)(0,-4.5)
\rput(1.6,0){
\psbezier[linecolor=blue,linewidth=1.5pt]{-}(-0.462, -0.191)(-0.785, -0.325)(-0.785, 0.325)(-1.109, 0.459)
\psbezier[linecolor=blue,linewidth=1.5pt]{-}(-0.191, -0.462)(-0.325, -0.785)(-0.785, -0.325)(-1.109, -0.459)
\psbezier[linecolor=blue,linewidth=1.5pt]{-}(0.191, -0.462)(0.325, -0.785)(-0.325, -0.785)(-0.459, -1.109)
\psbezier[linecolor=blue,linewidth=1.5pt]{-}(0.462, -0.191)(0.785, -0.325)(0.325, -0.785)(0.459, -1.109)
\psbezier[linecolor=blue,linewidth=1.5pt]{-}(0.462, 0.191)(0.785, 0.325)(0.785, -0.325)(1.109, -0.459)
\psbezier[linecolor=blue,linewidth=1.5pt]{-}(0.191, 0.462)(0.325, 0.785)(0.785, 0.325)(1.109, 0.459)
\psbezier[linecolor=blue,linewidth=1.5pt]{-}(-0.191, 0.462)(-0.325, 0.785)(0.325, 0.785)(0.459, 1.109)
\psbezier[linecolor=blue,linewidth=1.5pt]{-}(-0.462, 0.191)(-0.785, 0.325)(-0.325, 0.785)(-0.459, 1.109)
}
\rput(-1.6,0){
\psline[linecolor=blue,linewidth=1.5pt]{-}(0.191, 0.462)(0.459, 1.109)
\psline[linecolor=blue,linewidth=1.5pt]{-}(-0.191, 0.462)(-0.459, 1.109)
\psline[linecolor=blue,linewidth=1.5pt]{-}(-0.462, 0.191)(-1.109, 0.459)
\psline[linecolor=blue,linewidth=1.5pt]{-}(-0.462, -0.191)(-1.109, -0.459)
\psline[linecolor=blue,linewidth=1.5pt]{-}(-0.191, -0.462)(-0.459, -1.109)
\psline[linecolor=blue,linewidth=1.5pt]{-}(0.191, -0.462)(0.459, -1.109)
\psbezier[linecolor=blue,linewidth=1.5pt]{-}(1.109, 0.459)(0.785, 0.325)(0.785, -0.325)(1.109, -0.459)
}
\psline[linecolor=blue,linewidth=1.5pt]{-}(0.683, -3.433)(0.878, -4.414)
\psline[linecolor=blue,linewidth=1.5pt]{-}(1.944, -2.91)(2.5, -3.742)
\psline[linecolor=blue,linewidth=1.5pt]{-}(2.91, -1.944)(3.742, -2.5)
\psline[linecolor=blue,linewidth=1.5pt]{-}(3.433, -0.683)(4.414, -0.878)
\psline[linecolor=blue,linewidth=1.5pt]{-}(3.433, 0.683)(4.414, 0.878)
\psline[linecolor=blue,linewidth=1.5pt]{-}(2.91, 1.944)(3.742, 2.5)
\psline[linecolor=blue,linewidth=1.5pt]{-}(1.944, 2.91)(2.5, 3.742)
\psbezier[linecolor=blue,linewidth=1.5pt]{-}(0.683, 3.433)(0.813, 4.087)(-0.813, 4.087)(-0.683, 3.433)
\psline[linecolor=blue,linewidth=1.5pt]{-}(-1.944, 2.91)(-2.5, 3.742)
\psline[linecolor=blue,linewidth=1.5pt]{-}(-2.91, 1.944)(-3.742, 2.5)
\psline[linecolor=blue,linewidth=1.5pt]{-}(-3.433, 0.683)(-4.414, 0.878)
\psline[linecolor=blue,linewidth=1.5pt]{-}(-3.433, -0.683)(-4.414, -0.878)
\psline[linecolor=blue,linewidth=1.5pt]{-}(-2.91, -1.944)(-3.742, -2.5)
\psline[linecolor=blue,linewidth=1.5pt]{-}(-1.944, -2.91)(-2.5, -3.742)
\psline[linecolor=blue,linewidth=1.5pt]{-}(-0.683, -3.433)(-0.878, -4.414)
\psbezier[linecolor=blue,linewidth=1.5pt]{-}(0.683, -3.433)(0.546, -2.746)(-0.063, -0.689)(0.491, -0.459)
\psbezier[linecolor=blue,linewidth=1.5pt]{-}(1.944, -2.91)(1.556, -2.328)(0.911, -1.663)(1.141, -1.109)
\psbezier[linecolor=blue,linewidth=1.5pt]{-}(2.91, -1.944)(2.328, -1.556)(2.289, -1.663)(2.059, -1.109)
\psbezier[linecolor=blue,linewidth=1.5pt]{-}(3.433, -0.683)(3.089, -0.615)(2.93, -0.551)(2.709, -0.459)
\psbezier[linecolor=blue,linewidth=1.5pt]{-}(3.433, 0.683)(3.089, 0.615)(2.93, 0.551)(2.709, 0.459)
\psbezier[linecolor=blue,linewidth=1.5pt]{-}(2.91, 1.944)(2.328, 1.556)(2.289, 1.663)(2.059, 1.109)
\psbezier[linecolor=blue,linewidth=1.5pt]{-}(1.944, 2.91)(1.556, 2.328)(0.911, 1.663)(1.141, 1.109)
\psbezier[linecolor=blue,linewidth=1.5pt]{-}(0.683, 3.433)(0.546, 2.746)(-0.063, 0.689)(0.491, 0.459)

\psbezier[linecolor=blue,linewidth=1.5pt]{-}(-0.683, 3.433)(-0.546, 2.746)(0.063, 0.689)(-0.491, 0.459)
\psbezier[linecolor=blue,linewidth=1.5pt]{-}(-1.944, 2.91)(-1.556, 2.328)(-0.911, 1.663)(-1.141, 1.109)
\psbezier[linecolor=blue,linewidth=1.5pt]{-}(-2.91, 1.944)(-2.328, 1.556)(-2.289, 1.663)(-2.059, 1.109)
\psbezier[linecolor=blue,linewidth=1.5pt]{-}(-3.433, 0.683)(-3.089, 0.615)(-2.93, 0.551)(-2.709, 0.459)
\psbezier[linecolor=blue,linewidth=1.5pt]{-}(-3.433, -0.683)(-3.089, -0.615)(-2.93, -0.551)(-2.709, -0.459)
\psbezier[linecolor=blue,linewidth=1.5pt]{-}(-2.91, -1.944)(-2.328, -1.556)(-2.289, -1.663)(-2.059, -1.109)
\psbezier[linecolor=blue,linewidth=1.5pt]{-}(-1.944, -2.91)(-1.556, -2.328)(-0.911, -1.663)(-1.141, -1.109)
\psbezier[linecolor=blue,linewidth=1.5pt]{-}(-0.683, -3.433)(-0.546, -2.746)(0.063, -0.689)(-0.491, -0.459)
\psarc[linecolor=purple,linewidth=1.5pt]{-}(-1.6,0){0.5}{0}{360}
\psarc[linecolor=darkgreen,linewidth=1.5pt]{-}(1.6,0){0.5}{0}{360}
\psarc[linecolor=black,linewidth=0.5pt]{-}(0,0){3.5}{0}{360}
\end{pspicture}} \ ,
\ee
\end{subequations}
where we replaced $\lambda$, $\lambda_\pa$ and $\lambda_\pb$ by the identity for simplicity. Many more relations can be derived from \eqref{eq:Lambda.equations}, in particular 
\begin{subequations} \label{eq:Lambda.equations.extra}
\begin{alignat}{3}
&\lambda\, \Omega^{-1}(\lambda_\pa \otimes \lambda_\pb) = \lambda\, c_{N_\pa+2}\,(c_0^\dag \lambda_\pa \otimes \Omega^{-1} \lambda_\pb)\,,
\label{eq:Lambda.equations.f}
&\qquad& \textrm{for $N_\pb > 0$,} \\[0.15cm]
&\lambda\, c_0\,(\lambda_\pa \otimes \lambda_\pb) = \lambda\, c_{N_\pa} \,(\Omega\,\lambda_\pa \otimes \Omega^{-1} \lambda_\pb)\,,\quad
&& \textrm{for $N_\pa > 0$ and $N_\pb > 0$,}
\label{eq:Lambda.equations.g} \\[0.15cm]
\label{eq:Lambda.equations.h}
&\lambda\,\Omega\, (c_0^\dagger \lambda_\pa \otimes \lambda_\pb) = \lambda\, (\lambda_\pa \otimes c_0^\dagger\lambda_\pb)\,, \\[0.15cm]
\label{eq:Lambda.equations.i}
& \lambda\,c_0\, (\lambda_\pa \otimes c_0^\dagger \lambda_\pb) = \lambda\, (f\,\lambda_\pa \otimes \lambda_\pb) \,,
&&\textrm{for $N_\pa = 0$,} \\[0.15cm]
\label{eq:Lambda.equations.j}
& \lambda\,\Omega\, (\lambda_\pa \otimes \lambda_\pb) = \lambda\, c_1c_2\,(c_0^\dag\lambda_\pa \otimes c_0^\dag\lambda_\pb) \,,
&&\textrm{for $N_\pa = 0$ and $N_\pb>0$.}
\end{alignat}
\end{subequations}
For example, to prove \eqref{eq:Lambda.equations.g}, we replace $\lambda$ by $\lambda\, c_0\, \Omega^{-1}$ in \eqref{eq:Lambda.equations.e} and find
\begin{alignat}{2}
\lambda\, c_0 \,(\lambda_\pa \otimes \lambda_\pb) 
&= \lambda\, c_0\, \Omega^{-1}\, c_{N_\pa}\,(\Omega \,\lambda_\pa \otimes c_0^\dag \lambda_\pb) 
= \lambda\, c_{N_\pa+N_\pb-1}\, c_{N_\pa}\,(\Omega \,\lambda_\pa \otimes c_0^\dag \lambda_\pb)
\nn\\[0.1cm]&
= \lambda\, c_{N_\pa}\, c_{N_\pa+N_\pb+1}\,(\Omega \,\lambda_\pa \otimes c_0^\dag \lambda_\pb)
= \lambda\, c_{N_\pa}\, (\Omega \,\lambda_\pa \otimes c_{N_\pb+1}\,c_0^\dag \lambda_\pb)
\nn\\[0.1cm]&
= \lambda\, c_{N_\pa}\, (\Omega \,\lambda_\pa \otimes \Omega^{-1} \lambda_\pb)\,.
\end{alignat}
Similarly, to prove \eqref{eq:Lambda.equations.i}, we write
\begin{equation}
\lambda\,(f\,\lambda_\pa \otimes \lambda_\pb) 
= \lambda\,(c_1 c_0^\dag\lambda_\pa \otimes \lambda_\pb)
= \lambda \, c_1 (c_0^\dag\lambda_\pa \otimes \lambda_\pb)
= \lambda \, c_1 \Omega^{-1} (\lambda_\pa \otimes c_0^\dag\lambda_\pb)
= \lambda \, c_0 (\lambda_\pa \otimes c_0^\dag\lambda_\pb)\,,
\end{equation}
where we used \eqref{eq:Lambda.equations.a} and \eqref{eq:Lambda.equations.h}.
Notably, \eqref{eq:Lambda.equations.g} and \eqref{eq:Lambda.equations.h} describe the two ways that loop segments can, respectively, connect the holes $\pa$ to $\pb$, and pass between $\pa$ and $\pb$:
\be
\psset{unit=0.35cm}
\thispic{\begin{pspicture}[shift=-4.4](-4.5,-4.5)(4.5,4.5)
\psarc[linecolor=black,linewidth=0.5pt,fillstyle=solid,fillcolor=lightlightblue]{-}(0,0){4.5}{0}{360}
\psarc[linecolor=black,linewidth=0.5pt,fillstyle=solid,fillcolor=white]{-}(-1.6,0){0.5}{0}{360}
\psarc[linecolor=black,linewidth=0.5pt,fillstyle=solid,fillcolor=white]{-}(1.6,0){0.5}{0}{360}
\psarc[linecolor=black,linewidth=0.5pt]{-}(-1.6,0){1.2}{0}{360}
\psarc[linecolor=black,linewidth=0.5pt]{-}(1.6,0){1.2}{0}{360}
\psline[linestyle=dashed, dash= 1.5pt 1.5pt,linewidth=0.5pt]{-}(-1.1,0)(-0.4,0)
\psline[linestyle=dashed, dash= 1.5pt 1.5pt,linewidth=0.5pt]{-}(1.1,0)(0.4,0)
\psline[linestyle=dashed, dash= 1.5pt 1.5pt,linewidth=0.5pt]{-}(-0.4,0)(0,-3.5)
\psline[linestyle=dashed, dash= 1.5pt 1.5pt,linewidth=0.5pt]{-}(0.4,0)(0,-3.5)
\psline[linestyle=dashed, dash= 1.5pt 1.5pt,linewidth=0.5pt]{-}(0,-3.5)(0,-4.5)
\rput(1.6,0){
\psline[linecolor=blue,linewidth=1.5pt]{-}(0.191342, -0.46194)(0.45922, -1.10866)
\psline[linecolor=blue,linewidth=1.5pt]{-}(0.46194, -0.191342)(1.10866, -0.45922)
\psline[linecolor=blue,linewidth=1.5pt]{-}(0.46194, 0.191342)(1.10866, 0.45922)
\psline[linecolor=blue,linewidth=1.5pt]{-}(0.191342, 0.46194)(0.45922, 1.10866)
\psline[linecolor=blue,linewidth=1.5pt]{-}(-0.191342, 0.46194)(-0.45922, 1.10866)
\psline[linecolor=blue,linewidth=1.5pt]{-}(-0.46194, 0.191342)(-1.10866, 0.45922)
\psline[linecolor=blue,linewidth=1.5pt]{-}(-0.46194, -0.191342)(-1.10866, -0.45922)
\psline[linecolor=blue,linewidth=1.5pt]{-}(-0.191342, -0.46194)(-0.45922, -1.10866)
}
\rput(-1.6,0){
\psline[linecolor=blue,linewidth=1.5pt]{-}(0, -0.5)(0., -1.2)
\psline[linecolor=blue,linewidth=1.5pt]{-}(0.433013, -0.25)(1.03923, -0.6)
\psline[linecolor=blue,linewidth=1.5pt]{-}(0.433013, 0.25)(1.03923, 0.6)
\psline[linecolor=blue,linewidth=1.5pt]{-}(0, 0.5)(0., 1.2)
\psline[linecolor=blue,linewidth=1.5pt]{-}(-0.433013, 0.25)(-1.03923, 0.6)
\psline[linecolor=blue,linewidth=1.5pt]{-}(-0.433013, -0.25)(-1.03923, -0.6)
}
\psbezier[linecolor=blue,linewidth=1.5pt]{-}(-0.779, -3.412)(-0.927, -4.062)(0.927, -4.062)(0.779, -3.412)
\psline[linecolor=blue,linewidth=1.5pt]{-}(2.18221, -2.73641)(2.8057, -3.51824)
\psline[linecolor=blue,linewidth=1.5pt]{-}(3.15339, -1.51859)(4.05436,-1.95248)
\psline[linecolor=blue,linewidth=1.5pt]{-}(3.5, 0.)(4.5, 0)
\psline[linecolor=blue,linewidth=1.5pt]{-}(3.15339, 1.51859)(4.05436, 1.95248)
\psline[linecolor=blue,linewidth=1.5pt]{-}(2.18221, 2.73641)(2.8057, 3.51824)
\psline[linecolor=blue,linewidth=1.5pt]{-}(0.778823, 3.41225)(1.00134, 4.38718)
\psline[linecolor=blue,linewidth=1.5pt]{-}(-0.778823, 3.41225)(-1.00134, 4.38718)
\psline[linecolor=blue,linewidth=1.5pt]{-}(-2.18221, 2.73641)(-2.8057, 3.51824)
\psline[linecolor=blue,linewidth=1.5pt]{-}(-3.15339, 1.51859)(-4.05436, 1.95248)
\psline[linecolor=blue,linewidth=1.5pt]{-}(-3.5, 0.)(-4.5, 0)
\psline[linecolor=blue,linewidth=1.5pt]{-}(-3.15339, -1.51859)(-4.05436, -1.95248)
\psline[linecolor=blue,linewidth=1.5pt]{-}(-2.18221, -2.73641)(-2.8057, -3.5182)
\psbezier[linecolor=blue,linewidth=1.5pt]{-}(0.491345, -0.45922)(0.121793, -0.612293)(0.689815, -3.02228)(0.778823, -3.41225)
\psbezier[linecolor=blue,linewidth=1.5pt]{-}(1.14078, -1.10866)(0.987707, -1.47821)(1.93282, -2.42368)(2.18221, -2.73641)
\psbezier[linecolor=blue,linewidth=1.5pt]{-}(2.05922, -1.10866)(2.21229, -1.47821)(2.793, -1.34504)(3.15339, -1.51859)
\psbezier[linecolor=blue,linewidth=1.5pt]{-}(2.70866, -0.45922)(3.07821, -0.612293)(3.1, 0.)(3.5, 0.)
\psbezier[linecolor=blue,linewidth=1.5pt]{-}(2.70866, 0.45922)(3.07821, 0.612293)(2.793, 1.34504)(3.15339, 1.51859)
\psbezier[linecolor=blue,linewidth=1.5pt]{-}(2.05922, 1.10866)(2.21229, 1.47821)(1.93282, 2.42368)(2.18221, 2.73641)
\psbezier[linecolor=blue,linewidth=1.5pt]{-}(1.14078, 1.10866)(0.987707, 1.47821)(0.689815, 3.02228)(0.778823, 3.41225)
\psbezier[linecolor=blue,linewidth=1.5pt]{-}(0.491345, 0.45922)(0.121793, 0.612293)(-0.689815, 3.02228)(-0.778823, 3.41225)
\psbezier[linecolor=blue,linewidth=1.5pt]{-}(-2.182, 2.736)(-1.964, 2.463)(0.115, 0.99)(-0.561, 0.6)
\psbezier[linecolor=blue,linewidth=1.5pt]{-}(-3.153, 1.519)(-2.838, 1.367)(-1.6, 1.8)(-1.6, 1.2)
\psbezier[linecolor=blue,linewidth=1.5pt]{-}(-2.63923, 0.6)(-2.98564, 0.8)(-3.1, 0.)(-3.5, 0.)
\psbezier[linecolor=blue,linewidth=1.5pt]{-}(-2.63923, -0.6)(-2.98564, -0.8)(-2.793, -1.34504)(-3.15339, -1.51859)
\psbezier[linecolor=blue,linewidth=1.5pt]{-}(-1.6, -1.2)(-1.6, -1.6)(-1.93282, -2.42368)(-2.18221, -2.73641)
\psbezier[linecolor=blue,linewidth=1.5pt]{-}(-0.779, -3.412)(-0.701, -3.071)(-0.041, -0.9)(-0.561, -0.6)
\psarc[linecolor=purple,linewidth=1.5pt]{-}(-1.6,0){0.5}{0}{360}
\psarc[linecolor=darkgreen,linewidth=1.5pt]{-}(1.6,0){0.5}{0}{360}
\psarc[linecolor=black,linewidth=0.5pt]{-}(0,0){3.5}{0}{360}
\end{pspicture}}
\ = \
\thispic{\begin{pspicture}[shift=-4.4](-4.5,-4.5)(4.5,4.5)
\psarc[linecolor=black,linewidth=0.5pt,fillstyle=solid,fillcolor=lightlightblue]{-}(0,0){4.5}{0}{360}
\psarc[linecolor=black,linewidth=0.5pt,fillstyle=solid,fillcolor=white]{-}(-1.6,0){0.5}{0}{360}
\psarc[linecolor=black,linewidth=0.5pt,fillstyle=solid,fillcolor=white]{-}(1.6,0){0.5}{0}{360}
\psarc[linecolor=black,linewidth=0.5pt]{-}(-1.6,0){1.2}{0}{360}
\psarc[linecolor=black,linewidth=0.5pt]{-}(1.6,0){1.2}{0}{360}
\psline[linestyle=dashed, dash= 1.5pt 1.5pt,linewidth=0.5pt]{-}(-1.1,0)(-0.4,0)
\psline[linestyle=dashed, dash= 1.5pt 1.5pt,linewidth=0.5pt]{-}(1.1,0)(0.4,0)
\psline[linestyle=dashed, dash= 1.5pt 1.5pt,linewidth=0.5pt]{-}(-0.4,0)(0,-3.5)
\psline[linestyle=dashed, dash= 1.5pt 1.5pt,linewidth=0.5pt]{-}(0.4,0)(0,-3.5)
\psline[linestyle=dashed, dash= 1.5pt 1.5pt,linewidth=0.5pt]{-}(0,-3.5)(0,-4.5)
\rput(1.6,0){
\psbezier[linecolor=blue,linewidth=1.5pt]{-}(-0.462, -0.191)(-0.785, -0.325)(-0.785, 0.325)(-1.109, 0.459)
\psbezier[linecolor=blue,linewidth=1.5pt]{-}(-0.191, -0.462)(-0.325, -0.785)(-0.785, -0.325)(-1.109, -0.459)
\psbezier[linecolor=blue,linewidth=1.5pt]{-}(0.191, -0.462)(0.325, -0.785)(-0.325, -0.785)(-0.459, -1.109)
\psbezier[linecolor=blue,linewidth=1.5pt]{-}(0.462, -0.191)(0.785, -0.325)(0.325, -0.785)(0.459, -1.109)
\psbezier[linecolor=blue,linewidth=1.5pt]{-}(0.462, 0.191)(0.785, 0.325)(0.785, -0.325)(1.109, -0.459)
\psbezier[linecolor=blue,linewidth=1.5pt]{-}(0.191, 0.462)(0.325, 0.785)(0.785, 0.325)(1.109, 0.459)
\psbezier[linecolor=blue,linewidth=1.5pt]{-}(-0.191, 0.462)(-0.325, 0.785)(0.325, 0.785)(0.459, 1.109)
\psbezier[linecolor=blue,linewidth=1.5pt]{-}(-0.462, 0.191)(-0.785, 0.325)(-0.325, 0.785)(-0.459, 1.109)
}
\rput(-1.6,0){
\psbezier[linecolor=blue,linewidth=1.5pt]{-}(0.433, 0.25)(0.736, 0.425)(0., 0.85)(0., 1.2)
\psbezier[linecolor=blue,linewidth=1.5pt]{-}(0., 0.5)(0., 0.85)(-0.736, 0.425)(-1.039, 0.6)
\psbezier[linecolor=blue,linewidth=1.5pt]{-}(-0.433, 0.25)(-0.736, 0.425)(-0.736, -0.425)(-1.039, -0.6)
\psbezier[linecolor=blue,linewidth=1.5pt]{-}(-0.433, -0.25)(-0.736, -0.425)(0., -0.85)(0., -1.2)
\psbezier[linecolor=blue,linewidth=1.5pt]{-}(0., -0.5)(0., -0.85)(0.736, -0.425)(1.039, -0.6)
\psbezier[linecolor=blue,linewidth=1.5pt]{-}(0.433, -0.25)(0.736, -0.425)(0.736, 0.425)(1.039, 0.6)
}
\psline[linecolor=blue,linewidth=1.5pt]{-}(0.778823, -3.41225)(1.00134, -4.38718)
\psline[linecolor=blue,linewidth=1.5pt]{-}(2.18221, -2.73641)(2.8057, -3.51824)
\psline[linecolor=blue,linewidth=1.5pt]{-}(3.15339, -1.51859)(4.05436,-1.95248)
\psline[linecolor=blue,linewidth=1.5pt]{-}(3.5, 0.)(4.5, 0)
\psline[linecolor=blue,linewidth=1.5pt]{-}(3.15339, 1.51859)(4.05436, 1.95248)
\psline[linecolor=blue,linewidth=1.5pt]{-}(2.18221, 2.73641)(2.8057, 3.51824)
\psline[linecolor=blue,linewidth=1.5pt]{-}(0.778823, 3.41225)(1.00134, 4.38718)
\psbezier[linecolor=blue,linewidth=1.5pt]{-}(-0.779, 3.412)(-0.927, 4.062)(-2.598, 3.258)(-2.182, 2.736)
\psline[linecolor=blue,linewidth=1.5pt]{-}(-3.15339, 1.51859)(-4.05436, 1.95248)
\psline[linecolor=blue,linewidth=1.5pt]{-}(-3.5, 0.)(-4.5, 0)
\psline[linecolor=blue,linewidth=1.5pt]{-}(-3.15339, -1.51859)(-4.05436, -1.95248)
\psline[linecolor=blue,linewidth=1.5pt]{-}(-2.18221, -2.73641)(-2.8057, -3.5182)
\psline[linecolor=blue,linewidth=1.5pt]{-}(-0.778823, -3.41225)(-1.00134, -4.38718)
\psbezier[linecolor=blue,linewidth=1.5pt]{-}(0.491345, -0.45922)(0.121793, -0.612293)(0.689815, -3.02228)(0.778823, -3.41225)
\psbezier[linecolor=blue,linewidth=1.5pt]{-}(1.14078, -1.10866)(0.987707, -1.47821)(1.93282, -2.42368)(2.18221, -2.73641)
\psbezier[linecolor=blue,linewidth=1.5pt]{-}(2.05922, -1.10866)(2.21229, -1.47821)(2.793, -1.34504)(3.15339, -1.51859)
\psbezier[linecolor=blue,linewidth=1.5pt]{-}(2.70866, -0.45922)(3.07821, -0.612293)(3.1, 0.)(3.5, 0.)
\psbezier[linecolor=blue,linewidth=1.5pt]{-}(2.70866, 0.45922)(3.07821, 0.612293)(2.793, 1.34504)(3.15339, 1.51859)
\psbezier[linecolor=blue,linewidth=1.5pt]{-}(2.05922, 1.10866)(2.21229, 1.47821)(1.93282, 2.42368)(2.18221, 2.73641)
\psbezier[linecolor=blue,linewidth=1.5pt]{-}(1.14078, 1.10866)(0.987707, 1.47821)(0.689815, 3.02228)(0.778823, 3.41225)
\psbezier[linecolor=blue,linewidth=1.5pt]{-}(0.491345, 0.45922)(0.121793, 0.612293)(-0.689815, 3.02228)(-0.778823, 3.41225)
\psbezier[linecolor=blue,linewidth=1.5pt]{-}(-0.56077, 0.6)(-0.214359, 0.8)(-1.93282, 2.42368)(-2.18221, 2.73641)
\psbezier[linecolor=blue,linewidth=1.5pt]{-}(-1.6, 1.2)(-1.6, 1.6)(-2.793, 1.34504)(-3.15339, 1.51859)
\psbezier[linecolor=blue,linewidth=1.5pt]{-}(-2.63923, 0.6)(-2.98564, 0.8)(-3.1, 0.)(-3.5, 0.)
\psbezier[linecolor=blue,linewidth=1.5pt]{-}(-2.63923, -0.6)(-2.98564, -0.8)(-2.793, -1.34504)(-3.15339, -1.51859)
\psbezier[linecolor=blue,linewidth=1.5pt]{-}(-1.6, -1.2)(-1.6, -1.6)(-1.93282, -2.42368)(-2.18221, -2.73641)
\psbezier[linecolor=blue,linewidth=1.5pt]{-}(-0.56077, -0.6)(-0.214359, -0.8)(-0.689815, -3.02228)(-0.778823, -3.41225)
\psarc[linecolor=purple,linewidth=1.5pt]{-}(-1.6,0){0.5}{0}{360}
\psarc[linecolor=darkgreen,linewidth=1.5pt]{-}(1.6,0){0.5}{0}{360}
\psarc[linecolor=black,linewidth=0.5pt]{-}(0,0){3.5}{0}{360}
\end{pspicture}} \ \,,
\qquad
\thispic{\begin{pspicture}[shift=-4.4](-4.5,-4.5)(4.5,4.5)
\psarc[linecolor=black,linewidth=0.5pt,fillstyle=solid,fillcolor=lightlightblue]{-}(0,0){4.5}{0}{360}
\psarc[linecolor=black,linewidth=0.5pt,fillstyle=solid,fillcolor=white]{-}(-1.6,0){0.5}{0}{360}
\psarc[linecolor=black,linewidth=0.5pt,fillstyle=solid,fillcolor=white]{-}(1.6,0){0.5}{0}{360}
\psarc[linecolor=black,linewidth=0.5pt]{-}(-1.6,0){1.2}{0}{360}
\psarc[linecolor=black,linewidth=0.5pt]{-}(1.6,0){1.2}{0}{360}
\psline[linestyle=dashed, dash= 1.5pt 1.5pt,linewidth=0.5pt]{-}(-1.1,0)(-0.4,0)
\psline[linestyle=dashed, dash= 1.5pt 1.5pt,linewidth=0.5pt]{-}(1.1,0)(0.4,0)
\psline[linestyle=dashed, dash= 1.5pt 1.5pt,linewidth=0.5pt]{-}(-0.4,0)(0,-3.5)
\psline[linestyle=dashed, dash= 1.5pt 1.5pt,linewidth=0.5pt]{-}(0.4,0)(0,-3.5)
\psline[linestyle=dashed, dash= 1.5pt 1.5pt,linewidth=0.5pt]{-}(0,-3.5)(0,-4.5)
\rput(1.6,0){
\psline[linecolor=blue,linewidth=1.5pt]{-}(0.191342, -0.46194)(0.45922, -1.10866)
\psline[linecolor=blue,linewidth=1.5pt]{-}(0.46194, -0.191342)(1.10866, -0.45922)
\psline[linecolor=blue,linewidth=1.5pt]{-}(0.46194, 0.191342)(1.10866, 0.45922)
\psline[linecolor=blue,linewidth=1.5pt]{-}(0.191342, 0.46194)(0.45922, 1.10866)
\psline[linecolor=blue,linewidth=1.5pt]{-}(-0.191342, 0.46194)(-0.45922, 1.10866)
\psbezier[linecolor=blue,linewidth=1.5pt]{-}(-1.109, 0.459)(-0.785, 0.325)(-0.785, -0.325)(-1.109, -0.459)
\psline[linecolor=blue,linewidth=1.5pt]{-}(-0.191342, -0.46194)(-0.45922, -1.10866)
}
\rput(-1.6,0){
\psline[linecolor=blue,linewidth=1.5pt]{-}(0, -0.5)(0., -1.2)
\psline[linecolor=blue,linewidth=1.5pt]{-}(0.433013, -0.25)(1.03923, -0.6)
\psline[linecolor=blue,linewidth=1.5pt]{-}(0.433013, 0.25)(1.03923, 0.6)
\psline[linecolor=blue,linewidth=1.5pt]{-}(0, 0.5)(0., 1.2)
\psline[linecolor=blue,linewidth=1.5pt]{-}(-0.433013, 0.25)(-1.03923, 0.6)
\psline[linecolor=blue,linewidth=1.5pt]{-}(-0.433013, -0.25)(-1.03923, -0.6)
}
\psbezier[linecolor=blue,linewidth=1.5pt]{-}(0.779, -3.412)(0.89, -3.9)(-0.89, -3.9)(-1.001, -4.387)
\psbezier[linecolor=blue,linewidth=1.5pt]{-}(2.182, -2.736)(2.494, -3.127)(0.89, -3.9)(1.001, -4.387)
\psbezier[linecolor=blue,linewidth=1.5pt]{-}(3.153, -1.519)(3.604, -1.736)(2.494, -3.127)(2.806, -3.518)
\psbezier[linecolor=blue,linewidth=1.5pt]{-}(3.5, 0.)(4., 0.)(3.604, -1.736)(4.054, -1.952)
\psbezier[linecolor=blue,linewidth=1.5pt]{-}(3.153, 1.519)(3.604, 1.736)(4., 0.)(4.5, 0.)
\psbezier[linecolor=blue,linewidth=1.5pt]{-}(2.182, 2.736)(2.494, 3.127)(3.604, 1.736)(4.054, 1.952)
\psbezier[linecolor=blue,linewidth=1.5pt]{-}(0.779, 3.412)(0.89, 3.9)(2.494, 3.127)(2.806, 3.518)
\psbezier[linecolor=blue,linewidth=1.5pt]{-}(-0.779, 3.412)(-0.89, 3.9)(0.89, 3.9)(1.001, 4.387)
\psbezier[linecolor=blue,linewidth=1.5pt]{-}(-2.182, 2.736)(-2.494, 3.127)(-0.89, 3.9)(-1.001, 4.387)
\psbezier[linecolor=blue,linewidth=1.5pt]{-}(-3.153, 1.519)(-3.604, 1.736)(-2.494, 3.127)(-2.806, 3.518)
\psbezier[linecolor=blue,linewidth=1.5pt]{-}(-3.5, 0.)(-4., 0.)(-3.604, 1.736)(-4.054, 1.952)
\psbezier[linecolor=blue,linewidth=1.5pt]{-}(-3.153, -1.519)(-3.604, -1.736)(-4., 0.)(-4.5, 0.)
\psbezier[linecolor=blue,linewidth=1.5pt]{-}(-2.182, -2.736)(-2.494, -3.127)(-3.604, -1.736)(-4.054, -1.952)
\psbezier[linecolor=blue,linewidth=1.5pt]{-}(-0.779, -3.412)(-0.89, -3.9)(-2.494, -3.127)(-2.806, -3.518)
\psbezier[linecolor=blue,linewidth=1.5pt]{-}(0.491345, -0.45922)(0.121793, -0.612293)(0.689815, -3.02228)(0.778823, -3.41225)
\psbezier[linecolor=blue,linewidth=1.5pt]{-}(1.14078, -1.10866)(0.987707, -1.47821)(1.93282, -2.42368)(2.18221, -2.73641)
\psbezier[linecolor=blue,linewidth=1.5pt]{-}(2.05922, -1.10866)(2.21229, -1.47821)(2.793, -1.34504)(3.15339, -1.51859)
\psbezier[linecolor=blue,linewidth=1.5pt]{-}(2.70866, -0.45922)(3.07821, -0.612293)(3.1, 0.)(3.5, 0.)
\psbezier[linecolor=blue,linewidth=1.5pt]{-}(2.70866, 0.45922)(3.07821, 0.612293)(2.793, 1.34504)(3.15339, 1.51859)
\psbezier[linecolor=blue,linewidth=1.5pt]{-}(2.05922, 1.10866)(2.21229, 1.47821)(1.93282, 2.42368)(2.18221, 2.73641)
\psbezier[linecolor=blue,linewidth=1.5pt]{-}(1.14078, 1.10866)(0.987707, 1.47821)(0.689815, 3.02228)(0.778823, 3.41225)
\psbezier[linecolor=blue,linewidth=1.5pt]{-}(0.491345, 0.45922)(0.121793, 0.612293)(-0.689815, 3.02228)(-0.778823, 3.41225)
\psbezier[linecolor=blue,linewidth=1.5pt]{-}(-2.182, 2.736)(-1.964, 2.463)(0.115, 0.99)(-0.561, 0.6)
\psbezier[linecolor=blue,linewidth=1.5pt]{-}(-3.153, 1.519)(-2.838, 1.367)(-1.6, 1.8)(-1.6, 1.2)
\psbezier[linecolor=blue,linewidth=1.5pt]{-}(-2.63923, 0.6)(-2.98564, 0.8)(-3.1, 0.)(-3.5, 0.)
\psbezier[linecolor=blue,linewidth=1.5pt]{-}(-2.63923, -0.6)(-2.98564, -0.8)(-2.793, -1.34504)(-3.15339, -1.51859)
\psbezier[linecolor=blue,linewidth=1.5pt]{-}(-1.6, -1.2)(-1.6, -1.6)(-1.93282, -2.42368)(-2.18221, -2.73641)
\psbezier[linecolor=blue,linewidth=1.5pt]{-}(-0.779, -3.412)(-0.701, -3.071)(-0.041, -0.9)(-0.561, -0.6)
\psarc[linecolor=purple,linewidth=1.5pt]{-}(-1.6,0){0.5}{0}{360}
\psarc[linecolor=darkgreen,linewidth=1.5pt]{-}(1.6,0){0.5}{0}{360}
\psarc[linecolor=black,linewidth=0.5pt]{-}(0,0){3.5}{0}{360}
\end{pspicture}}
\ = \
\thispic{\begin{pspicture}[shift=-4.4](-4.5,-4.5)(4.5,4.5)
\psarc[linecolor=black,linewidth=0.5pt,fillstyle=solid,fillcolor=lightlightblue]{-}(0,0){4.5}{0}{360}
\psarc[linecolor=black,linewidth=0.5pt,fillstyle=solid,fillcolor=white]{-}(-1.6,0){0.5}{0}{360}
\psarc[linecolor=black,linewidth=0.5pt,fillstyle=solid,fillcolor=white]{-}(1.6,0){0.5}{0}{360}
\psarc[linecolor=black,linewidth=0.5pt]{-}(-1.6,0){1.2}{0}{360}
\psarc[linecolor=black,linewidth=0.5pt]{-}(1.6,0){1.2}{0}{360}
\psline[linestyle=dashed, dash= 1.5pt 1.5pt,linewidth=0.5pt]{-}(-1.1,0)(-0.4,0)
\psline[linestyle=dashed, dash= 1.5pt 1.5pt,linewidth=0.5pt]{-}(1.1,0)(0.4,0)
\psline[linestyle=dashed, dash= 1.5pt 1.5pt,linewidth=0.5pt]{-}(-0.4,0)(0,-3.5)
\psline[linestyle=dashed, dash= 1.5pt 1.5pt,linewidth=0.5pt]{-}(0.4,0)(0,-3.5)
\psline[linestyle=dashed, dash= 1.5pt 1.5pt,linewidth=0.5pt]{-}(0,-3.5)(0,-4.5)
\rput(1.6,0){
\psline[linecolor=blue,linewidth=1.5pt]{-}(-0.433, -0.25)(-1.039, -0.6)
\psline[linecolor=blue,linewidth=1.5pt]{-}(0., -0.5)(0., -1.2)
\psline[linecolor=blue,linewidth=1.5pt]{-}(0.433, -0.25)(1.039, -0.6)
\psline[linecolor=blue,linewidth=1.5pt]{-}(0.433, 0.25)(1.039, 0.6)
\psline[linecolor=blue,linewidth=1.5pt]{-}(0., 0.5)(0., 1.2)
\psline[linecolor=blue,linewidth=1.5pt]{-}(-0.433, 0.25)(-1.039, 0.6)
}
\rput(-1.6,0){
\psbezier[linecolor=blue,linewidth=1.5pt]{-}(1.109, -0.459)(0.785, -0.325)(0.785, 0.325)(1.109, 0.459)
\psline[linecolor=blue,linewidth=1.5pt]{-}(0.191, 0.462)(0.459, 1.109)
\psline[linecolor=blue,linewidth=1.5pt]{-}(-0.191, 0.462)(-0.459, 1.109)
\psline[linecolor=blue,linewidth=1.5pt]{-}(-0.462, 0.191)(-1.109, 0.459)
\psline[linecolor=blue,linewidth=1.5pt]{-}(-0.462, -0.191)(-1.109, -0.459)
\psline[linecolor=blue,linewidth=1.5pt]{-}(-0.191, -0.462)(-0.459, -1.109)
\psline[linecolor=blue,linewidth=1.5pt]{-}(0.191, -0.462)(0.459, -1.109)
}
\psline[linecolor=blue,linewidth=1.5pt]{-}(0.778823, -3.41225)(1.00134, -4.38718)
\psline[linecolor=blue,linewidth=1.5pt]{-}(2.18221, -2.73641)(2.8057, -3.51824)
\psline[linecolor=blue,linewidth=1.5pt]{-}(3.15339, -1.51859)(4.05436,-1.95248)
\psline[linecolor=blue,linewidth=1.5pt]{-}(3.5, 0.)(4.5, 0)
\psline[linecolor=blue,linewidth=1.5pt]{-}(3.15339, 1.51859)(4.05436, 1.95248)
\psline[linecolor=blue,linewidth=1.5pt]{-}(2.18221, 2.73641)(2.8057, 3.51824)
\psline[linecolor=blue,linewidth=1.5pt]{-}(0.778823, 3.41225)(1.00134, 4.38718)
\psline[linecolor=blue,linewidth=1.5pt]{-}(-0.778823, 3.41225)(-1.00134, 4.38718)
\psline[linecolor=blue,linewidth=1.5pt]{-}(-2.18221, 2.73641)(-2.8057, 3.51824)
\psline[linecolor=blue,linewidth=1.5pt]{-}(-3.15339, 1.51859)(-4.05436, 1.95248)
\psline[linecolor=blue,linewidth=1.5pt]{-}(-3.5, 0.)(-4.5, 0)
\psline[linecolor=blue,linewidth=1.5pt]{-}(-3.15339, -1.51859)(-4.05436, -1.95248)
\psline[linecolor=blue,linewidth=1.5pt]{-}(-2.18221, -2.73641)(-2.8057, -3.5182)
\psline[linecolor=blue,linewidth=1.5pt]{-}(-0.778823, -3.41225)(-1.00134, -4.38718)
\psbezier[linecolor=blue,linewidth=1.5pt]{-}(0.779, -3.412)(0.623, -2.73)(0.041, -0.9)(0.561, -0.6)
\psbezier[linecolor=blue,linewidth=1.5pt]{-}(2.182, -2.736)(1.746, -2.189)(1.6, -1.8)(1.6, -1.2)
\psbezier[linecolor=blue,linewidth=1.5pt]{-}(3.153, -1.519)(2.838, -1.367)(2.951, -0.78)(2.639, -0.6)
\psbezier[linecolor=blue,linewidth=1.5pt]{-}(3.5, 0.)(3.15, 0.)(2.951, 0.78)(2.639, 0.6)
\psbezier[linecolor=blue,linewidth=1.5pt]{-}(3.153, 1.519)(2.523, 1.215)(1.6, 2.16)(1.6, 1.2)
\psbezier[linecolor=blue,linewidth=1.5pt]{-}(2.182, 2.736)(1.746, 2.189)(-0.271, 1.08)(0.561, 0.6)
\psbezier[linecolor=blue,linewidth=1.5pt]{-}(0.779, 3.412)(0.623, 2.73)(0.063, 0.689)(-0.491, 0.459)
\psbezier[linecolor=blue,linewidth=1.5pt]{-}(-0.779, 3.412)(-0.623, 2.73)(-0.911, 1.663)(-1.141, 1.109)
\psbezier[linecolor=blue,linewidth=1.5pt]{-}(-2.182, 2.736)(-1.746, 2.189)(-2.289, 1.663)(-2.059, 1.109)
\psbezier[linecolor=blue,linewidth=1.5pt]{-}(-3.153, 1.519)(-2.838, 1.367)(-3.041, 0.597)(-2.709, 0.459)
\psbezier[linecolor=blue,linewidth=1.5pt]{-}(-3.5, 0.)(-3.15, 0.)(-3.041, -0.597)(-2.709, -0.459)
\psbezier[linecolor=blue,linewidth=1.5pt]{-}(-3.153, -1.519)(-2.838, -1.367)(-2.197, -1.441)(-2.059, -1.109)
\psbezier[linecolor=blue,linewidth=1.5pt]{-}(-2.182, -2.736)(-1.746, -2.189)(-0.842, -1.829)(-1.141, -1.109)
\psbezier[linecolor=blue,linewidth=1.5pt]{-}(-0.779, -3.412)(-0.623, -2.73)(0.063, -0.689)(-0.491, -0.459)
\psarc[linecolor=purple,linewidth=1.5pt]{-}(-1.6,0){0.5}{0}{360}
\psarc[linecolor=darkgreen,linewidth=1.5pt]{-}(1.6,0){0.5}{0}{360}
\psarc[linecolor=black,linewidth=0.5pt]{-}(0,0){3.5}{0}{360}
\end{pspicture}} \ \, .
\ee

We thus define $\Lambda(N,N_\pa,N_\pb)$ algebraically as the vector space \eqref{eq:Lambda.sumsLLL} endowed with the relations~\eqref{eq:Lambda.equations} and the relations \eqref{eq:c.relations} for $\cL(N,N'_{\pa\pb})$, $\cL(N'_\pa,N_\pa)$ and $\cL(N'_\pb,N_\pb)$. We note that an equivalent complete set of relations is obtained by selecting the four relations \eqref{eq:Lambda.equations.a} to \eqref{eq:Lambda.equations.d}, and one extra relation among \eqref{eq:Lambda.equations.e}, \eqref{eq:Lambda.equations.f}, \eqref{eq:Lambda.equations.g} and \eqref{eq:Lambda.equations.h}. Indeed, after choosing one such set of five relations, one can always combine them to prove the other relations. A proof of the following proposition is given in \cref{app:equiv.Lambda}.

\begin{Theorem}
\label{prop:Lambda.defs}
The diagrammatic and algebraic definitions of $\Lambda(N,N_\pa,N_\pb)$ are equivalent.
\end{Theorem}

\subsection{General definition of fusion}\label{sec:fusion.definition}

Let $\repM_\pa$ and $\repM_\pb$ be two families of modules. We now define the family of fusion modules 
\be
\label{eq:Ma*Mb.set}
(\repM_\pa \timesf \repM_\pb) = \{(\repM_\pa \timesf \repM_\pb)(N)\,|\,N=N_0, N_0+2, N_0+4, \dots\}\,,
\ee 
where $N_0=0$ or $1$ if $\repM_\pa$ and $\repM_\pb$ have the same or opposite parities, respectively. The corresponding modules are defined as
\be
\label{eq:fusion.of.modules}
(\repM_\pa \timesf \repM_\pb)(N) =
\sum_{N_\pa,N_\pb} \cL(N,N_{\pa\pb}) \cdot
\big(\repM_\pa(N_\pa) \otimes \repM_\pb(N_\pb) \big) \,,
\ee
subject to the relations
\begin{subequations} \label{eq:moving.arches}
\begin{alignat}{2} 
&c_j \cdot (u \otimes v) = (c_j u \otimes v)\,, \qquad 
&& 1 \le j \le N_\pa -1\,, \label{eq:move.cj.Ma}
\\[0.15cm]
&c_{j+N_\pa} \cdot (u \otimes v) = (u \otimes c_j v)\,, \qquad 
&& 1 \le j \le N_\pb-1\,, \label{eq:move.cj.Mb}
\\[0.15cm]
&c_j^\dag \cdot (u \otimes v) = (c_j^\dag u \otimes v)\,, \qquad 
&& 1 \le j \le N_\pa + 1\,, \label{eq:move.cjdag.Ma}
\\[0.15cm]
&c_{j+N_\pa}^\dag \cdot (u \otimes v) = (u \otimes c_j^\dag v)\,, \qquad 
&& 1 \le j \le N_\pb + 1\,, \label{eq:move.cjdag.Mb}
\\[0.15cm]
&\Omega \cdot (u \otimes v) = c_{N_\pa} \cdot (\Omega\, u \otimes c_0^\dag v)\,,
&\quad& \textrm{for }N_\pa >0 \,, \label{eq:move.c0dag}
\end{alignat}
\end{subequations}
for all $u \in \repM_\pa(N_\pa)$ and $v \in \repM_\pb(N_\pb)$. Here, $N_\pa$ and $N_\pb$ are admissible integers for the families $\repM_\pa$ and~$\repM_\pb$ respectively, $N_{\pa\pb}=N_\pa+N_\pb$, and the sum in \eqref{eq:fusion.of.modules} runs over all the corresponding admissible values.\medskip

The action of $\cL(N',N)$ on $(\repM_\pa \timesf \repM_\pb)(N)$, for all admissible integers $N'$, is defined from the associative action of $\cL(N',N)$ on $\cL(N,N_{\pa\pb})$, namely 
\begin{equation} \label{eq:L.on.fusion}
\lambda' \cdot \big(\lambda \cdot (u \otimes v)\big) = (\lambda' \lambda) \cdot (u \otimes v)\,,
\end{equation}
for all $\lambda' \in \cL(N',N)$, $\lambda \in \cL(N,N_{\pa\pb})$, $u \in \repM_\pa(N_\pa)$ and $v \in \repM_\pb(N_\pb)$. In the next section, we prove that this definition of $(\repM_\pa \timesf \repM_\pb)(N)$ indeed produces a family of modules.\medskip

Using \eqref{eq:moving.arches}, one can derive many more relations, and in particular
\begin{subequations}
\begin{alignat}{3}
& \Omega^{-1} \cdot (u \otimes v) = c_{N_\pa+2} \cdot (c_0^\dag u \otimes \Omega^{-1} v) \,,
&\qquad& \text{for $N_\pb > 0$,} \\[0.15cm]
& c_0 \cdot (u \otimes v) = c_{N_\pa} \cdot (\Omega\, u \otimes \Omega^{-1} v)\,,&&\text{for $N_\pa > 0$ and $N_\pb > 0$,} \label{eq:c0=cNa} \\[0.15cm]
& \Omega \cdot (c_0^\dag u \otimes v) = (u \otimes c_0^\dag v) \,, \\[0.15cm]
& c_0 \cdot (u \otimes c_0^\dag v) = (f\,u \otimes v) \,,
&&\text{for $N_\pa =0$,} \\[0.15cm]
& \Omega \cdot (u \otimes v) = c_1c_2\cdot(c_0^\dag u \otimes c_0^\dag v) \,,
&&\text{for $N_\pa =0$ and $N_\pb>0$.}
\end{alignat}
\end{subequations}

We remark that the relations \eqref{eq:moving.arches} defining the fusion of two modules are closely related to the relations \eqref{eq:Lambda.equations} for diagrams with two holes. Indeed, from the algebraic definition of $\Lambda(N,N_\pa,N_\pb)$ consisting of the vector space \eqref{eq:Lambda.sumsLLL} with the relations~\eqref{eq:Lambda.equations}, we see that, seen as a family of modules, it precisely corresponds to the fusion of the families of diagram spaces defined in \eqref{eq:LN}, namely
\begin{equation} \label{eq:Lambda=LxL}
\Lambda(N,N_\pa,N_\pb) = (\repL_{N_\pa} \timesf \repL_{N_\pb})(N) \,.
\end{equation}

\subsection{Fusion as a bifunctor}\label{sec:bifunctor}

In this section, we review the definition of bifunctors and show that the fusion of families of modules introduced in the previous section defines a bifunctor on the category $\mathcal C$ of families of modules of $\eptl_N(\beta)$. We first recall that, given a category $\mathcal C$, a bifunctor~$\mathcal F$ consists in
\begin{enumerate}
\item[(i)] a map $\cal F: (\repM_\pa, \repM_\pb) \to \repM_{\pa\pb}$ where $\repM_{\pa\pb} \in \mathcal C$, defined for all $\repM_\pa,\repM_\pb \in \mathcal C$,
\item[(ii)] a map $\cal F: (\phi_\pa, \phi_\pb) \to \phi_{\pa\pb}$ where $\phi_{\pa\pb}$ is a homomorphism from $\cal F(\repM_\pa,\repM_\pb)$ to $\cal F(\repM_\pa',\repM_\pb')$, defined for all homomorphisms $\phi_\pa:\repM_\pa\to\repM'_\pa$ and $\phi_\pb:\repM_\pb\to\repM'_\pb$,
and satisfying
\begin{equation} \label{eq:bifunctor.props}
\cal F(\id_{\repM_\pa}, \id_{\repM_\pb}) = \id_{\repM_{\pa\pb}} \quad \text{and} \quad
\cal F(\phi'_\pa \circ \phi_\pa, \phi'_\pb \circ \phi_\pb) = 
\cal F(\phi'_\pa, \phi'_\pb) \circ\cal F(\phi_\pa,\phi_\pb) \,,
\end{equation}
for all homomorphisms $\repM_\pa \overset{\phi_\pa}{\to} \repM'_\pa \overset{\phi'_\pa}{\to} \repM''_\pa$ and $\repM_\pb \overset{\phi_\pb}{\to} \repM'_\pb \overset{\phi'_\pb}{\to} \repM''_\pb$, and all families of modules $\repM_\pa, \repM'_\pa, \repM''_\pa, \repM_\pb, \repM'_\pb, \repM''_\pb \in \cal C$. Here, $\id_\repM$ denotes the trivial isomorphism $\repM \to \repM$.
\end{enumerate}
The following proposition states one useful property of bifunctors.

\begin{Proposition} \label{prop:fusion.iso}
Let $\repM_\pa, \repM'_\pa, \repM_\pb,\repM'_\pb \in \mathcal C$ and $
\phi_\pa: \repM_\pa \to \repM'_\pa$, $\phi_\pb: \repM_\pb \to \repM'_\pb$ be isomorphisms. Then $\phi_{\pa\pb} = \mathcal F(\phi_\pa,\phi_\pb)$ is an isomorphism from $\repM_{\pa\pb} = \mathcal F(\repM_\pa, \repM_\pb)$ to $\repM'_{\pa\pb} = \mathcal F(\repM'_\pa, \repM'_\pb)$.
\end{Proposition}
\proof Consider the maps $\psi_{\pa\pb}: \mathcal F(\repM'_\pa, \repM'_\pb) \to\mathcal F(\repM_\pa, \repM_\pb)$ defined as $\psi_{\pa\pb} = \mathcal F(\phi^{-1}_{\pa},\phi^{-1}_{\pb})$. By the second property in \eqref{eq:bifunctor.props}, we have
$\psi_{\pa\pb} \circ \phi_{\pa\pb} = \id_{\repM_{\pa\pb}}$ and $\phi_{\pa\pb} \circ \psi_{\pa\pb} = \id_{\repM'_{\pa\pb}}$.
\eproof

We now set out to prove that the fusion as defined in \cref{sec:fusion.definition} is a bifunctor. Let us first define the vector space
\be
\label{eq:Mab} 
\repL_{\repM_\pa,\repM_\pb}(N) = \bigoplus_{N_\pa,N_\pb} \cL(N,N_{\pa\pb}) \otimes \repM_\pa(N_\pa) \otimes \repM_\pb(N_\pb) \,,
\ee
endowed with the action of $\cL(N',N)$
\begin{equation} \label{eq:L.on.Mab}
\lambda' \cdot (\lambda \otimes u \otimes v) = (\lambda'\lambda) \otimes u \otimes v \,,
\end{equation}
for all $\lambda' \in \cL(N',N)$, $\lambda \in \cL(N,N_{\pa\pb})$, $u \in \repM_\pa(N_\pa)$ and $v \in \repM_\pb(N_\pa)$. Because $\repM_\pa(N_\pa)$ and $\repM_\pb(N_\pb)$ act as spectators in the above action and $\{\cL(N,N_{\pa\pb})\,|\,N=N_0, N_0+2, \dots\}$ is a family of modules from \cref{prop:L.family}, it is clear that $\repL_{\repM_\pa,\repM_\pb}$ is also a family of modules. Let us define
\begin{subequations} \label{eq:X(u,v)}
\begin{alignat}{2}
& X_j^{\tinyx a}(u,v) = c_j \otimes u\otimes v - \id \otimes c_j u \otimes v \,,
\qquad&& 1 \leq j \leq N_\pa-1 \,, \\[0.1cm]
& X_j^{\tinyx b}(u,v) = c_{j+N_\pa}\otimes u\otimes v - \id \otimes u \otimes c_j v \,,
\qquad&& 1 \leq j \leq N_\pb-1 \,, \\[0.1cm]
& X_j^{\tinyx c}(u,v) = c_j^\dag \otimes u \otimes v - \id \otimes c_j^\dag u \otimes v \,,
\qquad&& 1 \le j \le N_\pa + 1, \\[0.1cm]
& X_j^{\tinyx d}(u,v) = c_{j+N_\pa}^\dag \otimes u \otimes v
- \id \otimes u \otimes c_j^\dag v\,,
\qquad&& 1 \le j \le N_\pb + 1 \,, \\[0.1cm]
& X_0^{\tinyx e}(u,v) = \Omega \otimes u \otimes v
- c_{N_\pa} \otimes \Omega\,u \otimes c_0^\dag v \,,
\qquad&& \textrm{for }N_\pa>0 \,.
\end{alignat}
\end{subequations}
for $u \in \repM_\pa(N_\pa)$ and $v \in \repM_\pb(N_\pb)$.
These are elements of $\repL_{\repM_\pa,\repM_\pb}(N_{\pa\pb}^{\tinyx \alpha})$, where $N_{\pa\pb}^{\tinyx \alpha}$ is defined in \eqref{eq:N.alpha}. Let us also consider the subspace of $\repL_{\repM_\pa,\repM_\pb}(N)$ defined as
\begin{equation} \label{eq:Eab} 
\repE_{\repM_\pa,\repM_\pb}(N) =
\sum_{N_\pa,N_\pb,\alpha,j} \cL(N,N_{\pa\pb}^{\tinyx \alpha}) \cdot X_j^{\tinyx \alpha}(\repM_\pa(N_\pa),\repM_\pb(N_\pb))\,,
\end{equation}
where $\alpha$ runs over the set $\{a,b,c,d,e\}$, and $j$ takes all the $\alpha$-dependent values allowed in \eqref{eq:X(u,v)}. Because the action of the diagram spaces is well-defined,
it is easy to see that $\lambda \cdot v \in \repE_{\repM_\pa,\repM_\pb}(N')$, for all $\lambda \in \cL(N',N)$ and $v \in \repE_{\repM_\pa,\repM_\pb}(N)$.
Hence, $\repE_{\repM_\pa,\repM_\pb}$ is a subfamily of $\repL_{\repM_\pa,\repM_\pb}$. 
From this construction, it is clear that the definition \eqref{eq:fusion.of.modules} with the relations \eqref{eq:moving.arches} is equivalent to
\begin{equation} \label{eq:fusion=quotient}
(\repM_\pa \timesf \repM_\pb)(N)
= \repL_{\repM_\pa,\repM_\pb}(N) \big/ \repE_{\repM_\pa,\repM_\pb}(N) \,.
\end{equation}
The notation $\lambda \cdot (u \otimes v)$ that we use for the module $(\repM_\pa \timesf \repM_\pb)(N)$ stands for the equivalence class of $(\lambda \otimes u \otimes v)$ under the equivalence relation
\begin{equation} \label{eq:equivE}
x \equiv y \qquad 
\longleftrightarrow
\qquad x-y \in \repE_{\repM_\pa,\repM_\pb}(N) \,,
\end{equation}
for $x,y \in \repL_{\repM_\pa,\repM_\pb}(N)$.
Since $\repL_{\repM_\pa,\repM_\pb}$ is a family of modules and $\repE_{\repM_\pa,\repM_\pb}$ is a subfamily of $\repL_{\repM_\pa,\repM_\pb}$, we know from \cref{prop:sub.quot.families} that the quotient defined in the right-hand side of \eqref{eq:fusion=quotient} is also a family of modules. This readily proves the following proposition.
\begin{Proposition} \label{prop:fusion.stable}
The definition for $(\repM_\pa \timesf \repM_\pb)$ given by \eqref{eq:fusion.of.modules} with the relations \eqref{eq:moving.arches} yields a family of modules over the diagram spaces.
\end{Proposition}

\noindent This is item (i) in the definition of the bifunctor. We now address item (ii) related to homomorphisms.

\begin{Lemma} \label{lem:morphism.on.Ma.Mb}
Let $\repM_\pa,\repM_\pb,\repM$ be three families of modules, and $\phi = \{\phi_N \, | \, N_0, N_0+2, \dots\}$ be a set of linear maps from $\repL_{\repM_\pa,\repM_\pb}(N)$ to $\repM(N)$ satisfying
\begin{subequations}
\begin{alignat}{3} 
\label{eq:phi.on.Mab}
&\phi_{N'} (\lambda \cdot x) = \lambda \cdot \phi_N(x) \,,\qquad
&&\forall \lambda \in \cL(N',N) \,, \quad &&\forall x \in \repL_{\repM_\pa,\repM_\pb}(N)\,,
\\[0.15cm] 
\label{eq:phi(X)=0}
&\phi_{N^{(\alpha)}_{\pa\pb}} \big(X_j^{\tinyx \alpha}(u,v) \big) = 0\,, 
\qquad 
&&\forall u\in \repM_\pa(N_\pa) \,, \quad &&\forall v \in \repM_\pb(N_\pb) \,,
\end{alignat}
\end{subequations}
for all values of $\alpha$ and $j$. Then $\phi$ is a well-defined family of homomorphisms from $(\repM_\pa \timesf \repM_\pb)$ to $\repM$.
\end{Lemma}
\proof
It is clear from \eqref{eq:phi.on.Mab} that $\phi$ is a family of homomorphisms from $\repL_{\repM_\pa,\repM_\pb}(N)$ to $\repM(N)$. Let $x,y \in \repL_{\repM_\pa,\repM_\pb}(N)$ be two equivalent elements under the equivalence relation \eqref{eq:equivE}, namely $x \equiv y$. To show that $\phi_N$ is a well-defined map from $(\repM_\pa \timesf \repM_\pb)(N)$ to $\repM(N)$, it is sufficient to show that $\phi_N(x)=\phi_N(y)$ for all such $x$ and $y$. But clearly $x-y \in \repE_{\repM_\pa,\repM_\pb}(N)$, so we can decompose it as a sum of the form
\be
x-y = \sum_{k} \lambda_k \cdot X_{j_k}^{\tinyx {\alpha_k}}(u_k,v_k)\,,
\ee
for some $\lambda_k \in \cL(N,N_{\pa\pb}^{\tinyx {\alpha_k}})$, some indices $\alpha_k$ and $j_k$ and some states $u_k \in \repM_\pa(N_{\pa,k})$ and $ v_k \in \repM_\pb(N_{\pb,k})$, with sizes $N^{\tinyx {\alpha_k}}_{\pa\pb}$, $N_{\pa,k}$ and $N_{\pb,k}$ that may depend on $k$.
We then have
\be
\phi_N(x-y) = \sum_{k} \lambda_k \cdot \phi_{N_{\pa\pb}^{\tinyx{\alpha_k}}}\big(X_{j_k}^{\tinyx {\alpha_k}}(u_k,v_k)\big) = 0\,,
\ee
where we used \eqref{eq:phi.on.Mab} for the first equality and \eqref{eq:phi(X)=0} for the second. 
\eproof

In the specific case when $\repM$ is of the form $\repM=\repM'_\pa \timesf \repM'_\pb$, the above result allows us to construct a family of homomorphisms between pairs of families of fusion modules.
\begin{Corollary} \label{cor:morphism.on.Ma.Mb}
Let $(\repM_\pa,\repM_\pb)$ and $(\repM'_\pa,\repM'_\pb)$ be two pairs of families of modules, and $\phi$ be a set of maps from $\repL_{\repM_\pa,\repM_\pb}$ to $\repL_{\repM'_\pa,\repM'_\pb}$ such that 
\begin{subequations}
\begin{alignat}{3} 
&\phi_{N'} (\lambda \cdot x) = \lambda \cdot \phi_N(x) \,,\qquad
&&\forall \lambda \in \cL(N',N) \,, \quad &&\forall x \in \repL_{\repM_\pa,\repM_\pb}(N)\,,
\\[0.1cm] 
&\phi_{N_{\pa\pb}^{(\alpha)}} \big(X_j^{\tinyx \alpha}(u,v) \big) \equiv 0\,, 
\qquad 
&&\forall u\in \repM_\pa(N_\pa) \,, \quad &&\forall v \in \repM_\pb(N_\pb) \,,
\end{alignat}
\end{subequations}
for all values of $\alpha$ and $j$. Let also $\psi$ be the family of homomorphisms from $\repL_{\repM'_\pa,\repM'_\pb}$ to $(\repM'_\pa \timesf \repM'_\pb)$ defined as $\psi_N(\lambda \otimes u \otimes v) = \lambda \cdot(u \otimes v)$ for all $\lambda \in \cL(N',N)$, $u\in \repM'_\pa(N_\pa)$ and $v\in \repM'_\pb(N_\pb)$. Then $\psi \circ \phi$ is a family of homomorphisms from $(\repM_\pa \timesf \repM_\pb)$ to $(\repM'_\pa \timesf \repM'_\pb)$.
\end{Corollary}

We use these results to construct the map $\cal F: (\phi_\pa, \phi_\pb) \to \phi_{\pa\pb}$, namely the item (ii) defining the bifunctor.
\begin{Proposition} \label{prop:phi_(ab)}
Let $(\repM_\pa,\repM_\pb)$ and $(\repM'_\pa,\repM'_\pb)$ be two pairs of families of modules, and
\begin{equation}
\phi_{\pa} : \repM_\pa \to \repM'_\pa \,,
\qquad \phi_{\pb} : \repM_\pb \to \repM'_\pb \,,
\end{equation}
be two families of homomorphisms. Then the map from $\repL_{\repM_\pa,\repM_\pb}$ to $\repL_{\repM'_\pa,\repM'_\pb}$ defined as
\begin{equation} \label{eq:phiab.def}
\phi_{\pa\pb}(\lambda \otimes u \otimes v) = \lambda \otimes \phi_{\pa}(u) \otimes \phi_{\pb}(v)
\end{equation}
yields a well-defined family of homomorphisms $\mathcal{F}(\phi_\pa,\phi_\pb)$ from $(\repM_\pa \timesf \repM_\pb)$ to $(\repM'_\pa \timesf \repM'_\pb)$.
\end{Proposition}
\proof
For all $\lambda' \in \cL(N',N)$, we have
\begin{alignat}{2}
\phi_{\pa\pb}\big(\lambda' \cdot(\lambda \otimes u \otimes v)\big)
&= \phi_{\pa\pb}\big((\lambda'\lambda) \otimes u \otimes v\big)
= \lambda'\lambda \otimes \phi_\pa(u) \otimes \phi_\pb(v) \nonumber \\[0.1cm]
&= \lambda' \cdot \big(\lambda \otimes \phi_\pa(u) \otimes \phi_\pb(v)\big)
= \lambda' \cdot \phi_{\pa\pb}(\lambda \otimes u \otimes v) \,.
\end{alignat}
This confirms that $\phi_{\pa\pb}$ acts as a homomorphism. We now show that $\phi_{\pa\pb}$ preserves the equivalence relations $X_j^{\tinyx \alpha}(u,v)\equiv 0$. To proceed, we note that each $X_j^{\tinyx \alpha}(u,v)$ is of the form
\begin{equation}\label{eq:X.general}
X(u,v) = \lambda \otimes u \otimes v - \lambda' \otimes \lambda_\pa u \otimes \lambda_\pb v\,,
\end{equation}
for some $\lambda, \lambda', \lambda_\pa, \lambda_\pb$ specific to each $X_j^{\tinyx \alpha}(u,v)$. We then have
\begin{alignat}{2}
\phi_{\pa\pb}\big(X(u,v)\big)
&= \phi_{\pa\pb}(\lambda \otimes u \otimes v)- \phi_{\pa\pb}(\lambda' \otimes \lambda_\pa u \otimes \lambda_\pb v) \nn \\[0.1cm]
&= \lambda \otimes \phi_\pa(u) \otimes \phi_\pb(v)
- \lambda' \otimes \phi_\pa(\lambda_\pa u) \otimes \phi_\pb(\lambda_\pb v) \nn \\[0.1cm]
&= \lambda \otimes \phi_\pa(u) \otimes \phi_\pb(v)
- \lambda' \otimes \lambda_\pa \phi_\pa(u) \otimes \lambda_\pb\phi_\pb(v) \nn \\[0.1cm]
&= X\big(\phi_\pa(u),\phi_\pb(v)\big) \equiv 0 \,.
\end{alignat}
At the last step, we used the equivalence relation $X(u',v')\equiv 0$ in $(\repM'_\pa \timesf \repM'_\pb)$, which holds for all $u'\in\repM'_\pa$ and $v'\in\repM'_\pb$. Thus, \cref{cor:morphism.on.Ma.Mb} applies and allows us to construct the family of homomorphisms $\psi \circ \phi_{\pa\pb}$ from $(\repM_\pa \timesf \repM_\pb)$ to $(\repM'_\pa \timesf \repM'_\pb)$.
\eproof

We now discuss the properties \eqref{eq:bifunctor.props}. Let $\repM_\pa$, $\repM'_\pa$, $\repM''_\pa$, $\repM_\pb$, $\repM'_\pb$, $\repM''_\pb$ be families of modules with homomorphisms 
\be
\phi_\pa: \repM_\pa \to \repM'_\pa\,, \qquad
\phi'_\pa: \repM'_\pa \to \repM''_\pa\,, \qquad
\phi_\pb: \repM_\pb \to \repM'_\pb\,, \qquad 
\phi'_\pb: \repM'_\pb \to \repM''_\pb\,.
\ee 
Moreover let 
\be
\phi_{\pa\pb}:(\repM_\pa \timesf \repM_\pb) \to (\repM'_\pa \timesf \repM'_\pb)\,, \qquad
\phi'_{\pa\pb}:(\repM'_\pa \timesf \repM'_\pb) \to (\repM''_\pa \timesf \repM''_\pb)
\ee 
be the homomorphisms defined in \cref{prop:phi_(ab)} in terms of $(\phi_\pa, \phi_\pb)$ and $(\phi'_\pa, \phi'_\pb)$, respectively. Then it readily follows from the definition \eqref{eq:phiab.def} of the homomorphisms $\phi_{\pa\pb}$ and $\phi'_{\pa\pb}$ that
\be
(\phi'_{\pa\pb} \circ \phi_{\pa\pb})(\repM_\pa \timesf \repM_\pb) = \big(\phi'_\pa\circ \phi_\pa(\repM_\pa) \timesf \phi'_\pb \circ \phi_\pb (\repM_\pb)\big)\,,
\ee
thus confirming the second property in \eqref{eq:bifunctor.props}. Moreover, if $\phi_\pa: \repM_\pa \to \repM_\pa$ and $\phi_\pb: \repM_\pb \to \repM_\pb$ are families of isomorphisms, then it is easy to check that the family $\phi_{\pa\pb}$ constructed from $\phi_\pa$ et $\phi_\pb$ is an isomorphism from $(\repM_\pa \times \repM_\pb)$ to itself. This is the first property in \eqref{eq:bifunctor.props}. We have thus shown the following proposition.
\begin{Proposition}
The fusion of families of modules defined from \eqref{eq:fusion.of.modules} and \eqref{eq:moving.arches}, with the definition~\eqref{eq:phiab.def} for the homomorphisms, defines a bifunctor on the category $\mathcal C$ of families of $\eptl_N(\beta)$-modules.
\end{Proposition}

\subsection{Properties of the fusion functor}

In this section, we investigate various properties of the families of fusion modules. The first property, stated in the following corollary, follows directly from \cref{prop:fusion.iso}.

\begin{Corollary}
Let $\repM_\pa$, $\repM'_\pa$ and $\repM_\pb$ be families of modules, such that $\repM_{\pa}$ and $\repM'_{\pa}$ are isomorphic. Then, the families of modules $(\repM_\pa \timesf \repM_\pb)$ and $(\repM'_\pa \timesf \repM_\pb)$ are also isomorphic.
\end{Corollary}

\noindent The following two propositions state that fusion is distributive and commutative.
\begin{Proposition}\label{prop:distributivity}
Let $\repM_\pa, \repM'_\pa, \repM''_\pa, \repM_\pb$ be families of modules, with $\repM_\pa$ decomposing as $\repM_\pa = \repM'_\pa \oplus \repM''_\pa$. Then $(\repM'_\pa \timesf \repM_\pb)$ and $(\repM''_\pa \timesf \repM_\pb)$ are isomorphic to subfamilies of $(\repM_\pa \timesf \repM_\pb)$, and
\begin{equation}\label{eq:distrib}
\repM_\pa \timesf \repM_\pb
= (\repM'_\pa \oplus \repM''_\pa)\timesf \repM_\pb
\simeq (\repM'_\pa \timesf \repM_\pb) \oplus (\repM''_\pa \timesf \repM_\pb) \,.
\end{equation}
\end{Proposition}
\proof Let us first prove that $(\repM'_\pa \timesf \repM_\pb)$ is isomorphic to a subfamily of $(\repM_\pa \timesf \repM_\pb)$. From the decomposition of $\repM_\pa$, we know that there exist families of homomorphisms 
\be
\phi'_\pa:\repM'_\pa\to\repM_\pa\,, \qquad 
\varphi'_\pa:\repM_\pa\to\repM'_\pa\,, \qquad 
\phi''_\pa:\repM''_\pa\to\repM_\pa\,, \qquad 
\varphi''_\pa:\repM_\pa\to\repM''_\pa\,.
\ee
Let $N$ be an admissible integer for $\repM_\pa$. Any element $u \in \repM_\pa(N)$ decomposes uniquely into a sum of the form $u=\phi_\pa'(u')+\phi_\pa''(u'')$ for some $u'\in\repM'_\pa(N)$ and $u''\in \repM''_\pa(N)$. The homomorphisms can be chosen such that $\varphi_\pa'\circ\phi'_\pa(u') = u'$ and $\varphi''_\pa\circ\phi''_\pa(u'') = u''$, for all $u'\in\repM'_\pa(N)$ and $u''\in \repM''_\pa(N)$. Equivalently, we have 
\begin{equation}
\varphi'_\pa\circ \phi'_\pa = \id_{\repM'_\pa}\,,\qquad
\varphi''_\pa\circ \phi''_\pa = \id_{\repM''_\pa}\,,\qquad 
\varphi'_\pa\circ \phi''_\pa = \varphi''_\pa\circ \phi'_\pa =0\,.
\end{equation}
Let us define
\begin{equation}
\phi'_{\pa\pb}={\cal F}(\phi'_\pa,\id_{\repM_\pb}) \,,
\qquad \varphi'_{\pa\pb}={\cal F}(\varphi'_\pa,\id_{\repM_\pb}) \,,
\qquad \phi''_{\pa\pb}={\cal F}(\phi''_\pa,\id_{\repM_\pb}) \,,
\qquad \varphi''_{\pa\pb}={\cal F}(\varphi''_\pa,\id_{\repM_\pb}) \,.
\end{equation}
From \eqref{eq:bifunctor.props}, we find $\varphi'_{\pa\pb}\circ \phi'_{\pa\pb}=\id_{\repM'_\pa \timesf \repM_\pb}$, $\varphi''_{\pa\pb}\circ \phi''_{\pa\pb}=\id_{\repM''_\pa \timesf \repM_\pb}$, $\varphi'_{\pa\pb}\circ \phi''_{\pa\pb}=0$, and $\varphi''_{\pa\pb}\circ \phi'_{\pa\pb}=0$. This implies that $\phi'_{\pa\pb}$ is a family of injective homomorphisms from $(\repM'_\pa \timesf \repM_\pb)$ to $(\repM_\pa \timesf \repM_\pb)$, and therefore that $(\repM'_\pa \timesf \repM_\pb)$ is isomorphic to $\phi'_{\pa\pb}(\repM'_\pa \timesf \repM_\pb)$. The same argument is used to show that $(\repM''_\pa \timesf \repM_\pb) \simeq \phi''_{\pa\pb}(\repM''_\pa \timesf \repM_\pb)$.\medskip

Next, to prove the decomposition \eqref{eq:distrib}, we start from the identity $\phi'_\pa\circ \varphi'_\pa + \phi''_\pa\circ \varphi''_\pa = \id_{\repM_\pa}$. Using~\eqref{eq:bifunctor.props}, we find $\phi'_{\pa\pb}\circ \varphi'_{\pa\pb} + \phi''_{\pa\pb}\circ \varphi''_{\pa\pb} = \id_{\repM_\pa \timesf \repM_\pb}$. Hence, for all $x \in (\repM_\pa \timesf \repM_\pb)(N)$ with $N$ admissible for $(\repM_\pa \timesf \repM_\pb)$, one can write
\begin{equation}
x = \phi'_{\pa\pb} \circ \varphi'_{\pa\pb}(x) + \phi''_{\pa\pb} \circ \varphi''_{\pa\pb}(x) \,.
\end{equation}
This state is of the form $x=\phi'_{\pa\pb}(y') + \phi''_{\pa\pb}(y'')$ for some $y' \in (\repM'_\pa \timesf \repM_\pb)(N)$ and $y'' \in (\repM''_\pa \timesf \repM_\pb)(N)$. Moreover, $y'$ and $y''$ are uniquely fixed by $x$, namely $\varphi'_{\pa\pb}(x)=y'$ and $\varphi''_{\pa\pb}(x)=y''$. This proves that
\begin{equation}
\repM_\pa \timesf \repM_\pb
= \phi'_{\pa\pb}(\repM'_\pa \timesf \repM_\pb) \oplus \phi''_{\pa\pb}(\repM''_\pa \timesf \repM_\pb) \,,
\end{equation}
ending the proof.
\eproof

\begin{Proposition} \label{prop:commutativity}
Let $\repM_\pa,\repM_\pb$ be families of modules. Then $(\repM_\pa \timesf \repM_\pb) \simeq (\repM_\pb \timesf \repM_\pa)$.
\end{Proposition}
\proof We first introduce the set of linear maps $\phi = \{\phi_N \, | \, N=N_0, N_0+2, \dots\}$ from $\repL_{\repM_\pa,\repM_\pb}$ to $\repL_{\repM_\pb,\repM_\pa}$, defined as
\begin{equation} \label{eq:swap.Ma.Mb}
\phi_N(\lambda \otimes u \otimes v) =\lambda\,\Omega^{N_\pb} \otimes v \otimes u \,,
\end{equation}
for all $\lambda \in \cL(N,N_{\pa\pb})$, $u \in \repM_\pa(N_\pa)$ and $v \in \repM_\pb(N_\pb)$.
Using \eqref{eq:L.on.fusion}, we first observe that
\begin{subequations}
\begin{alignat}{2}
\phi_N\big(\lambda' \cdot (\lambda \otimes u\otimes v)\big)
&= \phi_N\big((\lambda'\lambda) \otimes u\otimes v\big)
= \lambda'\lambda\,\Omega^{N_\pb} \otimes v \otimes u
\nonumber\\[0.1cm]
&= \lambda' \cdot \big(\lambda\,\Omega^{N_\pb} \otimes v \otimes u\big)
= \lambda' \cdot \phi_N(\lambda \otimes u \otimes v) \,,
\label{eq:swap.is.morphism}
\end{alignat}
\end{subequations}
for all $\lambda' \in\cL(N',N)$. This shows that $\phi$ is a family of homomorphisms from $\repL_{\repM_\pa,\repM_\pb}$ to $\repL_{\repM_\pb,\repM_\pa}$. Second, we prove that $\phi$ preserves the equivalence relations $X^{\tinyx \alpha}_j(u,v) \equiv 0$:
\begingroup
\allowdisplaybreaks
\begin{subequations}
\begin{alignat}{3}
\phi_{N_{\pa\pb}-2}(c_j \otimes u\otimes v)
& = c_j\,\Omega^{N_\pb} \otimes v \otimes u
= \Omega^{N_\pb} \,c_{j+N_\pb}\otimes v \otimes u 
\nn \\[0.15cm]
& \equiv \Omega^{N_\pb} \otimes v \otimes c_j u
= \phi_{N_{\pa\pb}-2}(\id \otimes c_j u \otimes v) \,, 
&&\hspace{-0.5cm}1\le j \le N_\pa-1,
\\[0.15cm]
\phi_{N_{\pa\pb}-2}(c_{j+N_\pa} \otimes u\otimes v)
& = c_{j+N_\pa}\,\Omega^{N_\pb} \otimes v \otimes u
= \Omega^{N_\pb-2} \,c_j \otimes v \otimes u
\nn \\[0.15cm]
&\equiv \Omega^{N_\pb-2} \otimes c_j v\otimes u
= \phi_{N_{\pa\pb}-2}(\id \otimes u \otimes c_j v) \,, 
\qquad&&\hspace{-0.5cm}1\le j \le N_\pb-1,
\\[0.15cm]
\phi_{N_{\pa\pb}+2}(c^\dag_j \otimes u \otimes v)
&= c^\dag_j\,\Omega^{N_\pb} \otimes v \otimes u
= \Omega^{N_\pb} \,c^\dag_{j+N_\pb} \otimes v \otimes u
\nn \\[0.15cm]
&\equiv \Omega^{N_\pb}\otimes v \otimes c^\dag_j u 
= \phi_{N_{\pa\pb}+2}(\id \otimes c^\dag_j u \otimes v) \,, 
\qquad&&\hspace{-0.5cm}1\le j \le N_\pa+1,
\\[0.15cm]
\phi_{N_{\pa\pb}+2}(c^\dag_{j+N_\pa} \otimes u\otimes v)
& = c^\dag_{j+N_\pa}\,\Omega^{N_\pb}\otimes v \otimes u
= \Omega^{N_\pb+2} \,c^\dag_j \otimes v \otimes u
\nn \\[0.15cm]
&\equiv \Omega^{N_\pb+2} \otimes c^\dag_j v\otimes u
= \phi_{N_{\pa\pb}+2}(\id \otimes u \otimes c^\dag_j v) \,, 
\qquad&&\hspace{-0.5cm}1\le j \le N_\pb+1,
\\[0.15cm]
\phi_{N_{\pa\pb}}(\Omega \otimes u \otimes v\big)
&= \Omega^{N_\pb+1} \otimes v \otimes u 
= \Omega^{N_\pb+1}\, c_{N_\pb+2}\, c^\dag_{N_\pb+1}\otimes v \otimes u
\nn \\[0.15cm]
&\equiv \Omega^{N_\pb+1} c_{N_\pb+2} \otimes c^\dag_{N_\pb+1} v \otimes u
= \Omega^{N_\pb+1} c_{N_\pb+2} \otimes \Omega\, c^\dag_0 v \otimes u
\nn \\[0.15cm]
&\equiv \Omega^{N_\pb+1} c_0 \otimes c^\dag_0 v \otimes \Omega\,u
= c_{N_\pa}\, \Omega^{N_\pb+2} \otimes c^\dag_0 v \otimes \Omega\,u
\nn \\[0.15cm]&
= \phi_{N_{\pa\pb}}(c_{N_\pa} \otimes \Omega\,u \otimes c_0^\dag v) \,,
\qquad&&\hspace{-0.5cm} \textrm{for }N_\pa>0 \,.
\end{alignat}
\end{subequations}
where we used the identities \eqref{eq:Omega.cj}. Using \cref{cor:morphism.on.Ma.Mb}, we deduce that $\phi_N$ yields a well-defined family of homomorphisms from $(\repM_\pa \timesf \repM_\pb)$ to $(\repM_\pb \timesf \repM_\pa)$.
With similar arguments, we find that the maps 
\begin{equation}
\psi_N(\lambda \otimes v \otimes u) = \lambda\,\Omega^{-N_\pb} \otimes u \otimes v
\end{equation}
for $\lambda \in \cL(N,N_{\pa\pb})$, $u \in \repM_\pa(N_\pa)$ and $v \in \repM_\pb(N_\pb)$,
define a family of homomorphisms from $\repL_{\repM_\pb,\repM_\pa}$ to $\repL_{\repM_\pa,\repM_\pb}$ that preserves the equivalence relations, and thus extends to $(\repM_\pb \timesf \repM_\pa) \to (\repM_\pa \timesf \repM_\pb)$. Because
\begin{subequations}
\begin{alignat}{3}
\psi_N \circ \phi_N(x)&= x\,,
\qquad &&\forall x \in \repL_{\repM_\pa,\repM_\pb}(N) \,, 
\\[0.15cm]
\phi_N \circ \psi_N(y)&= y\,,
\qquad &&\forall y \in \repL_{\repM_\pb,\repM_\pa}(N) \,, 
\end{alignat}
\end{subequations}
\endgroup
we conclude that $\psi_N = (\phi_N)^{-1}$, and thus that $\phi$ is a family of isomorphisms.
\eproof

\begin{Proposition}
\label{prop:Fmaps}
Let $\repM_\pa$ and $\repM_\pb$ be pairs of families of modules. The maps
\begin{subequations}
\begin{alignat}{4}
F_\pa&:\lambda \cdot(u \otimes v) \mapsto \lambda \cdot(F\,u \otimes v) \,, 
\qquad &\Fb_\pa&&:\lambda \cdot(u \otimes v) \mapsto \lambda \cdot(\Fb\,u \otimes v) \,, \\
F_\pb&:\lambda \cdot(u \otimes v) \mapsto \lambda \cdot(u \otimes F\,v) \,,
\qquad &\Fb_\pb&&:\lambda \cdot(u \otimes v) \mapsto \lambda \cdot(u \otimes \Fb\,v) \,, \\
F_{\pa\pb}&:\lambda \cdot(u \otimes v) \mapsto F \,\lambda \cdot(u \otimes v) \,,
\qquad &\Fb_{\pa\pb}&&:\lambda \cdot(u \otimes v) \mapsto \Fb \,\lambda \cdot(u \otimes v)\,,\
\end{alignat}
\end{subequations}
define families of endomorphisms on $(\repM_\pa \timesf \repM_\pb)$. Moreover, these six maps commute pairwise.
\end{Proposition}
\proof
The braid transfer matrices $F,\Fb$ commute with the action of $\cL(N',N)$. Thus, in any module~$\repM(N)$, the maps $u \mapsto F u$ and $u \mapsto \Fb u$ define families of endomorphisms over $\repM$, and $F_{\pa\pb}$ and $\Fb_{\pa\pb}$ indeed define families of endomorphisms.\medskip

Applying \cref{prop:phi_(ab)} with $(\phi_\pa,\phi_\pb)=(F,\id)$, we find that $F_\pa$ is well defined on $(\repM_\pa \timesf \repM_\pb)(N)$ and yields a family of endomorphisms on $(\repM_\pa \timesf \repM_\pb)$. A similar argument works for the maps $F_\pb$, $\Fb_\pa$ and $\Fb_\pb$.
\medskip

Proving the commutativity of these maps is straightforward. For instance, we have
\begin{subequations}
\begin{alignat}{2} 
F_\pa \circ F_\pb \big( \lambda \cdot(u \otimes v)\big)
&= F_\pb \circ F_\pa \big( \lambda \cdot(u \otimes v)\big)
= \lambda \cdot(F\,u \otimes F\,v) \,, \\[0.15cm]
F_\pa \circ \Fb_\pa \big( \lambda \cdot(u \otimes v) \big)
&= \Fb_\pa \circ F_\pa \big( \lambda \cdot(u \otimes v) \big) 
= \lambda \cdot(F\, \Fb\,u \otimes v) \,, \\[0.15cm]
F_\pa \circ F_{\pa\pb} \big( \lambda \cdot(u \otimes v)\big)
&= F_{\pa\pb} \circ F_\pa \big( \lambda \cdot(u \otimes v)\big)
= F\,\lambda \cdot(F\, u \otimes v) \,, \\[0.15cm]
F_{\pa\pb} \circ \Fb_{\pa\pb} \big( \lambda \cdot(u \otimes v)\big)
&= \Fb_{\pa\pb} \circ F_{\pa\pb}\big( \lambda \cdot(u \otimes v)\big)
= F\, \Fb\, \lambda \cdot(u \otimes v) \,,
\end{alignat}
\end{subequations}
and similarly for the other cases.
\eproof

We now investigate fusion in the case where one of the families is a quotient family. 

\begin{Proposition}
Let $\repM_\pa$ and $\repM_\pb$ be families of modules, $\repM'_\pa$ be a subfamily of $\repM_\pa$, and $\repQ_\pa$ be the quotient family $\repQ_\pa=\repM_\pa/ \repM'_\pa$. Then the fused family $(\repQ_\pa \timesf \repM_\pb)$ is isomorphic to a quotient family of $(\repM_\pa \timesf \repM_\pb)$.
\end{Proposition}
\proof
For $u\in \repM_\pa(N_\pa)$, we denote by $[u]$ the equivalence class of $u$ modulo $\repM'_\pa(N_\pa)$. The map $u\mapsto [u]$ yields a family $\eta_\pa$ of homomorphisms from $\repM_\pa$ to $\repQ_\pa$. Applying \cref{prop:phi_(ab)} with $(\phi_\pa,\phi_\pb)=(\eta_\pa,\id_{\repM_\pb})$, we find that the map defined as
\begin{equation} \label{eq:MxM->QxM}
\eta_N\big(\lambda \cdot (u\otimes v) \big) = \lambda \cdot ([u]\otimes v)\,,
\end{equation}
for $\lambda \in \cL(N,N_\pa+N_\pb), \ u \in \repM_\pa(N_\pa), \ v \in \repM_\pb(N_\pb)$, yields a well-defined family of homomorphisms $\eta = \{\eta_N\,|\, N=N_0, N_0+2, \dots\}$ from $(\repM_\pa \timesf \repM_\pb)$ to $(\repQ_\pa \timesf \repM_\pb)$. Moreover, since $(\repQ_\pa \timesf \repM_\pb)(N)$ is spanned by states of the form $\lambda \cdot ([u]\otimes v)$, we can construct a state in the pre-image of $\eta_N$ for any state in this module, implying that the map \eqref{eq:MxM->QxM} yields a surjective homomorphism from $(\repM_\pa \timesf \repM_\pb)(N)$ to $(\repQ_\pa \timesf \repM_\pb)(N)$, for all admissible~$N$. As a consequence, the fused module $(\repQ_\pa \timesf \repM_\pb)(N)$ is isomorphic to a quotient of $(\repM_\pa \timesf \repM_\pb)(N)$:
\begin{equation}
(\repQ_\pa \timesf \repM_\pb)(N)
= \eta_N\big((\repM_\pa \timesf \repM_\pb)(N)\big)
\simeq (\repM_\pa \timesf \repM_\pb)(N) \Big/ \ker \eta_N \,.
\end{equation}
\eproof

Finally, we show that the fused modules can be generated by the repeated action of $c_0$ on states of the form $(u\otimes v)$ with $u\in \repM_\pa(N_\pa)$ and $v\in \repM_\pb(N_\pb)$, for some admissible $N_\pa$ and $N_\pb$ restricted to the following shaded region:
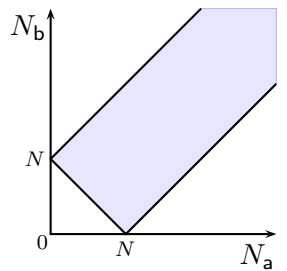
\begin{equation*}
\thispic{\begin{pspicture}(0,0)(3,3)
\psline{<->}(0,3)(0,0)(3,0)
\rput(2.75,-0.3){$N_\pa$}
\rput(-0.3,2.75){$N_\pb$}
\rput(-0.1,-0.1){$_0$}
\rput(1,-0.2){$_N$}
\rput(-0.2,1){$_N$}
\pspolygon[fillstyle=solid,fillcolor=lightlightblue,linewidth=0pt,linecolor=lightblue](1,0)(3,2)(3,3)(2,3)(0,1)
\psline{-}(2,3)(0,1)(1,0)(3,2)
\end{pspicture}}\ .
\end{equation*}

\begin{Proposition} \label{prop:MxM=c.(uxv)}
Let $\repM_\pa$ and $\repM_\pb$ be two families of modules. The fused module $(\repM_\pa \timesf \repM_\pb)(N)$ is given by
\begin{equation}
(\repM_\pa \timesf \repM_\pb)(N) =
\sum_{\substack{N_\pa,N_\pb\\[0.05cm]|N_\pa-N_\pb|\leq N \leq N_{\pa\pb}}}
c_0^{(N_{\pa\pb}-N)/2} \cdot \big(\repM_\pa(N_\pa)\otimes \repM_\pb(N_\pb)\big)\,,
\end{equation}
where $N_{\pa\pb}=N_\pa+N_\pb$, and $N_\pa$ and $N_\pb$ are admissible integers for the families $\repM_\pa$ and $\repM_\pb$, respectively.
\end{Proposition}
\proof
From its definition \eqref{eq:fusion.of.modules}, we know that the module $(\repM_\pa \timesf \repM_\pb)(N)$ is spanned by states of the form $\lambda \cdot (u'\otimes v')$, where $\lambda$ is a diagram in $\cL(N,N'_{\pa\pb})$, $u' \in \repM_\pa(N'_\pa)$, and $v'\in \repM_\pb(N'_\pb)$, for some admissible integers $N'_\pa$ and $N'_\pb$. We want to show that we always have
\begin{equation} \label{eq:reduced-form}
\lambda \cdot (u' \otimes v') = c_0^{(N_{\pa\pb}-N)/2}\cdot(u\otimes v)
\end{equation}
for some $u \in \repM_\pa(N_\pa)$, $v\in \repM_\pb(N_\pb)$, where $N_{\pa\pb} = N_\pa + N_\pb$, and $N_\pa$ and $N_\pb$ are admissible integers subject to the constraint $|N_\pa-N_\pb|\leq N \leq N_\pa+N_\pb$. We write $n=\frac12(N_{\pa\pb}-N)$, and remark that this condition is equivalent to $0\leq n\leq \mathrm{min}(N_\pa,N_\pb)$.
\medskip

To prove \eqref{eq:reduced-form}, we decompose $\lambda$ as a word in the generators $c_j$ and $c_j^\dag$, and show that the statement holds iteratively on the length $\ell$ of the word $\lambda$. For $\lambda = \id$, the length is $\ell = 0$ and the statement holds trivially with $n = 0$. The inductive assumption is that \eqref{eq:reduced-form} holds for all words of length~$\ell$. Each word $\lambda$ of length $\ell+1$ can be written as either $c^\dag_j \lambda'$ or $c_j \lambda'$, where $\lambda'$ has length $\ell$, and $\lambda'\in \cL(N',N_{\pa\pb})$, with $N'=N-2$ if $\lambda=c^\dag_j \lambda'$, and $N'=N+2$ if $\lambda=c_j \lambda'$. Using the inductive assumption, we write
\begin{equation}
\lambda' \cdot (u' \otimes v') = c_0^n \cdot (u \otimes v)
\end{equation}
for some $u\in \repM_\pa(N_\pa)$ and $v\in \repM_\pb(N_\pb)$, with $n=\frac12(N_{\pa\pb}-N')$ satisfying $0\leq n\leq \mathrm{min}(N_\pa,N_\pb)$.
\medskip

For $\lambda = c^\dag_j \lambda'$, we have
\begin{equation} \label{eq:cjdag.c0n.u.v}
\lambda\cdot(u'\otimes v') = c_j^\dag\, c_0^n \cdot (u\otimes v) = 
\left\{\begin{array}{ll}
c_0^{n+1} \cdot (c_{n+1}^\dag u\otimes c_{N_\pb-n+1}^\dag v)
\quad & j=0 \,,\\[0.1cm]
c_0^n \cdot (c_{j+n}^\dag u\otimes v)
& 1\leq j\leq N_\pa-n \,, \\[0.1cm]
c_0^n \cdot (u\otimes c_{j-N_\pa+n'}^\dag v)
& N_\pa-n+1\leq j\leq N_{\pa\pb}-2n+1 \,. \\[0.1cm]
\end{array}\right.
\end{equation}
For $j=0$, we used the identity $c_0^\dag =c_0 c_1^\dag c_{N-1}^\dag$ in $\cL(N,N-2)$. The expressions on the right side have $(n'',N''_\pa, N''_\pb)$ equal to $(n+1,N_\pa+2, N_\pb+2)$, $(n,N_\pa+2, N_\pb)$ and $(n,N_\pa, N_\pb+2)$, respectively. These triples all satisfy the inequality $0\leq n''\leq \mathrm{min}(N''_\pa,N''_\pb)$. We have thus obtained expressions for $\lambda \cdot (u'\otimes v')$ of the desired form for $\lambda = c^\dag_j \lambda'$, for all indices $j$.
\medskip

For $\lambda = c_j \lambda'$, we similarly use the inductive assumption and find
\begin{equation} \label{eq:cj.c0n.u.v}
\lambda\cdot(u'\otimes v') =
c_j\, c_0^n \cdot (u\otimes v) = \left\{\begin{array}{ll}
c_0^{n+1} \cdot (u \otimes v)
& j=0 \,,\\[0.1cm]
c_0^n \cdot (c_{j+n} u\otimes v)
& 1\leq j\leq N_\pa-n-1 \,, \\[0.1cm]
c_0^{n+1} \cdot (\Omega^{-1} u \otimes \Omega\, v)
& j=N_\pa-n \,, \ \quad n < \min(N_\pa,N_\pb)\,,
\\[0.1cm]
c_0^n \cdot (u\otimes c_{j-N_\pa+n} v)
\quad & N_\pa-n+1\leq j\leq N_{\pa\pb}-2n-1 \,.
\end{array}\right.
\end{equation}
where $0 \le n < \frac12N_{\pa\pb}$. The inequality $n<\frac12N_{\pa\pb}$ ensures that $\lambda$ is in $\cL(N,N_{\pa\pb})$ with $N \ge 0$. For the case $j=N_\pa-n$, we used the relation \eqref{eq:c0=cNa}. The value $n = \min(N_\pa,N_\pb)$ is excluded in this case, since the corresponding generator $c_j$ is equal to $c_0$ and its evaluation is instead given by the first line. We find that the values of $(n'',N''_\pa, N''_\pb)$ are $(n+1,N_\pa, N_\pb)$, $(n,N_\pa-2, N_\pb)$, $(n+1,N_\pa, N_\pb)$, and $(n,N_\pa, N_\pb-2)$, respectively. For the second line, we must have $n\le N_\pa-2$ in order to have at least one value of~$j$, and this implies that $0 \le n'' \le \min(N''_\pa,N''_\pb)$. Similarly, for the fourth line, we have $n\le N_\pb-2$, also implying that $0 \le n'' \le \min(N''_\pa,N''_\pb)$. For the third line, the constraint $0 \le n< \min(N_\pa,N_\pb)$ also ensures that $0 \le n'' \le \min(N''_\pa,N''_\pb)$. In all three cases, this yields expressions for $\lambda \cdot (u' \otimes v')$ of the desired form, for $\lambda = c_j \lambda'$ for all $j>0$.\medskip

Only the case $j=0$ remains to be resolved. For $n\leq \mathrm{min}(N_\pa,N_\pb)-1$, we readily observe that $c_0^{n+1} \cdot (u\otimes v)$ is of the desired form with $0 \le n''\le \min(N''_\pa,N''_\pb)$. For $n= \mathrm{min}(N_\pa,N_\pb)$, we have either $n=N_\pa \le N_\pb-2$ or $n=N_\pb \le N_\pa -2$. Cases with $|N_\pa-N_\pb| = 1$ are excluded because they would violate the inequality $n < \frac12N_{\pa\pb}$. We then have
\begin{equation}
u\otimes v = \left\{\begin{array}{ll}
c_{N_\pa+2} \cdot (c_{N_\pa+1}^\dag u\otimes v)
= c_0 \cdot (\Omega^{-1} c_{N_\pa+1}^\dag u\otimes \Omega\, v)
= c_0 \cdot (c_0^\dag u\otimes \Omega\, v) & \ \ n=N_\pa \le N_\pb - 2 \,, \\[0.1cm]
c_{N_\pa} \cdot (u\otimes c_1^\dag v)
= c_0 \cdot (\Omega^{-1} u\otimes \Omega\, c_1^\dag v)
= c_0 \cdot (\Omega^{-1} u\otimes c_0^\dag v) & \ \ n=N_\pb \le N_\pa - 2 \,,
\end{array}\right.
\end{equation}
where we used \eqref{eq:c0=cNa}.
This yields
\begin{equation}
\lambda \cdot (u' \otimes v') = 
c_0^{n+1} \cdot (u\otimes v)
= \left\{\begin{array}{ll}
c_0^{N_\pa+2} \cdot (c_0^\dag u\otimes \Omega \,v)
& n=N_\pa\le N_\pb -2 \,, \ \ 
\\[0.1cm]
c_0^{N_\pb+2} \cdot (\Omega^{-1} u\otimes c_0^\dag v)
\quad & n=N_\pb\le N_\pa -2 \,. \ \ \end{array}\right.
\end{equation}
The resulting states have $(n'',N''_\pa, N''_\pb)$ equal to $(N_\pa+2, N_\pa+2, N_\pb)$ and $(N_\pb+2, N_\pa, N_\pb+2)$, respectively, both of which respect the inequality $0 \le n'' \le \min(N''_\pa,N''_\pb)$. We thus obtained expressions of the desired form for $\lambda \cdot (u' \otimes v')$, for all indices $j$. This ends the proof of the inductive hypothesis.
\eproof

\subsection{Fusion of transformed families of modules}

\begin{Proposition}\label{prop:change-sign}
Let $\repM_\pa$ and $\repM_\pb$ be families of modules. Then
\be
(\repM^-_\pa \timesf \repM_\pb) \simeq (\repM_\pa \timesf \repM_\pb)^-\,.
\ee
\end{Proposition}
\proof
We denote the elements of $\repL_{\repM^-_\pa,\repM_\pb}(N)$ by $(\lambda \otimes u^- \otimes v)$ to distinguish them from the identical elements $(\lambda \otimes u \otimes v)$ of $\repL_{\repM_\pa,\repM_\pb}(N)$. Similarly, we denote by $(\lambda \otimes u \otimes v)^-$ the elements of $\repL^-_{\repM_\pa,\repM_\pb}(N)$. We define the linear map $\phi_N:\repL_{\repM^-_\pa,\repM_\pb}(N) \to \repL^-_{\repM_\pa,\repM_\pb}(N)$ as
\begin{equation}
\phi_N(\lambda \otimes u^- \otimes v) = \sigma(\lambda) \, (\lambda \otimes u \otimes v)^- 
\end{equation}
for all $\lambda \in \cL(N,N_{\pa\pb})$, $u \in \repM_\pa(N_\pa)$ and $v \in \repM_\pb(N_\pb)$.
We then have
\begin{alignat}{2}
\phi_{N'}\big(\lambda' \cdot (\lambda\otimes u^- \otimes v)\big)
&= \phi_{N'}\big((\lambda'\lambda) \otimes u^- \otimes v\big)
= \sigma(\lambda'\lambda) \, (\lambda'\lambda \otimes u \otimes v)^-
\nonumber\\[0.15cm]
&= \sigma(\lambda) \sigma(\lambda') \big(\lambda' \cdot (\lambda \otimes u \otimes v)\big)^-
= \sigma(\lambda) \, \lambda' \cdot (\lambda \otimes u \otimes v)^-
\nonumber\\[0.15cm]
&= \lambda' \cdot \phi_N(\lambda \otimes u^- \otimes v) \,,
\end{alignat}
for all $\lambda' \in \cL(N',N)$, $\lambda \in \cL(N,N_{\pa\pb})$, $u \in \repM_\pa(N_\pa)$ and $v \in \repM_\pb(N_\pb)$. As a result, we conclude that $\phi = \{\phi_N\,|\, N=N_0, N_0+2, \dots\}$ is a family of homomorphisms. Moreover, $\phi$ preserves the equivalence relations $X^{\tinyx \alpha}_j(u,v) \equiv 0$. Indeed, using the general form \eqref{eq:X.general} for these relations, we find
\begin{alignat}{2}
\phi_{N^{\tinyx \alpha}_{\pa\pb}}(\lambda \otimes u^-\otimes v)
&= \sigma(\lambda)(\lambda \otimes u \otimes v)^-
\nn\\[0.15cm]&
\equiv \sigma(\lambda)(\lambda' \otimes \lambda_\pa u \otimes \lambda_\pb v)^-
= \sigma(\lambda)\sigma(\lambda')\phi_{N_{\pa\pb}^{\tinyx \alpha}}(\lambda' \otimes (\lambda_\pa u)^- \otimes \lambda_\pb v)
\nn\\[0.15cm]&
= \sigma(\lambda)\sigma(\lambda')\sigma(\lambda_\pa)\phi_{N^{\tinyx \alpha}_{\pa\pb}}(\lambda \otimes \lambda_\pa u^- \otimes \lambda_\pb v)\,.
\end{alignat}
It is then easy to check from \eqref{eq:X(u,v)} that $\sigma(\lambda)\sigma(\lambda')\sigma(\lambda_\pa)=1$ in all cases. From \cref{cor:morphism.on.Ma.Mb}, we find that $\phi$ is a well-defined family of homomorphisms from $(\repM^-_\pa \timesf \repM_\pb)$ to $(\repM_\pa \timesf \repM_\pb)^-$.
\medskip

Similarly, one can prove that the map $\psi_N:\repL^-_{\repM_\pa,\repM_\pb}(N) \to \repL_{\repM^-_\pa,\repM_\pb}(N)$ defined by
\begin{equation}
\psi_N\big((\lambda \otimes u \otimes v)^-\big)
= \sigma(\lambda)\, (\lambda \otimes u^- \otimes v)
\end{equation}
is a homomorphism that preserves the equivalence relations. It is thus well-defined from $(\repM^-_\pa \timesf \repM_\pb)(N)$ to $(\repM_\pa \timesf \repM_\pb)^-(N)$. These maps satisfy $\psi_N \circ \phi_N(x) = x$ and $\phi_N \circ \psi_N(y) = y$ for all $x \in \repL_{\repM^-_\pa,\repM_\pb}(N)$ and $y \in \repL^-_{\repM_\pa,\repM_\pb}(N)$, thus proving that the two families of modules are isomorphic.
\eproof

\begin{Proposition} \label{prop:reverse}
Let $\repM_\pa$ and $\repM_\pb$ be families of modules. Then
\begin{equation}
(\repM^r_\pa \timesf \repM^r_\pb) \simeq (\repM_\pb \timesf \repM_\pa)^r\,.
\end{equation}
\end{Proposition}
\proof
We denote by $(\lambda \otimes u^r \otimes v^r)$ and $(\lambda \otimes v \otimes u)^r$ the elements of the modules $\repL_{\repM^r_\pa,\repM^r_\pb}(N)$ and $\repL^r_{\repM_\pb,\repM_\pa}(N)$, respectively. We define the linear maps $\phi_N: \repL_{\repM^r_\pa,\repM^r_\pb}(N)\to \repL^r_{\repM_\pb,\repM_\pa}(N)$ by
\begin{equation}
\phi_N(\lambda \otimes u^r \otimes v^r) = \big(R(\lambda) \otimes v\otimes u\big)^r \,.
\end{equation}
We then have
\begin{alignat}{2}
\phi_N\big(\lambda' \cdot (\lambda \otimes u^r \otimes v^r)\big)
&= \phi_N(\lambda' \lambda \otimes u^r \otimes v^r)
= \big(R(\lambda' \lambda) \otimes v \otimes u\big)^r
\nn\\[0.1cm]
&= \big(R(\lambda') R(\lambda) \otimes v \otimes u\big)^r
= \lambda' \cdot \big(R(\lambda) \otimes v\otimes u\big)^r
\nn\\[0.1cm]
&= \lambda' \cdot \phi_N(\lambda \otimes u^r \otimes v^r) \,.
\end{alignat}
Moreover, these maps preserve the equivalence relations $X^{\tinyx \alpha}_j(u,v) \equiv 0$:
\begingroup
\allowdisplaybreaks
\begin{subequations}
\begin{alignat}{2}
\phi_{N_{\pa\pb}-2}(c_j \otimes u^r \otimes v^r)
&= (c_{N_{\pa\pb}-j} \otimes v \otimes u)^r
\equiv (\id \otimes v \otimes c_{N_\pa-j}u)^r 
\nn\\[0.15cm]\nn
&= \phi_{N_{\pa\pb}-2}\big(\id \otimes (c_{N_\pa-j}u)^r \otimes v^r\big)
\nn\\[0.15cm]
&= \phi_{N_{\pa\pb}-2}(\id \otimes c_j u^r \otimes v^r)\,, 
\quad &&1 \le j \le N_\pa-1,
\\[0.15cm]
\phi_{N_{\pa\pb}-2}(c_{j+N_\pa} \otimes u^r\otimes v^r)
&= (c_{N_\pb-j} \otimes v \otimes u)^r
\equiv (\id \otimes c_{N_\pb-j}v \otimes u)^r
\nn\\[0.15cm]
&= \phi_{N_{\pa\pb}-2}\big(\id \otimes u^r\otimes (c_{N_\pb-j}v)^r\big)
\nn\\[0.15cm]
&= \phi_{N_{\pa\pb}-2}(\id \otimes u^r \otimes c_jv^r) \,, 
\quad&& 1 \le j \le N_\pb-1,
\\[0.15cm]
\phi_{N_{\pa\pb}+2}(c^\dag_j \otimes u^r\otimes v^r) 
&= (c^\dag_{N_{\pa\pb}+2-j} \otimes v \otimes u)^r
 \equiv (\id \otimes v \otimes c^\dag_{N_\pa+2-j}u)^r 
\nn\\[0.15cm]
&= \phi_{N_{\pa\pb}+2}\big(\id \otimes (c^\dag_{N_\pa+2-j}u)^r \otimes v^r\big)
\nn\\[0.15cm]
&= \phi_{N_{\pa\pb}+2}(\id \otimes c^\dag_j u^r \otimes v^r)\,, 
\quad&&1 \le j \le N_\pa+1,
\label{eq:r.phi.c}
\\[0.15cm]
\phi_{N_{\pa\pb}+2}(c^\dag_{j+N_\pa} \otimes u^r\otimes v^r)
&= (c^\dag_{N_\pb+2-j}\otimes v \otimes u)^r
\equiv (\id \otimes c^\dag_{N_\pb+2-j}v \otimes u)^r 
\nn\\[0.15cm]
&= \phi_{N_{\pa\pb}+2}\big(\id \otimes u^r\otimes (c^\dag_{N_\pb+2-j}v)^r\big)
\nn\\[0.15cm]
&= \phi_{N_{\pa\pb}+2}(\id \otimes u^r \otimes c^\dag_j v^r) \,, 
\quad&&1 \le j \le N_\pb+1,
\label{eq:r.phi.d}
\\[0.15cm]
\phi_{N_{\pa\pb}}(\Omega \otimes u^r\otimes v^r)
& = (\Omega^{-1} \otimes v\otimes u)^r
\equiv (c_{N_\pb+2} \otimes c_0^\dagger v\otimes \Omega^{-1} u)^r
\nn\\[0.15cm] 
&= \phi_{N_{\pa\pb}} \big(c_{N_\pa} \otimes (\Omega^{-1} u)^r \otimes (c_0^\dagger v)^r\big)
\nn\\[0.15cm] 
&= \phi_{N_{\pa\pb}} \big(c_{N_\pa} \otimes \Omega\, u^r \otimes c_0^\dagger v^r\big)
 \,,\quad&& \textrm{for }N_\pa>0 \,.
\end{alignat}
\end{subequations}
\endgroup
Using \cref{cor:morphism.on.Ma.Mb}, we deduce that $\phi =\{\phi_N\,|\, N = N_0, N_0+2, \dots\}$ is a family of homomorphisms from $(\repM^r_\pa \timesf \repM^r_\pb)$ to $(\repM_\pb \timesf \repM_\pa)^r$.\medskip

Similarly, one can construct the maps $\psi_N: \repL^r_{\repM_\pb,\repM_\pa}(N) \to \repL_{\repM^r_\pa,\repM^r_\pb}(N)$, defined by
\begin{equation}
\psi_N\big((\lambda \otimes v \otimes u)^r\big)
= R(\lambda) \otimes u^r \otimes v^r.
\end{equation}
With the same arguments as above, we find that these maps define a family $\psi$ of homomorphisms that preserve the equivalence relations, and is thus well-defined from $(\repM_\pb \timesf \repM_\pa)^r$ to $(\repM^r_\pa \timesf \repM^r_\pb)$. It is also easy to see that $\psi_N$ is the inverse $\phi_N$. This ends the proof that $(\repM^r_\pa \timesf \repM^r_\pb)$ and $(\repM_\pb \timesf \repM_\pa)^r$ are isomorphic.
\eproof

\subsection{Fusion with the vacuum} 

In this section, we show that the family $\repV$ is the neutral element for the fusion of families of modules.
\begin{Proposition}\label{prop:vacuum.fusion}
Let $\repM$ be a family of modules. The fusion of $\repM$ with $\repV$ yields
\begin{equation}
\repM \timesf \repV \simeq \repM \,.
\end{equation}
\end{Proposition}
\proof
Let $N$ be a non-negative even integer and $v_0$ be the unique empty link state of $\repV(0)$. We recall from \cref{sec:definitions-EPTL} that $\cL_0(N,N')$ is the subspace of $\cL(N,N')$ spanned by the diagrams without loop segments intersecting the dashed line. For each link state $v \in \repV(N)$, there exists a unique diagram $\lambda \in \cL_0(N,0)$ such that $v=\lambda \cdot v_0$. Thus, the states of the form $\lambda \cdot v_0$ with $\lambda \in \cL_0(N,0)$ form a basis of $\repV(N)$.\medskip

Let $\lambda_\pa$ and $\lambda_\pb$ be diagrams in $\cL_0(N_\pa,N'_\pa)$ and $\cL_0(N_\pb,N'_\pb)$ respectively. We define the concatenated diagram $(\lambda_\pa \otimes \lambda_\pb) \in \cL_0(N_{\pa\pb},N'_{\pa\pb})$, obtained by assigning to the nodes $1,2, \dots, N_\pa$ and $1,2, \dots, N'_\pa$ on the inner and outer boundaries of the disc the same connections as $\lambda_\pa$, and similarly to the nodes $N_\pb+1, N_\pb+2, \dots, N_{\pa\pb}$ and $1,2, \dots, N'_\pa$ on the same boundary the connections of $\lambda_\pb$. For example, we have
\be
\psset{unit=0.6cm}
\thispic{\begin{pspicture}[shift=-1.5](-1.6,-1.6)(1.6,1.6)
\psarc[linecolor=black,linewidth=0.5pt,fillstyle=solid,fillcolor=lightlightblue]{-}(0,0){1.5}{0}{360}
\psarc[linecolor=black,linewidth=0.5pt,fillstyle=solid,fillcolor=white]{-}(0,0){0.7}{0}{360}
\psbezier[linecolor=blue,linewidth=1.5pt]{-}(0.666, 0.216)(1.046, 0.34)(0., 1.1)(0., 0.7)
\psbezier[linecolor=blue,linewidth=1.5pt]{-}(0.651, -1.351)(0.477, -0.991)(1.072, -0.245)(1.462, -0.334)
\psbezier[linecolor=blue,linewidth=1.5pt]{-}(0.411, -0.566)(0.764, -1.052)(0.86, 0.686)(1.173, 0.935)
\psbezier[linecolor=blue,linewidth=1.5pt]{-}(-0.411, -0.566)(-0.647, -0.89)(-0.477, -0.991)(-0.651, -1.351)
\psbezier[linecolor=blue,linewidth=1.5pt]{-}(-1.173, 0.935)(-0.86, 0.686)(-1.072, -0.245)(-1.462, -0.334)
\psbezier[linecolor=blue,linewidth=1.5pt]{-}(-0.666, 0.216)(-1.046, 0.34)(0., 1.1)(0., 1.5)
\psline[linestyle=dashed, dash= 1.5pt 1.5pt,linewidth=0.5pt]{-}(0,-1.5)(0,-0.7)
\end{pspicture}}
\ \otimes \
\thispic{\begin{pspicture}[shift=-1.5](-1.6,-1.6)(1.6,1.6)
\psarc[linecolor=black,linewidth=0.5pt,fillstyle=solid,fillcolor=lightlightblue]{-}(0,0){1.5}{0}{360}
\psarc[linecolor=black,linewidth=0.5pt,fillstyle=solid,fillcolor=white]{-}(0,0){0.7}{0}{360}
\psbezier[linecolor=blue,linewidth=1.5pt]{-}(0.7, 0.)(1.1, 0.)(0.778, -0.778)(1.061, -1.061)
\psbezier[linecolor=blue,linewidth=1.5pt]{-}(-0.7, 0.)(-1.1, 0.)(-0.778, -0.778)(-1.061, -1.061)
\psbezier[linecolor=blue,linewidth=1.5pt]{-}(1.061, 1.061)(0.778, 0.778)(-0.778, 0.778)(-1.061, 1.061)
\psline[linestyle=dashed, dash= 1.5pt 1.5pt,linewidth=0.5pt]{-}(0,-1.5)(0,-0.7)
\end{pspicture}}
\ = \
\thispic{\begin{pspicture}[shift=-1.5](-1.6,-1.6)(1.6,1.6)
\psarc[linecolor=black,linewidth=0.5pt,fillstyle=solid,fillcolor=lightlightblue]{-}(0,0){1.5}{0}{360}
\psarc[linecolor=black,linewidth=0.5pt,fillstyle=solid,fillcolor=white]{-}(0,0){0.7}{0}{360}
\psbezier[linecolor=blue,linewidth=1.5pt]{-}(0.423, -1.439)(0.31, -1.055)(0.831, -0.72)(1.134, -0.982)
\psbezier[linecolor=blue,linewidth=1.5pt]{-}(0.304, -0.631)(0.477, -0.991)(1.089, -0.157)(1.485, -0.213)
\psbezier[linecolor=blue,linewidth=1.5pt]{-}(0.682, -0.156)(1.072, -0.245)(0.86, 0.686)(0.547, 0.436)
\psbezier[linecolor=blue,linewidth=1.5pt]{-}(0., 0.7)(0., 1.1)(1.001, 0.457)(1.364, 0.623)
\psbezier[linecolor=blue,linewidth=1.5pt]{-}(0.811, 1.262)(0.595, 0.925)(0., 1.1)(0., 1.5)
\psbezier[linecolor=blue,linewidth=1.5pt]{-}(-0.547, 0.436)(-0.86, 0.686)(-0.595, 0.925)(-0.811, 1.262)
\psbezier[linecolor=blue,linewidth=1.5pt]{-}(-1.485, -0.213)(-1.089, -0.157)(-0.831, -0.72)(-1.134, -0.982)
\psbezier[linecolor=blue,linewidth=1.5pt]{-}(-0.682, -0.156)(-1.072, -0.245)(-1.001, 0.457)(-1.364, 0.623)
\psbezier[linecolor=blue,linewidth=1.5pt]{-}(-0.304, -0.631)(-0.477, -0.991)(-0.31, -1.055)(-0.423, -1.439)
\psline[linestyle=dashed, dash= 1.5pt 1.5pt,linewidth=0.5pt]{-}(0,-1.5)(0,-0.7)
\end{pspicture}}\ \ .
\ee

We introduce the maps $\phi_N:\repL_{\repM,\repV}(N) \to \repM(N)$ defined by
\begin{equation}
\phi_N\big( \lambda \otimes u \otimes \lambda_\pb v_0 \big) = \lambda \, (\id_{N_\pa} \otimes \lambda_\pb) \cdot u \,,
\end{equation}
for $\lambda \in \cL(N,N_{\pa\pb})$, $\lambda_\pb \in \cL_0(N_\pb,0)$ and $u \in \repM(N_\pa)$. First, we show that $\phi =\{\phi_N\,|\,N=N_0, N_0+2, \dots\}$ is a family of homomorphisms. Indeed, we have
\begin{equation}
\phi_{N'}\big(\lambda'\cdot ( \lambda \otimes u \otimes \lambda_\pb\, v_0)\big)
= \phi_{N'} \big(\lambda'\lambda \otimes u \otimes \lambda_\pb v_0 \big)
= \lambda' \, \lambda \, (\id_{N_\pa} \otimes \lambda_\pb ) \cdot u
= \lambda'\cdot \phi_N \big( \lambda \otimes u \otimes \lambda_\pb v_0 \big) \,,
\end{equation}
for all $\lambda' \in \cL(N',N)$, $\lambda \in \cL(N,N_{\pa\pb})$, $\lambda_\pb \in \cL_0(N_\pb,0)$ and $u \in \repM(N_\pa)$. Second, we check that $\phi$ preserves the equivalence relations $X^{\tinyx\alpha}_j(u,v)\equiv0$:
\begingroup
\allowdisplaybreaks
\begin{subequations}
\begin{alignat}{3}
\phi_{N_{\pa\pb}-2}(c_j \otimes u \otimes \lambda_\pb v_0)
&= c_j \, (\id_{N_\pa} \otimes \lambda_\pb) \cdot u
= (\id_{N_\pa} \otimes \lambda_\pb)\, c_j \cdot u
\nn\\[0.1cm]
&= \phi_{N_{\pa\pb}-2}(\id \otimes c_j u \otimes \lambda_\pb v_0) \,, 
&&\quad1\le j \le N_\pa-1,
\\[0.1cm]
\phi_{N_{\pa\pb}-2}(c_{j+N_\pa} \otimes u \otimes \lambda_\pb v_0)
&= c_{j+N_\pa} \, (\id_{N_\pa} \otimes \lambda_\pb) \cdot u
= (\id_{N_\pa} \otimes c_j\lambda_\pb) \cdot u
\nn\\[0.1cm]
&= \phi_{N_{\pa\pb}-2}(\id \otimes u \otimes c_j \lambda_\pb v_0) \,, 
&&\quad1\le j \le N_\pb-1,
\\[0.1cm]
\phi_{N_{\pa\pb}+2}(c^\dag_j \otimes u \otimes \lambda_\pb v_0)
&= c^\dag_j \, (\id_{N_\pa} \otimes \lambda_\pb) \cdot u
= (\id_{N_\pa} \otimes \lambda_\pb) \, c^\dag_j\,\cdot u
\nn\\[0.1cm]
&= \phi_{N_{\pa\pb}+2}(\id \otimes c^\dag_j u \otimes \lambda_\pb v_0) \,,
&&\quad1\le j \le N_\pa+1, 
\\[0.1cm]
\phi_{N_{\pa\pb}+2}(c^\dag_{j+N_\pa} \otimes u \otimes \lambda_\pb v_0)
&= c^\dag_{j+N_\pa} \, (\id_{N_\pa} \otimes \lambda_\pb) \cdot u
= (\id_{N_\pa} \otimes c^\dag_j\lambda_\pb) \cdot u
\nn\\[0.1cm]
&= \phi_{N_{\pa\pb}+2}(\id \otimes u \otimes c^\dag_j\lambda_\pb v_0) \,, 
&&\quad1\le j \le N_\pb+1, 
\\[0.1cm]
\phi_{N_{\pa\pb}}(\Omega \otimes u \otimes \lambda_\pb v_0)
&= \Omega\,(\id_{N_\pa} \otimes \lambda_\pb) \cdot u 
= (\id_{N_\pa-1} \otimes \lambda_\pb \otimes \id_1)\, \Omega \cdot u 
\nn\\[0.1cm]
& = c_{N_\pa}(\id_{N_\pa} \otimes \big(\id_1 \otimes \lambda_\pb \otimes \id_1)c^\dag_1\big) \,\Omega \cdot u 
\nn\\[0.1cm]
&= \phi_{N_{\pa\pb}}\big(c_{N_\pa} \otimes \Omega\, u \otimes \big(\id_1 \otimes \lambda_\pb \otimes \id_1)\,c_1^\dag v_0\big)
\nn\\[0.1cm]
&= \phi_{N_{\pa\pb}}(c_{N_\pa} \otimes \Omega\, u \otimes c_0^\dag \lambda_\pb \, v_0) \,,
&&\quad \textrm{for }N_\pa>0 \,,
\end{alignat}
\end{subequations}
\endgroup
for all $u\in\repM_\pa(N_\pa)$ and $\lambda_\pb \in \cL_0(N_\pb,0)$. At the last step, we used the identity $c^\dag_1 v_0 = c^\dag_0 v_0$ in $\repV$. From \cref{lem:morphism.on.Ma.Mb}, we deduce that $\phi$ is also a family of homomorphisms from $(\repM \timesf \repV)$ to $\repM$. Its action on $(\repM \timesf \repV)$ is given by
\begin{equation}
\phi_N \big(\lambda\cdot(u\otimes \lambda_\pb v_0) \big)
= \lambda\, (\id_{N_\pa}\otimes \lambda_\pb) \cdot u \,.
\end{equation}

Next, we consider the map $\psi_N$ from $\repM(N)$ to $(\repM \timesf \repV)(N)$ defined by
\begin{equation}
\psi_N(u) = u \otimes v_0 \,,
\end{equation}
for $u \in \repM(N)$. To prove that $\psi = \{\psi_N\,|\,N=N_0,N_0+2, \dots\}$ is a family of homomorphisms, it is sufficient to prove that
\begin{equation}
\psi_{N-2}(c_j \cdot u) = c_j \cdot \psi_N(u)
\qquad \text{and} \qquad
\psi_{N+2}(c^\dag_j \cdot u) = c^\dag_j \cdot \psi_N(u) \,,
\end{equation}
for all indices $j$ and all $u \in \repM(N)$.
First, we have
\begin{subequations}
\begin{alignat}{3}
\psi_{N-2}(c_j \cdot u)
&= c_j \, u \otimes v_0 = c_j \cdot (u \otimes v_0)
= c_j \cdot \psi_N(u) \,,
\qquad && 1 \le j \le N-1 \,, \\[0.1cm]
\psi_{N+2}(c^\dag_j \cdot u)
&= c^\dag_j \, u \otimes v_0
= c^\dag_j \cdot (u \otimes v_0)
= c^\dag_j \cdot \psi_N(u) \,, \qquad &&1 \le j \le N+1 \,.
\end{alignat}
\end{subequations}
For $c_0$ and $c_0^\dag$, we write
\begin{subequations}
\begin{alignat}{2}
\psi_{N-2}(c_0 \cdot u)
&= c_0 \, u \otimes v_0
= c_1 c_{N-1} c^\dag_0 u \otimes v_0
= c_1 c_{N-1} \cdot (c^\dag_0 u \otimes v_0)
= c_1 c_{N-1} \Omega^{-1} \cdot (u \otimes c^\dag_0 v_0)
\nn \\[0.15cm]&
= c_1 c_{N-1} \Omega^{-1} \cdot (u \otimes c^\dag_1 v_0)
= c_1 c_{N-1} \Omega^{-1} \,c^\dag_{N+1} \cdot (u \otimes v_0)
= c_0 \cdot \psi_N(u) \,, 
\\[0.15cm]
\psi_{N+2}(c^\dag_0 \cdot u) &= c^\dag_0 u \otimes v_0 
= \Omega^{-1} \cdot (u \otimes c^\dag_0 v_0)
= \Omega^{-1} \cdot (u \otimes c^\dag_1 v_0)
= \Omega^{-1} c^\dag_{N+1} \cdot (u \otimes v_0)
\nn\\[0.15cm]&
= c^\dag_0 \cdot \psi_N(u) \,,
\end{alignat}
\end{subequations}
where we used again the identity $c^\dag_1 v_0 = c^\dag_0 v_0$ in $\repV$.
\medskip

Finally, we prove that $\phi$ is an isomorphism as follows. For all $u \in \repM(N)$, we have 
\begin{equation}
\phi_N \circ \psi_N(u)= \phi_N(u \otimes v_0) = u \,.
\end{equation}
The module $(\repM\timesf \repV)(N)$ is spanned by states of the form $x=\lambda \cdot (u \otimes v)$ with $\lambda \in \cL(N,N_{\pa\pb})$, $u \in \repM(N_\pa)$ and $v \in \repV(N_\pb)$, for some integer $N_\pa$ admissible in $\repM$ and some even integer $N_\pb$.
Moreover, $v$ can be written as $v=c^\dag_{j_1}c^\dag_{j_2} \cdots c^\dag_{j_{N_\pb/2}} \cdot v_0$, for some indices $j_i \ge 1$ for each $i$. Using~\eqref{eq:move.cjdag.Mb}, we can rewrite $x$ as
\be
x=\lambda'\cdot (u \otimes v_0)\,,
\qquad \textrm{where}\qquad
\lambda'=\lambda\,c^\dag_{N_\pb+j_1}c^\dag_{N_\pb+j_2} \cdots c^\dag_{N_\pb+j_{N_\pb/2}} \in \cL(N,N_\pa)\,.
\ee 
Hence, we have
\begin{equation}
\psi_N \circ \phi_N(x) = \psi_N (\lambda' \cdot u)
= \lambda' \cdot \psi_N(u) 
= \lambda' \cdot (u \otimes v_0) = x \,,
\end{equation}
which confirms that $\psi \circ \phi = \id_{(\repM\timesf \repV)}$.
\eproof

Let us recall from \cref{sec:automorphs} that $\repV$ has a partner family $\repV^-$ not isomorphic to $\repV$, wherein the action of the diagrams $\lambda$ differs only by the minus signs $\sigma(\lambda)$. The following corollary follows directly from \cref{prop:change-sign,prop:vacuum.fusion}.

\begin{Corollary}
Let $\repM$ be a family of modules. Then
\begin{equation}
\repM \timesf \repV^- \simeq \repM^-\,.
\end{equation}
\end{Corollary}

\subsection{Fusion with adjoint families of modules}

Up to now, we only considered fusion of families of left modules. We now consider fusion products involving one family $\repM^\dag_\pa$ of right modules and one family $\repM_\pb$ of left modules. The family of fused modules $(\repM^\dag_\pa \timesfh \repM_\pb)$ consists of the right modules defined as 
\begin{equation}
(\repM^\dag_\pa \timesfh \repM_\pb)(N)
= \sum_{N_\pa \geq N_\pb} \repM^\dag_\pa(N_\pa) \cdot 
\big(\cL(N_\pa-N_\pb,N) \otimes \repM_\pb(N_\pb)\big) \,,
\end{equation}
subject to the relations 
\begin{subequations} \label{eq:rel.MdagxM}
\begin{alignat}{3}
&u^\dag c_j \cdot (\id \otimes v) = 
u^\dag \cdot (c_j \otimes v)\,, \qquad 
&& 1 \le j \le N_\pa-N_\pb + 1,
\\[0.15cm]
&u^\dag c_{N_\pa-N_\pb+j+2} \cdot (\id \otimes v) = 
u^\dag \cdot(\id \otimes c_j\,v)\,, \qquad 
&& 1 \le j \le N_\pb -1,
\\[0.15cm]
&u^\dag c_j^\dagger \cdot(\id \otimes v) = 
u^\dag \cdot(c_j^\dagger \otimes v)\,, \qquad 
&& 1 \le j \le N_\pa-N_\pb -1,
\\[0.15cm]
&u^\dag c^\dagger_{N_\pa-N_\pb+j-2} \cdot(\id \otimes v) = 
u^\dag \cdot(\id \otimes c^\dagger_jv)\,, \qquad 
&& 1 \le j \le N_\pb +1,
\\[0.15cm]
&u^\dag \Omega\cdot(\id \otimes v) = u^\dag c_{N_\pa-N_\pb}\cdot(\Omega \, \otimes c_0^\dag v)\,,
&& \textrm{for } N_\pa>N_\pb \,.
\end{alignat}
\end{subequations}
for $u^\dag \in \repM_\pa(N_\pa)$ and $v \in \repM_\pb(N_\pb)$. The right action of $\lambda' \in \cL(N,N')$ on $u^\dag \cdot (\lambda \otimes v)$ is defined as
\begin{equation}
\big(u^\dag \cdot (\lambda \otimes v)\big) \lambda' = u^\dag \cdot (\lambda \lambda' \otimes v) \,.
\end{equation}
The following proposition shows that this construction is equivalent to the adjoint family of $(\repM_\pa \timesf \repM_\pb^r)$.

\begin{Proposition} \label{prop:Ma+.Mb=(Ma.Mbr)+}
Let $\repM_\pa$ and $\repM_\pb$ be families of left modules. Then
the families of fusion modules $(\repM_\pa^\dag \timesfh \repM_\pb)$ and $(\repM_\pa \timesf \repM_\pb^r)^\dag$ are isomorphic.
\end{Proposition}
\proof
For this proof, we construct a family of anti-homomorphisms, namely a sequence $\phi$ of linear maps from $(\repM_\pa^\dag \timesfh \repM_\pb)(N)$ to $(\repM_\pa \timesf \repM_\pb^r)(N)$ satisfying $\phi(x\cdot \lambda) = \lambda^\dag \cdot\phi(x)$, for all $\lambda \in \cL(N,N')$ and $x \in (\repM_\pa^\dag \timesfh \repM_\pb)(N)$. We will then prove that $\phi$ is bijective.\medskip

Let $\repL_{\repM_\pa^\dag,\repM_\pb}(N)$ be the right module
\begin{equation}
\repL_{\repM_\pa^\dag,\repM_\pb}(N) = \bigoplus_{N_\pa\geq N_\pb} 
\repM^\dag_\pa(N_\pa)
\otimes \cL(N_\pa-N_\pb,N) \otimes \repM_\pb(N_\pb) \,,
\end{equation}
with the right action of $\lambda' \in \cL(N,N')$ defined as
\begin{equation}
(u^\dag \otimes \lambda \otimes v)\,\lambda'
= (u^\dag \otimes \lambda\lambda' \otimes v)\,.
\end{equation}
We introduce the linear map from $\repL_{\repM_\pa^\dag,\repM_\pb}(N)$ to $(\repM_\pa \timesf \repM_\pb^r)(N)$, defined as
\begin{equation}
\phi_N(u^\dag \otimes \lambda \otimes v) =
\lambda^\dag \, c_{N_\pa-N_\pb+1}\cdots c_{N_\pa-1}c_{N_\pa} \cdot (u \otimes v^r) \,,
\end{equation}
for $u\in \repM_\pa(N_\pa)$, $v\in\repM_\pb(N_\pb)$ and $\lambda\in \cL(N_\pa-N_\pb,N)$.
For all $\lambda' \in \cL(N,N')$, we have
\begin{equation}
\phi_N\big((u^\dag \otimes \lambda \otimes v)\lambda'\big)
= \phi_N(u^\dag \otimes \lambda\lambda' \otimes v)
= (\lambda')^\dag\lambda^\dag \,c_{N_\pa-N_\pb+1}\cdots c_{N_\pa-1}c_{N_\pa} \cdot (u \otimes v^r)
= (\lambda')^\dag \phi_N(u^\dag \otimes \lambda \otimes v) \,.
\end{equation}
This confirms that $\phi_N$ defines a family of anti-homomorphisms. It moreover respects the relations in~\eqref{eq:rel.MdagxM}:
\begingroup
\allowdisplaybreaks
\begin{subequations}
\begin{alignat}{2}
\phi_{N_\pa-N_\pb+2}(u^\dag c_j \otimes \id\otimes v)
&= c_{N_\pa-N_\pb+3}\cdots c_{N_\pa+1}c_{N_\pa+2} \cdot (c_j^\dag u \otimes v^r)
\nn \\&
= c_{N_\pa-N_\pb+3}\cdots c_{N_\pa+1}c_{N_\pa+2}\,c_j^\dag \cdot (u \otimes v^r) 
\nn \\&
= c_j^\dag \,c_{N_\pa-N_\pb+1}\cdots c_{N_\pa-1}c_{N_\pa} \cdot (u \otimes v^r)
\nn \\&
= \phi_{N_\pa-N_\pb+2}(u^\dag \otimes c_j \otimes v) \,, 
&\hspace{-2.0cm} 1 \le j \le N_\pa-N_\pb+1\,,
\label{eq:phi.antih.a}
\\[0.1cm]
\phi_{N_\pa-N_\pb+2}(u^\dag c_{N_\pa-N_\pb+j+2} \otimes \id\otimes v)
&= c_{N_\pa-N_\pb+3}\cdots c_{N_\pa+1} c_{N_\pa+2} \cdot (c_{N_\pa-N_\pb+j+2}^\dag u \otimes v^r) 
\nn \\&
= c_{N_\pa-N_\pb+3}\cdots c_{N_\pa+1} c_{N_\pa+2} \,c_{N_\pa-N_\pb+j+2}^\dag \cdot (u \otimes v^r) 
\nn \\&
= c_{N_\pa-N_\pb+3}\cdots c_{N_\pa-1} c_{N_\pa} \,c_{N_\pa+N_\pb-j} \cdot (u \otimes v^r) \nn \nn \\&
= c_{N_\pa-N_\pb+3}\cdots c_{N_\pa-1} c_{N_\pa} \cdot (u \otimes c_{N_\pb-j}v^r)
\nn \\&
= \phi_{N_\pa-N_\pb+2}(u^\dag \otimes \id \otimes c_j v) \,, 
&\hspace{-2.0cm}1 \le j \le N_\pb-1\,,
\label{eq:phi.antih.b}\\[0.1cm]
\phi_{N_\pa-N_\pb-2}(u^\dag c^\dag_j \otimes \id\otimes v)
&= c_{N_\pa-N_\pb-1}\cdots c_{N_\pa-3}c_{N_\pa-2} \cdot (c_j u \otimes v^r)
\nn \\&
= c_{N_\pa-N_\pb-1}\cdots c_{N_\pa-3}c_{N_\pa-2}\,c_j \cdot (u \otimes v^r) 
\nn \\&
= c_j \,c_{N_\pa-N_\pb+1}\cdots c_{N_\pa-1}c_{N_\pa} \cdot (u \otimes v^r)
\nn \\&
= \phi_{N_\pa-N_\pb-2}(u^\dag \otimes c^\dag_j \otimes v) \,, 
&\hspace{-2.0cm}1 \le j \le N_\pa-N_\pb-1\,,
\label{eq:phi.antih.c}\\[0.1cm]
\phi_{N_\pa-N_\pb-2}(u^\dag c^\dag_{N_\pa-N_\pb+j-2} \otimes \id\otimes v)
&= c_{N_\pa-N_\pb-1}\cdots c_{N_\pa-3}c_{N_\pa-2} \cdot (c_{N_\pa-N_\pb+j-2} u \otimes v^r) 
\nn \\&
= c_{N_\pa-N_\pb-1}\cdots c_{N_\pa-3}c_{N_\pa-2} \,c_{N_\pa-N_\pb+j-2} \cdot (u \otimes v^r) 
\nn \\&
= c_{N_\pa-N_\pb-1}\cdots c_{N_\pa-1}c_{N_\pa} \,c^\dag_{N_\pa+N_\pb-j+2} \cdot (u \otimes v^r) 
\nn \\&
= c_{N_\pa-N_\pb-1}\cdots c_{N_\pa-1}c_{N_\pa} \cdot (u \otimes c^\dag_{N_\pb-j+2}v^r) 
\nn \\&
= \phi_{N_\pa-N_\pb-2}(u^\dag \otimes \id \otimes c^\dag_j v) \,, 
&\hspace{-2.0cm}1 \le j \le N_\pb+1\,,
\label{eq:phi.antih.d}\\[0.1cm]
\label{eq:phi.antih.e}
\phi_{N_\pa-N_\pb}(u^\dag \Omega \otimes \id\otimes v)
&= c_{N_\pa-N_\pb+1}\cdots c_{N_\pa-1}c_{N_\pa} \cdot (\Omega^{-1}u \otimes v^r) 
\nn \\&
= \Omega^{-1} c_{N_\pa-N_\pb}\cdots c_{N_\pa-2}c_{N_\pa-1}\Omega \cdot (\Omega^{-1}u \otimes v^r) 
\nn \\&
= \Omega^{-1} c_{N_\pa-N_\pb}\cdots c_{N_\pa-2}c_{N_\pa-1}c_{N_\pa} \cdot (u \otimes c_0^\dag v^r) 
\nn \\&
= \Omega^{-1} c_{N_\pa-N_\pb+1}\cdots c_{N_\pa+1}c_{N_\pa+2}c^\dag_{N_\pa-N_\pb} \cdot (u \otimes c_0^\dag v^r) 
\nn \\&
= \Omega^{-1} c_{N_\pa-N_\pb+1}\cdots c_{N_\pa+1}c_{N_\pa+2} \cdot (c^\dag_{N_\pa-N_\pb}u \otimes c_0^\dag v^r)
\nn \\&
= \phi_{N_\pa-N_\pb}(u^\dag c_{N_\pa-N_\pb} \otimes \Omega \otimes c_0^\dag v) \,,
&\hspace{-2.0cm}\textrm{for } N_\pa>N_\pb\,.
\end{alignat}
\end{subequations}
\endgroup
For the proofs of \eqref{eq:phi.antih.b}, \eqref{eq:phi.antih.d} and \eqref{eq:phi.antih.e}, we respectively used the identities
\begin{subequations}
\begin{alignat}{2}
&c_{N_\pa-N_\pb+3}\cdots c_{N_\pa+1}\,c_{N_\pa+2}\,c_{N_\pa-N_\pb+j+2}^\dag
= c_{N_\pa-N_\pb+3}\cdots c_{N_\pa-1}\,c_{N_\pa}\,c_{N_\pa+N_\pb-j}
&\quad 1\leq j\leq N_\pb-1 \,, 
\\ 
& c_{N_\pa-N_\pb-1}\cdots c_{N_\pa-1}\,c_{N_\pa}\,c_{N_\pa+N_\pb-j+2}^\dag
= c_{N_\pa-N_\pb-1}\cdots c_{N_\pa-3}\,c_{N_\pa-2}\,c_{N_\pa-N_\pb+j-2}
&\quad 1\leq j\leq N_\pb+1 \,,
\\
&c_{N_\pa-N_\pb+1}\cdots c_{N_\pa+1}\,c_{N_\pa+2}\,c^\dag_{N_\pa-N_\pb}
= c_{N_\pa-N_\pb}\cdots c_{N_\pa-2}\,c_{N_\pa-1}\,c_{N_\pa}
&\quad\textrm{for } N_\pa>N_\pb\,,
\end{alignat}
\end{subequations}
which are easily proven by writing the left and right sides as diagrams.
\medskip

Hence, by adapting the argument of \cref{cor:morphism.on.Ma.Mb}, one can show that $\phi_N$ yields a family of anti-homomorphisms from $(\repM^\dag_\pa \timesfh \repM_\pb)$ to $(\repM_\pa \timesf \repM_\pb^r)$, defined as
\begin{equation}
\label{eq:phi.dual}
\phi_N\big(u^\dag \cdot( \lambda \otimes v) \big) =
\lambda^\dag \, c_{N_\pa-N_\pb+1}\cdots c_{N_\pa-1}c_{N_\pa} \cdot (u \otimes v^r) \,.
\end{equation}
Similarly, one can show that the map
\begin{equation}
\label{eq:psi.dual}
\psi_N\big(\lambda \cdot( u \otimes v^r) \big) =
u^\dag \, c_{N_\pa+1}c_{N_\pa+2}\cdots c_{N_\pa+N_\pb} \cdot (\lambda^\dag \otimes v)
\end{equation}
for $\lambda \in \cL(N,N_\pa+N_\pb)$, $u \in \repM_\pa(N_\pa)$ and $v \in \repM_\pb(N_\pb)$, 
is well-defined on $(\repM_\pa \timesf \repM_\pb^r)(N)$, and yields a family of anti-homomorphisms from $(\repM_\pa \timesf \repM_\pb^r)$ to $(\repM^\dag_\pa \timesfh \repM_\pb)$.
Finally, using the property
\begin{equation}
c_{j+1}c_{j+2} \cdots c_{j+k} \, c^\dag_j c^\dag_{j-1} \cdots c^\dag_{j-k+1}= \id \,,
\qquad 0 \leq k \leq j \,,
\end{equation}
we compute the compositions
\begin{subequations}
\begin{alignat}{1}
(\phi_N \circ \psi_N)\big(\lambda \cdot( u \otimes v^r) \big) 
&= \phi_N\big(u^\dag \, c_{N_\pa+1}c_{N_\pa+2}\cdots c_{N_\pa+N_\pb} \cdot (\lambda^\dag \otimes v) \big) 
\nn \\&
= \lambda\, c_{N_\pa+N_\pb+1}c_{N_\pa+N_\pb+2} \cdots c_{N_\pa+2N_\pb}
\cdot (c^\dag_{N_\pa+N_\pb}\dots c^\dag_{N_\pa+2}c^\dag_{N_\pa+1} u \otimes v^r) \nn \\&
= \lambda\, c_{N_\pa+N_\pb+1}c_{N_\pa+N_\pb+2} \cdots c_{N_\pa+2N_\pb}
\, c^\dag_{N_\pa+N_\pb}\cdots c^\dag_{N_\pa+2}c^\dag_{N_\pa+1} \cdot(u \otimes v^r) 
\nn \\&
= \lambda \cdot(u \otimes v^r) \,, 
\\[0.15cm]
(\psi_N \circ \phi_N)\big( u^\dag \cdot( \lambda \otimes v)\big) 
&= \psi_N\big( \lambda^\dag \, c_{N_\pa-N_\pb+1}\cdots c_{N_\pa-1}c_{N_\pa} \cdot (u \otimes v^r) \big) \nn \\
&= u^\dag c_{N_\pa+1}c_{N_\pa+2} \cdots c_{N_\pa+N_\pb} \cdot 
(c^\dag_{N_\pa}c^\dag_{N_\pa+1}\cdots c^\dag_{N_\pa-N_\pb+1} \lambda \otimes v) \nn \\
&= u^\dag c_{N_\pa+1}c_{N_\pa+2} \cdots c_{N_\pa+N_\pb} c^\dag_{N_\pa}c^\dag_{N_\pa+1}\dots c^\dag_{N_\pa-N_\pb+1} \cdot 
(\lambda \otimes v) \nn \\
&= u^\dag \cdot( \lambda \otimes v) \,.
\end{alignat}
\end{subequations}
This shows that $\phi$ and $\psi$ are families of inverse anti-isomorphisms, ending the proof.
\eproof

We note that, by setting $\repM_\pa = \repL_{N_\pa'}$ and $\repM_\pb = \repL_{N_\pb'}$ for some non-negative integers $N'_\pa$ and $N'_\pb$, the map $\psi_N$ defined in \eqref{eq:psi.dual} is in fact a map from $\Lambda(N,N'_\pa,N'_\pb)$ to $\Lambda(N'_\pa,N,N'_\pb)$. It is defined as
\begin{equation}
\psi_N\big(\lambda \cdot( \lambda_\pa \otimes \lambda_\pb^r) \big) =
\lambda_\pa^\dag \, c_{N_\pa+1}c_{N_\pa+2}\cdots c_{N_\pa+N_\pb} \cdot (\lambda^\dag \otimes \lambda_\pb)
\end{equation}
for $\lambda \in \cL(N,N_\pa+N_\pb)$, $\lambda_\pa \in \cL(N_\pa,N_\pa')$ and $\lambda_\pb \in \cL(N_\pb,N_\pb')$, and satisfies
\begin{subequations}
\begin{alignat}{2}
\psi_N\big(\lambda'\lambda \cdot( \lambda_\pa \otimes \lambda_\pb^r) \big)
&= \psi_N\big(\lambda \cdot( \lambda_\pa \otimes \lambda_\pb^r) \big)\cdot\big((\lambda')^\dag\otimes \id \big)
&& \qquad \forall \lambda' \in \cL(N',N)\,,
\\[0.1cm]
\psi_N\big(\lambda \cdot( \lambda_\pa \lambda_\pa' \otimes \lambda_\pb^r) \big)
&= (\lambda_\pa')^\dag \cdot \psi_N\big(\lambda \cdot( \lambda_\pa \otimes \lambda_\pb^r) \big)
&& \qquad \forall \lambda_\pa' \in \cL(N_\pa',N_\pa'')\,,
\\[0.1cm]
\psi_N\big(\lambda \cdot( \lambda_\pa \otimes \lambda_\pb^r (\lambda'_\pb)^r) \big)
&=\psi_N\big(\lambda \cdot( \lambda_\pa \otimes \lambda_\pb^r) \big)\cdot \big(\id \otimes (\lambda_\pb')^\dag \big)
&& \qquad \forall \lambda_\pb' \in \cL(N_\pb',N_\pb'')\,,
\end{alignat}
\end{subequations}
for all admissible integers $N'$, $N_\pa''$ and $N_\pb''$. Diagrammatically, the action of $\psi_N$ is similar to the adjoint operation for diagrams with one hole, in the sense that it flips the diagram inside out, interchanging the role of hole $\pa$ and of the outer circle of the disc. The hole $\pb$ however remains in its position, with its diagram $\lambda_\pb$ reflected.

%
\section{Diagrams with three holes and associativity}
\label{sec:associativity}
%

In this section, we introduce diagrams with three holes and investigate the associativity of the fusion of families of modules.

\subsection{Diagrams with three holes}

\paragraph{Diagrammatic definition of $\boldsymbol{\Lambda(N,N_\pa,N_\pb,N_\pc)}$.}

Let $N$, $N_\pa$, $N_\pb$ and $N_\pc$ be non-negative integers such that $N$ and $N_{\pa\pb\pc}=N_{\pa}+N_{\pb}+N_{\pc}$ have the same parity. We first give the diagrammatic definition of $\Lambda(N,N_\pa,N_\pb,N_\pc)$ as the vector space of connectivity diagrams with three holes $\pa$, $\pb$ and $\pc$. For each diagram $\nu \in \Lambda(N,N_\pa,N_\pb,N_\pc)$, we align the three holes horizontally in the disc, with $\pa$ on the right, $\pb$~in the middle and $\pc$ on the left and drawn in orange. The diagram $\nu$ consists of a set of non-intersecting loop segments connecting $N$ nodes on the outside of the disc, and $N_\pa$, $N_\pb$ and $N_\pc$ nodes on the holes $\pa$, $\pb$ and $\pc$, respectively. There can also be non-contractible loops encircling either one, two or three holes. There are therefore seven types of non-contractible loops, namely those encircling $\pa$, $\pb$, $\pc$, $\pa\pb$, $\pa\pc$, $\pb\pc$ and $\pa\pb\pc$. Here is an example of an element of $\Lambda(13,2,4,5)$:
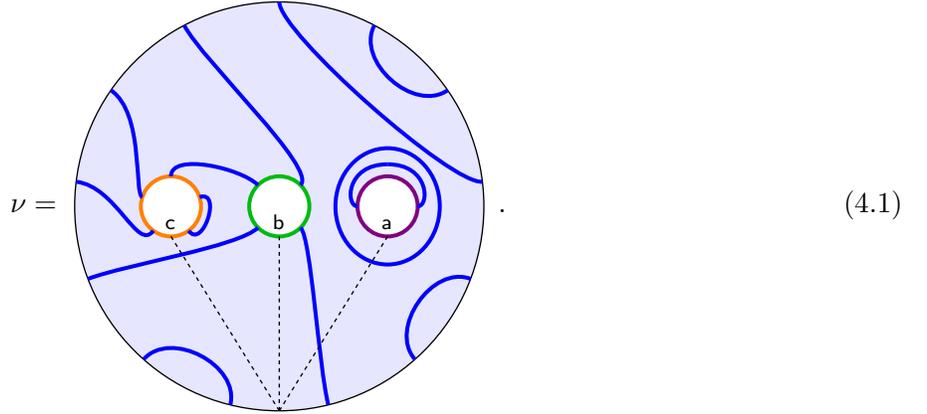
\be
\psset{unit=0.8cm}
\nu = \
\thispic{\begin{pspicture}[shift=-3.3](-3.4,-3.4)(3.4,3.4)
\psarc[linecolor=black,linewidth=0.5pt,fillstyle=solid,fillcolor=lightlightblue]{-}(0,0){3.4}{0}{360}
\psarc[linecolor=black,linewidth=0.5pt,fillstyle=solid,fillcolor=white]{-}(-1.8,0){0.5}{0}{360}
\psarc[linecolor=black,linewidth=0.5pt,fillstyle=solid,fillcolor=white]{-}(0,0){0.5}{0}{360}
\psarc[linecolor=black,linewidth=0.5pt,fillstyle=solid,fillcolor=white]{-}(1.8,0){0.5}{0}{360}
\psarc[linecolor=orange,linewidth=1.5pt]{-}(-1.8,0){0.5}{0}{360}
\psarc[linecolor=darkgreen,linewidth=1.5pt]{-}(0,0){0.5}{0}{360}
\psarc[linecolor=purple,linewidth=1.5pt]{-}(1.8,0){0.5}{0}{360}
\psbezier[linecolor=blue,linewidth=1.5pt]{-}(0.354, -0.354)(0.53, -0.53)(0.651, -2.641)(0.814, -3.301)
\psbezier[linecolor=blue,linewidth=1.5pt]{-}(2.3,0)(2.5,0)(2.5,0.7)(1.8,0.7)
\psbezier[linecolor=blue,linewidth=1.5pt]{-}(1.3,0)(1.1,0)(1.1,0.7)(1.8,0.7)
\psellipse[linecolor=blue,linewidth=1.5pt](1.8,0)(0.9,1.0)
\psbezier[linecolor=blue,linewidth=1.5pt]{-}(2.255, -2.545)(1.804, -2.036)(2.543, -0.965)(3.179, -1.206)
\psbezier[linecolor=blue,linewidth=1.5pt]{-}(2.798, 1.931)(2.239, 1.545)(1.264, 2.408)(1.58, 3.011)
\psbezier[linecolor=blue,linewidth=1.5pt]{-}(3.375, 0.41)(2.869, 0.348)(0., 2.89)(0., 3.4)
\psbezier[linecolor=blue,linewidth=1.5pt]{-}(-2.255, -2.545)(-1.804, -2.036)(-0.651, -2.641)(-0.814, -3.301)
\psbezier[linecolor=blue,linewidth=1.5pt]{-}(-1.506, -0.405)(-1.271, -0.728)(-0.944, 0.278)(-1.324, 0.155)
\psbezier[linecolor=blue,linewidth=1.5pt]{-}(-0.354, 0.354)(-0.636, 0.636)(-1.8, 0.9)(-1.8, 0.5)
\psbezier[linecolor=blue,linewidth=1.5pt]{-}(0.354, 0.354)(0.636, 0.636)(-1.264, 2.408)(-1.58, 3.011)
\psbezier[linecolor=blue,linewidth=1.5pt]{-}(-2.276, 0.155)(-2.418, 0.201)(-2.239, 1.545)(-2.798, 1.931)
\psbezier[linecolor=blue,linewidth=1.5pt]{-}(-2.094, -0.405)(-2.358, -0.769)(-2.7, 0.328)(-3.375, 0.41)
\psbezier[linecolor=blue,linewidth=1.5pt]{-}(-0.354, -0.354)(-0.672, -0.672)(-2.543, -0.965)(-3.179, -1.206)
\psline[linestyle=dashed, dash= 1.5pt 1.5pt,linewidth=0.5pt]{-}(0,-3.4)(-1.8,-0.5)
\psline[linestyle=dashed, dash= 1.5pt 1.5pt,linewidth=0.5pt]{-}(0,-3.4)(1.8,-0.5)
\psline[linestyle=dashed, dash= 1.5pt 1.5pt,linewidth=0.5pt]{-}(0,-3.4)(0,-0.5)
\rput(1.8,-0.3){$_\pa$}
\rput(0,-0.25){$_\pb$}
\rput(-1.8,-0.3){$_\pc$}
\end{pspicture}}\ \,.
\ee

The space $\Lambda(N,N_\pa,N_\pb,N_\pc)$ is endowed with a natural diagrammatic left action of $\cL(N',N)$, whereby $\lambda\, \nu$ is obtained by drawing $\nu$ inside $\lambda$, for all $\lambda \in \cL(N',N)$ and $\nu \in\Lambda(N,N_\pa,N_\pb,N_\pc)$. It also has a diagrammatic right action of $\cL(N_\pa, N'_\pa) \otimes \cL(N_\pb, N'_\pb) \otimes \cL(N_\pc, N'_\pc)$, whereby $\nu(\lambda_\pa \otimes \lambda_\pb \otimes \lambda_\pc)$ is obtained by drawing $\lambda_\pa$, $\lambda_\pb$ and $\lambda_\pc$ in the holes $\pa$, $\pb$ and $\pc$ respectively. Each contractible loop is removed and replaced by a weight $\beta$. We write this action as
\begin{subequations}
\begin{alignat}{2}
&\cL(N',N) \Lambda(N,N_\pa,N_\pb,N_\pc) \subseteq \Lambda(N',N_\pa,N_\pb,N_\pc)\,,
\label{eq:L.LLL.in.LLL}\\[0.15cm]
&\Lambda(N,N_\pa,N_\pb,N_\pc)\big(\cL(N_\pa,N'_\pa)\otimes\cL(N_\pb,N'_\pb)\otimes\cL(N_\pc,N'_\pc)\big) \subseteq \Lambda(N,N'_\pa,N'_\pb,N'_\pc)\,.
\end{alignat}
\end{subequations}

\paragraph{Algebraic definition of $\boldsymbol{\Lambda(N,N_\pa,N_\pb,N_\pc)}$.}

We now give the algebraic definition of the space $\Lambda(N,N_\pa,N_\pb,N_\pc)$ of the diagram with three holes. Let us note that any diagram $\nu$ in $\Lambda(N,N_\pa,N_\pb,N_\pc)$ can be expressed as $\nu = \lambda(\lambda_\pa \otimes \lambda_\pb \otimes \lambda_\pc)$ for some $\lambda \in \cL(N,N'_{\pa\pb\pc})$, $\lambda_\pa \in \cL(N'_\pa,N_\pa)$, $\lambda_\pb \in \cL(N'_\pb,N_\pb)$ and $\lambda_\pc \in \cL(N'_\pc,N_\pc)$, and some integers $N'_\pa$, $N'_\pb$, $N'_\pc$ and $N'_{\pa\pb\pc}=N'_\pa+N'_\pb+N'_\pc$. This diagram takes the form
\be
\label{eq:nu.diag}
\psset{unit=0.45cm}
\nu = \lambda(\lambda_\pa \otimes \lambda_\pb\otimes \lambda_\pc) = \
\thispic{\begin{pspicture}[shift=-5.9](-6,-6)(6,6)
\psarc[linecolor=black,linewidth=0.5pt,fillstyle=solid,fillcolor=lightlightblue]{-}(0,0){6}{0}{360}
\psarc[linecolor=black,linewidth=0.5pt,fillstyle=solid,fillcolor=white]{-}(-3.2,0){0.5}{0}{360}
\psarc[linecolor=black,linewidth=0.5pt,fillstyle=solid,fillcolor=white]{-}(0,0){0.5}{0}{360}
\psarc[linecolor=black,linewidth=0.5pt,fillstyle=solid,fillcolor=white]{-}(3.2,0){0.5}{0}{360}
\psarc[linecolor=black,linewidth=0.5pt]{-}(-3.2,0){1.2}{0}{360}
\psarc[linecolor=black,linewidth=0.5pt]{-}(0,0){1.2}{0}{360}
\psarc[linecolor=black,linewidth=0.5pt]{-}(3.2,0){1.2}{0}{360}
\psarc[linecolor=orange,linewidth=1.5pt]{-}(-3.2,0){0.5}{0}{360}
\psarc[linecolor=purple,linewidth=1.5pt]{-}(0,0){0.5}{0}{360}
\psarc[linecolor=darkgreen,linewidth=1.5pt]{-}(3.2,0){0.5}{0}{360}
\psarc[linecolor=black,linewidth=0.5pt]{-}(0,0){5}{0}{360}
\rput(3.2,0){\psline[linestyle=dashed, dash= 1.5pt 1.5pt,linewidth=0.5pt]{-}(-1.1,0)(-0.55,0)}
\rput(-3.2,0){\psline[linestyle=dashed, dash= 1.5pt 1.5pt,linewidth=0.5pt]{-}(1.1,0)(0.55,0)}
\psline[linestyle=dashed, dash= 1.5pt 1.5pt,linewidth=0.5pt]{-}(0,-0.55)(0,-5)
\psline[linestyle=dashed, dash= 1.5pt 1.5pt,linewidth=0.5pt]{-}(-2.0,0)(0,-5)
\psline[linestyle=dashed, dash= 1.5pt 1.5pt,linewidth=0.5pt]{-}(2.0,0)(0,-5)
\psline[linestyle=dashed, dash= 1.5pt 1.5pt,linewidth=0.5pt]{-}(0,-5)(0,-6)
\psbezier[linecolor=blue,linewidth=1.5pt]{-}(2.161, -0.6)(1.537, -0.96)(0.78, -3.923)(0.975, -4.904)
\psbezier[linecolor=blue,linewidth=1.5pt]{-}(3.2, -1.2)(3.2, -1.92)(2.222, -3.326)(2.778, -4.157)
\psbezier[linecolor=blue,linewidth=1.5pt]{-}(4.239, -0.6)(4.863, -0.96)(3.326, -2.222)(4.157, -2.778)
\psbezier[linecolor=blue,linewidth=1.5pt]{-}(4.239, 0.6)(5.175, 1.14)(4.414, -0.878)(4.904, -0.975)
\psbezier[linecolor=blue,linewidth=1.5pt]{-}(3.2, 1.2)(3.2, 2.28)(4.414, 0.878)(4.904, 0.975)
\psbezier[linecolor=blue,linewidth=1.5pt]{-}(2.161, 0.6)(1.537, 0.96)(3.326, 2.222)(4.157, 2.778)
\psbezier[linecolor=blue,linewidth=1.5pt]{-}(0.849, -0.849)(1.527, -1.527)(2.222, 3.326)(2.778, 4.157)
\psbezier[linecolor=blue,linewidth=1.5pt]{-}(0.849, 0.849)(1.527, 1.527)(0.78, 3.923)(0.975, 4.904)
\psbezier[linecolor=blue,linewidth=1.5pt]{-}(-0.849, 0.849)(-1.527, 1.527)(-0.78, 3.923)(-0.975, 4.904)
\psbezier[linecolor=blue,linewidth=1.5pt]{-}(-0.849, -0.849)(-1.527, -1.527)(-2.222, 3.326)(-2.778, 4.157)
\psbezier[linecolor=blue,linewidth=1.5pt]{-}(-2.161, 0.6)(-1.537, 0.96)(-3.326, 2.222)(-4.157, 2.778)
\psbezier[linecolor=blue,linewidth=1.5pt]{-}(-3.2, 1.2)(-3.2, 2.28)(-4.414, 0.878)(-4.904, 0.975)
\psbezier[linecolor=blue,linewidth=1.5pt]{-}(-4.239, 0.6)(-5.175, 1.14)(-4.414, -0.878)(-4.904, -0.975)
\psbezier[linecolor=blue,linewidth=1.5pt]{-}(-4.239, -0.6)(-4.863, -0.96)(-3.326, -2.222)(-4.157, -2.778)
\psbezier[linecolor=blue,linewidth=1.5pt]{-}(-3.2, -1.2)(-3.2, -1.92)(-2.222, -3.326)(-2.778, -4.157)
\psbezier[linecolor=blue,linewidth=1.5pt]{-}(-2.161, -0.6)(-1.537, -0.96)(-0.78, -3.923)(-0.975, -4.904)
\rput(0.3,5.5){$_\lambda$}
\rput(3.2,0.85){$_{\lambda_\pa}$}
\rput(0,0.85){$_{\lambda_\pb}$}
\rput(-3.2,0.85){$_{\lambda_\pc}$}
\end{pspicture}}\ \, .
\ee
For a given diagram, this decomposition is in general not unique. The vector space of diagrams with three holes can be written as
\begin{equation} \label{eq:Lambda.sumsLLLL}
\Lambda(N,N_\pa,N_\pb, N_\pc) =
\sum_{N'_\pa,N'_\pb,N'_\pc} \cL(N,N'_{\pa\pb\pc})
\big(\cL(N'_\pa,N_\pa) \otimes \cL(N'_\pb,N_\pb) \otimes \cL(N'_\pc,N_\pc) \big) \,,
\end{equation}
where $N'_\pa$, $N'_\pb$ and $N'_\pc$ run over all non-negative integers with the same parities as $N_\pa$, $N_\pb$ and $N_\pc$, respectively, and $N'_{\pa\pb\pc}=N'_\pa+N'_\pb+N'_\pc$. This vector space is endowed with the relations
\begingroup
\allowdisplaybreaks
\begin{subequations} \label{eq:3hole.equations}
\begin{alignat}{2}
&\lambda\, c_j (\lambda_\pa \otimes \lambda_\pb \otimes \lambda_\pc) = \lambda\, (c_j \lambda_\pa \otimes \lambda_\pb \otimes \lambda_\pc)\,, \qquad 
&& 1 \le j \le N_\pa -1,
\label{eq:3hole.equations.a}\\[0.15cm]
&\lambda\, c_{j+N_\pa} (\lambda_\pa \otimes \lambda_\pb \otimes \lambda_\pc) =
\lambda\, (\lambda_\pa \otimes c_j\lambda_\pb \otimes \lambda_\pc)\,, \qquad 
&& 1 \le j \le N_\pb -1,
\label{eq:3hole.equations.b}\\[0.15cm]
&\lambda\, c_{j+N_{\pa\pb}} (\lambda_\pa \otimes \lambda_\pb \otimes \lambda_\pc) = 
\lambda\, (\lambda_\pa \otimes \lambda_\pb \otimes c_j\lambda_\pc)\,, \qquad 
&& 1 \le j \le N_\pc -1,
\label{eq:3hole.equations.c}\\[0.15cm]
&\lambda\, c^\dagger_{j} (\lambda_\pa \otimes \lambda_\pb \otimes \lambda_\pc) 
= \lambda\, (c^\dagger_j\lambda_\pa \otimes \lambda_\pb \otimes \lambda_\pc)\,, 
&& 1 \le j \le N_\pa +1,
\label{eq:3hole.equations.d}\\[0.15cm]
&\lambda\, c^\dagger_{j+N_\pa} (\lambda_\pa \otimes \lambda_\pb \otimes \lambda_\pc) 
= \lambda\, (\lambda_\pa \otimes c^\dagger_j\lambda_\pb \otimes \lambda_\pc)\,, \qquad 
&& 1 \le j \le N_\pb +1,
\label{eq:3hole.equations.e}\\[0.15cm]
&\lambda\, c^\dagger_{j+N_{\pa\pb}} (\lambda_\pa \otimes \lambda_\pb \otimes \lambda_\pc) 
= \lambda\, (\lambda_\pa \otimes \lambda_\pb \otimes c^\dagger_j\lambda_\pc)\,, \qquad 
&& 1 \le j \le N_\pc +1,
\label{eq:3hole.equations.f}\\[0.15cm]
&\lambda\,\Omega\, (\lambda_\pa \otimes \lambda_\pb \otimes \lambda_\pc)
= \lambda\, c_{N_{\pa\pb}}c_{N_\pa} (\Omega\, \lambda_\pa \otimes c_0^\dagger\lambda_\pb \otimes c_0^\dagger\lambda_\pc)\,,
&\qquad& \textrm{for }N_\pa>0 \,,
\label{eq:3hole.equations.g}
\end{alignat}
\end{subequations}
\endgroup
for all $\lambda \in \cL(N',N^{\tinyx \alpha}_{\pa\pb\pc})$, $\lambda_\pa \in \cL(N_\pa,N'_\pa)$, $\lambda_\pb \in \cL(N_\pb,N'_\pb)$ and $\lambda_\pc \in \cL(N_\pc,N'_\pc)$, where $N_{\pa\pb}=N_\pa+N_\pb$, and $N_{\pa\pb\pc}^{\tinyx \alpha}$ in (\ref{eq:3hole.equations}$\alpha$) takes the values
\begin{equation} \label{eq:N.beta}
N^{\tinyx \alpha}_{\pa\pb\pc} = \left\{\begin{array}{ll}
N_\pa+N_\pb+N_\pc-2 & \alpha \in \{a,b,c\}\,,\\[0.15cm]
N_\pa+N_\pb+N_\pc+2 & \alpha \in \{d,e,f\}\,,\\[0.15cm]
N_\pa+N_\pb+N_\pc & \alpha \in \{g\}\,.\\[0.15cm]
\end{array} \right.
\end{equation}
The algebraic definition of $\Lambda(N,N_\pa,N_\pb,N_\pc)$ thus consists of the vector space \eqref{eq:Lambda.sumsLLLL} subject to the relations~\eqref{eq:3hole.equations}.
\begin{Theorem}
\label{prop:LLL.equiv}
The diagrammatic and algebraic definitions of $\Lambda(N,N_\pa,N_\pb,N_\pc)$ are equivalent.
\end{Theorem}
\noindent The proof is given at the end of \cref{sec:assoc.proof}.\medskip 

We note that one can use \eqref{eq:3hole.fusion} to prove many more relations, and in particular
\begin{subequations} \label{eq:3hole.equations.extra}
\begin{alignat}{3}
&\lambda\,\Omega^{-1}\cdot (\lambda_\pa \otimes \lambda_\pb \otimes \lambda_\pc) =
\lambda\, c_{N_\pa+2}\, c_{N_{\pa\pb}+4}\cdot (c_0^\dag \lambda_\pa \otimes c_0^\dag \lambda_\pb \otimes \Omega^{-1} \lambda_\pc)\,,
&\qquad&\textrm{for }N_\pc>0 \,,
\label{eq:3hole.equations.extra.a}\\[0.15cm]
&\lambda\,(c_0^\dag \lambda_\pa \otimes \lambda_\pb \otimes \lambda_\pc) = \lambda\,\Omega^{-1} c_{N_{\pa\pb}+2}\cdot (\lambda_\pa \otimes c^\dag_0 \lambda_\pb \otimes c^\dag_0 \lambda_\pc)\,,
\label{eq:3hole.equations.extra.b}\\[0.15cm]
&\lambda\,(\lambda_\pa \otimes c_0^\dag \lambda_\pb \otimes \lambda_\pc) = \lambda\,c_0\cdot (c^\dag_0 \lambda_\pa \otimes \lambda_\pb \otimes c^\dag_0 \lambda_\pc)\,,
\label{eq:3hole.equations.extra.c}\\[0.15cm]
&\lambda\,(\lambda_\pa \otimes \lambda_\pb \otimes c_0^\dag \lambda_\pc) = \lambda\,\Omega\, c_{N_\pa+2}\cdot (c^\dag_0 \lambda_\pa \otimes c^\dag_0 \lambda_\pb \otimes \lambda_\pc)\,,
\label{eq:3hole.equations.extra.d}\\[0.15cm]
&\lambda\,c_0 \cdot (\lambda_\pa \otimes \lambda_\pb \otimes \lambda_\pc) = \lambda\,c_{N_\pa}\, c_{N_{\pa\pb}+2}\cdot (\Omega \lambda_\pa \otimes c^\dag_0 \lambda_\pb \otimes \Omega^{-1} \lambda_\pc)\,,
&&\textrm{for }N_\pa,N_\pc>0 \,,
\label{eq:3hole.equations.extra.e}\\[0.15cm]
&\lambda\,c_{N_\pa} \cdot (\lambda_\pa \otimes \lambda_\pb \otimes \lambda_\pc) = \lambda\,c_0\, c_{N_{\pa\pb}}\cdot (\Omega^{-1} \lambda_\pa \otimes \Omega \lambda_\pb \otimes c^\dag_0 \lambda_\pc)\,,
&&\textrm{for }N_\pa,N_\pb>0 \,,
\label{eq:3hole.equations.extra.f}\\[0.15cm]
&\lambda\,c_{N_{\pa\pb}} \cdot (\lambda_\pa \otimes \lambda_\pb \otimes \lambda_\pc) = \lambda\,c_0\, c_{N_\pa+2}\cdot (c^\dag_0 \lambda_\pa \otimes \Omega^{-1} \lambda_\pb \otimes \Omega \lambda_\pc)\,,
&&\textrm{for }N_\pb,N_\pc>0 \,.
\label{eq:3hole.equations.extra.g}
\end{alignat}
\end{subequations}
One can in fact equivalently define $\Lambda(N,N_\pa,N_\pb,N_\pc)$ algebraically with the relations \eqref{eq:3hole.equations.a}$-$\eqref{eq:3hole.equations.f} and one of the relations in \eqref{eq:3hole.equations.extra}.

\subsection{Triple fusion products and associativity}

\paragraph{Definitions of the triple products.}

To show that fusion is associative, we must show that the families $((\repM_\pa \timesf \repM_\pb) \timesf \repM_\pc)$ and $(\repM_\pa \timesf (\repM_\pb \timesf \repM_\pc))$ are isomorphic. The first family is defined as
\be
\label{eq:first.fam}
\big((\repM_\pa \timesf \repM_\pb) \timesf \repM_\pc\big)(N) =
\sum_{N_\pa,N_\pb,N_\pc,N'_{\pa\pb}} \cL(N,N'_{\pa\pb,\pc}) \cdot
\bigg(\Big(\cL(N'_{\pa\pb},N_{\pa\pb})\cdot\big(\repM_\pa(N_\pa) \otimes \repM_\pb(N_\pb)\big)\Big)\otimes \repM_\pc(N_\pc)\bigg)\,,
\ee
with $N'_{\pa\pb,\pc} = N'_{\pa\pb}+N_\pc$,
subject to the relations
\begingroup
\allowdisplaybreaks
\begin{subequations} \label{eq:2+1hole.equations}
\begin{alignat}{2}
&c_j \cdot \big((u \otimes v) \otimes w\big) = 
\big(c_j(u \otimes v) \otimes w\big)\,,&& 1 \le j \le N_{\pa\pb} -1\,,
\label{eq:2+1hole.equations.a}\\[0.15cm]
&\big(c_j(u \otimes v) \otimes w\big) = 
\big((c_j u \otimes v) \otimes w\big)\,,
&& 1 \le j \le N_\pa -1\,,
\label{eq:2+1hole.equations.aa}\\[0.15cm]
&\big(c_{j+N_\pa}(u \otimes v) \otimes w\big) = 
\big((u \otimes c_j v) \otimes w\big)\,, 
&& 1 \le j \le N_\pb -1\,,
\label{eq:2+1hole.equations.b}\\[0.15cm]
&c_{j+N_{\pa\pb}}\cdot \big((u \otimes v) \otimes w\big) = 
\big((u \otimes v) \otimes c_jw\big)\,,
&& 1 \le j \le N_\pc -1\,,
\label{eq:2+1hole.equations.c}\\[0.15cm]
&c^\dag_j \cdot \big((u \otimes v) \otimes w\big) = 
\big(c^\dag_j(u \otimes v) \otimes w\big)\,,
&& 1 \le j \le N_{\pa\pb} +1\,,
\label{eq:2+1hole.equations.d}\\[0.15cm]
&\big(c^\dag_j(u \otimes v) \otimes w\big) = 
\big((c^\dag_j u \otimes v) \otimes w\big)\,,
&& 1 \le j \le N_\pa +1\,,
\label{eq:2+1hole.equations.dd}\\[0.15cm]
&\big(c^\dag_{j+N_\pa}(u \otimes v) \otimes w\big) = 
\big((u \otimes c^\dag_j v) \otimes w\big)\,,
&& 1 \le j \le N_\pb +1\,,
\label{eq:2+1hole.equations.e}\\[0.15cm]
&c^\dag_{j+N_{\pa\pb}}\cdot \big((u \otimes v) \otimes w\big) = 
\big((u \otimes v) \otimes c^\dag_j w\big)\,,
&& 1 \le j \le N_\pc +1\,,
\label{eq:2+1hole.equations.f}\\[0.15cm]
&\Omega\cdot \big((u \otimes v) \otimes w\big) = c_{N_{\pa\pb}}\cdot\big(\Omega(u \otimes v) \otimes c_0^\dag w\big)\,,
&\qquad& \textrm{for }N_\pa+N_\pb>0 \,,
\label{eq:2+1hole.equations.g}\\[0.15cm]
\label{eq:2+1hole.equations.h}
&\big(\Omega(u \otimes v) \otimes w\big) = \big(c_{N_\pa}(\Omega u \otimes c_0^\dag v) \otimes w\big)\,,
&&\textrm{for }N_\pa>0 \,,
\end{alignat}
\end{subequations}
\endgroup
for all $u\in \repM_\pa(N_\pa)$, $v \in \repM_\pb(N_\pb)$ and $w \in \repM_\pc(N_\pc)$. Similarly, the second family is defined as
\be
\big(\repM_\pa \timesf (\repM_\pb \timesf \repM_\pc)\big)(N) =
\sum_{N_\pa,N_\pb,N_\pc,N'_{\pb\pc}} \cL(N,N'_{\pa,\pb\pc}) \cdot
\bigg(\repM_\pa(N_\pa)\otimes \Big(\cL(N'_{\pb\pc},N_{\pb\pc})\cdot \big(\repM_\pb(N_\pb)\otimes \repM_\pc(N_\pc)\big)\Big)\bigg) \,,
\ee
with $N'_{\pa,\pb\pc} = N_\pa+N'_{\pb\pc}$,
subject to the relations
\begingroup
\allowdisplaybreaks
\begin{subequations}
\label{eq:1+2hole.equations}
\begin{alignat}{2}
&c_j \cdot \big(u \otimes (v \otimes w)\big) = 
\big(c_j u \otimes (v \otimes w)\big)\,,
\qquad
&& 1 \le j \le N_\pa -1\,,
\label{eq:1+2hole.equations.a}\\[0.15cm]
&c_{j+N_\pa} \cdot \big(u \otimes (v \otimes w)\big) = 
\big(u \otimes c_j(v \otimes w)\big)\,,
\qquad 
&& 1 \le j \le N_{\pb\pc} -1\,,
\label{eq:1+2hole.equations.b}\\[0.15cm]
&
\big(u \otimes c_j(v \otimes w)\big) = 
\big(u \otimes (c_j v \otimes w)\big)\,, 
\qquad 
&& 1 \le j \le N_\pb -1\,,
\label{eq:1+2hole.equations.bb}\\[0.15cm]
&\big(u \otimes c_{j+N_\pb}(v \otimes w)\big) =
\big(u \otimes (v \otimes c_jw)\big)\,,
\qquad
&& 1 \le j \le N_\pc -1\,,
\label{eq:1+2hole.equations.c}\\[0.15cm]
&c^\dag_j \cdot \big(u \otimes (v \otimes w)\big) = 
\big(c^\dag_j u \otimes (v \otimes w)\big)
\,,\qquad
&& 1 \le j \le N_\pa +1\,,
\label{eq:1+2hole.equations.d}\\[0.15cm]
&c^\dag_{j+N_\pa} \cdot \big(u \otimes (v \otimes w)\big) = 
\big(u \otimes c^\dag_j(v \otimes w)\big)\,, \qquad
&& 1 \le j \le N_{\pb\pc} +1\,,
\label{eq:1+2hole.equations.e}\\[0.15cm]
&\big(u \otimes c^\dag_j(v \otimes w)\big) = 
\big(u \otimes (c^\dag_j v \otimes w)\big)\,,\qquad
&& 1 \le j \le N_\pb +1\,,
\label{eq:1+2hole.equations.ee}\\[0.15cm]
&\big(u \otimes c^\dag_{j+N_\pb}(v \otimes w)\big) =
\big(u \otimes (v \otimes c^\dag_jw)\big)\,,\qquad
&& 1 \le j \le N_\pc +1\,,
\label{eq:1+2hole.equations.f}\\[0.15cm]
&\Omega \cdot \big(u \otimes (v \otimes w)\big) = c_{N_{\pa}}\cdot\big(\Omega u \otimes c_0^\dag (v \otimes w)\big)\,,
&\qquad&\textrm{for }N_\pa>0 \,,
\label{eq:1+2hole.equations.g}\\[0.15cm]
\label{eq:1+2hole.equations.h}
&\big(u \otimes \Omega(v \otimes w)\big) = \big(u \otimes c_{N_\pa} (\Omega v \otimes c_0^\dag w)\big)\,,
&& \textrm{for }N_\pb>0 \,,
\end{alignat}
\end{subequations}
\endgroup
for all $u\in \repM_\pa(N_\pa)$, $v \in \repM_\pb(N_\pb)$ and $w \in \repM_\pc(N_\pc)$.\medskip 

We note that elements of the first and second fusion families generally take the form of linear combinations involving terms of the form $\lambda\cdot\big(\lambda_{\pa\pb}(u\otimes v) \otimes w\big)$ and $\lambda\cdot\big(u\otimes \lambda_{\pb\pc}(v \otimes w)\big)$, respectively. However, using the relations \eqref{eq:2+1hole.equations} and \eqref{eq:1+2hole.equations}, one can rewrite these as linear combinations of terms of the form $\lambda\cdot\big((u\otimes v) \otimes w\big)$ and $\lambda\cdot\big(u\otimes (v \otimes w)\big)$. This inspires us to introduce a third family, denoted by $(\repM_\pa \timesf \repM_\pb \timesf \repM_\pc)$ and whose modules are defined as 
\begin{equation} \label{eq:triple.fusion}
(\repM_\pa \timesf \repM_\pb \timesf \repM_\pc)(N) =
\sum_{N_\pa,N_\pb,N_\pc} \cL(N,N_{\pa\pb\pc}) \cdot
\big(\repM_\pa(N_\pa) \otimes\repM_\pb(N_\pb) \otimes \repM_\pc(N_\pc)\big)\,,
\end{equation}
with $N_{\pa\pb\pc}=N_\pa+N_\pb+N_\pc$, and subject to the relations
\begingroup
\allowdisplaybreaks
\begin{subequations} \label{eq:3hole.fusion}
\begin{alignat}{2}
&c_j \cdot (u \otimes v \otimes w) = 
(c_j u \otimes v \otimes w)\,,\qquad
&& 1 \le j \le N_\pa -1\,,
\label{eq:3hole.fusion.a}\\[0.15cm]
&c_{j+N_\pa} \cdot (u \otimes v \otimes w) = 
(u \otimes c_j v \otimes w)\,, \qquad 
&& 1 \le j \le N_\pb -1\,,
\label{eq:3hole.fusion.b}\\[0.15cm]
&c_{j+N_{\pa\pb}}\cdot (u \otimes v \otimes w) = 
(u \otimes v \otimes c_j w)\,,\qquad
&& 1 \le j \le N_\pc -1\,,
\label{eq:3hole.fusion.c}\\[0.15cm]
&c^\dag_j \cdot (u \otimes v \otimes w) = 
(c^\dag_j u \otimes v \otimes w)\,,\qquad
&& 1 \le j \le N_\pa +1\,,
\label{eq:3hole.fusion.d}\\[0.15cm]
&c^\dag_{j+N_\pa} \cdot (u \otimes v \otimes w) = 
(u \otimes c^\dag_j v \otimes w)\,, \qquad 
&& 1 \le j \le N_\pb +1\,,
\label{eq:3hole.fusion.e}\\[0.15cm]
&c^\dag_{j+N_{\pa\pb}}\cdot (u \otimes v \otimes w) = 
(u \otimes v \otimes c^\dag_j w)\,,\qquad
&& 1 \le j \le N_\pc +1\,,
\label{eq:3hole.fusion.f}\\[0.15cm]
&\Omega \cdot (u \otimes v \otimes w) = c_{N_\pa+N_\pb} c_{N_\pa} \cdot (\Omega\, u \otimes c_0^\dag v \otimes c_0^\dag w)\,,
&\qquad&\textrm{for }N_\pa>0 \,.
\label{eq:3hole.fusion.g}\end{alignat}
\end{subequations}
\endgroup
We can use these to prove a number of relations, in particular
\begingroup
\allowdisplaybreaks
\begin{subequations} \label{eq:3hole.fusion.extra}
\begin{alignat}{3}
&\Omega^{-1}\cdot (u \otimes v \otimes w) =
c_{N_\pa+2}\, c_{N_{\pa\pb}+4}\cdot (c_0^\dag u \otimes c_0^\dag v \otimes \Omega^{-1} w)\,,
&\qquad&\textrm{for }N_\pc>0 \,,
\label{eq:3hole.fusion.extra.a}\\[0.15cm]
&(c_0^\dag u \otimes v \otimes w) = \Omega^{-1} c_{N_{\pa\pb}+2}\cdot (u \otimes c^\dag_0 v \otimes c^\dag_0 w)\,,
\label{eq:3hole.fusion.extra.b}\\[0.15cm]
&(u \otimes c_0^\dag v \otimes w) = c_0\cdot (c^\dag_0 u \otimes v \otimes c^\dag_0 w)\,,
\label{eq:3hole.fusion.extra.c}\\[0.15cm]
&(u \otimes v \otimes c_0^\dag w) = \Omega\, c_{N_\pa+2}\cdot (c^\dag_0 u \otimes c^\dag_0 v \otimes w)\,,
\label{eq:3hole.fusion.extra.d}\\[0.15cm]
&c_0 \cdot (u \otimes v \otimes w) = c_{N_\pa}\, c_{N_{\pa\pb}+2}\cdot (\Omega u \otimes c^\dag_0 v \otimes \Omega^{-1} w)\,,
&&\textrm{for }N_\pa,N_\pc>0 \,,
\label{eq:3hole.fusion.extra.e}\\[0.15cm]
&c_{N_\pa} \cdot (u \otimes v \otimes w) = c_0\, c_{N_{\pa\pb}}\cdot (\Omega^{-1} u \otimes \Omega v \otimes c^\dag_0 w)\,,
&&\textrm{for }N_\pa,N_\pb>0 \,,
\label{eq:3hole.fusion.extra.f}\\[0.15cm]
&c_{N_{\pa\pb}} \cdot (u \otimes v \otimes w) = c_0\, c_{N_\pa+2}\cdot (c^\dag_0 u \otimes \Omega^{-1} v \otimes \Omega w)\,,
&&\textrm{for }N_\pb,N_\pc>0 \,.
\label{eq:3hole.fusion.extra.g}
\end{alignat}
\end{subequations}
\endgroup

Lastly, we define another triple fusion product, wherein the first operand is a family of right modules and the two others are families of left modules
\begin{equation} 
(\repM^\dag_\pa \timesfh \repM_\pb \timesf \repM_\pc)(N) =
\sum_{N_\pa \geq N_\pb+N_\pc} \repM^\dag_\pa(N_\pa) \cdot 
\big(\cL(N_\pa-N_\pb-N_\pc,N) \otimes \repM_\pb(N_\pb) \otimes \repM_\pc(N_\pc)\big)\,,
\end{equation}
subject to the relations 
\begin{subequations} 
\begin{alignat}{3}
&u^\dag c_j \cdot (\id \otimes v \otimes w) = 
u^\dag \cdot (c_j \otimes v \otimes w)\,, \qquad 
&& 1 \le j \le N_\pa-N_\pb -N_\pc + 1,
\\[0.15cm]
&u^\dag c_{N_\pa-N_\pb-N_\pc+j+2} \cdot (\id \otimes v \otimes w) = 
u^\dag \cdot(\id \otimes c_j\,v\otimes w)\,, \qquad 
&& 1 \le j \le N_\pb -1,
\\[0.15cm]
&u^\dag c_{N_\pa-N_\pc+j+2} \cdot (\id \otimes v \otimes w) = 
u^\dag \cdot(\id \otimes v \otimes c_j\, w)\,, \qquad 
&& 1 \le j \le N_\pc -1,
\\[0.15cm]
&u^\dag c_j^\dagger \cdot(\id \otimes v \otimes w) = 
u^\dag \cdot(c_j^\dagger \otimes v \otimes w)\,, \qquad 
&& 1 \le j \le N_\pa-N_\pb-N_\pc -1,
\\[0.15cm]
&u^\dag c^\dagger_{N_\pa-N_\pb-N_\pc+j-2} \cdot(\id \otimes v \otimes w) = 
u^\dag \cdot(\id \otimes c^\dagger_j v \otimes w)\,, \qquad 
&& 1 \le j \le N_\pb +1,
\\[0.15cm]
&u^\dag c^\dagger_{N_\pa-N_\pc+j-2} \cdot(\id \otimes v \otimes w) = 
u^\dag \cdot(\id \otimes v \otimes c^\dagger_j w)\,, \qquad 
&& 1 \le j \le N_\pc +1,
\\[0.15cm]
&u^\dag \Omega\cdot(\id \otimes v \otimes w) = u^\dag c_{N_\pa-N_\pc}c_{N_\pa-N_\pb-N_\pc}\cdot(\Omega \, \otimes c_0^\dag v \otimes c_0^\dag w)\,, \qquad
&& \textrm{for } N_\pa>N_\pb+N_c \,.
\end{alignat}
\end{subequations}
for $u^\dag \in \repM_\pa(N_\pa)$, $v \in \repM_\pb(N_\pb)$ and $w \in \repM_\pb(N_\pc)$.

\paragraph{Properties of the triple product.}

Using the same arguments as those presented in \cref{sec:bifunctor}, we obtain the following proposition.
\begin{Proposition}
The fusion of three familes of modules defined from \eqref{eq:triple.fusion} with the relations \eqref{eq:3hole.fusion} defines a trifunctor $\mathcal F$ on the category $\mathcal C$ of families of modules, namely it defines
\begin{enumerate}
\item[(i)] a map $\cal F: (\repM_\pa, \repM_\pb, \repM_\pc) \to \repM_{\pa\pb\pc}$ for some $\repM_{\pa\pb\pc} \in \mathcal C$, defined for all $\repM_\pa,\repM_\pb, \repM_\pc \in \mathcal C$,
\item[(ii)] a map $\cal F: (\phi_\pa, \phi_\pb, \phi_\pc) \to \phi_{\pa\pb\pc}$ for some family of homomorphisms $\phi_{\pa\pb\pc}$ on $\mathcal C$, defined for all families of homomorphisms $\phi_\pa$, $\phi_\pb$, $\phi_\pc$ on $\mathcal C$,
and satisfying
\begin{subequations} \label{eq:trifunctor.props}
\begin{alignat}{2}
& \cal F(\id_{\repM_\pa}, \id_{\repM_\pb}, \id_{\repM_\pc}) = \id_{\cal F(\repM_\pa,\repM_\pb,\repM_\pc)}\,, \\[0.1cm]
& \cal F(\phi'_\pa \circ \phi_\pa, \phi'_\pb \circ \phi_\pb, \phi'_\pc \circ \phi_\pc) = 
\cal F(\phi'_\pa, \phi'_\pb, \phi'_\pc) \circ\cal F(\phi_\pa,\phi_\pb, \phi_\pc) \,,
\end{alignat}
\end{subequations}
for all $\repM_\pa, \repM_\pb, \repM_\pc \in \mathcal C$ and all families of homomorphisms $\phi_\pa$, $\phi'_\pa$, $\phi_\pb$, $\phi'_\pb$, $\phi_\pc$ and $\phi'_\pc$.
\end{enumerate}
\end{Proposition}

\noindent The following theorem asserts that fusion is associative. Its proof is given at the end of \cref{sec:assoc.proof}.

\begin{Theorem} \label{prop:assoc}
Let $\repM_\pa$, $\repM_\pb$ and $\repM_\pc$ be families of modules. Then
\begin{equation}
(\repM_\pa \timesf \repM_\pb) \timesf \repM_\pc \simeq
\repM_\pa \timesf \repM_\pb \timesf \repM_\pc \simeq
\repM_\pa \timesf (\repM_\pb \timesf \repM_\pc)\,.
\end{equation}
\end{Theorem}

\noindent Let us remark that \eqref{eq:2+1hole.equations} and \eqref{eq:1+2hole.equations} each consist in ten types of relations, whereas \eqref{eq:3hole.fusion} has only seven. It is thus remarkable that the three corresponding families of modules are isomorphic. 

\begin{Proposition}
\label{prop:3hole.props}
Let $\repM_\pa$, $\repM'_\pa$, $\repM_\pb$ and $\repM_\pc$ be families of modules. Then
\begin{subequations}
\begin{alignat}{2}
(\repM_\pa \oplus \repM'_\pa) \timesf \repM_\pb \timesf \repM_\pc &\simeq
\repM_\pa \timesf \repM_\pb \timesf \repM_\pc \oplus
\repM'_\pa \timesf \repM_\pb \timesf \repM_\pc\,,
\label{eq:(Ma+Ma').Mb.Mc=Ma.Mb.Mc+Ma'.Mb.Mc}
\\[0.15cm]
\repM_\pa \timesf \repM_\pb \timesf \repM_\pc &\simeq 
\repM_\pb \timesf \repM_\pa \timesf \repM_\pc \simeq
\repM_\pa \timesf \repM_\pc \timesf \repM_\pb\,,
\label{eq:Ma.Mb.Mc=Mb.Ma.Mc}
\\[0.15cm]
\repM_\pa^\dag \timesfh \repM_\pb \timesf \repM_\pc&\simeq 
(\repM_\pa \timesf \repM_\pc^r \timesf \repM_\pb^r)^\dag\,.
\label{eq:Ma+.Mb.Mc=(Ma.Mcr.Mbr)+}
\end{alignat}
\end{subequations}
\end{Proposition} 
\proof
The isomorphisms \eqref{eq:(Ma+Ma').Mb.Mc=Ma.Mb.Mc+Ma'.Mb.Mc} are obtained by combining \cref{prop:distributivity} and \cref{prop:assoc}. Similarly, the isomorphisms \eqref{eq:Ma.Mb.Mc=Mb.Ma.Mc} are obtained by combining \cref{prop:commutativity} and \cref{prop:assoc}. Lastly, the isomorphism~\eqref{eq:Ma+.Mb.Mc=(Ma.Mcr.Mbr)+} can be proven with an argument similar to the one used to prove \cref{prop:Ma+.Mb=(Ma.Mbr)+}, namely by constructing an explicit family of anti-isomorphisms from $\repM_\pa^\dag \timesfh \repM_\pb \timesf \repM_\pc$ to $\repM_\pa \timesf \repM_\pc^r \timesf \repM_\pb^r$.
\eproof

\begin{Corollary}
\label{cor:assoc}
Let $\repM_\pa$, $\repM_\pb$ and $\repM_\pc$ be families of modules. Then
\be
(\repM^\dag_\pa \timesfh \repM_\pb) \timesfh \repM_\pc \simeq
\repM^\dag_\pa \timesfh \repM_\pb \timesf \repM_\pc \simeq
\repM^\dag_\pa \timesfh (\repM_\pb \timesf \repM_\pc)\,.
\ee
\end{Corollary}
The proof of the first isomorphism uses \cref{prop:Ma+.Mb=(Ma.Mbr)+,prop:3hole.props,prop:assoc}:
\begin{alignat}{2}
\repM^\dag_\pa \timesfh \repM_\pb \timesf \repM_\pc 
&\simeq (\repM_\pa \timesf \repM^r_\pc \timesf \repM^r_\pb)^\dag
\simeq (\repM_\pa \timesf \repM^r_\pb \timesf \repM^r_\pc)^\dag
\simeq \big((\repM_\pa \timesf \repM^r_\pb) \timesf \repM^r_\pc\big)^\dag
\nn\\[0.2cm]& 
\simeq (\repM_\pa \timesf \repM^r_\pb)^\dag \timesfh \repM_\pc
\simeq (\repM^\dag_\pa \timesfh \repM_\pb) \timesfh \repM_\pc\,.
\end{alignat}
The proof of the second isomorphism is similar, and uses also \cref{prop:reverse}.

\subsection{Homomorphisms and proof of \cref{prop:assoc}}
\label{sec:assoc.proof}

In this section, we define two families of homomorphisms and use them to give a proof of \cref{prop:LLL.equiv}. We start however by studying equivalence formulas for elements in $(\repM_\pa \timesf \repM_\pb)(N)$.

\paragraph{Equivalences in $\boldsymbol{(\repM_\pa \timesf \repM_\pb)}$.}

Let $\repM_\pa$ and $\repM_\pb$ be two families of modules. Using \eqref{eq:moving.arches}, we derive the relations
\begin{subequations} \label{eq:map-diagrams}
\begin{alignat}{2}
(c_j u \otimes v) &= c_j \cdot(u \otimes v) && 1\leq j \leq N_\pa-1 \,, \\
(c_j^\dag u \otimes v) &= c_j^\dag \cdot(u \otimes v) && 1\leq j \leq N_\pa+1 \,, \\
(c_0 u \otimes v) &= c_0 c_{N_\pa} \cdot (u \otimes c_0^\dag v) \,,\\
(c_0^\dag u \otimes v) &= \Omega^{-1} \cdot (u \otimes c_0^\dag v) \,, 
\end{alignat}
\end{subequations}
which hold for all $u \in \repM_\pa(N_\pa)$ and $v \in \repM_\pb(N_\pb)$.\medskip

Let $\lambda$ be a diagram in $\cL(N'_\pa,N_\pa)$. It can be decomposed as $\lambda_1\lambda_2\cdots \lambda_m$, where each $\lambda_i$ is an operator $c_j$ or $c_j^\dag$. For $m \ge 1$, we associate to $\lambda$ the diagram $\rho(\lambda)$ and the integer $r(\lambda)$ defined as
\begin{equation}
\rho(\lambda)=\rho(\lambda_1)\rho(\lambda_2)\cdots \rho(\lambda_m) \,,
\qquad r(\lambda)=r(\lambda_1)+r(\lambda_2)+\dots + r(\lambda_m) \,,
\end{equation}
where
\begin{subequations} 
\begin{alignat}{3}
\rho(c_j)&=c_j\,, \qquad &&r(c_j)=0\,, \qquad && 1\leq j \leq N_\pa-1 \,, \\
\rho(c^\dag_j)&=c^\dag_j\,, \qquad && r(c^\dag_j)=0\,, \qquad&& 1\leq j \leq N_\pa+1 \,, \\
\rho(c_0) &= c_0 c_{N_\pa}\,, \qquad && r(c_0) = 1 \,,\\
\rho(c_0^\dag)&= \Omega^{-1}\,, \qquad && r(c_0^\dag) = 1 \,.
\end{alignat}
\end{subequations}
For $m=0$, we set $\rho(\id)=\id$ and $r(\id)=0$. The relations \eqref{eq:map-diagrams} then ensure that
\begin{equation} \label{eq:prop.rho}
(\lambda\, u \otimes v) = \rho(\lambda) \cdot \big( u \otimes (c_0^\dag)^{r(\lambda)} v\big) \,,
\end{equation}
for all $u \in \repM_\pa(N_\pa)$ and $v \in \repM_\pb(N_\pb)$. The element $\rho(\lambda)$ in \eqref{eq:prop.rho} is a diagram in $\cL\big(N'_\pa + N_\pb, N_\pa + N_\pb + 2 r(\lambda)\big)$. This formula gives equivalent expressions for elements in $\repM_\pa \timesf \repM_\pb$.\medskip

Let $\lambda$ and $\lambda'$ be two words in $\cL(N'_\pa,N_\pa)$. These words are equivalent if and only if $\lambda$ can be transformed into $\lambda'$ using the relations \eqref{eq:c.relations}. These may however be associated to two distinct pairs $(\rho(\lambda),r(\lambda))$ and $(\rho(\lambda'),r(\lambda'))$. In this case, we nonetheless have
\be
\rho(\lambda) \cdot (u\otimes (c_0^\dag)^{r(\lambda)}v)
= (\lambda\,u \otimes v)
= (\lambda'\,u \otimes v)
= \rho(\lambda') \cdot (u\otimes (c_0^\dag)^{r(\lambda')}v) \,,
\ee
for all $u\in\repM_\pa(N_\pa)$ and $v\in\repM_\pb(N_\pb)$.
Moreover, for each diagram $\lambda$ in $\cL(N_\pa',N_\pa)$, we have
\begin{subequations}
\label{eq:commutation.rho}
\begin{alignat}{2}
c_{N_\pa'+j} \, \rho(\lambda) = \rho(\lambda) \, c_{N_\pa+j+r(\lambda)} \,, 
\qquad &&
j = 1,2, \dots, N_\pb-1\,,
\label{eq:rho.c=c.rho} \\[0.1cm]
\qquad 
c_{N_\pa'+j}^\dag \, \rho(\lambda) = \rho(\lambda) \, c_{N_\pa+j+r(\lambda)}^\dag \,,
\qquad &&
j = 1,2, \dots, N_\pb+1\,.
\end{alignat}
\end{subequations}
These identities are easily proved for $\lambda=c_k,c^\dag_k,c_0,c^\dag_0$ from the relations \eqref{eq:c.relations}. To prove that they also hold for arbitrary words $\lambda$, we decompose them as $\lambda=\lambda_1\lambda_2 \cdots \lambda_m$ and apply iteratively \eqref{eq:commutation.rho} on each~$\rho(\lambda_i)$.

\paragraph{Homomorphisms.}

We define two linear maps, $\psi_N:\repL_{\repM_\pa,\repM_\pb,\repM_\pc}(N)\to \big((\repM_\pa \timesf\repM_\pb) \timesf \repM_\pc\big)(N)$ and $\phi_N: \repL_{\repL_{\repM_\pa,\repM_\pb},\repM_\pc}(N) \to (\repM_\pa \timesf \repM_\pb \timesf\repM_\pc)(N)$, as
\begin{equation}
\psi_N(\lambda \otimes u \otimes v \otimes w)=
\lambda \cdot \big((u \otimes v) \otimes w \big) \,,
\end{equation}
for $\lambda \in \cL(N,N_{\pa\pb\pc}), \ u \in \repM_\pa(N_\pa), \ v \in \repM_\pb(N_\pb), \ w\in \repM_\pc(N_\pc)$, and
\begin{equation}
\phi_N\big(\lambda \otimes (\lambda_{\pa\pb} \otimes u \otimes v) \otimes w \big)
= \lambda \,\rho(\lambda_{\pa\pb}) \cdot \big(u \otimes v \otimes (c_0^\dag)^{r(\lambda_{\pa\pb})}w \big) \,,
\end{equation}
for $\lambda \in \cL(N,N'_{\pa\pb}+N_\pc), \ \lambda_{\pa\pb} \in \cL(N'_{\pa\pb},N_{\pa\pb}), \ u \in \repM_\pa(N_\pa), \ v \in \repM_\pb(N_\pb), \ w\in \repM_\pc(N_\pc)$.
The following propositions establish that $\psi_N$ and $\phi_N$ define families of homomorphisms, and that they are also well defined when their domains are replaced by the corresponding families of fused modules.

\begin{Proposition}
The maps $\psi_N$ define a family of homomorphisms, which is also well defined from $(\repM_\pa \timesf \repM_\pb \timesf\repM_\pc)$ to $\big((\repM_\pa \timesf\repM_\pb) \timesf \repM_\pc\big)$.
\end{Proposition}
\proof
It is easy to check that $\psi_N$ satisfies
\be
\psi_{N'}\big(\lambda' \cdot (\lambda \otimes u \otimes v \otimes w)\big)
= \lambda' \cdot \psi_N(\lambda \otimes u \otimes v \otimes w) \,,
\ee
for all $\lambda'\in\cL(N',N)$, so we deduce that $\psi_N$ indeed defines a family of homomorphisms. To show that it also defines a family of homomorphisms from $(\repM_\pa \timesf \repM_\pb \timesf\repM_\pc)$ to $\big((\repM_\pa \timesf\repM_\pb) \timesf \repM_\pc\big)$, we prove that $\psi_N$ preserves the relations~\eqref{eq:3hole.fusion}. These relations are all of the form
\be
\lambda \cdot (u \otimes v \otimes w) = \lambda' \cdot(\lambda_\pa u \otimes \lambda_\pb v \otimes \lambda_\pc w)\,,
\ee
for some diagrams $\lambda$, $\lambda'$, $\lambda_\pa$, $\lambda_\pb$, $\lambda_\pc$ specific to each case. It is easy to check from \eqref{eq:2+1hole.equations} that each such relation has a counterpart in $\big((\repM_\pa \timesf\repM_\pb) \timesf \repM_\pc\big)$ of the form
\be
\lambda \cdot \big((u \otimes v) \otimes w\big) = \lambda' \cdot\big((\lambda_\pa u \otimes \lambda_\pb v) \otimes \lambda_\pc w\big)\,.
\ee
We then have
\be
\psi_{N_{\pa\pb\pc}^{\tinyx\alpha}}(\lambda \otimes u \otimes v \otimes w)
= \lambda \cdot \big((u \otimes v) \otimes w\big)
= \lambda'\cdot\big((\lambda_\pa u \otimes \lambda_\pb v) \otimes \lambda_\pc w\big)
=\psi_{N_{\pa\pb\pc}^{\tinyx\alpha}}(\lambda' \otimes \lambda_\pa u \otimes \lambda_\pb v \otimes \lambda_\pc w)\,,
\ee
ending the proof.
\eproof

\noindent Proving the same result for the maps $\phi_N$ is more tedious.

\begin{Proposition}
The maps $\phi_N$ define a family of homomorphisms, which is also well defined from $((\repM_\pa \timesf \repM_\pb) \timesf\repM_\pc)$ to $\big(\repM_\pa \timesf\repM_\pb \timesf \repM_\pc\big)$.
\end{Proposition}
\proof
We first check that $\phi_N$ is a well-defined map. Let $\lambda,\lambda'$ be two words in $\cL(N'_{\pa\pb},N_{\pa\pb})$ that take different expressions in terms of products of $c_j$ and $c_j^\dag$ generators, but are equivalent under the relations~\eqref{eq:c.relations}. We then have to show that
\begin{equation} \label{eq:rho.welldef}
\rho(\lambda) \cdot \big(u \otimes v \otimes (c_0^\dag)^{r(\lambda)}w \big) = \rho(\lambda') \cdot \big(u \otimes v \otimes (c_0^\dag)^{r(\lambda')}w \big) \,,
\end{equation}
for all $u \in \repM_\pa(N_\pa)$, $v \in \repM_\pb(N_\pb)$ and $w \in \repM_\pc(N_\pc)$. The definitions of $\rho(\lambda)$ as a product and $r(\lambda)$ as a sum allow us to prove this using the relations \eqref{eq:c.relations}. Indeed, if \eqref{eq:rho.welldef} holds for two pairs $(\lambda_1, \lambda'_1)$ and $(\lambda_2, \lambda'_2)$, then it is straightforward to show that it holds for $(\lambda_1\lambda_2, \lambda'_1\lambda'_2)$. For any pair $(\lambda,\lambda')$ of equivalent words, there exists a finite number of transformations that one can perform using \eqref{eq:c.relations} that produces $\lambda'$ starting from $\lambda$. We thus show that \eqref{eq:rho.welldef} is satisfied at each step of the transformation. In other words, we check that \eqref{eq:rho.welldef} holds for each relation in \eqref{eq:c.relations}, with $\lambda$ and $\lambda'$ fixed to the words arising in the left and right sides of the relation considered. The proof divides into four cases.
\begin{enumerate}
\item[1)] The relations \eqref{eq:c.relations.a}$-$\eqref{eq:c.relations.d} do not involve $c_0$ nor $c_0^\dag$. As a result, we have $\rho(\lambda) = \lambda = \lambda' = \rho(\lambda')$ as well as $r(\lambda) = r(\lambda') = 0$, so clearly \eqref{eq:rho.welldef} holds in this case.
\item[2)] For the relations \eqref{eq:c.relations.e}$-$\eqref{eq:c.relations.g}, and \eqref{eq:c.relations.h} with $j\neq 0$, we find with a similar computation that $\rho(\lambda)=\rho(\lambda')$, and $r(\lambda)=r(\lambda')=1$, confirming \eqref{eq:rho.welldef} in this case as well.
\item[3)] For the relation \eqref{eq:c.relations.h} with $j=0$, namely $c_0c_0^\dag=\beta \,\id$, we use the relations \eqref{eq:3hole.fusion} and find
\begin{alignat}{2}
\rho(c_0c_0^\dag) & \cdot \big(u\otimes v\otimes (c_0^\dag)^{r(c_0c_0^\dag)}w \big) 
= c_0\,c_{N_{\pa\pb}+2}\,\Omega^{-1} \cdot \big(u\otimes v\otimes (c_0^\dag)^2w \big) \nn\\
&= c_0\,\Omega^{-1} c_{N_{\pa\pb}+1} \cdot \big(u\otimes v\otimes (c_0^\dag)^2w \big)
= c_0\,\Omega^{-1}\cdot \big(u\otimes v\otimes c_1(c_0^\dag)^2w \big)\nn\\
&= c_{N_{\pa\pb\pc}+1} \cdot \big(u\otimes v\otimes c^\dag_{N_\pc+1}w \big)
= c_{N_{\pa\pb\pc}+1}\, c_{N_{\pa\pb\pc}+1}^\dag \cdot (u\otimes v\otimes w)
= \beta \, (u\otimes v\otimes w) \,,
\end{alignat}
where $N_{\pa\pb} = N_\pa+N_\pb$, and $N_{\pa\pb\pc} = N_\pa+N_\pb+N_\pc$.
\item[4)] For the relations \eqref{eq:c.relations.i}, we first compute
\be
\rho(\Omega_N)=\rho(c_1c_0^\dag) = c_0 \,,
\qquad \rho(\Omega_N^{-1})=\rho(c_0c_1^\dag) = \Omega_N^{-1}
c_{N} \,,
\qquad r(\Omega_N)=r(\Omega_N^{-1})=1 \,.
\ee
We write
\begin{alignat}{2}
\rho(c_1 c_0^\dag c_0 c_1^\dag)&\cdot \big(u\otimes v\otimes (c_0^\dag)^{r(c_1 c_0^\dag c_0 c_1^\dag)}w \big) 
= c_0\,\Omega^{-1}c_{N_{\pa\pb}} \cdot \big(u\otimes v\otimes (c_0^\dag)^2w \big)
\nn\\[0.1cm]&
= c_{N_{\pa\pb}}c_{N_{\pa\pb\pc}+3} \cdot \big(u\otimes v\otimes (c_0^\dag)^2w \big)
= c_{N_{\pa\pb}}\cdot \big(u\otimes v\otimes c_{N_\pc+3}(c_0^\dag)^2w \big)
\nn\\[0.1cm]&
= c_{N_{\pa\pb}}\cdot (u\otimes v\otimes c_1^\dag w)
= c_{N_{\pa\pb}}c^\dag_{N_{\pa\pb}+1}\cdot (u\otimes v\otimes w)
= (u\otimes v\otimes w) \,,
\end{alignat}
and
\begin{alignat}{2}
\rho(c_0 c_1^\dag c_1 c_0^\dag) &\cdot \big(u\otimes v\otimes (c_0^\dag)^{r(c_0 c_1^\dag c_1 c_0^\dag)}w \big) 
= \Omega^{-1} c_{N_{\pa\pb}} c_0 \cdot \big(u\otimes v\otimes (c_0^\dag)^2w \big)
\nn\\[0.1cm]&
= \Omega^{-1} c_0 c_{N_{\pa\pb}+1} \cdot \big(u\otimes v\otimes (c_0^\dag)^2w \big) 
= \Omega^{-1} c_0 \cdot \big(u\otimes v\otimes c_1(c_0^\dag)^2w \big)
\nn\\[0.1cm]&
= \Omega^{-1} c_0 \cdot \big(u\otimes v\otimes c^\dag_{N_\pc+1}w \big) 
= \Omega^{-1} c_0 c^\dag_{N_{\pa\pb\pc}+1}\cdot \big(u\otimes v\otimes w \big) 
\nn\\[0.1cm]&
= \Omega^{-1} \Omega\cdot \big(u\otimes v\otimes w \big) 
= \big(u\otimes v\otimes w \big) \,.
\end{alignat}
\end{enumerate}
Hence we have proven that $\phi_N$ is a well-defined map from $\repL_{\repL_{\repM_\pa,\repM_\pb},\repM_\pc}(N)$ to $(\repM_\pa \timesf \repM_\pb \timesf\repM_\pc)(N)$.
\medskip

It is straightforward to show that
\be
\phi_{N'}\big(\lambda' \cdot (\lambda \otimes (\lambda_{\pa\pb} \otimes u \otimes v) \otimes w)\big)
= \lambda' \cdot \phi_N\big(\lambda \otimes (\lambda_{\pa\pb} \otimes u \otimes v) \otimes w\big) \,,
\ee
for all $\lambda'\in \cL(N',N)$, so that $\phi_N$ indeed yields a family of homomorphisms from $\repL_{\repL_{\repM_\pa,\repM_\pb},\repM_\pc}(N)$ to $(\repM_\pa \timesf \repM_\pb \timesf\repM_\pc)(N)$.
Proving that it is well defined on $((\repM_\pa \timesf \repM_\pb) \timesf\repM_\pc)$ amounts to showing the relations
\begin{subequations}
\begin{alignat}{2}
&\phi_{N'_{\pa\pb}+N_\pc}\big(\id \otimes \lambda\, X_j^{\tinyx \alpha}(u,v) \otimes w\big) = 0 \,,
&&\qquad \lambda\in\cL(N'_{\pa\pb},N_{\pa\pb}^{\tinyx\alpha}) \,, \\
&\phi_{N_{(\pa\pb)',c}^{\tinyx\alpha}}\big(X_j^{\tinyx \alpha}(\lambda \otimes u \otimes v, w)\big) = 0 \,,
&&\qquad \lambda\in\cL(N'_{\pa\pb},N_{\pa\pb}) \,,
\end{alignat}
\end{subequations}
for all $\alpha$ and $j$, where
\be
N_{(\pa\pb)',c}^{\tinyx\alpha} = \left\{\begin{array}{cl}
N'_{\pa\pb}+N_\pc-2 & \alpha \in \{a,b\}\,,\\[0.15cm]
N'_{\pa\pb}+N_\pc+2 & \alpha \in \{c,d\}\,,\\[0.15cm]
N'_{\pa\pb}+N_\pc & \alpha \in \{e\}\,.
\end{array}\right.
\ee
We also recall that $X_j^{\tinyx \alpha}$ is defined in \eqref{eq:X(u,v)}. The first set of conditions reads
\begin{subequations} \label{eq:phi(X1)=0}
\begin{alignat}{3}
& \rho(\lambda\, c_j) \cdot \big(u \otimes v \otimes (c_0^\dag)^{r(\lambda\, c_j)}w\big)
= \rho(\lambda) \cdot \big(c_j u \otimes v \otimes (c_0^\dag)^{r(\lambda)}w\big) \,,
&& \quad 1 \leq j \leq N_\pa-1 \,, \label{eq:phi(X1)=0.a} 
\\[0.1cm]
& \rho(\lambda\, c_{N_\pa+j}) \cdot \big(u \otimes v \otimes (c_0^\dag)^{r(\lambda\, c_{N_\pa+j})}w\big)
= \rho(\lambda) \cdot \big(u \otimes c_j v \otimes (c_0^\dag)^{r(\lambda)}w\big) \,,
&& \quad 1 \leq j \leq N_\pb-1 \,, \label{eq:phi(X1)=0.b} 
\\[0.1cm]
& \rho(\lambda\, c^\dag_j) \cdot \big(u \otimes v \otimes (c_0^\dag)^{r(\lambda\, c^\dag_j)}w\big)
= \rho(\lambda) \cdot \big(c^\dag_j u \otimes v \otimes (c_0^\dag)^{r(\lambda)}w\big) \,,
&& \quad 1 \leq j \leq N_\pa+1 \,, \label{eq:phi(X1)=0.c}
\\[0.1cm]
& \rho(\lambda\, c^\dag_{N_\pa+j}) \cdot \big(u \otimes v \otimes (c_0^\dag)^{r(\lambda \,c^\dag_{N_\pa+j})}w\big)
= \rho(\lambda) \cdot \big(u \otimes c^\dag_j v \otimes (c_0^\dag)^{r(\lambda)}w\big) \,,
&& \quad 1 \leq j \leq N_\pb+1 \,, \label{eq:phi(X1)=0.d}
\\[0.1cm]
& \rho(\lambda\, \Omega) \cdot \big(u \otimes v \otimes (c_0^\dag)^{r(\lambda\, \Omega)}w\big)
= \rho(\lambda\, c_{N_\pa}) \cdot \big(\Omega\, u \otimes c_0^\dag v \otimes (c_0^\dag)^{r(\lambda\,c_{N_\pa})}w\big) \,,
&& \quad\textrm{for }N_\pa>0 \,, \label{eq:phi(X1)=0.e}
\end{alignat}
\end{subequations}
where $\lambda\, c_j,\lambda\, c_j^\dag$ and $\lambda\, \Omega$ are diagrams in $\cL(N'_{\pa\pb},N_{\pa\pb})$.
The second set of conditions reads
\begin{subequations} \label{eq:phi(X2)=0}
\begin{alignat}{3}
& c_j \rho(\lambda) \cdot \big(u \otimes v \otimes (c_0^\dag)^{r(\lambda)}w\big)
= \rho(c_j \lambda) \cdot \big(u \otimes v \otimes (c_0^\dag)^{r(c_j \lambda)}w\big) \,,
&& 1 \leq j \leq N_{\pa\pb}'-1 \,, 
\label{eq:phi(X2)=0.a} \\[0.1cm]
& c_{N'_{\pa\pb}+j} \rho(\lambda) \cdot \big(u \otimes v \otimes (c_0^\dag)^{r(\lambda)}w\big)
= \rho(\lambda) \cdot \big(u \otimes v \otimes (c_0^\dag)^{r(\lambda)} c_j w\big) \,, 
&& 1 \leq j \leq N_{\pc} -1 \,, 
\label{eq:phi(X2)=0.b}\\[0.1cm]
& c^\dag_j \rho(\lambda) \cdot \big(u \otimes v \otimes (c_0^\dag)^{r(\lambda)}w\big)
= \rho(c^\dag_j \lambda) \cdot \big(u \otimes v \otimes (c_0^\dag)^{r(c^\dag_j \lambda)}w\big) \,,
&&1 \leq j \leq N'_{\pa\pb}+1\,, 
\label{eq:phi(X2)=0.c}\\[0.1cm]
& c_{N'_{\pa\pb}+j}^\dag \rho(\lambda) \cdot \big(u \otimes v \otimes (c_0^\dag)^{r(\lambda)}w\big)
= \rho(\lambda) \cdot \big(u \otimes v \otimes (c_0^\dag)^{r(\lambda)} c_j^\dag w\big) \,,
&& 1 \leq j \leq N_{\pc}+1 \,, 
\label{eq:phi(X2)=0.d}\\[0.1cm]
& \Omega\, \rho(\lambda) \cdot \big(u \otimes v \otimes (c_0^\dag)^{r(\lambda)}w\big)
= c_{N'_{\pa\pb}} \rho(\Omega\, \lambda) \cdot \big(u \otimes v \otimes (c_0^\dag)^{r(\Omega\, \lambda)+1}w\big) \,,
&\quad&\textrm{for }N'_{\pa\pb}>0 \,,
\label{eq:phi(X2)=0.e}
\end{alignat}
\end{subequations}
for all diagrams $\lambda$ in $\cL(N'_{\pa\pb},N_{\pa\pb})$.\medskip

The proofs of \eqref{eq:phi(X1)=0.a}$-$\eqref{eq:phi(X1)=0.d} use \eqref{eq:3hole.fusion} and are straightforward. For instance, we have
\be
\rho(\lambda\, c_j) \cdot \big(u \otimes v \otimes (c_0^\dag)^{r(\lambda\, c_j)}w\big)
= \rho(\lambda)\, c_j\cdot \big(u \otimes v \otimes (c_0^\dag)^{r(\lambda)}w\big) 
= \rho(\lambda) \cdot \big(c_j u \otimes v \otimes (c_0^\dag)^{r(\lambda)}w\big) \,,
\ee
for $1 \leq j \leq N_\pa-1$. Similarly, \eqref{eq:phi(X2)=0.a} and \eqref{eq:phi(X2)=0.c} directly follow from the definition of $\rho(\lambda)$, and \eqref{eq:phi(X2)=0.b} and \eqref{eq:phi(X2)=0.d} follow from \eqref{eq:commutation.rho}.\medskip

To prove \eqref{eq:phi(X1)=0.e}, we use the identity in $(\repM_\pa \timesf\repM_\pb \timesf\repM_\pc)$
\begin{equation}
c_0 \cdot (u\otimes v \otimes c_0^\dag w)
= c_{N_\pa} \cdot (\Omega u \otimes c_0^\dag v \otimes w) \,,
\end{equation}
which is easily derived using \eqref{eq:3hole.fusion.extra.e} and subsequently \eqref{eq:3hole.fusion.f}. We then have
\begin{alignat}{2}
\rho(\lambda\, \Omega) &\cdot \big(u \otimes v \otimes (c_0^\dag)^{r(\lambda\, \Omega)}w\big)
= \rho(\lambda)\,c_0 \cdot \big(u \otimes v \otimes (c_0^\dag)^{r(\lambda)+1}w\big) \nn \\
&= \rho(\lambda)\,c_{N_\pa}\cdot \big(\Omega u \otimes c_0^\dag v \otimes (c_0^\dag)^{r(\lambda)}w\big) 
= \rho(\lambda\, c_{N_\pa})\cdot \big(\Omega u \otimes c_0^\dag v \otimes (c_0^\dag)^{r(\lambda\, c_{N_\pa})}w\big) \,.
 \end{alignat}
To prove \eqref{eq:phi(X2)=0.e}, we write its left side as
\begin{alignat}{2}
\Omega\, &\rho(\lambda) \cdot \big(u \otimes v \otimes (c_0^\dag)^{r(\lambda)}w\big)
= c_0\, c^\dag_{N'_{\pa\pb}+N_\pc+1} \rho(\lambda) \cdot \big(u \otimes v \otimes (c_0^\dag)^{r(\lambda)}w\big) \nn \\[0.1cm]
&= c_0\, \rho(\lambda)\, c^\dag_{N_\pa+N_\pb+N_\pc+r(\lambda)+1}\cdot \big(u \otimes v \otimes (c_0^\dag)^{r(\lambda)}w\big) 
= c_0 \,\rho(\lambda) \cdot \big(u \otimes v \otimes c^\dag_{N_\pc+r(\lambda)+1}(c_0^\dag)^{r(\lambda)}w\big) \,.
\end{alignat}
Similarly, the right side of \eqref{eq:phi(X2)=0.e} reads
\begin{alignat}{2}
c_{N'_{\pa\pb}} \rho(\Omega\, \lambda) &\cdot \big(u \otimes v \otimes (c_0^\dag)^{r(\Omega \,\lambda)+1}w\big)
= c_{N'_{\pa\pb}} c_0\,\rho(\lambda) \cdot \big(u \otimes v \otimes (c_0^\dag)^{r(\lambda)+2}w\big) 
\nn \\[0.1cm]&
= c_0\, c_{N'_{\pa\pb}+1} \,\rho(\lambda) \cdot \big(u \otimes v \otimes (c_0^\dag)^{r(\lambda)+2}w\big) 
= c_0 \,\rho(\lambda)\, c_{N_\pa+N_\pb+r(\lambda)+1}\cdot \big(u \otimes v \otimes (c_0^\dag)^{r(\lambda)+2}w\big) \nn \\[0.1cm]
&= c_0 \,\rho(\lambda) \cdot \big(u \otimes v \otimes c_{r(\lambda)+1}(c_0^\dag)^{r(\lambda)+2}w\big) \,.
\end{alignat}
For all nonnegative integer $r$ and all states $w \in \repM_\pc(N_\pc)$, we have
\begin{equation}
c^\dag_{N_\pc+r+1} (c_0^\dag)^{r}w = c_{r+1}(c_0^\dag)^{r+2}w\,,
\end{equation}
and hence \eqref{eq:phi(X2)=0.e} is proven.
\eproof

\paragraph{Proof of \cref{prop:assoc}.}

We now consider the maps $\psi_N$ and $\phi_N$ as acting on the fusion modules, namely
$\psi_N:(\repM_\pa \timesf \repM_\pb \timesf\repM_\pc)(N) \to \big((\repM_\pa \timesf\repM_\pb) \timesf \repM_\pc\big)(N)$ 
and 
$\phi_N: \big((\repM_\pa \timesf\repM_\pb) \timesf \repM_\pc\big)(N) \to (\repM_\pa \timesf \repM_\pb \timesf\repM_\pc)(N)$
defined as
\begin{subequations}
\begin{alignat}{2}
&\psi_N \big(\lambda \cdot( u \otimes v \otimes w) \big)=
\lambda \cdot \big((u \otimes v) \otimes w \big) \,,
\\[0.1cm]
&\phi_N \Big(\lambda \cdot \big((\lambda_{\pa\pb} \cdot( u \otimes v)) \otimes w\big) \Big)=
\lambda \,\rho(\lambda_{\pa\pb}) \cdot \big(u \otimes v \otimes (c_0^\dag)^{r(\lambda_{\pa\pb})}w \big) \,.
\end{alignat}
\end{subequations}
We proved above that these maps define families of homomorphisms.
We also have
\begin{subequations}
\begin{alignat}{2}
& \psi_N \circ \phi_N
\big(\lambda \cdot ((\lambda_{\pa\pb} \cdot( u \otimes v)) \otimes w)\big)
= \lambda \,\rho(\lambda_{\pa\pb}) \cdot \big((u \otimes v) \otimes (c_0^\dag)^{r(\lambda_{\pa\pb})}w \big)
= \lambda \cdot \big((\lambda_{\pa\pb} \cdot( u \otimes v)) \otimes w\big) \,, \\[0.1cm]
& \phi_N \circ \psi_N
\big(\lambda \cdot(u \otimes v \otimes w)\big)
= \lambda \,\rho(\id_{N_{\pa\pb}}) \cdot \big(u \otimes v \otimes (c_0^\dag)^{r(\id_{N_{\pa\pb}})}w \big)
= \lambda \cdot( u \otimes v \otimes w) \,.
\end{alignat}
\end{subequations}
This shows that $\psi_N$ and $\phi_N$ are inverse maps, and therefore
that the families of modules $\big((\repM_\pa \timesf \repM_\pb) \timesf \repM_\pc\big)$ and $(\repM_\pa \timesf \repM_\pb \timesf\repM_\pc)$ are isomorphic. With a similar argument, one proves that $\big(\repM_\pa \timesf (\repM_\pb \timesf \repM_\pc)\big)$ and $(\repM_\pa \timesf \repM_\pb \timesf\repM_\pc)$ are isomorphic families, ending the proof.\eproof

\subsection{Equivalence of definitions for diagram spaces with three holes}

We distinguish between the diagrammatic and algebraic definitions of the spaces of diagrams with three holes, and denote them as $\cLad(N,N_\pa,N_\pb,N_\pc)$ and $\cLaa(N,N_\pa,N_\pb,N_\pc)$ respectively. From \cref{prop:equivalence.L,prop:Lambda.defs}, we have 
\begin{equation}
\cLd(N,N') \simeq \cLa(N,N') \,,
\qquad
\cLad(N,N_\pa,N_\pb) \simeq \cLaa(N,N_{\pa},N_\pb) \,.
\end{equation}
Moreover, using the definition \eqref{eq:LN} of the family $\repL_N$, we find from 
\eqref{eq:Lambda=LxL}, \eqref{eq:Lambda.sumsLLLL} and \eqref{eq:triple.fusion} that
\begin{equation}
\cLaa(N,N_\pa,N_\pb)= (\repL_{N_\pa}\timesf \repL_{N_\pb})(N) \,,
\qquad
\cLaa(N,N_\pa,N_\pb,N_\pc)= (\repL_{N_\pa}\timesf \repL_{N_\pb} \timesf \repL_{N_\pc} )(N) \,,
\end{equation}
as family of modules.

\begin{Proposition}
\label{prop:LxLxLequiv}
We have the isomorphism of spaces 
\be
\label{eq:LxLxLequiv}
\big((\repL_{N_\pa}\timesf \repL_{N_\pb}) \timesf \repL_{N_\pc}\big)(N)
\simeq (\repL_{N_\pa}\timesf \repL_{N_\pb} \timesf \repL_{N_\pc})(N)
\simeq \big(\repL_{N_\pa}\timesf (\repL_{N_\pb} \timesf \repL_{N_\pc})\big)(N)\,.
\ee
\end{Proposition}
\proof
By setting $\repM_\pa = \repL_{N_\pa}$, $\repM_\pb = \repL_{N_\pb}$ and $\repM_\pc = \repL_{N_\pc}$ in \cref{prop:assoc}, we readily find that \eqref{eq:LxLxLequiv} holds as isomorphisms of left modules. This implies that (i) there exist bijections between the bases of the three modules, and (ii) the action of $\cL(N',N)$ from the left is identical in all three modules.\medskip 

To prove that \eqref{eq:LxLxLequiv} holds as isomorphisms of diagram spaces, one must additionally show that the action of $\cL(N_\pa,N'_\pa) \otimes \cL(N_\pb,N'_\pb) \otimes \cL(N_\pc,N'_\pc) $ from the right is identical in the three modules. For the action of $\cL(N_\pa,N'_\pa) \otimes \id \otimes \id$, we write
\begin{alignat}{2}
\Big(\big(\repL_{N_\pa}\timesf \repL_{N_\pb} &\timesf \repL_{N_\pc}\big)(N)\big(\cL(N_\pa,N'_\pa) \otimes \id \otimes \id\big)\Big)^\dag
\simeq 
\big(\cL(N'_\pa,N_\pa) \otimes \id \otimes \id\big)\big(\repL_{N_\pa}^\dag\timesfh \repL_{N_\pc}^r \timesf \repL_{N_\pb}^r\big)(N)
\nn\\[0.15cm]&
\simeq \big(\cL(N'_\pa,N_\pa) \otimes \id \otimes \id\big)\big(\repL_{N_\pa}^\dag\timesfh \repL_{N_\pb}^r \timesf \repL_{N_\pc}^r\big)(N)
\nn\\[0.15cm]&
\simeq \big(\cL(N'_\pa,N_\pa) \otimes \id \otimes \id\big)\big((\repL_{N_\pa}^\dag\timesfh \repL_{N_\pb}^r) \timesfh \repL_{N_\pc}^r\big)(N)
\nn\\[0.15cm]&
\simeq \big(\cL(N'_\pa,N_\pa) \otimes \id \otimes \id\big)\big((\repL_{N_\pa}\timesf \repL_{N_\pb})^\dag \timesfh \repL_{N_\pc}^r\big)(N)
\nn\\[0.15cm]&
\simeq \big(\cL(N'_\pa,N_\pa) \otimes \id \otimes \id\big)\big((\repL_{N_\pa}\timesf \repL_{N_\pb}) \timesf \repL_{N_\pc}\big)^\dag(N)
\nn\\[0.15cm]&
\simeq \Big(\big((\repL_{N_\pa}\timesf \repL_{N_\pb}) \timesf \repL_{N_\pc}\big)(N)\big(\cL(N_\pa,N'_\pa) \otimes \id \otimes \id\big)\Big)^\dag\,,
\end{alignat}
where we used \cref{cor:assoc} and \cref{prop:3hole.props}. Similarly,
\begin{alignat}{2}
\Big(\big(\repL_{N_\pa}\timesf \repL_{N_\pb} &\timesf \repL_{N_\pc}\big)(N)\big(\cL(N_\pa,N'_\pa) \otimes \id \otimes \id\big)\Big)^\dag
\simeq \big(\cL(N'_\pa,N_\pa) \otimes \id \otimes \id\big)\big(\repL_{N_\pa}^\dag\timesfh \repL_{N_\pb}^r \timesf \repL_{N_\pc}^r\big)(N)
\nn\\[0.15cm]&
\simeq \big(\cL(N'_\pa,N_\pa) \otimes \id \otimes \id\big)\big(\repL_{N_\pa}^\dag\timesfh (\repL_{N_\pb}^r \timesf \repL_{N_\pc}^r)\big)(N)
\nn\\[0.15cm]&
\simeq \big(\cL(N'_\pa,N_\pa) \otimes \id \otimes \id\big)\big(\repL_{N_\pa}^\dag\timesf (\repL_{N_\pb} \timesf \repL_{N_\pc}\big)^r)(N)
\nn\\[0.15cm]&
\simeq \big(\cL(N'_\pa,N_\pa) \otimes \id \otimes \id\big)\big(\repL_{N_\pa}\timesf (\repL_{N_\pb} \timesf \repL_{N_\pc})\big)^\dag(N)
\nn\\[0.15cm]&
\simeq \Big(\big(\repL_{N_\pa}\timesf (\repL_{N_\pb} \timesf \repL_{N_\pc})\big)(N)\big(\cL(N_\pa,N'_\pa) \otimes \id \otimes \id\big)\Big)^\dag\,,
\end{alignat}
where we also used \cref{prop:reverse}. Using the commutativity of the fusion product, we can show with similar arguments that the actions of $\id \otimes \cL(N_\pb,N'_\pb) \otimes \id$ and $\id \otimes \id \otimes \cL(N_\pc,N'_\pc)$ are identical in the three modules.
\eproof

\noindent We now have all the ingredients needed to prove \cref{prop:LLL.equiv}.\medskip

\noindent {\scshape Proof of \cref{prop:LLL.equiv}.\ }
For the proof, we first note that, for all non-negative integers $N$, $N_\pa$, $N_\pb$ and $N_\pc$ such that $N+N_\pa+N_\pb+N_\pc$ is even, and for all non-negative integer $M$, we have
\begin{equation}
\cLad(N,N_\pa,N_\pb,N_\pc) \simeq
\sum_{N_{\pa\pb}\ge M}\cLad(N,N_{\pa\pb},N_\pc)
\big(\cLad(N_{\pa\pb},N_\pa,N_\pb)\otimes \cLd(N_\pc,N_\pc)\big) \,.
\end{equation}
Indeed, any diagram $\nu \in \Lambda(N,N_\pa,N_\pb,N_\pc)$ can be decomposed as
\be
\psset{unit=0.8cm}
\thispic{\begin{pspicture}[shift=-2.5](-2.6,-2.6)(2.6,2.6)
\psarc[linecolor=black,linewidth=0.5pt,fillstyle=solid,fillcolor=lightlightblue]{-}(0,0){2.6}{0}{360}
\psarc[linecolor=black,linewidth=0.5pt,fillstyle=solid,fillcolor=white]{-}(-1.6,0){0.5}{0}{360}
\psarc[linecolor=black,linewidth=0.5pt,fillstyle=solid,fillcolor=white]{-}(0,0){0.5}{0}{360}
\psarc[linecolor=black,linewidth=0.5pt,fillstyle=solid,fillcolor=white]{-}(1.6,0){0.5}{0}{360}
\psarc[linecolor=orange,linewidth=1.5pt]{-}(-1.6,0){0.5}{0}{360}
\psarc[linecolor=darkgreen,linewidth=1.5pt]{-}(0,0){0.5}{0}{360}
\psarc[linecolor=purple,linewidth=1.5pt]{-}(1.6,0){0.5}{0}{360}
\psline[linestyle=dashed, dash= 1.5pt 1.5pt,linewidth=0.5pt]{-}(0,-2.6)(-1.6,-0.5)
\psline[linestyle=dashed, dash= 1.5pt 1.5pt,linewidth=0.5pt]{-}(0,-2.6)(1.6,-0.5)
\psline[linestyle=dashed, dash= 1.5pt 1.5pt,linewidth=0.5pt]{-}(0,-2.6)(0,-0.5)
\rput(1.6,-0.3){$_\pa$}
\rput(0,-0.25){$_\pb$}
\rput(-1.6,-0.3){$_\pc$}
\rput(0,1.5){$\nu$}
\end{pspicture}} 
\ = \
\thispic{\begin{pspicture}[shift=-2.5](-2.6,-2.6)(2.6,2.6)
\psarc[linecolor=black,linewidth=0.5pt,fillstyle=solid,fillcolor=lightlightblue]{-}(0,0){2.6}{0}{360}
\psarc[linecolor=black,linewidth=0.5pt,fillstyle=solid,fillcolor=lightlightblue]{-}(0.8,0){1.6}{0}{360}
\psarc[linecolor=black,linewidth=0.5pt,fillstyle=solid,fillcolor=white]{-}(-1.6,0){0.5}{0}{360}
\psarc[linecolor=black,linewidth=0.5pt,fillstyle=solid,fillcolor=white]{-}(0,0){0.5}{0}{360}
\psarc[linecolor=black,linewidth=0.5pt,fillstyle=solid,fillcolor=white]{-}(1.6,0){0.5}{0}{360}
\psarc[linecolor=orange,linewidth=1.5pt]{-}(-1.6,0){0.5}{0}{360}
\psarc[linecolor=darkgreen,linewidth=1.5pt]{-}(0,0){0.5}{0}{360}
\psarc[linecolor=purple,linewidth=1.5pt]{-}(1.6,0){0.5}{0}{360}
\psline[linestyle=dashed, dash= 1.5pt 1.5pt,linewidth=0.5pt]{-}(0,-2.6)(-1.6,-0.5)
\psline[linestyle=dashed, dash= 1.5pt 1.5pt,linewidth=0.5pt]{-}(0.8,-1.6)(1.6,-0.5)
\psline[linestyle=dashed, dash= 1.5pt 1.5pt,linewidth=0.5pt]{-}(0.8,-1.6)(0,-0.5)
\psline[linestyle=dashed, dash= 1.5pt 1.5pt,linewidth=0.5pt]{-}(0,-2.6)(0.8,-1.6)
\rput(1.6,-0.3){$_\pa$}
\rput(0,-0.25){$_\pb$}
\rput(-1.6,-0.3){$_\pc$}
\rput(0.8,1.0){$\mu_{\pa\pb}$}
\rput(-1,1.5){$\mu_\pc$}
\end{pspicture}} 
\ee
for some $\mu_{\pa\pb} \in \Lambda(N_{\pa\pb},N_\pa,N_\pb)$ and $\mu_\pc \in \Lambda(N,N_{\pa\pb},N_\pc)$. By deforming the contour around the holes $\pa$ and $\pb$, the number $N_{\pa\pb}$ of times it is intersected by loop segments can always be increased to be larger or equal to $M$. Moreover, it is easy to see that the action of diagrams on the outside or inside the holes are precisely the same for both vector spaces.\medskip

We then find
\begin{alignat}{2}
\cLad&(N,N_\pa,N_\pb,N_\pc) \simeq \sum_{N_{\pa\pb}\ge M}\cLaa(N,N_{\pa\pb},N_\pc)\big(\cLaa(N_{\pa\pb},N_\pa,N_\pb)\otimes \cLa(N_\pc,N_\pc)\big) 
\nn\\[0.1cm]&
= \sum_{N'_\pc,N'_{\pa\pb}} \sum_{N_{\pa\pb}\ge M}\cLa(N,N'_{\pa\pb}+N'_\pc)\big(\cLa(N'_{\pa\pb},N_{\pa\pb})\otimes \cLa(N'_\pc, N_\pc)\big)\big(\cLaa(N_{\pa\pb},N_\pa,N_\pb)\otimes \cLa(N_\pc,N_\pc)\big)
\nn\\[0.1cm]\label{eq:Lambdad.intermediate}&
= \sum_{N'_\pc,N'_{\pa\pb}} \cLa(N,N'_{\pa\pb}+N'_\pc)\bigg(\sum_{N_{\pa\pb}\ge M}\cLa(N'_{\pa\pb},N_{\pa\pb})\cLaa(N_{\pa\pb},N_\pa,N_\pb)\otimes \cLa(N'_\pc, N_\pc)\bigg)
\end{alignat}
where we used \cref{prop:LL=L}. For $N_{\pa\pb} \ge N'_{\pa\pb}$, we have
\begin{alignat}{2}
\cLa(N'_{\pa\pb},N_{\pa\pb})\cLaa(N_{\pa\pb},N_\pa,N_\pb) &= 
\sum_{N'_\pa,N'_\pb}\cLa(N'_{\pa\pb},N_{\pa\pb})\cLa(N_{\pa\pb},N'_\pa+N'_\pb)\big(\cLa(N'_\pa,N_\pa)\otimes \cLa(N'_\pb,N_\pb)\big)
\nn\\[0.15cm]&
=\sum_{N'_\pa,N'_\pb}\cLa(N'_{\pa\pb},N'_\pa+N'_\pb)\big(\cLa(N'_\pa,N_\pa)\otimes \cLa(N'_\pb,N_\pb)\big)
\nn\\[0.15cm]\label{eq:LLa=La}
&
= \cLaa(N'_{\pa\pb},N_\pa,N_\pb)\,,
\end{alignat}
which is independent of $N_{\pa\pb}$. Thus, setting $M = N'_{\pa\pb}$ in \eqref{eq:Lambdad.intermediate}, we find that all the terms in the sum over $N_{\pa\pb}$ are identical vector spaces, and hence that this sum can be removed. We note that \eqref{eq:LLa=La} is a generalisation of \cref{prop:LL=L} to diagrams with two holes. 
We then find
\begin{alignat}{2}
\cLad&(N,N_\pa,N_\pb,N_\pc) \simeq
\sum_{N'_\pc,N'_{\pa\pb}} \cLa(N,N'_{\pa\pb}+N'_\pc)\big(\cLaa(N'_{\pa\pb},N_\pa,N_\pb)\otimes \cLa(N'_\pc, N_\pc)\big)
\nn\\[0.1cm]&
=\sum_{N'_\pa, N'_\pb, N'_\pc,N'_{\pa\pb}} \cLa(N,N'_{\pa\pb}+N'_\pc)\Big(\cLa(N'_{\pa\pb},N'_{\pa}+N'_{\pb})\big(\cLa(N'_\pa, N_\pa)\otimes \cLa(N'_\pb, N_\pb)\big)\otimes \cLa(N'_\pc, N_\pc)\Big)
\nn\\[0.1cm]&
= \big((\repL_{N_\pa}\timesf \repL_{N_\pb}) \timesf \repL_{N_\pc}\big)(N) 
\simeq (\repL_{N_\pa}\timesf \repL_{N_\pb} \timesf \repL_{N_\pc})(N) 
= \cLaa(N,N_\pa,N_\pb,N_\pc)\,.
\end{alignat}
Here, we used \eqref{eq:first.fam} and \cref{prop:LxLxLequiv} at the third and fourth step, respectively. We have thus proved that $\cLad(N,N_\pa,N_\pb,N_\pc)$ and $\cLaa(N,N_\pa,N_\pb,N_\pc)$ are isomorphic.
\eproof

%
\section{Conclusion}\label{sec:conclusion}
%

In this paper, we first illustrated the notion of families of modules on a few physical examples, and subsequently constructed a new fusion bifunctor, which takes as input a pair of families of modules over $\eptl_N(\beta)$ and outputs a new family of modules. We established some important properties of this fusion product, including distributivity, commutativity, and associativity. All of the constructions and results hold for $\beta \in \Cbb$.\medskip

A possible extension of these results would be to consider diagrams with more than three holes, and to generalise the result of \cref{sec:associativity} to the corresponding multiply fused modules. Let $\repM_1,\repM_2,\dots,\repM_m$ be families of modules, with $m \ge 2$. We conjecture that the repeated two-fold fusion of these families $((((\repM_1 \otimes \repM_2) \otimes \repM_3) \otimes \repM_4) \dots \otimes \repM_m)$ yields a module isomorphic to the following generalisation of \eqref{eq:triple.fusion}$-$\eqref{eq:3hole.fusion}:
\begin{equation}
(\repM_1 \timesf \repM_2 \times \dots \timesf \repM_m)(N)=\sum_{N_1,N_2,\dots,N_m} \cL(N,\widehat{N}_m)
\cdot \big(\repM_1(N_1) \otimes \repM_2(N_2) \otimes \dots \otimes \repM_m(N_m) \big) \,,
\end{equation}
subject to the relations
\begin{subequations}
\begin{alignat}{1}
& c_{j+\widehat{N}_{i-1}} \cdot (u_1 \otimes u_2 \otimes \dots \otimes u_m)
= u_1 \otimes \dots \otimes u_{i-1} \otimes c_j u_i \otimes u_{i+1}\otimes \dots \otimes u_m
\qquad 1\leq j \leq N_i-1 \,, \\
& c_{j+\widehat{N}_{i-1}}^\dag \cdot (u_1 \otimes u_2 \otimes \dots \otimes u_m)
= u_1 \otimes \dots \otimes u_{i-1} \otimes c_j^\dag u_i \otimes u_{i+1} \otimes \dots \otimes u_m
\qquad 1\leq j \leq N_i+1 \,,
\end{alignat}
for $1 \le i \le m$, as well as
\be
\Omega \cdot (u_1 \otimes u_2 \otimes \dots \otimes u_m)
= c_{\widehat{N}_{m-1}} \cdots\, 
c_{\widehat{N}_2}\,c_{\widehat{N}_1}
\cdot (\Omega\, u_1 \otimes c_0^\dag u_2 \otimes c_0^\dag u_3 \otimes \dots \otimes c_0^\dag u_m)
\qquad \text{for } N_1>0 \,,
\ee
\end{subequations}
where $\widehat N_{i}=\sum_{k=1}^i N_k$.\medskip

Computing explicity the fusion of the families associated to the physically interesting models requires substantially more work, and is beyond the scope of the paper. For instance, one can show that, if both $\repM_\pa$ and $\repM_\pb$ are families of XXZ modules, the resulting fused family $\repM_\pa \timesf\repM_\pb$ consists of more complicated modules that are not of XXZ type, with two species of spins associated to the two original chains, and are infinite dimensional for all $N$. As explained in \cite{RP07,NS07,GV13,GS16,GJS18,BSA18}, the general purpose of the fusion of (periodic or ordinary) Temperley--Lieb modules is to understand the structure of correlation functions of lattice models defined with nonlocal geometric observables. One relevant aspect pertains to the description of the space of states in the transfer matrix formulation and to the insertion of connectivity operators. The fusion product for the Temperley--Lieb algebras is then a lattice analog of the Operator Product Expansion (OPE) in conformal field theory. In further work in preparation, we will investigate specific applications of the fusion functor defined in this paper, namely the fusion of the families $\repW_{k,x}$ and their quotients, and describe the impact on the correlation functions of lattice connectivity operators.

\subsection*{Acknowledgments}

The authors thank Th\'eo Pinet for valuable discussions in early stages of this project. They also thank Alexis Langlois-R\'emillard, Th\'eo Pinet and David Ridout for their comments on the manuscript.
\bigskip

\appendix

%
\section{Equivalences of algebraic and diagrammatic definitions}
%

\subsection[Equivalent definitions for $\cL(N,N')$]{Equivalent definitions for $\boldsymbol{\cL(N,N')}$}
\label{app:equiv.L}

In this appendix, we prove \cref{prop:equivalence.L} about the equivalent definitions of $\cL(N,N')$. To proceed, we distinguish between its algebraic and diagrammatic definitions given in \cref{sec:definitions-EPTL}, and denote them by $\cLa(N,N')$ and $\cLd(N,N')$.\medskip

To show that the two definitions are equivalent, we construct bases for these two spaces. For $\cLd(N,N')$, we use the fact that, as a vector space, it decomposes as
\begin{alignat}{1}
\cLd(N,N') &= \bigoplus_{k = N_0/2}^{\min(N,N')/2} \cLkd{k}(N,N')\nn\\
&= \bigoplus_{k = N_0/2}^{\min(N,N')/2} \cLkd{k,0}(N,2k)\, \cLkd{k}(2k,2k)\, \cLkd{k,0}(2k,N')\,, \qquad
N_0 = \left\{\begin{array}{ll}
0 & N \textrm{ even,}\\[0.1cm]
1 & N \textrm{ odd,}
\end{array}\right.
\end{alignat}
where $\cLkd{k}(N,N')$ and $\cLkd{k,0}(N,N')$ are the subspaces of $\cLd(N,N')$ defined 
at the end of \cref{sec:definitions-EPTL}. This allows us to construct a basis for $\cLd(N,N')$. Indeed, let us first note that, for $k>0$, $\cLkd{k}(2k,2k)$ is spanned by the identity $\id_{2k}$ and powers of $\Omega_{2k}$ and $\Omega_{2k}^{-1}$. For $k=0$, it is instead spanned by $\id$ and positive powers of~$f$, which we denote here as $\Omega_0$ for simplicity. Second, the elements of $\cLkd{k,0}(2k,N)$ consist of $k$ bridges that do not intersect the dashed line $\pc\pc'$, and $\frac N2-k$ arches connecting nodes on the inner boundary. We denote such an arch by the pair $(i,j)$ of nodes that it connects, with $1 \le i<j \le N$ if the arch does not intersect the dashed line $\pc\pc'$, and $1\le j<i \le N$ otherwise. It is easy to see that a diagram in $\cLkd{k,0}(2k,N)$ is uniquely specified by the tuple of integers $\boldsymbol i = (i_1, i_2, \dots, i_{N/2-k})$ of its arches $(i_1, j_1)$, $(i_2, j_2)$, \dots, $(i_{N/2-k}, j_{N/2-k})$, with $1 \le i_1 < i_2< \dots < i_{N/2-k} \le N$. Indeed, for each such tuple, there is a unique choice for the nodes $j_1,j_2,\dots, j_{N/2-k}$. Let us denote the corresponding diagram as $\lambda_{\boldsymbol i}$, and the set of possible tuples as
\begin{equation} \label{eq:Ik(N)}
\mathcal I_k(N) = 
\big\{(i_1, i_2, \dots, i_{\frac N2-k}) \ \big| \ 1 \le i_1 < i_2< \dots < i_{\frac N2-k} \le N\big\}\,.
\end{equation}
There is in fact a simple bijection between the set of diagrams spanning $\cLkd{k,0}(N,2k)$ and the link states spanning $\repW_{k,x}(N)$. This bijection uses the same ideas as in \eqref{eq:Wk.vs.L(N,2k)}. The two vector spaces are finite-dimensional, with $\dim\cLkd{k,0}(N,2k)=\dim \repW_{k,x}(N) = \binom{N}{N/2-k}$.\medskip

Similarly, $\cLkd{k,0}(N,2k)$ has a basis made of the diagrams $\lambda^\dag_{\boldsymbol i}$ with $\boldsymbol i \in \mathcal I_k(N)$. Then a basis of $\cLd(N,N')$ is given by
\begin{equation}
\repB\textrm{d}(N,N') = \bigcup_{k=N_0/2}^{\min(N,N')/2} \repBd_k(N,N') \,,
\end{equation}
where
\begin{equation}
\repBd_k(N,N') = \bigg\{ 
\lambda^\dag_{\boldsymbol i} \,\Omega^\ell_{2k}\, \lambda_{\boldsymbol {i'}}\ \Big|
\ \ell \in \Zbb_k\,,\,
\begin{array}{l}
\boldsymbol i \in \mathcal I_k(N)
\\[0.15cm]
\boldsymbol {i'} \in \mathcal I_k(N')
\end{array}
\bigg\}\,,
\qquad
\Zbb_k = \left\{\begin{array}{ll}
\Zbb_{\ge 0} & k = 0\,,\\[0.1cm]
\Zbb & k>0\,. \end{array}\right.
\end{equation}

We now proceed to construct a basis for $\cLa(N,N')$. For this, let us introduce the notations 
\be
\boldsymbol j = (j_1,j_2, \dots, j_n) \in \mathcal J_n(N)\,, \qquad
w^{\tinyx s}_{\boldsymbol j} = (c_0)^s \, c_{j_1} \, c_{j_2} \cdots c_{j_n}\,,
\ee 
where
\be
\mathcal J_{n}(N) = \big\{(j_1, j_2, \dots, j_n)\ \big| \ 1 \le j_1<j_2<\dots <j_n \le N-1\,,\, 1 \le j_m \le N -2n+2m-1\big\}\,.
\ee
We also define the subset of words of $\cLa(2k,N)$
\be
\repS(2k,N) = \bigcup_{n=0}^{N/2-k}\big\{
w^{\tinyx {N/2\!-\!k\!-\!n}}_{\boldsymbol j}
 \ \big| \ \boldsymbol j \in \mathcal J_{n}(N)\big\}\,, \qquad 0 \le k<\tfrac N2\,,
\ee
and $\repS(N,N)=\{\id\}$. Before constructing a basis for $\cLa(N,N')$, we establish two preparatory lemmas.
\begin{Lemma}
\label{lem:word.in.c}
Any word in $\cLa(2k,N)$ that contains no operators $c^\dag_j$ is equal to a unique element of $\repS(2k,N)$.
\end{Lemma}
\proof
Let $w \in \cLa(2k,N)$ be a word without operators $c^\dag_j$, of length $L = \frac N2-k$. The proof is inductive for increasing values of $L$. For $L = 0$ and $L=1$, the proof is trivial. For $L \ge 2$, we can write $w = w' c_j$ for some index $j$ satisfying $0 \le j \le N-1$, and some word $w' \in \repS(2k,N-2)$ of length $L-1$. By the inductive assumption, $w'$ can be written as $w' = w^{\tinyx {N/2\!-\!k\!-\!n\!-\!1}}_{\boldsymbol j}$ for some $n\in\{0,1,\dots, \frac N2-k-1\}$ and some $\boldsymbol j \in \mathcal J_n(N-2)$. In the case $n=0$, the word $w=(c_0)^{L-1}c_j$ is an element of $\repS(2k,N)$. Next, we discuss the case $n \geq 1$ with $j \geq 1$. If $j>j_n$, then $w$ is already in the desired form. If $j \le j_n$, then from~\eqref{eq:c.relations.a}, we have $c_{j_n} c_j = c_{j} c_{j_n+2}$. If $j > j_{n-1}$, the resulting expression for $w$ is of the desired form. If $j \leq j_{n-1}$, we repeat the process, and write $c_{j_{n-1}} c_j = c_j c_{j_{n-1}+2}$.
This iterative process is repeated and terminates when we find an integer $m\ge 0$ such that $j_m <j \le j_{m+1}$, with the convention $j_0 = 0$ for $m=0$. The resulting expression for $w$ is 
\be
w = \left\{\begin{array}{ll}
(c_0)^{N/2-k-n-1} c_j\, c_{j_1+2}\,c_{j_2+2} \cdots c_{j_n+2} & m=0\,,\\[0.15cm]
(c_0)^{N/2-k-n-1} c_{j_1}\,c_{j_2} \cdots c_{j_m}\, c_j\, c_{j_{m+1}+2}\, c_{j_{m+2}+2}\cdots c_{j_n+2}&m\ge1\,,
\end{array}\right.
\ee
both of which are elements of $\repS(2k,N)$. Finally, in the case $n \geq 1$ with $j=0$, we use \eqref{eq:c.relations.e} and find after a similar iterative argument
\begin{equation}
w = (c_0)^{N/2-k-n} c_{j_1+1}\,c_{j_2+1} \cdots c_{j_n+1} \,,
\end{equation}
which is also in $\repS(2k,N)$.
\eproof

\begin{Lemma}
\label{lem:w'cdagj}
Let $w'$ be a word in $\repS(2k,N)$ and $w=w'\, c^\dag_j$ for some index $j$ satisfying $0 \le j \le N-1$. Then $w$ takes one of the following forms
\begin{enumerate}
\item[$\mathrm{(i)}$] $w$ is proportional to an element of $\repS(2k,N-2)$;
\item[$\mathrm{(ii)}$] $w = c^\dag_{j'}\, w''$ for some $w''\in \repS(2k+2,N-2)$ and some index $j'$ satisfying $0 \le j' \le 2k-1$;
\item[$\mathrm{(iii)}$] $w = c_1 \, c_0^\dagger\, w''$ for some $w'' \in \repS(2k,N-2)$;
\item[$\mathrm{(iv)}$] $w = c_0 \, c_1^\dagger\, w''$ for some $w'' \in \repS(2k,N-2)$.
\end{enumerate}
\end{Lemma}
\proof
Let $w' = w^{\tinyx s}_{\boldsymbol j}$ with $s = \frac N2-k-n$ for some $n\in\{0,1,\dots, \frac N2-k\}$, and $\boldsymbol j \in \mathcal J_n(N)$. To evaluate $w=w'\, c^\dag_j$, we apply \eqref{eq:c.relations.d} or \eqref{eq:c.relations.g} several times to commute $c^\dag_j$ to the left. Let us first discuss the case where $1 \le j \le N-1$. In this commutation process, it may happen that, for some integer $m \ge 1$, we reach a point where there are two adjacent letters $c_{j_m}c^\dagger_{j'}$ for some $m$, with $j'\in \{j_m-1,j_m,j_m+1\}$. In this case, the product of these two letters is proportional to the identity and the resulting expression for $w$ proportional to an element in $\repS(2k,N-2)$, as in form~(i). Otherwise, the process terminates when $c^\dag_j$ is commuted across all the other operators, and the resulting expression for $w$ reads
\be
w = (c_0)^s \, c^\dag_{j'} \, c_{j'_1} \, c_{j'_2} \cdots c_{j'_n}
\ee
for some $j'$ satisfying $1\le j' \le 2s+2k-1$ and $(j'_1, j'_2, \dots, j'_n) \in \mathcal J_n(N-2)$. For $s=0$, $w$ is of the form~(ii). For $s\ge 1$, we use \eqref{eq:c.relations.d} to show that
\be
\label{eq:c0scdagj}
(c_0)^s \, c^\dag_{j} = 
\left\{\begin{array}{ll}
(c_0)^{s-2}\,c_{2s+2k-j-2} &1 \le j \le s-1\,,\\[0.15cm]
c_0 \, c^\dag_1 \, (c_0)^{s-1}& j=s\,,\\[0.15cm]
c^\dag_{j-s} \, (c_0)^s& s+1 \le j \le s+2k-1\,,\\[0.15cm]
c_1 \, c^\dag_0 \,(c_0)^{s-1}& j = s+2k\,,\\[0.15cm]
(c_0)^{s-2}\,c_{2s+2k-j} & s+2k+1 \le j \le 2s+2k-1\,. 
\end{array}\right.
\ee
This is easily checked for $s=1$ and then proven inductively on increasing values of $s$. For $1 \le j \le s-1$ and $s+2k+1 \le j \le 2s+2k-1$, the resulting expression for $w$ contains no $c^\dag_j$ operators, so we use \cref{lem:word.in.c} to rewrite it as a word in $\repS(2k,N-2)$. For the other three cases, the resulting word is of the form (ii), (iii) or (iv). This ends the proof for $1 \le j \le N-1$.\medskip

Similar ideas apply for $j=0$. If $j_1 \neq 1$ and $j_n \neq N-1$, we commute $c^\dag_0$ to the left using \eqref{eq:c.relations.g}. For $s=0$, the resulting word is $c^\dag_0 c_{j_1+1} \, c_{j_2+1} \cdots c_{j_n+1}$ and is of the form (ii). For $s \ge 1$, we find that $w = \beta\, (c_0)^{s-1}\, c_{j_1+1} \, c_{j_2+1} \cdots c_{j_n+1}$, which is proportional to a word in $\repS(2k,N-2)$, as in form (i). If $j_n = N-1$, then using \eqref{eq:c.relations.h}, we find $w = (c_0)^{s} c_{j_1}\,c_{j_2} \cdots c_{j_{n-1}} c_0 c_1^\dag$. Using \cref{lem:word.in.c}, this can be rewritten as $w = w'' c_1^\dag$ for some $w'' \in \repS(2k,N)$. The proof that $w$ is of one of the desired form is already given above for $j=1$. Finally, if $j_n \neq N-1$ and $j_1 = 1$, then we find 
\begin{alignat}{2}
w &= (c_0)^s\, c_1 c^\dag_0 \, c_{j_2+1}\,c_{j_3+1} \cdots c_{j_n+1} = (c_0)^{s+1}\, c^\dag_{2s+2k+1} \, c_{j_2+1}\,c_{j_3+1} \cdots c_{j_n+1}
\nn \\[0.1cm]&
= c_1 \, (c_0)^{s-1} \, c_{j_2+1}\,c_{j_3+1} \cdots c_{j_n+1}\,,
\end{alignat}
where we used \eqref{eq:c.relations.h} and \eqref{eq:c0scdagj}. We then use \cref{lem:word.in.c} to reexpress $w$ as an element of $\repS(2k,N-2)$, as in form (i).
\eproof

We note that the forms (iii) and (iv) of the above propositions are related, since $c_1\,c^\dag_0 = (c_0\,c^\dag_1)^{-1}$ for $k>0$, and $c_1\,c^\dag_0 = c_0\,c^\dag_1$ for $k=0$. We now introduce the set
\begin{equation}
\repBa(N,N') = \bigcup_{k=N_0/2}^{\min(N,N')/2} \repBa_k(N,N')
\end{equation}
with
\begin{equation}
\repBa_k(N,N') = \bigcup_{n=0}^{N/2-k}\bigcup_{n'=0}^{N'/2-k}
\bigg\{ (w^{\tinyx{N/2\!-\!k\!-\!n}}_{\boldsymbol j})^\dag \,(c_1 c_0^\dag)^\ell\, w^{\tinyx{N'\!/2\!-\!k\!-\!n'}}_{\boldsymbol {j'}} \ \bigg| \ \ell \in \Zbb_k\,, 
\begin{array}{l}
\boldsymbol j \in \mathcal J_n(N)\\[0.1cm]
\boldsymbol j' \in \mathcal J_{n'}(N')
\end{array}
\bigg\}\,.
\end{equation}

\begin{Proposition}
\label{prop:La.basis}
Any word in $\cLa(N,N')$ is equal or proportional to an element in $\repBa(N,N')$.
\end{Proposition}
\proof
The proof is inductive on the length $L$ of the words. Clearly, all words of lengths $L=0$ and $L=1$ are elements of the above sets. Let us therefore assume that $w \in \cLa(N,N')$ has length $L\ge 2$. It can be decomposed as $w = w' c_j$ or $w = w' c^\dag_j$ for some shorter word $w'$ and some index~$j$. By the induction hypothesis, we know that $w'$ is proportional to an element of $\repBa(N,N'-2)$ or $\repBa(N,N'+2)$, respectively. For $w = w' c_j$, we deduce from \cref{lem:word.in.c} that $w$ is in $\repBa(N,N')$ for all $j$. For $w = w' c^\dag_j$, we write $w' = (w^{\tinyx{s}}_{\boldsymbol j})^\dag \,(c_1 c_0^\dag)^\ell\, w^{\tinyx{s'}}_{\boldsymbol {j'}}$ for $s = \frac N2-k-n$ and $s = \frac{N'}2-k-n'+1$, for some $k,\ell, n,n',\boldsymbol j, \boldsymbol j'$. We now use \cref{lem:w'cdagj}. If $w^{\tinyx{s'}}_{\boldsymbol {j'}} c^\dag_j$ is in the forms (i), (iii) or (iv), then we readily find that $w$ is equal or proportional to an element in $\repBa(N,N')$. If it is of the form (ii), then 
$w = (w^{\tinyx{s}}_{\boldsymbol j})^\dag \,(c_1 c_0^\dag)^\ell c^\dag_{j'} w''$
for some index $j$ and some word $w''$ in $\repS(2k+2,N')$. We then use
\be
c_1\, c^\dag_0\, c^\dag_j = 
\left\{
\begin{array}{ll}
c^\dag_{j-1} & j = 0,1\,,\\[0.1cm]
c^\dag_{j-1}\,c_1\, c_0^\dag & j \ge 2\,,
\end{array}\right.
\ee 
multiple times to find $w = (w^{\tinyx{s}}_{\boldsymbol j})^\dag c^\dag_{j''} \,(c_1 c_0^\dag)^{\ell'} w''$ for some integers $j''$ and $\ell'$. Finally, using \cref{lem:word.in.c} with all operators replaced by their adjoints, we find that $(w^{\tinyx{s}}_{\boldsymbol j})^\dag c^\dag_{j''}$ can be written as $(w''')^\dag$ for some $w''' \in \repS(2k+2,N)$. This ends the proof that all words of length $L$ are equal or proportional to elements in the sets $\repBa(N,N')$.
\eproof
\medskip

\begin{Proposition}\label{prop:L.equiv.defs}
We have the isomorphism of diagram spaces
\begin{equation} 
\cLa(N,N') \simeq \cLd(N,N') \,.
\end{equation} 
Moreover, $\repBa(N,N')$ is a basis of $\cLa(N,N')$.
\end{Proposition}
\proof
\cref{prop:La.basis} showed that $\repBa(N,N')$ is a generating set for $\cLa(N,N')$. However, it could be that some of its elements are linearly dependent, which would imply that $\repBa(N,N')$ is not a basis. To show that $\repBa(N,N')$ is indeed a basis and that $\cLa(N,N')$ and $\cLd(N,N')$ are isomorphic, we turn to \eqref{eq:def.cj} and view it as a map from $\cLa(N,N')$ to $\cLd(N,N')$. This map preserves the relations~\eqref{eq:c.relations}, so it is a homomorphism of diagram spaces. We must prove that it is also an isomorphism, by showing that the map is one-to-one from $\repBd(N,N')$ to $\repBa(N,N')$.
\medskip

An element of $\cLkd{k,0}(2k,N)$ is specified by a tuple $\boldsymbol i = (i_1,i_2, \dots, i_{N/2-k})$. Let $n$ be the number of its arcs that do not cross the dashed line, and $s = \frac N2-k-n$ be the number of arcs that do cross the dashed line. Let also $\boldsymbol j = (j_1,j_2, \dots, j_n)$ be the sub-tuple of $\boldsymbol i$ of the starting positions of the arcs that do not cross the dashed line. With this data, it is easy to see that \eqref{eq:def.cj} maps $w^{\tinyx s}_{\boldsymbol j}$ to $\lambda_{\boldsymbol i}$. It is also straightforward to construct the unique pre-image of $\lambda_{\boldsymbol i}$ and show that it is indeed $w^{\tinyx s}_{\boldsymbol j}$. In other words, the map is one-to-one from $\repS(2k,N)$ to $\cLkd{k,0}(2k,N)$.\medskip 

An element of $\repBd(N,N')$ is specified by two numbers $k$ and $\ell$ and two tuples $\boldsymbol i \in \mathcal I_k(N)$ and $\boldsymbol {i'} \in \mathcal I_k(N')$. We denote the corresponding sub-tuples as $\boldsymbol j \in \mathcal J_n(N)$ and $\boldsymbol {j'} \in \mathcal J_{n'}(N')$. Because $(c_1c_0^\dag)^\ell \mapsto \Omega_{2k}^\ell$ under the map, we find the one-to-one correspondence
\begin{equation}
(w^{\tinyx{N/2\!-\!k\!-\!n}}_{\boldsymbol j})^\dag \,(c_1 c_0^\dag)^\ell\, w^{\tinyx{N'\!/2\!-\!k\!-\!n'}}_{\boldsymbol{j'}}
\quad \mapsto \quad \lambda^\dag_{\boldsymbol i} \,\Omega^\ell_{2k}\, \lambda_{\boldsymbol{i'}}\,.
\end{equation}
This is a bijective map from the generating set $\repBa(N,N')$ of $\cLa(N,N')$, to the basis $\repBd(N,N')$ of $\cLd(N,N')$. Because the elements of $\repBd(N,N')$ are independent, this must also be true for the elements of $\repBa(N,N')$, thus confirming that the latter is a basis $\cLa(N,N')$. This also proves that the homomorphism \eqref{eq:def.cj} is an isomorphism, ending the proof.
\eproof

\subsection[Equivalent definitions for $\Lambda(N,N_\pa, N_\pb)$]{Equivalent definitions for $\boldsymbol{\Lambda(N,N_\pa, N_\pb)}$}
\label{app:equiv.Lambda}

In this appendix, we prove \cref{prop:Lambda.defs} about the equivalent definitions of $\Lambda(N,N_\pa, N_\pb)$. We consider separately its algebraic and diagrammatic definitions, denoted by $\cLaa(N,N_\pa, N_\pb)$ and $\cLad(N,N_\pa, N_\pb)$ respectively, and subsequently prove that they are equivalent. For this appendix, we refer to the outer boundary of the disc with two holes as $\pc$, and write $N=N_\pc$. Then $\cLaa(N_\pc,N_\pa, N_\pb)$ is defined as
\be
\cLaa(N_\pc,N_\pa,N_\pb) =
\sum_{N'_\pa,N'_\pb} \cLa(N_\pc,N'_{\pc})
\big(\cLa(N'_\pa,N_\pa) \otimes \cLa(N'_\pb,N_\pb) \big) \,, \qquad N'_{\pc} = N'_{\pa\pb}\,,
\ee
subject to the relations \eqref{eq:Lambda.equations} with $\Omega\to c_1 c_0^\dag$ and $\Omega^{-1}\to c_0 c_1^\dag$, and the relations \eqref{eq:c.relations} for $\cLa(N_\pc,N'_{\pc})$, $\cLa(N'_\pa,N_\pa)$ and $\cLa(N'_\pb,N_\pb)$.\medskip

In contrast, $\cLad(N_\pc,N_\pa, N_\pb)$ is the vector space spanned by diagrams on a disc with two holes with $N_\pa$, $N_\pb$ and $N_\pc$ nodes on $\pa$, $\pb$ and $\pc$ respectively, as described in the beginning of \cref{sec:diagrams.2holes}. Let $\mu$ be a diagram in $\cLad(N_\pc,N_\pa, N_\pb)$. We define a subdomain including $\pa$ but not $\pb$, whose boundary intersects only $2k_\pa$ bridges connecting $\pa$ to $\pb$, $\pc$, or to itself after encircling $\pb$, but no other loop segments. Similarly, we define a subdomain including $\pb$ but not $\pa$, whose boundary intersects only the $2k_\pb$ bridges of $\pb$. We also define a third larger subdomain containing the subdomains around $\pa$ and $\pb$, whose boundary intersects only the $2k_\pc$ bridges of $\pc$, connecting it to $\pa$, $\pb$, or to itself by through-lines passing between $\pa$ and $\pb$. These numbers $k_\pa$, $k_\pb$ and $k_\pc$ satisfy 
\be
\label{eq:ki.constraints}
0 \le k_\ppi \le \tfrac{N_\ppi}2\,, \qquad 
\tfrac{N_\ppi}2-k_\ppi \in \Zbb\,,
\qquad \ppi \in \{\pa,\pb,\pc\}.
\ee
There is a freedom in choosing the path followed by the dashed lines $\pa\pc$ and $\pa\pb$ and how they intersect the boundaries of the subdomains. Here we make use of this freedom, so that the region between the larger subdomain and the two smaller ones takes the form of the diagram $\mu_{k_\pc, k_\pb, k_\pc} \in \Lambda(2k_\pc, 2k_\pa, 2k_\pb)$ defined as
\be
\mu_{k_\pc,k_\pa,k_\pb} = 
\left\{\begin{array}{llll}
\psset{unit=0.4cm}
\thispic{\begin{pspicture}[shift=-3.8](-3.5,-3.9)(3.5,3.9)
\psarc[linecolor=black,linewidth=0.5pt,fillstyle=solid,fillcolor=lightlightblue]{-}(0,0){3.5}{0}{360}
\psline[linestyle=dashed, dash= 1.5pt 1.5pt,linewidth=0.5pt]{-}(1.46, 0.65)(0,-3.5)
\psline[linestyle=dashed, dash= 1.5pt 1.5pt,linewidth=0.5pt]{-}(-1.46,-0.65)(0,-3.5)
\rput{24}(0,0){
\psbezier[linecolor=blue,linewidth=1.5pt]{-}(1.86, -0.43)(1.94, -0.55)(2.42, -1.4)(3.03, -1.75)
\psbezier[linecolor=blue,linewidth=1.5pt]{-}(2.09, -0.12)(2.23, -0.15)(2.78, -0.29)(3.48, -0.37)
\psbezier[linecolor=blue,linewidth=1.5pt]{-}(2.03, 0.26)(2.15, 0.34)(2.66, 0.87)(3.33, 1.08)
\psbezier[linecolor=blue,linewidth=1.5pt]{-}(1.72, 0.49)(1.75, 0.63)(2.08, 1.87)(2.6, 2.34)
\psbezier[linecolor=blue,linewidth=1.5pt]{-}(-0.73, -3.42)(-0.5, -0.68)(-0.53, 0.64)(-1.42, 3.2)
\psbezier[linecolor=blue,linewidth=1.5pt]{-}(0.73, -3.42)(0., -0.68)(-0.15, 0.7)(0., 3.5)
\psbezier[linecolor=blue,linewidth=1.5pt]{-}(2.06, -2.83)(0.41, -0.57)(0.23, 0.64)(1.37, 3.2)
\psbezier[linecolor=blue,linewidth=1.5pt]{-}(-1.69, 0.49)(-1.67, 0.39)(-2.08, 1.87)(-2.6, 2.34)
\psbezier[linecolor=blue,linewidth=1.5pt]{-}(-1.98, 0.32)(-1.91, 0.26)(-2.66, 0.87)(-3.33, 1.08)
\psbezier[linecolor=blue,linewidth=1.5pt]{-}(-2.1, 0.)(-2., 0.)(-2.78, -0.29)(-3.48, -0.37)
\psbezier[linecolor=blue,linewidth=1.5pt]{-}(-1.98, -0.32)(-1.91, -0.26)(-2.42, -1.4)(-3.03, -1.75)
\psbezier[linecolor=blue,linewidth=1.5pt]{-}(-1.69, -0.49)(-1.67, -0.39)(-1.65, -2.27)(-2.06, -2.83)
\psarc[linecolor=purple,linewidth=1.5pt,fillstyle=solid,fillcolor=white]{-}(-1.6,0){0.5}{0}{360}
\psarc[linecolor=darkgreen,linewidth=1.5pt,fillstyle=solid,fillcolor=white]{-}(1.6,0){0.5}{0}{360}
}
\end{pspicture}}
& \quad k_\pc>k_\pa + k_\pb \,,
\qquad &
\psset{unit=0.4cm}
\thispic{\begin{pspicture}[shift=-3.8](-3.5,-3.9)(3.5,3.9)
\psarc[linecolor=black,linewidth=0.5pt,fillstyle=solid,fillcolor=lightlightblue]{-}(0,0){3.5}{0}{360}
\psbezier[linestyle=dashed, dash= 1.5pt 1.5pt,linewidth=0.5pt]{-}(-1.6,0)(-0.5,1.6)(0,-1)(0,-3.5)
\psbezier[linestyle=dashed, dash= 1.5pt 1.5pt,linewidth=0.5pt]{-}(1.6,0)(0.5,1.6)(0,-1)(0,-3.5)
\psbezier[linecolor=blue,linewidth=1.5pt]{-}(1.39, -0.45)(1.33, -0.59)(0.58, -2.74)(0.73, -3.42)
\psbezier[linecolor=blue,linewidth=1.5pt]{-}(1.67, -0.49)(1.69, -0.64)(1.65, -2.27)(2.06, -2.83)
\psbezier[linecolor=blue,linewidth=1.5pt]{-}(1.93, -0.38)(2.03, -0.49)(2.42, -1.4)(3.03, -1.75)
\psbezier[linecolor=blue,linewidth=1.5pt]{-}(2.08, -0.14)(2.22, -0.18)(2.78, -0.29)(3.48, -0.37)
\psbezier[linecolor=blue,linewidth=1.5pt]{-}(2.08, 0.14)(2.22, 0.18)(2.66, 0.87)(3.33, 1.08)
\psbezier[linecolor=blue,linewidth=1.5pt]{-}(1.93, 0.38)(2.03, 0.49)(2.08, 1.87)(2.6, 2.34)
\psbezier[linecolor=blue,linewidth=1.5pt]{-}(1.67, 0.49)(1.69, 0.64)(1.14, 2.56)(1.42, 3.2)
\psbezier[linecolor=blue,linewidth=1.5pt]{-}(1.39, 0.45)(1.33, 0.59)(0., 2.8)(0., 3.5)
\psline[linecolor=blue,linewidth=1.5pt]{-}(1.1, 0)(-1.1, 0)
\psbezier[linecolor=blue,linewidth=1.5pt]{-}(1.18, 0.27)(1.05, 0.35)(-1.07, 0.38)(-1.2, 0.29)
\psbezier[linecolor=blue,linewidth=1.5pt]{-}(1.18, -0.27)(1.05, -0.35)(-1.07, -0.38)(-1.2, -0.29)
\psbezier[linecolor=blue,linewidth=1.5pt]{-}(-1.45, 0.48)(-1.4, 0.62)(-1.14, 2.56)(-1.42, 3.2)
\psbezier[linecolor=blue,linewidth=1.5pt]{-}(-1.75, 0.48)(-1.8, 0.62)(-2.08, 1.87)(-2.6, 2.34)
\psbezier[linecolor=blue,linewidth=1.5pt]{-}(-2., 0.29)(-2.13, 0.38)(-2.66, 0.87)(-3.33, 1.08)
\psbezier[linecolor=blue,linewidth=1.5pt]{-}(-2.1, 0.)(-2.25, 0.)(-2.78, -0.29)(-3.48, -0.37)
\psbezier[linecolor=blue,linewidth=1.5pt]{-}(-2., -0.29)(-2.13, -0.38)(-2.42, -1.4)(-3.03, -1.75)
\psbezier[linecolor=blue,linewidth=1.5pt]{-}(-1.75, -0.48)(-1.8, -0.62)(-1.65, -2.27)(-2.06, -2.83)
\psbezier[linecolor=blue,linewidth=1.5pt]{-}(-1.45, -0.48)(-1.4, -0.62)(-0.58, -2.74)(-0.73, -3.42)
\psarc[linecolor=purple,linewidth=1.5pt,fillstyle=solid,fillcolor=white]{-}(-1.6,0){0.5}{0}{360}
\psarc[linecolor=darkgreen,linewidth=1.5pt,fillstyle=solid,fillcolor=white]{-}(1.6,0){0.5}{0}{360}
\end{pspicture}}
& \quad |k_\pa-k_\pb|\leq k_\pc \leq k_\pa+k_\pb \,,
\\
\psset{unit=0.4cm}
\thispic{\begin{pspicture}[shift=-3.8](-3.5,-3.9)(3.5,3.9)
\psarc[linecolor=black,linewidth=0.5pt,fillstyle=solid,fillcolor=lightlightblue]{-}(0,0){3.5}{0}{360}
\psbezier[linecolor=blue,linewidth=1.5pt]{-}(1.78, -0.47)(1.83, -0.61)(0.72, -2.7)(0.91, -3.38)
\psbezier[linecolor=blue,linewidth=1.5pt]{-}(1.97, -0.33)(2.09, -0.43)(1.98, -1.98)(2.47, -2.47)
\psbezier[linecolor=blue,linewidth=1.5pt]{-}(2.09, -0.12)(2.23, -0.16)(2.7, -0.72)(3.38, -0.91)
\psbezier[linecolor=blue,linewidth=1.5pt]{-}(2.09, 0.12)(2.23, 0.16)(2.7, 0.72)(3.38, 0.91)
\psbezier[linecolor=blue,linewidth=1.5pt]{-}(1.97, 0.33)(2.09, 0.43)(1.98, 1.98)(2.47, 2.47)
\psbezier[linecolor=blue,linewidth=1.5pt]{-}(1.78, 0.47)(1.83, 0.61)(0.72, 2.7)(0.91, 3.38)
\psline[linecolor=blue,linewidth=1.5pt]{-}(1.1, 0.)(-1.1, 0.)
\psbezier[linecolor=blue,linewidth=1.5pt]{-}(1.16, -0.23)(1.02, -0.3)(-1.02, -0.3)(-1.16, -0.23)
\psbezier[linecolor=blue,linewidth=1.5pt]{-}(1.16, 0.23)(1.02, 0.3)(-1.02, 0.3)(-1.16, 0.23)
\psbezier[linecolor=blue,linewidth=1.5pt]{-}(1.32, -0.41)(1.17, -0.62)(-2.7,-1.4)(-2.7,0)
\psbezier[linecolor=blue,linewidth=1.5pt]{-}(1.32, 0.41)(1.17, 0.62)(-2.7,1.4)(-2.7,0)
\psbezier[linecolor=blue,linewidth=1.5pt]{-}(1.54, -0.5)(1.51, -0.74)(-3.0,-1.9)(-3.0,0)
\psbezier[linecolor=blue,linewidth=1.5pt]{-}(1.54, 0.5)(1.51, 0.74)(-3.0,1.9)(-3.0,0)
\psbezier[linestyle=dashed, dash= 1.5pt 1.5pt,linewidth=0.5pt]{-}(-1.6,0)(-0.5,1.6)(0,-1)(0,-3.5)
\psbezier[linestyle=dashed, dash= 1.5pt 1.5pt,linewidth=0.5pt]{-}(1.4,0)(0.5,1.5)(0,-1)(0,-3.5)
\psarc[linecolor=purple,linewidth=1.5pt,fillstyle=solid,fillcolor=white]{-}(-1.6,0){0.5}{0}{360}
\psarc[linecolor=darkgreen,linewidth=1.5pt,fillstyle=solid,fillcolor=white]{-}(1.6,0){0.5}{0}{360}
\end{pspicture}}
& \quad k_\pc < k_\pa-k_\pb \,,
&
\psset{unit=0.4cm}
\thispic{\begin{pspicture}[shift=-3.8](-3.5,-3.9)(3.5,3.9)
\psarc[linecolor=black,linewidth=0.5pt,fillstyle=solid,fillcolor=lightlightblue]{-}(0,0){3.5}{0}{360}
\psbezier[linecolor=blue,linewidth=1.5pt]{-}(-1.78, -0.47)(-1.83, -0.61)(-0.72, -2.7)(-0.91, -3.38)
\psbezier[linecolor=blue,linewidth=1.5pt]{-}(-1.97, -0.33)(-2.09, -0.43)(-1.98, -1.98)(-2.47, -2.47)
\psbezier[linecolor=blue,linewidth=1.5pt]{-}(-2.09, -0.12)(-2.23, -0.16)(-2.7, -0.72)(-3.38, -0.91)
\psbezier[linecolor=blue,linewidth=1.5pt]{-}(-2.09, 0.12)(-2.23, 0.16)(-2.7, 0.72)(-3.38, 0.91)
\psbezier[linecolor=blue,linewidth=1.5pt]{-}(-1.97, 0.33)(-2.09, 0.43)(-1.98, 1.98)(-2.47, 2.47)
\psbezier[linecolor=blue,linewidth=1.5pt]{-}(-1.78, 0.47)(-1.83, 0.61)(-0.72, 2.7)(-0.91, 3.38)
\psline[linecolor=blue,linewidth=1.5pt]{-}(1.1, 0.)(-1.1, 0.)
\psbezier[linecolor=blue,linewidth=1.5pt]{-}(1.16, -0.23)(1.02, -0.3)(-1.02, -0.3)(-1.16, -0.23)
\psbezier[linecolor=blue,linewidth=1.5pt]{-}(1.16, 0.23)(1.02, 0.3)(-1.02, 0.3)(-1.16, 0.23)
\psbezier[linecolor=blue,linewidth=1.5pt]{-}(-1.32, -0.41)(-1.17, -0.62)(2.7,-1.4)(2.7,0)
\psbezier[linecolor=blue,linewidth=1.5pt]{-}(-1.32, 0.41)(-1.17, 0.62)(2.7,1.4)(2.7,0)
\psbezier[linecolor=blue,linewidth=1.5pt]{-}(-1.54, -0.5)(-1.51, -0.74)(3.0,-1.9)(3.0,0)
\psbezier[linecolor=blue,linewidth=1.5pt]{-}(-1.54, 0.5)(-1.51, 0.74)(3.0,1.9)(3.0,0)
\psbezier[linestyle=dashed, dash= 1.5pt 1.5pt,linewidth=0.5pt]{-}(1.6,0)(0.5,1.6)(0,-1)(0,-3.5)
\psbezier[linestyle=dashed, dash= 1.5pt 1.5pt,linewidth=0.5pt]{-}(-1.4,0)(-0.5,1.5)(0,-1)(0,-3.5)
\psarc[linecolor=purple,linewidth=1.5pt,fillstyle=solid,fillcolor=white]{-}(-1.6,0){0.5}{0}{360}
\psarc[linecolor=darkgreen,linewidth=1.5pt,fillstyle=solid,fillcolor=white]{-}(1.6,0){0.5}{0}{360}
\end{pspicture}}
& \quad k_\pc < k_\pb - k_\pa \,.
\end{array}\right.
\ee
This construction of subdomains implies that we can decompose $\cLad(N_\pc,N_\pa, N_\pb)$ as
\begin{alignat}{1}
\cLad(N_\pc,N_\pa, N_\pb) = \ \sum_{k_\pa,k_\pb,k_\pc}
\cLkd{k_\pc}(N_\pc,2k_\pc) \, \mu_{k_\pc, k_\pa, k_\pb}
\big(\cLkd{k_\pa}(2k_\pa,N_\pa) \otimes \cLkd{k_\pb}(2k_\pb,N_\pb)\big) \,,
\end{alignat}
where $k_\pa,k_\pb,k_\pc$ run over the values allowed in \eqref{eq:ki.constraints}.
Moreover, we have the identities
\begin{subequations}
\begin{alignat}{2}
\cLkd{k}(N,2k) &= \cLkd{k,0}(N,2k) \, \cLkd{k}(2k,2k) \,,
\\[0.1cm]
\cLkd{k}(2k,N) &= \cLkd{k}(2k,2k) \, \cLkd{k,0}(2k,N) \,, 
\\[0.1cm]
\cLkd{k}(2k,2k) &= \mathrm{span}\big((\Omega_{2k})^n \,, n \in \Zbb_k \big) \,,
\end{alignat}
\end{subequations}
for $N \geq 2k$, where we use the same notations as in \cref{app:equiv.L}, namely
\begin{equation}
\Omega_N = \left\{\begin{array}{cl} 
f & N=0\,, \\[0.1cm] \Omega & N>0\,,
\end{array}\right. \qquad
\Omega^{-1}_N = \left\{\begin{array}{cl} 
f & N=0\,, \\[0.1cm] \Omega^{-1} & N>0\,,
\end{array}\right. \qquad
\Zbb_k = \left\{\begin{array}{cl}
\Zbb_{\geq 0} & k=0\,, \\[0.1cm] \Zbb & k>0\,.
\end{array}\right.
\end{equation}
We thus obtain a basis of $\cLad(N_\pc,N_\pa, N_\pb)$
\begin{subequations}
\be
\repBd(N_\pc,N_\pa, N_\pb) = \bigcup_{k_\pa,k_\pb,k_\pc} \repBd_{k_\pa,k_\pb,k_\pc}(N_\pc,N_\pa, N_\pb)
\ee
\be
\repBd_{k_\pa,k_\pb,k_\pc}(N_\pc,N_\pa, N_\pb) = 
\left\{ 
\lambda^\dag_{\boldsymbol i_\pc} \, \Omega^{-\ell_\pc}_{2k_\pc}\, \mu_{k_\pa, k_\pb, k_\pc}\, (\Omega^{\ell_\pa}_{2k_\pa}\,\lambda_{\boldsymbol i_\pa} \otimes \Omega^{\ell_\pb}_{2k_\pb}\,\lambda_{\boldsymbol i_\pb})\ \Bigg|
\begin{array}{ll}
\ell_\pa \in \Zbb_{k_\pa}\,, & \boldsymbol i_\pa \in \mathcal I_{k_\pa}(N_\pa)\\[0.1cm]
\ell_\pb \in \Zbb_{k_\pb}\,, & \boldsymbol i_\pb \in \mathcal I_{k_\pb}(N_\pb)\\[0.1cm]
\ell_\pc \in \Zbb_{k_\pc}\,, & \boldsymbol i_\pc \in \mathcal I_{k_\pc}(N_\pc)
\end{array}
\right\}
\ee 
\end{subequations}
where
\be
{\boldsymbol i}_\pa = (i_1, i_2, \dots, i_{N_\pa/2-k_\pa})\,,\qquad
{\boldsymbol i}_\pb = (i'_1, i'_2, \dots, i'_{N_\pb/2-k_\pb})\,,\qquad
{\boldsymbol i}_\pc = (i''_1, i''_2, \dots, i''_{N_\pc/2-k_\pc})\,.
\ee
The set $\mathcal{I}_k(N)$ and the diagram $\lambda_{\boldsymbol i} \in \cLd^{\tinyx{k,0}}(N,2k)$ associated to $\boldsymbol{i} \in \mathcal{I}_k(N)$ are defined in \cref{app:equiv.L}.\medskip

We now define the set
\begin{subequations}
\label{eq:canonical.form}
\begin{equation}
\repBa(N_\pc,N_\pa,N_\pc) =
\bigcup_{k_\pa,k_\pb,k_\pc}\repBa_{k_\pc,k_\pa,k_\pb}(N_\pc,N_\pa,N_\pc)
\end{equation}
with
\begin{equation}
\repBa_{k_\pc,k_\pa,k_\pb} = 
\bigcup_{n_\pa,n_\pb,n_\pc}
\left\{ (w^{\tinyx{s_\pc}}_{\boldsymbol j_\pc})^\dag 
(c_0c_1^\dag)^{\ell_\pc}
\big((c_0^\dag)^{k_\pc-k_\pa-k_\pb}
(c_1c_0^\dag)^{\ell_\pa}w^{\tinyx{s_\pa}}_{\boldsymbol j_\pa} \otimes
(c_1c_0^\dag)^{\ell_\pb}w^{\tinyx{s_\pb}}_{\boldsymbol j_\pb}
\big) \ \Bigg| 
\begin{array}{l}
\ell_\pa \in \Zbb_{k_\pa}\,,\, \boldsymbol j_\pa \in \mathcal J_{n_\pa}(N_\pa)\\[0.1cm]
\ell_\pb \in \Zbb_{k_\pb}\,,\, \boldsymbol j_\pb \in \mathcal J_{n_\pb}(N_\pb)\\[0.1cm]
\ell_\pc \in \Zbb_{k_\pc}\,,\, \boldsymbol j_\pc \in \mathcal J_{n_\pc}(N_\pc)
\end{array}
\right\}
\end{equation}
for $k_\pa+k_\pb < k_\pc$, and
\begin{equation}
\repBa_{k_\pc,k_\pa,k_\pb} = 
\bigcup_{n_\pa,n_\pb,n_\pc}
\left\{ (w^{\tinyx{s_\pc}}_{\boldsymbol j_\pc})^\dag 
\,(c_0c_1^\dag)^{\ell_\pc}
(c_0)^{k_\pa+k_\pb-k_\pc}
\big(
(c_1c_0^\dag)^{\ell_\pa} w^{\tinyx{s_\pa}}_{\boldsymbol j_\pa} \otimes
(c_1c_0^\dag)^{\ell_\pb} w^{\tinyx{s_\pb}}_{\boldsymbol j_\pb}
\big) \ \Bigg| 
\begin{array}{l}
\ell_\pa \in \Zbb_{k_\pa}\,,\, \boldsymbol j_\pa \in \mathcal J_{n_\pa}(N_\pa)\\[0.1cm]
\ell_\pb \in \Zbb_{k_\pb}\,,\, \boldsymbol j_\pb \in \mathcal J_{n_\pb}(N_\pb)\\[0.1cm]
\ell_\pc \in \Zbb_{k_\pc}\,,\, \boldsymbol j_\pc \in \mathcal J_{n_\pc}(N_\pc)
\end{array}
\right\}
\end{equation}
\end{subequations}
for $k_\pa+k_\pb \ge k_\pc$, where $k_\ppi$ runs over the values allowed in \eqref{eq:ki.constraints}, $n_\ppi$ runs over $\{0,1,\dots, N_\ppi/2-k_\ppi\}$, $s_\ppi = N_\ppi/2-k_\ppi-n_\ppi$, and $\boldsymbol j_\ppi = (j_1, j_2, \dots, j_{n_\ppi})$.

\begin{Proposition} \label{prop:Lambda.basis}
Any word in $\cLaa(N_\pc,N_\pa,N_\pb)$ is equal or proportional to an element in the set $\repBa(N_\pc,N_\pa,N_\pc)$.
\end{Proposition}
\proof
Let $w = w_\pc(w_\pa \otimes w_\pb)$ be a word in $\cLaa(N_\pc,N_\pa,N_\pb)$ for some $w_\pc \in \cLa(N_\pc,N'_{\pc})$, $w_\pa \in \cLa(N'_\pa,N_\pa)$, $w_\pb \in \cLa(N'_\pb,N_\pb)$, with $N'_\pc = N'_{\pa\pb}$. The length of $w$ is $L=L_\pa+L_\pb+L_\pc$ where $L_\ppi$ is the length of $w_\ppi$. For $L = 0$, the proposition holds trivially. For $L=1$, it is also easy to see that the proposition holds using the relations \eqref{eq:Lambda.equations} and \eqref{eq:Lambda.equations.extra}. For $L \ge 2$, we use an inductive argument. Let us suppose that $L_\pa > 0$. In this case, we can write $w =\kappa\, w'(c_j \otimes \id)$ or $w = \kappa\, w'(c_j^\dag \otimes \id)$ for some scalar constant $\kappa$, for some index~$j$, and some element $w'$ of length strictly smaller than $L$, which by the induction assumption we can write as
\be
w' = 
\left\{\begin{array}{ll}
(w'_\pc)^\dag \big((c_0^\dag)^{k_\pc-k_\pa-k_\pb}w'_\pa \otimes w'_\pb\big) & k_\pa+k_\pb<k_\pc\,,
\\[0.15cm]
(w'_\pc)^\dag (c_0)^{k_\pa+k_\pb-k_\pc}(w'_\pa \otimes w'_\pb) & k_\pa+k_\pb \ge k_\pc\,,
\end{array}\right.
\qquad \textrm{with}\qquad
\left\{\begin{array}{l}
w'_\pa = (c_1 c_0^\dag)^{\ell_\pa}w^{\tinyx{s_\pa}}_{\boldsymbol j_\pa}\,, \\[0.15cm]
w'_\pb = (c_1 c_0^\dag)^{\ell_\pb}w^{\tinyx{s_\pb}}_{\boldsymbol j_\pb}\,, \\[0.15cm]
w'_\pc = (c_1 c_0^\dag)^{\ell_\pc}w^{\tinyx{s_\pc}}_{\boldsymbol j_\pc}\,,
\end{array}\right.
\ee
for some $k_\pa$, $k_\pb$, $k_\pc$, $\ell_\pa$, $\ell_\pb$, $\ell_\pc$, $\boldsymbol j_\pa,\boldsymbol j_\pb, \boldsymbol j_\pc$. For the case $w \simeq \kappa\, w'(c_j \otimes \id)$, we deduce from \cref{lem:word.in.c} that $w'_\pa c_j \in \repS(2k_\pa, N_\pa)$ and that $w$ is in $\repBa_{k_\pc, k_\pa, k_\pb}$. For the case $w = \kappa\, w'(c_j^\dag \otimes \id)$, \cref{prop:L.equiv.defs} tells us that $w'_\pa$ can be equivalently described as an element of $\cLd^{\tinyx{k_\pa}}(2k_\pa,N_\pa+2)$, and similarly for $w'_\pb \in \cLd^{\tinyx{k_\pb}}(2k_\pb,N_\pb)$ and $(w'_\pc)^\dag \in \cLd^{\tinyx{k_\pc}}(N_\pc,2k_\pc)$. With this description, we find that $w'_\pa c^\dag_j$ is either in $\cLd^{\tinyx{k_\pa}}(2k_\pa,N_\pa)$, or equal to $c^\dag_{j'} w''_{\pa}$ for some $w''_{\pa} \in \cLd^{\tinyx{k_\pa}}(2k_\pa-2,N_\pa)$ and some index $0 \le j' \le 2k_{\pa}-1$. In the former case, $w$ is proportional to an element in $\repB_{k_\pc, k_\pa, k_\pb}$. In the latter case, we have
\be
w'(c_j^\dag \otimes \id) =
\left\{\begin{array}{ll}
(w'_\pc)^\dag \big((c_0^\dag)^{k_\pc-k_\pa-k_\pb}c_{j'}^\dag w''_\pa \otimes w'_\pb\big) & k_\pa+k_\pb<k_\pc\,,
\\[0.15cm]
(w'_\pc)^\dag (c_0)^{k_\pa+k_\pb-k_\pc}(c_{j'}^\dag w''_\pa \otimes w'_\pb) & k_\pa+k_\pb \ge k_\pc\,.
\end{array}\right.
\ee
We first discuss the case $k_\pa+k_\pb<k_\pc$. If $j' = 0$, then it is clear that $w'(c_j^\dag \otimes \id)$ is in $\repB_{k_\pc,k_\pa-1,k_\pb}$. If $1 \le j'\le 2k_\pa-1$, then we use \eqref{eq:c.relations.f} and \eqref{eq:Lambda.equations.c} and find
\be
(w'_\pc)^\dag \big((c_0^\dag)^{k_\pc-k_\pa-k_\pb}c_{j'}^\dag w''_\pa \otimes w'_\pb\big) =
(w'_\pc)^\dag c_{j'+k_\pc-k_\pa-k_\pb}^\dag \big((c_0^\dag)^{k_\pc-k_\pa-k_\pb} w''_\pa \otimes w'_\pb\big)\,.
\ee
Using the description of $(w'_\pc)^\dag$ as an element of $\cLd^{\tinyx{k_\pc}}(N_\pc, 2k_\pc)$, we find that $(w'_\pc)^\dag c^\dag_{j'+k_\pc-k_\pa-k_\pb}$ is in $\cLkd{k_\pc\!-\!1}(N_\pc,2k_\pc-2)$. This shows that $w'(c_j^\dag \otimes \id)$ is in $\repBa_{k_\pc-1, k_\pa-1, k_\pb}$ in this case.\medskip

Second we discuss the case $k_\pa+k_\pb\ge k_\pc$. For $j'=0$, we use \eqref{eq:Lambda.equations.g} with $\lambda_\pa \to c^\dag_0 \lambda_{\pa}$ to derive the relation
\be
\lambda \, c_0\,(c_0^\dag \lambda_\pa \otimes \lambda_\pb) = \lambda\,(\lambda_\pa \otimes c_0^\dag c_1 \lambda_\pb)\,.
\ee
This yields
\be
w'(c_j^\dag \otimes \id) = (w'_\pc)^\dag (c_0)^{k_\pa+k_\pb-k_\pc}(c_{0}^\dag w''_\pa \otimes w'_\pb) = (w'_\pc)^\dag (c_0)^{k_\pa+k_\pb-k_\pc-1}(w''_\pa \otimes (c_0^\dag c_1)w'_\pb)
\ee
which is an element of $\repB_{k_\pc,k_\pa-1,k_\pb}$. For $1 \le j' \le 2k_\pa-1$, we use \eqref{eq:Lambda.equations.c}
and find
\be
w'(c_j^\dag \otimes \id) = (w'_\pc)^\dag (c_0)^{k_\pa+k_\pb-k_\pc}c_{j'}^\dag(w''_\pa \otimes w'_\pb)\,.
\ee
We now compute $(c_0)^{k_\pa+k_\pb-k_\pc}c^\dag_{j'}$ using \eqref{eq:c0scdagj}. If the result equals $(c_0 c_1^\dag)^{\pm 1}(c_0)^{k_\pa+k_\pb-k_\pc-1}$, the resulting word is in $\repB_{k_\pc,k_\pa,k_\pb}$ with $\ell_\pc$ shifted by $\pm1$. If the result is $c^\dag_{j'-k_\pa-k_\pb+k_\pc}(c_0)^{k_\pa+k_\pb-k_\pc}$, we know that $(w'_\pc)^\dag c^\dag_{j'-k_\pa-k_\pb+k_\pc} $ is in $\cLd(N_\pc,2k_\pc-2)$, confirming that $w'(c_j^\dag \otimes \id)$ is in $\repB_{k_\pc-1,k_\pa-1,k_\pb}$. The final case is if the result is $(c_0)^{k_\pa+k_\pb-k_\pc-2}c_{j''}$ for some $j''$ satisfying $1 \le j'' \le 2k_\pa+2k_\pb-3$. We then find
\begin{alignat}{2}
w'(c_j^\dag \otimes \id) &= (w'_\pc)^\dag (c_0)^{k_\pa+k_\pb-k_\pc-2} c_{j''} (w''_\pa \otimes w'_\pb) 
\nn\\[0.2cm]&= 
\left\{\begin{array}{ll}
(w'_\pc)^\dag (c_0)^{k_\pa+k_\pb-k_\pc-2} (c_{j''} w''_\pa \otimes w'_\pb) & 1 \le j'' \le 2k_\pa-3\,,
\\[0.15cm]
(w'_\pc)^\dag (c_0)^{k_\pa+k_\pb-k_\pc-1}\big((c^\dag_1 c_0)^{-1}w''_\pa \otimes c_1^\dag c_0\, w'_\pb\big) & j'' = 2k_\pa-2\,,
\\[0.15cm]
(w'_\pc)^\dag (c_0)^{k_\pa+k_\pb-k_\pc-2} (w''_\pa \otimes c_{j''-2k_\pa-2} w'_\pb) & 2k_\pa-1 \le j \le 2k_\pa + 2k_\pb-3\,.
\end{array}\right.
\end{alignat}
In these three cases, the result is in $\repBa_{k_\pc,k_\pa-2,k_\pb}$, $\repBa_{k_\pc,k_\pa-1,k_\pb}$ and $\repBa_{k_\pc,k_\pa-1,k_\pb-1}$, respectively.\medskip

The above arguments are presented in the case where $L_\pa > 0$. If $L_\pa=0$, then either $L_\pb>0$ or $L_\pc>0$, in which cases we can respectively write $w_\pb$ and $w_\pc$ as an element in the sets \eqref{eq:canonical.form} times an operator $c_j$ or $c^\dag_j$ for some $j$, and the rest of the proof follows the same arguments.
\eproof

\begin{Proposition}\label{prop:La.equiv.defs}
We have the isomorphism of diagram spaces
\begin{equation}
\cLaa(N_\pc,N_\pa,N_\pb) \simeq \cLad(N_\pc,N_\pa,N_\pb) \,.
\end{equation}
Moreover, $\repBa(N_\pc,N_\pa,N_\pb)$ is a basis of $\cLaa(N_\pc,N_\pa,N_\pb)$.
\end{Proposition}
\proof
We use the same arguments as in the proof of \cref{prop:L.equiv.defs}. The map from $\cLaa(N_\pc,N_\pa,N_\pb)$ to $\cLad(N_\pc,N_\pa,N_\pb)$ described in \cref{sec:fusion.modules} preserves all the relations \eqref{eq:Lambda.equations}, and is therefore a homomorphism of diagram spaces. With the identification
\begin{equation}
\mu_{k_\pc,k_\pa, k_\pb} = \left\{\begin{array}{ll}
\id_{2 k_\pc}\big((c_0^\dag)^{k_\pc-k_\pa-k_\pb}\id_{2k_\pa}\otimes \id_{2k_\pb} \big)
& k_\pa + k_\pb < k_\pc\,, \\[0.15cm]
\id_{2 k_\pc}(c_0)^{k_\pc-k_\pa-k_\pb}\big(\id_{2k_\pa} \otimes \id_{2k_\pb} \big)
& k_\pa + k_\pb \ge k_\pc\,,
\end{array}\right.
\end{equation}
it is clear that the map is one-to-one, from the generating set $\repBa(N_\pc,N_\pa,N_\pb)$ of $\cLaa(N_\pc,N_\pa,N_\pb)$ to the basis $\repBd(N_\pc,N_\pa,N_\pb)$ of $\cLad(N_\pc,N_\pa,N_\pb)$, with the same relations tying $\boldsymbol i_\pa$, $\boldsymbol i_\pb$, $\boldsymbol i_\pc$ to $\boldsymbol j_\pa$, $\boldsymbol j_\pb$, $\boldsymbol j_\pc$ and $s_\pa$, $s_\pb$, $s_\pc$ given in the proof of \cref{prop:L.equiv.defs}. This proves both that $\repBa(N_\pc,N_\pa,N_\pb)$ is a basis of $\cLaa(N_\pc,N_\pa,N_\pb)$ and that the map is an isomorphism of diagram spaces.
\eproof

%
\section{Transformations of families of modules}\label{app:mod.transf}
%

In this appendix, we prove the transformations formulas \eqref{eq:transformed} for the families of modules introduced in \cref{sec:example.fams}.\medskip

\paragraph{The families $\boldsymbol{\repW_k}$.}

The vector space for $\repW_k(N)$ is $\cL^{\tinyx k}(N,2k)$. For diagrams $\lambda \in \cL^{\tinyx k}(N,2k)$, we denote by $u_\lambda$ the corresponding link state in $\repW_k(N)$ obtained from the bijection illustrated in \eqref{eq:Wk.vs.L(N,2k)}. The map $\lambda \mapsto u_\lambda$ is extended linearly, and we write
\begin{equation}
\repW_k(N) = \mathrm{span} \big[u_\lambda \,|\, \lambda \in \cL^{\tinyx k}(N,2k) \big] \,.
\end{equation}
The action of $\cL(N',N)$ on $\repW_k(N)$ is defined as
\begin{equation}
\lambda' \cdot u_\lambda = \left\{\begin{array}{cl}
u_{\lambda'\lambda} & \lambda'\lambda \in \cL^{\tinyx k}(N',2k) \,, \\[0.15cm]
0 & \text{otherwise.}
\end{array}\right.
\end{equation}

To investigate the relation between $\repW_k$ and $\repW_k^-$, we consider the linear map $\phi_N: \repW_k(N) \to \repW^-_k(N)$ defined as
\begin{equation}
\phi_N: u_\lambda \mapsto \sigma(\lambda) \, u_\lambda^- \,.
\end{equation}
For all diagrams $\lambda'$ in $\cL(N',N)$ and link states $u_\lambda \in \repW_k(N)$, we have
\be
\phi_{N'}(\lambda' \cdot u_\lambda) = \left\{\begin{array}{cl}
\sigma(\lambda' \lambda) \, u^-_{\lambda'\lambda} & \lambda'\lambda \in \cL^{\tinyx k}(N',2k) \,, 
\\[0.15cm]
0 & \text{otherwise.}
\end{array}\right.
\ee
On the other hand, we have
\be
\lambda' \cdot \phi_N(u_\lambda) = \sigma(\lambda) \, \lambda' \cdot u_\lambda^-
= \sigma(\lambda')\,\sigma(\lambda) \,(\lambda'\cdot u_\lambda)^-
= \left\{\begin{array}{cl}
\sigma(\lambda')\,\sigma(\lambda)\, u_{\lambda'\lambda}^- 
&\lambda'\lambda \in \cL^{\tinyx k}(N',2k) \,, \\[0.15cm]
0 
& \text{otherwise.}
\end{array}\right.
\ee
Because $\sigma(\lambda)\,\sigma(\lambda') = \sigma(\lambda' \lambda)$, we find that $\phi_{N'}(\lambda' \cdot u_\lambda)=\lambda' \cdot \phi_N(u_\lambda)$, and thus that $\phi$ is a family of homomorphisms. Moreover, it is easy to see that $\phi$ is bijective. We conclude that $\repW_k$ and $\repW_k^-$ are isomorphic families of modules.
\medskip

To investigate the relation between $\repW_k$ and $\repW_k^r$, we consider the linear map $\psi_N: \repW_k(N) \to \repW^r_k(N)$ defined as
\begin{equation}
\psi_N: u_\lambda \mapsto u_{R(\lambda)}^r \,.
\end{equation}
For all diagrams in $\lambda' \in \cL(N',N)$ and link states $u_\lambda \in \repW_k(N)$, we have
\begin{equation}
\psi_{N'}(\lambda' \cdot u_\lambda) = \left\{
\begin{array}{cl}
u^r_{R(\lambda'\lambda)} & \lambda'\lambda \in \cL^{\tinyx k}(N',2k) \,, 
\\[0.15cm]
0 & \text{otherwise.}
\end{array}\right.
\end{equation}
We also have
\be
\lambda' \cdot \psi_N(u_\lambda) = \lambda' \cdot u_{R(\lambda)}^r
= \big(R(\lambda')\cdot u_{R(\lambda)}\big)^r
= \left\{\begin{array}{cl}
u^r_{R(\lambda')R(\lambda)} & R(\lambda')R(\lambda) \in \cL^{\tinyx k}(N',2k) \,, \\[0.15cm]
0 & \text{otherwise.}
\end{array}\right.
\ee
Using the property $R(\lambda'\lambda)=R(\lambda')R(\lambda)$, we obtain $\psi_{N'}(\lambda' \cdot u_\lambda) = \lambda' \cdot \psi_N(u_\lambda)$. It is also easy to see that $\psi_N$ is bijective. We deduce that the families of modules $\repW_k$ and $\repW_k^r$ are isomorphic.

\paragraph{The families $\boldsymbol{\repW_{k,x}}$.}

The vector space of the module $\repW_{k,x}(N)$ is 
\begin{equation}
\repW_{k,x}(N) = \mathrm{span} \big[ u_\lambda \,|\, \lambda \in \cL^{\tinyx{k,0}}(N,2k) \big] \,.
\end{equation}
Let $\lambda$ be a diagram in $\cL^{\tinyx k}(N,2k)$. For $k=0$, there exists a unique $n(\lambda) \in \Zbb_{\ge 0}$ and a unique diagram $\tau(\lambda) \in \cL^{\tinyx{0,0}}(N,0)$ such that $\lambda=\tau(\lambda) f^{n(\lambda)}$. For $k>0$, there exists a unique $n(\lambda) \in \Zbb$ and a unique diagram $\tau(\lambda) \in \cL^{\tinyx{k,0}}(N,2k)$ such that $\lambda=\tau(\lambda) \, \Omega^{n(\lambda)}$. We have
\be
\label{eq:sigma.R.tau} 
\sigma\big(\tau(\lambda)\big) = (-1)^{n(\lambda)} \, \sigma(\lambda) \,,
\qquad
R\big(\tau(\lambda)\big)=\tau\big(R(\lambda)\big)\,,
\qquad 
n\big(R(\lambda)\big) = \left\{\begin{array}{cl}
n(\lambda) & k=0 \,, \\[0.1cm]
-n(\lambda) & k>0 \,.
\end{array}\right.
\ee
The action of $\cL(N',N)$ on $\repW_{k,x}(N)$ reads
\begin{equation}
\lambda' \cdot u_\lambda = \left\{\begin{array}{cl}
\gamma(x)^{n(\lambda'\lambda)} \, u_{\tau(\lambda'\lambda)} 
& \lambda'\lambda \in \cL^{\tinyx k}(N',2k) \,, \\[0.15cm]
0 & \text{otherwise,}
\end{array}\right.
\quad \textrm{where} \qquad
\gamma(x) = 
\left\{\begin{array}{cl}
x+x^{-1} & k=0 \,, \\[0.15cm]
x & k>0 \,.
\end{array}\right.
\end{equation}
We now consider the linear map $\phi_N: \repW_{k,-x}(N) \to \repW^-_{k,x}(N)$ defined as
\begin{equation}
\phi_N: u_\lambda \mapsto \sigma(\lambda) \, u_\lambda^- \,.
\end{equation}
For all diagrams $\lambda' \in \cL(N',N)$ and link states 
$u_\lambda \in \repW_{k,-x}(N)$, we have
\be
\phi_{N'}(\lambda' \cdot u_\lambda) = \left\{\begin{array}{cl}
\sigma\big(\tau(\lambda' \lambda)\big) \, \gamma(-x)^{n(\lambda'\lambda)} \, u_{\tau(\lambda'\lambda)}^- 
& \lambda'\lambda \in \cL^{\tinyx k}(N',2k) \,, \\[0.15cm]
0 & \text{otherwise.}
\end{array}\right.
\ee
We also have
\be
\lambda' \cdot \phi_N(u_\lambda) = \sigma(\lambda) \, \lambda' \cdot u_\lambda^-
= \sigma(\lambda')\,\sigma(\lambda) \,(\lambda'\cdot u_\lambda)^-
= \left\{\begin{array}{cl}
\sigma(\lambda')\,\sigma(\lambda)\, \gamma(x)^{n(\lambda'\lambda)} \, u^-_{\tau(\lambda'\lambda)} & 
\lambda'\lambda \in \cL^{\tinyx k}(N',2k) \,, \\[0.15cm]
0 & \text{otherwise.}
\end{array}\right.
\ee
Using the identity $\sigma(\lambda'\lambda)=\sigma(\lambda')\,\sigma(\lambda)$ and the first relation in \eqref{eq:sigma.R.tau}, we find that $\phi_{N'}(\lambda' \cdot u_\lambda)=\lambda' \cdot \phi_N(u_\lambda)$. Moreover, it is easy to see that $\phi_N$ is bijective. This implies that the families of modules $\repW_{k,x}^-$ and $\repW_{k,-x}$ are isomorphic.
\medskip

Similarly, to study $\repW_{k,x}^r$, we define the map $\psi_N:\repW_{k,x^{-1}} \to \repW_{k,x}^r$ as
\be
\psi_N: u_\lambda \mapsto u^r_{R(\lambda)}\,.
\ee
For all diagrams $\lambda' \in \cL(N',N)$ and link states $u_\lambda \in \repW_{k,x^{-1}}(N)$, we have
\be
\psi_{N'}(\lambda' \cdot u_\lambda) = \left\{\begin{array}{cl}
\gamma(x^{-1})^{n(\lambda'\lambda)} \, u_{R(\tau(\lambda'\lambda))}^r 
& \lambda'\lambda \in \cL^{\tinyx k}(N',2k) \,. \\[0.15cm]
0 & \text{otherwise.}
\end{array}\right.
\ee
We also have
\be
\lambda' \cdot \psi_N(u_\lambda) = \lambda' \cdot u^r_{R(\lambda)}
= \big(R(\lambda') \cdot u_{R(\lambda)}\big)^r
= \left\{\begin{array}{cl}
\gamma(x)^{n(R(\lambda')R(\lambda))} \, u^r_{\tau(R(\lambda')R(\lambda))} & 
R(\lambda')R(\lambda) \in \cL^{\tinyx k}(N',2k) \,, \\[0.15cm]
0 & \text{otherwise.}
\end{array}\right.
\ee
Using $R(\lambda')R(\lambda) = R(\lambda'\lambda)$ and the two rightmost relations in \eqref{eq:sigma.R.tau}, we find that $\psi_{N'}(\lambda' \cdot u_\lambda) = \lambda' \cdot \psi_N(u_\lambda)$, implying that $\psi$ defines a family of homomorphisms. We can easily see that $\psi_N$ is bijective and hence conclude that $\repW_{k,x^{-1}}$ and $\repW_{k,x}^r$ are isomorphic.
\medskip

\paragraph{The family $\boldsymbol{\repV}$.}

To show that $\repV(N) \simeq \repV^r(N)$, we use the map $\psi_N: \repV(N) \to \repV^r(N)$ defined as
\be
\psi_N: u \mapsto \big(R(u)\big)^r\,,
\ee
for all $u$ in $\repV(N)$, where $R(u)$ is obtained from $u$ by changing each arch connecting the pair of nodes $(i,j)$ to an arch connecting the pair $(N+1-i,N+1-j)$. We have $R(\lambda \cdot u)=R(\lambda)\cdot R(u)$ for all diagrams $\lambda \in \cL(N',N)$ and link states $u \in \repV(N)$. We can use this to show that $\psi_N$ is a homomorphism. Because it is bijective, we conclude that $\repV \simeq \repV^r$.
\medskip

In contrast, $\repV^-$ is in general \emph{not} isomorphic to $\repV$. Indeed, the central element $F$ acts as $\beta \id$ on $\repV(N)$, whereas, because $\sigma(F)=-1$, $F$ instead acts as $-\beta \id$ on $\repV^-(N)$, so $\repV$ and $\repV^-$ cannot be isomorphic for $\beta \neq 0$.

\paragraph{The families $\boldsymbol{\repM_{\g,K}}$.}

To understand how the reflection automorphism transforms the RSOS module $\repM_{\g,K}$, we introduce the linear map $\psi_N:\repM_{\g,K^{-1}}(N) \to \repM_{\g,K}^r(N)$ defined as
\begin{equation}
\psi_N: \ket{a_0,a_1,\dots,a_N} \mapsto \ket{a_N,a_{N-1},\dots,a_0}^r \,.
\end{equation}
With an argument similar to the above cases, we readily find that $\psi_N$ defines a family of isomorphisms.\medskip

To study the action of the parity sign flip on $\repM_{\g,K}(N)$, we recall that the Dynkin diagrams for all Lie algebras in the ADE series are bipartite. We define $\theta(a) = 1$ if $a \in \mathcal G$ is black, and $\theta(a) = -1$ if it is white. We introduce the operator $\phi_N$ that acts on $\repM_{\g,K}(N)$ as
\begin{equation}
\phi_N \, \ket{a_0,a_1,\dots,a_N} = \theta(a_0) \,\ket{a_0,a_1,\dots,a_N} \,.
\end{equation}
From the definition \eqref{eq:def.RSOS} and the bipartiteness of $\mathcal G$, it is easy to check that
\begin{subequations}
\begin{alignat}{2}
c_{N,j}\,\phi_N &= \phi_N \, c_{N,j} \,,
\qquad 
&&c^\dag_{N,j}\,\phi_N = \phi_N \, c^\dag_{N,j} \,, 
\qquad 
j=1,2,\dots,N-1,
\\[0.15cm]
c_{N,0}\,\phi_N &= -\phi_N \, c_{N,0} \,, 
\qquad 
&&c^\dag_{N,0}\,\phi_N = -\phi_N \, c^\dag_{N,0} \,.
\end{alignat}
\end{subequations}
This implies that $\phi_N\, \lambda = \sigma(\lambda) \,\lambda \,\phi_N$, for all diagrams $\lambda \in \cL(N,N')$. It is then straightforward to show that the map $\ket{a_0,a_1,\dots,a_N}^-\mapsto \phi_N\ket{a_0,a_1,\dots,a_N}$ yields a family of isomorphisms from $\repM_{\g,K}^-$ to $\repM_{\g,K}$.

\bibliography{biblio}
\bibliographystyle{unsrt}

\end{document}